\documentclass[a4paper,12pt]{article}
\usepackage[utf8]{inputenc}
\usepackage[english]{babel}
\usepackage{amsmath, amssymb,graphicx}
\usepackage{braket}
\usepackage{caption}
\usepackage{subcaption}
\usepackage{ulem}
\pdfoutput=1
\usepackage{jheppub}
\newcommand{\be}{\begin{equation}}
\newcommand{\ee}{\end{equation}}
\newcommand{\bea}{\begin{eqnarray}}
\newcommand{\eea}{\end{eqnarray}}
\newcommand{\nn}{\nonumber}

\newcommand{\fg}{\mathfrak{g}}

\newcommand{\e}{\mathrm{e}}

\author{Irina Ya. Aref'eva$^a$, Alexander Patrushev$^b$ and  Pavel Slepov$^a$}
\affiliation{$^a$Steklov Mathematical Institute, Russian Academy of Sciences,\\ Gubkina str. 8, 119991, Moscow, Russia\\$^b$Bauman Moscow State Technical University,\\  2-ya Baumanskaya str. 5/1, 105005, Moscow, Russia}

\emailAdd{arefeva@mi-ras.ru}
\emailAdd{apatrush@mi-ras.ru}
\emailAdd{slepov@mi-ras.ru}

\title{Holographic Entanglement Entropy in Anisotropic Background with Confinement-Deconfinement Phase Transition}
\abstract{We discuss
 a general five-dimensional completely anisotropic holographic model with three different spatial scale factors, characterized by a Van der Waals-like phase transition between small and large black holes. A peculiar feature of the model is the relation between anisotropy of the background and anisotropy of the colliding heavy ions geometry. We calculate
the holographic entanglement entropy (HEE) of the slab-shaped region, the orientation of which relatively to the beams line and the impact parameter is characterized by the Euler angles.
We  study the dependences of the HEE and its density on the thermodynamic (temperature, chemical potential) and geometric (parameters of anisotropy, thickness, and orientation of entangled regions) parameters.
 As a particular case the model with two equal transversal scaling factors is considered.
 This model is supported by the dilaton and two Maxwell fields. In this case we discuss the HEE and its density in detail: interesting 
 features of this model are jumps of the entanglement entropy and its density near the line of the small/large black hole phase transition. These jumps depend on the anisotropy parameter, chemical potential, and orientation. We also discuss different definitions and  behavior of c-functions in this model. The c-function calculated in the Einstein frame decreases while $\ell$ is increasing for all  $\ell$  in the isotropic case (in regions of $(\mu,T)$-plane far away from the line of the phase transition). We find the non-monotonicity of the c-functions for several anisotropic configurations, which however does not contradict with any of the existing c-theorems since they all are based on Lorentz invariance.}

\begin{document}

\maketitle
\newpage
\section{Introduction}
Fundamental  questions addressed in studies of high energy heavy ions  collisions (HIC) at RHIC and LHC, and future experiments NICA and FAIR,   concern  understanding of quark-gluon plasma (QGP) formation, i.e.  thermalization of media produced in HIC,
thermodynamic entropy production, and its characteristics such as  quantum entanglement, decoherence etc.
Most of our knowledge on the formation and properties of QGP resulting from HIC  is obtained from measurements of the yields and spectra of particles in the final state of colliding heavy ions and their thermo/hydrodynamic interpretation.
According to our common  understanding, within
a short time of order $1-2$ fm/$c$ collision systems reach a
state that  can be approximated by a  thermal medium located in an expanding ball. This medium is characterized
by local thermodynamic parameters
including temperature and entropy density.\\

HIC experiments at RHIC and LHC have provided strong evidence that this medium is a QGP at large temperatures and densities. There is also a strong evi\-dence for the existence of the confinement/deconfinement phase transition  in the $(\mu,T)$-plane, i.e. temperature on chemical potential. The experimental search for the QCD phase transition is nowadays one of the central goals for current and future collider facilities \cite{MCHIC}.The experimental search is mainly related to the measurement of fluctuations  of net-proton or net-charge multiplicity \cite{Aggarwal:2010wy,Luo:2012kja,Adamczyk:2013dal,Luo:2015ewa} which are expected to exhibit non-monotonic behavior near the phase transition.
The proper understanding of the experimental results requires careful theoretical analysis of the dynamical processes taking place near the phase transition lines, especially near the critical endpoint (CEP)\footnote{There are indications that  CEP's location on the phase diagram is  determined solely through chiral symmetry breaking, see for example \cite{Ayala:2014jla}} \cite{Berdnikov:1999ph}.
 There are several theoretical approaches to searches  of the  QCD phase transitions. One of the promising theoretical approaches is based on lattice calculations, but this approach encounters difficulties with non-zero chemical potential \cite{Lattice,Bazavov:2017dus}.\\

Holographic duality provides an alternative approach to the study of the QCD phase transitions \cite {Witten:1998zw,GK,GKN,Gubser:2008yx,Gubser:2008ny}  (and for review see \cite{Kiritsis,Solana,IA-UFN,DeWolf,Arefeva-conf,IAQ18} and refs therein).
Holography allows to define thermal entropy and free energy as functions of temperature and chemical potential of the gravitational background.
With this approach, the thermalization process is dual to the black hole formation, linking the study of thermalization in the QCD with the study of the black hole formation
in 5-dimensional gravity. The thermalization process is conveniently monitored, at least theoretically, by tracking the evolution of entanglement entropy  of a selected region\footnote{Note that in phenomenological discussions the entropy production in HIC is used  as a signal for
  quark gluon plasma phase transition \cite{Reiter:1998uq,Dumitru:2007qr}}. The entanglement entropy of the selected domain  $A$ is defined as the von Neumann
entropy of the reduced density matrix obtained by tracing out the degrees of freedom located out of domain $A$.
The entanglement entropy is hard to compute in QCD, but it is suitable to compute it in the lattice \cite{Buividovich:2008kq,Velytsky:2008rs,Itou:2015cyu}, see also 
\cite{Kunihiro:2010tg}. The entanglement entropy can be also evaluated at the  gravity side. It is called  the holographic entanglement entropy (HEE) and is defined as the area
of a minimal surface extending from some predefined surface $A$ on the boundary into the
bulk \cite{RT1,RT2,HRT}.  The HEE during thermalization  usually evolves to the thermal entanglement entropy \cite{Bala,Liu:2013iza, 1506.02658,1110.5035,Arefeva:2015jkr,AGG,Narayan:2012hk,Ageev:2017wet,AKT,Wondrak:2017kgp,Mishra:2015cpa,Ghosh:2017ygi}. With this approach, there is a natural possibility of studying the evolution of entropy in HIC (thermalization) and phase transitions for the obtained thermal media in the framework of the same holographic model.\\

Starting from the Landau thermodynamic approach to high energy collision \cite{LL} one connects the entropy of the ball of QGP produced in HIC
with the
total multiplicity of particle production with HIC. The ideal hydrodynamics preserves the total entropy, but dissipative  one does not. One can relate  the entanglement entropy, associated with a given region of the ball, to the multiplicity of particles  produced in this region  during the HIC 
\cite{HEE1}.  After thermalization the entanglement entropy  of the area  depends on whether this area  belongs to  Glauber's participant area
of ions collisions or does not.  It occurs that the dependence of the entanglement entropy on geometrical size of entangled areas
is closely related to energy loss and jet quenching.\\

The HEE has been also used to study phase transitions in equilibrium. In particular, in \cite{IK} it was proposed to use the HEE as a probe of confinement. The HEE  
has been extensively studied and applied  in the investigation of the phase transitions in various  holographic QCD (HQCD) models \cite{1208.2937,Lewkowycz:2012mw,Kim:2013ysa,Kol:2014nqa, Ling:2015dma,Ghodrati:2015rta,Kundu:2016dyk,Ling:2016wyr,Zhang:2016rcm,Zeng:2016fsb,Dudal:2016joz,Mahapatra:2019uql,Knaute:2017lll,Ali-Akbari:2017vtb,Dudal:2017max,1805.02938,Rahimi:2018ica,Baggioli:2018afg,Liu:2019npm,Ebrahim:2020qif,Narayan:2012ks}. \\

We start with the most general anisotropic holographic model and consider the general orientation of the slab-shaped entangled region with respect to the geometry of HIC.
The natural coordinate system defined by the HIC is such that the first axis (the longitudinal axis) is directed along the line of collision and the second (the transversal axis) is determined by the direction of the impact parameter. The orientation of the entangled region can be set by Euler angles.
Then, we calculate the HEE for an anisotropic model with symmetry in transversal directions studied in \cite{AG,AR}. 
The choice of this anisotropic model  is motivated by its previous use for the holographic description of HIC \cite{AG}. 
It is this model that for special parameter of anisotropy $\nu=4.5$ gives the dependence of the produced entropy on energy in accordance with the experimental data for the energy dependence of the total multiplicity of particles produced
in heavy ion collisions \cite{Alice}. Isotropic holographic models had not been able to reproduce the experimental  multiplicity dependence on energy \cite{Gubser,Gubser:2009sx,Grumiller:2008va,AlvarezGaume:2008fx,Albacete:2008vs,Lin:2009pn,Albacete:2009ji,ABG,ABJ,Kovchegov:2009du,1111.1931,APP-AA,1409.7558}. As shown in \cite{1808.05596}, the model \cite{AR} describes smeared confinement/deconfinement phase transitions. As we will see, the model indicates the relations of the fluctuations of the multiplicity, i.e. the entanglement entropy, with the phase transitions.\\

There are more reasons to use the anisotropic models in context of HIC \cite{Strickland:2013uga} and QCD itself \cite{Arefeva:1993rz}.
Other anisotropic holographic models have been actively studied in recent years \cite{Giataganas, Finazzo, Giataganas2, Mateos:2011ix, Mateos:2011tv, Brehm:2017dmt, Bohra:2019ebj}, and the main motivation for these studies was also the anisotropic nature of HIC.\\

We select an entangled slab-shaped area that has a finite
extent in one direction and infinite extent in the other two directions\footnote{See few refs with calculations of the HEE for more complicated entanglement region \cite{Tonii}}. We implicitly assume that the slab of the interest is located   inside the overlapping area of two ions. In this case of the spacetimes with
the fully anisotropic  metric (see  \eqref{Gmetric} below) the problem of finding the extremal
area functional  effectively reduces to finding the geodesics in some auxiliary 2-dimensional Euclidean space. For the model \cite{AR} we find that varying the angle between the axis of collisions and the direction of the smallest side of the slab-shaped entanglement area changes the slab HEE.   Increasing this angle we enhance the HEE, and according to conjecture \cite{HEE1}, this means the enhancement of the multiplicity of particle production from this area.
 This  enhancement depends on geometrical (length), and thermodynamical (temperature $T$ and chemical potential $\mu$) parameters. The HEE density \cite{NNT,BHRT,1708.09376,1709.07016} is a more convenient object for study 
 since it does not suffer from ultraviolet divergencies.
 The HEE and its density undergo jumps and these jumps increase while the angle is increasing. Moreover, the values of the HEE density and its jumps increase with the  anisotropy  increasing.\\

We also discuss various definitions and behavior of c-functions in this model. In the isotropic case we use different frames, the Einstein frame or the string frame, as well as different types of renormalizations. Here we mainly use the geometric renormalization scheme, which consists in subtracting the disconnected configuration from the connected one
 \cite{IK,Kol:2014nqa}. 
For the isotropic case in regions of $(\mu,T)$-plane far away from the phase transition line, the c-function calculated in the Einstein frame decreases while $\ell$ is increasing for all  $\ell$.
We find the non-monotonicity of c-functions in the string frame, that is related to dynamics of the dilaton field in UV in our model.
 There are several proposals in the literature how to define the holographic  c-function in the anisotropic backgrounds \cite{Liu:2012wf,Swingle:2013zla,Cremonini:2013ipa,Bea:2015fja,1806.09072,Chu:2019uoh,Ghasemi:2019xrl,Hoyos:2020zeg}. Here we use prescription \cite{Chu:2019uoh}
adapted to our renormalization schemes.
We find the non-monotonicity of c-functions for several anisotropic configurations and discuss their origin. Note, that generally speaking, 
the non-monotonicity of c-functions does not contradict any of the existing c-theorems 
\cite{Zamolodchikov:1986gt,Freedman:1999gp,Girardello:1998pd,Cardy:1988cwa,Casini:2006es,Myers:2010tj,Myers:2010xs,Komargodski:2011xv,Myers:2012ed,Kolekar:2018chf} since they all base on Lorentz invariance.\\

The paper is organized as follows. In Sect. \ref{Sec:AHM} we briefly describe anisotropic holographic models. In Sect. \ref{Sec:GA}
we present  the most general anisotropic holographic model.  
We present the action 
and the ansatz that solves the EOM for the anisotropic model with symmetry in transversal directions in Sect. \ref{Sec:AHM2t} and  thermodynamics of the background in Sect. \ref{Sec:Thermo}. In  Sect. \ref{Sec: EE} we present an expression for the HEE for a slab-shaped entangling 
region oriented differently with respect to the HIC axes. In Sect. \ref{Sect:longitudinal} and Sect. \ref{Sect:transversal}   we present its special forms 
corresponding to transversal and longitudinal orientations in respect to the collision axis, respectively, and discuss    the regularization procedure
used for the transversal and longitudinal orientations. Sect. \ref{Sect:density1} is devoted to entanglement entropy density and definitions of c-functions.
In Sect. \ref{Sect:NR} we display and discuss our main numerical results. In Sect. \ref{Sect:HEE-TL} we demonstrate the dependence of the HEE 
on the geometrical parameters -- length $\ell$,  anisotropy $\nu$ and orientations, and on thermodynamical parameters -- temperature $T$ and chemical potential $\mu$.
In Sect. \ref{Sect:density} we show  dependencies  of the HEE density  on the thickness, anisotropy and orientation of the slab as well as on  temperature and chemical potential. In Sect. \ref{Sect:c-function} we discuss the scaling behavior of the modified c-function.
 In Sect. \ref{sect:origine} we discuss  possible origins of a non-monotonic behavior of  c-functions.
In Sect. \ref{nearBB} we present behavior of different c-functions near the background phase transition. In Sect. \ref{sect: summary} we present a table  of the dependencies of various  c-functions on $\ell$. In Sect. \ref{Sect:PT}  we compare the position of the phase transition for HEE with the positions of phase transitions related to the background instability.

Finally, we end the paper with the conclusion 
and discussion of future directions of  research on the subject.

\section{Setup}\label{Sec:Hol-mod}
\subsection{ Anisotropic Holographic Models }
\label{Sec:AHM}

\subsubsection{ General Anisotropic Model}\label{Sec:GA}
We start with a  general anisotropic holographic model 

 \bea\label{Gmetric}
ds^2&=& \frac{L^2 b_s(z)}{z^2}\sum _{M=0}^{4}G_{M}(z)(dx^{M})^2,\\
G_{0}&=&-g(z), \,\,\,\,G_{i}=\fg_i(z), \,\,i=1,2,3,\,\,G_{4}=\frac{1}{g(z)}.
\eea
Here $b_s(z)=b(z)\e^{\sqrt{\frac{2}{3}}\phi(z)}$ is the AdS deformation factor (in the string frame in the presence of the dilaton), $g(z)$ is a blackening function and  $\fg_i(z)$ are anisotropy factors.

\subsubsection{Anisotropic Model with Symmetry in Transversal Directions  }\label{Sec:AHM2t}
Taking in the previous formula $\fg_1(z)=1$ and $\fg_2(z)=\fg_3(z)=\fg(z)$ we get
\bea
  ds^2 = \frac{L^2 b_{s}(z)}{z^2} \left[ - \, g(z) dt^2 + dx^2 +
   \fg(z)\left( (dy^1)^2 + (dy^2)^2 \right) +
    \cfrac{dz^2}{g(z)} \right].\,\,\label{ARG}
\eea
Metric \eqref{ARG} with a special form of anisotropy factor $\fg(z)$ as in \cite{AGG} and a particular case of 
$b(z)$ describing the holographic model for heavy quarks \cite{AR} present a special interest for us:
\be
  ds^2 = \frac{L^2 b(z)}{z^2} \left[ - \, g(z) dt^2 + dx^2 +
    \left(\frac{z}{L}\right)^{2-\frac{2}{\nu}} \left(  (dy^1)^2 + (dy^2)^2 \right) +
    \cfrac{dz^2}{g(z)} \right],\label{ARmetric}\ee
where $L$ is the characteristic length scale of the geometry and $b(z)=e^{cz^2/2}$. In all our calculations we set $c=-1$. Here the metric is in the Einstein frame.  In the next section to perform calculation of the HEE  we also switch  to the string frame adding an extra dilaton-dependent exponential prefactor
 \cite{RT1,IK}. The metric \eqref{ARmetric} is supported  by the Einstein-Dilaton-two-Maxwell
 action with special potential $V$  for the dilaton field $\phi$ and strength potentials  $f_1$ and $f_2$
for two Maxwell fields:
\be
  S = \frac{1}{16\pi G_5} \int d^5x \sqrt{-g} \left[ R -
    \frac{f_1(\phi)}{4} F^{(1)\,2} -  \frac{f_2(\phi)}{4}
    F^{(2)\,2} - \frac{1}{2} (\partial \phi )^2
    - V(\phi) \right], \label{action}
\ee
where $F^{(1)}\,\!^2$ and $F^{(2)\,2}$ are the squares of the Maxwell
fields.

The ansatz with  the Maxwell
fields  $F^{(1)}_{\,\mu\nu} = \partial_{\mu} A^{(1)}_{\nu} - \partial_{\nu}
A^{(1)}_{\mu}$, $A^{(1)}_\mu = A_t (z)\delta_\mu^0,$ and $F^{(2)} = q \ dy^1 \wedge dy^2$,
$\phi = \phi(z)$ and metric \eqref{ARmetric}
satisfies the equations of motion under relations  between the warp factor
$b(z)$, dilaton and Maxwell potentials $V(\phi),\,f_1(\phi), \,f_2(\phi)$ \cite{AR}.

The ansatz \eqref{ARmetric} breaks isotropy while preserving translation and $(t,x)$-boost  invariances.   The advantage of this model is that it allows to find  potentials $V$, $f_1$
and $f_2$ and the blackening function $g$ explicitly. Note that in this case the dilaton potential  can be approximated by the sum of two exponents:
\be
  V(\phi,\mu,\nu) = V_0(\nu) + C_1(\mu,\nu) e^{k_1(\nu) \phi} +
  C_2(\mu,\nu) e^{k_2(\nu) \phi}.
\ee
Note, that in \cite{AGP} an explicit isotropic solution for the
dilaton potential as a sum of two exponents and zero chemical
potential has been constructed. The isotropic version of the model has beed considered earlier in \cite{yang2015}.

 There is another holographic anisotropic  model that has symmetry in transversal directions, but breaks the boost invariance in $(t,x^1)$ plane, keeping the $(t,x^2)$  and $(t,x^3)$ planes.  It is  based on  the Mateos-Trancanelli metric \cite{Mateos:2011tv, Mateos:2011ix}

\be \label{MTmetric}
ds^2=\frac{L^2 b(z)}{z^2} \left[-g(z) dt^2+e^{2h(z)}(dx^{1})^{2}+(dx^{2})^{2}+(dx^{3})^{2}+\frac{dz^2}{g(z)}\right].
\ee
This metric is supported by the  Einstein-Axion-Dilaton action.

\subsection{Thermodynamics of the Background \eqref{ARmetric}}\label{Sec:Thermo}

The thermodynamical properties of the anisotropic holographic model were studied in  \cite{AR}. The temperature of the model is given as $T = \left|g'(z_h)\right|/4\pi$, where $g(z_h)$ is given by the incomplete gamma
function (see (2.37) in \cite{AR}).

\begin{figure}[b!]
  \centering
  \includegraphics[scale=0.35]{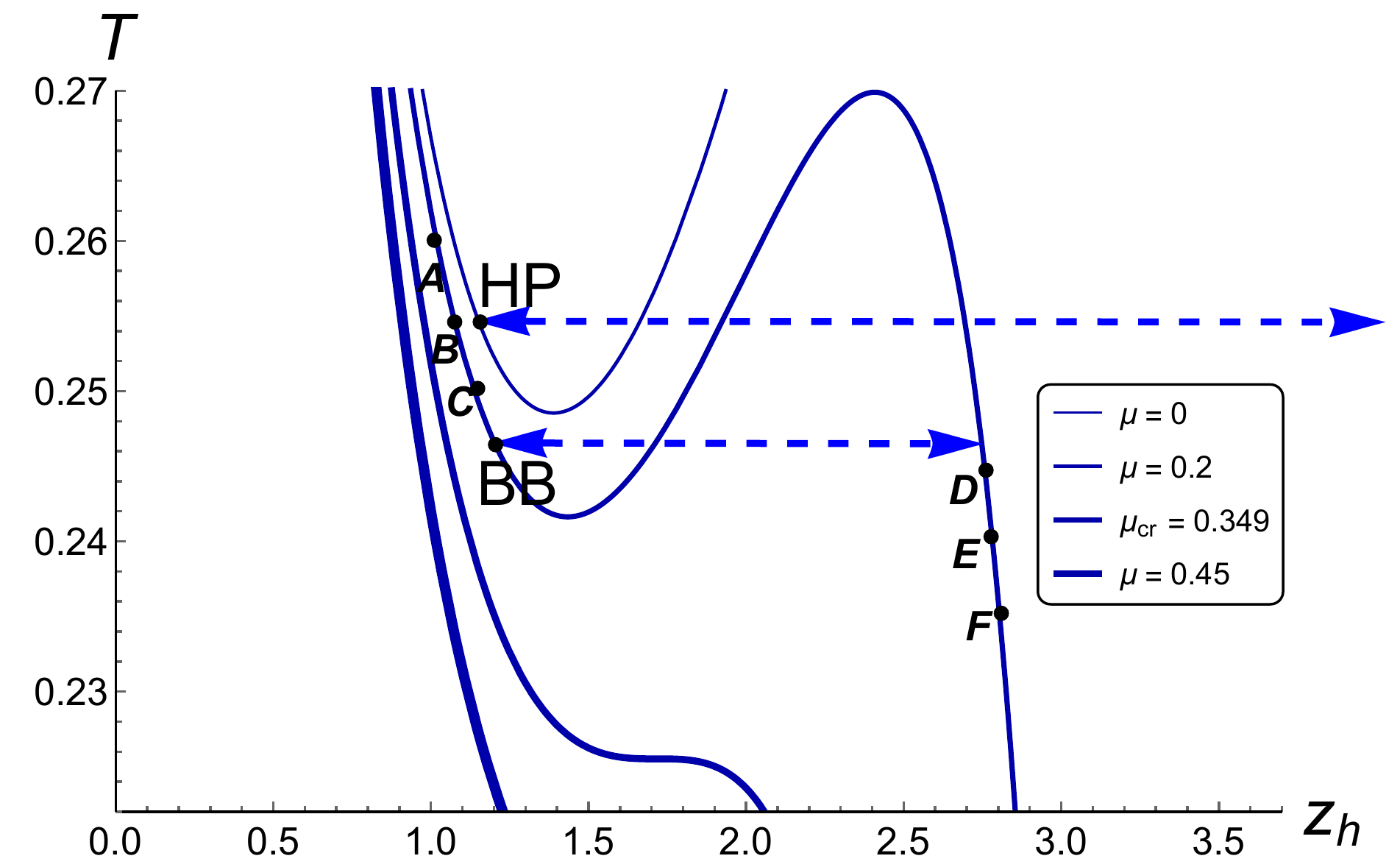} \quad
  \includegraphics[scale=0.35]{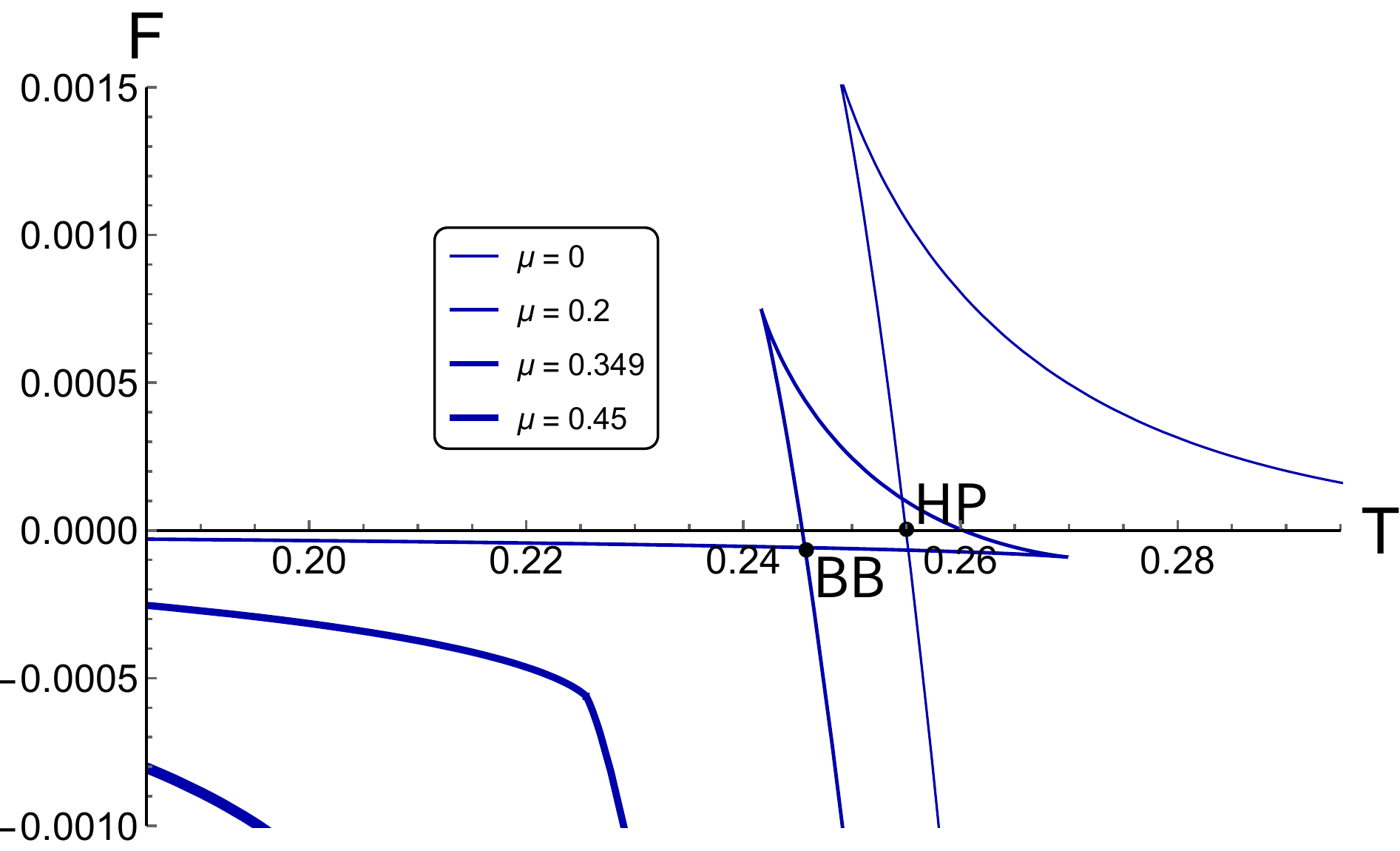}\\{\bf A)}\qquad\qquad\qquad\qquad\qquad\qquad\qquad{\bf B)}

  \caption{{\bf A)} The temperature $T(z_h,\mu,c,\nu)$ dependence  on the horizon size $z_h$ for  $\nu=4.5$ and various values
of the chemical potential $\mu$.  {\bf B)} The black hole free energy $F(T)$ for various values
of  $\mu$ in anisotropic case, $\nu = 4.5$.  The intersection with the horizontal axis gives the value of the Hawking-Page horizon $z_{h,HP}(\nu)$.}
  \label{T_zh.pdf}
\end{figure}

The expressions for thermal entropy and the free energy are the following:
\bea
  s(z_h,c,\nu) = \frac{e^{\frac{3}{4} c z_h^2}}{4  z_h^{1+2/\nu}}, ~~~~ F(z_h,c,\nu,\mu)  = \int_{z_h}^{z_{{h}_{2}}}
  s(z_h,c,\nu) \, T^\prime (z_h,c,\nu,\mu) \, dz_h,
\eea
where  $z_{h_2}$ is the second horizon, i.e. at this point  $T(z_{h_2})=0$.  For the considered blackening function the function $T(z_h)$ is three-valued i.e. has the Van der Waals type of behavior at $0 < \mu < \mu_{cr}(\nu)$  (Fig.\ref{T_zh.pdf}.{\bf A)}) and  free energy shows the swallow-tailed dependence on the temperature in this range of chemical potentials (Fig.\ref{T_zh.pdf}.{\bf B)}). The loop of the swallow-tailed shape
disappears at $(\mu, T)=(\mu_{cr}(\nu),T_{cr}(\nu))$. For $\mu \geq\mu_{cr}$, the function of free energy increases smoothly while temperature is decreasing. The lowest values of free energy correspond to the thermodynamically stable phases.

\begin{figure}[b!]
  \centering
  \includegraphics[scale=0.4]{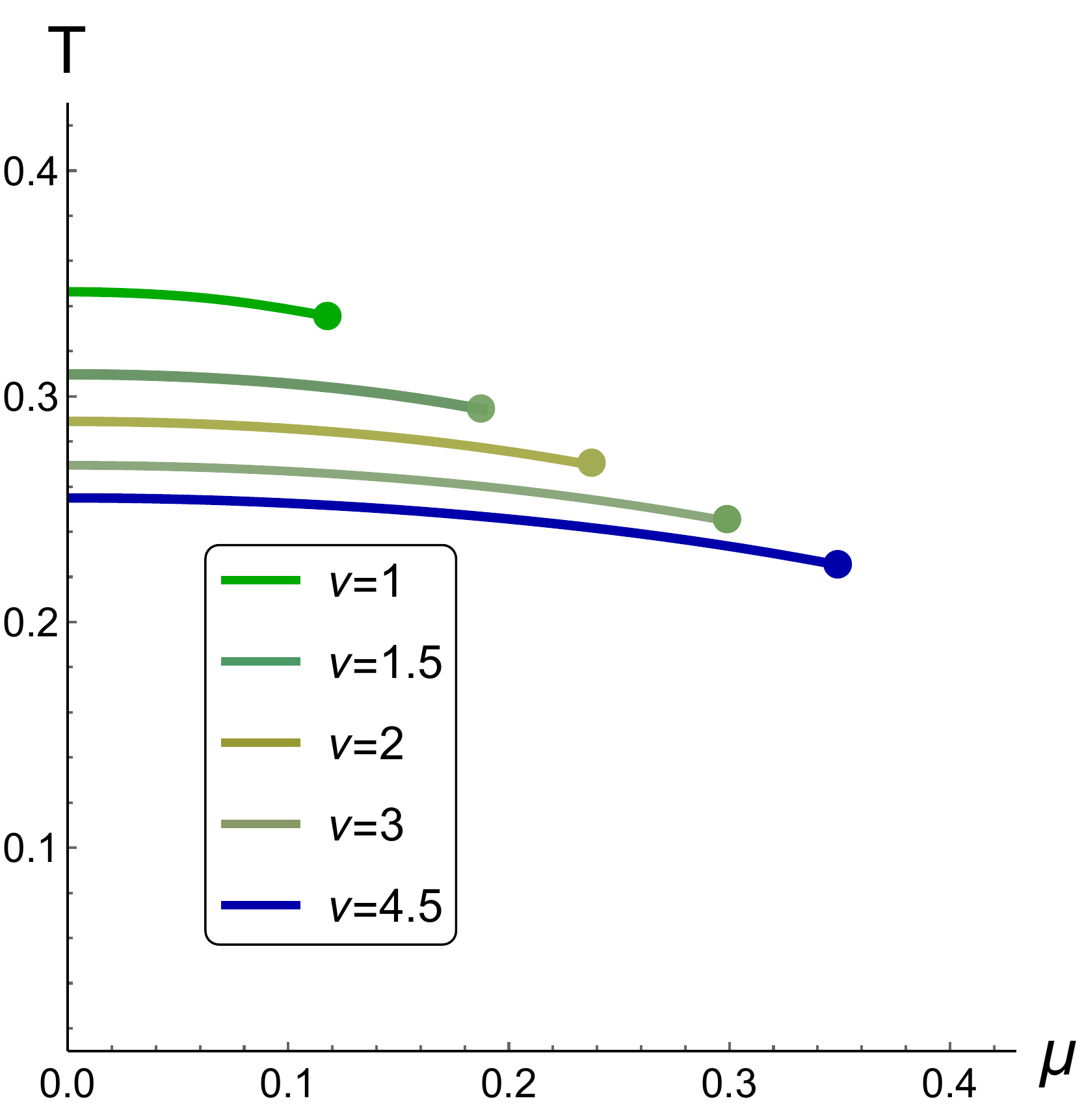} \quad\includegraphics[scale=0.36]{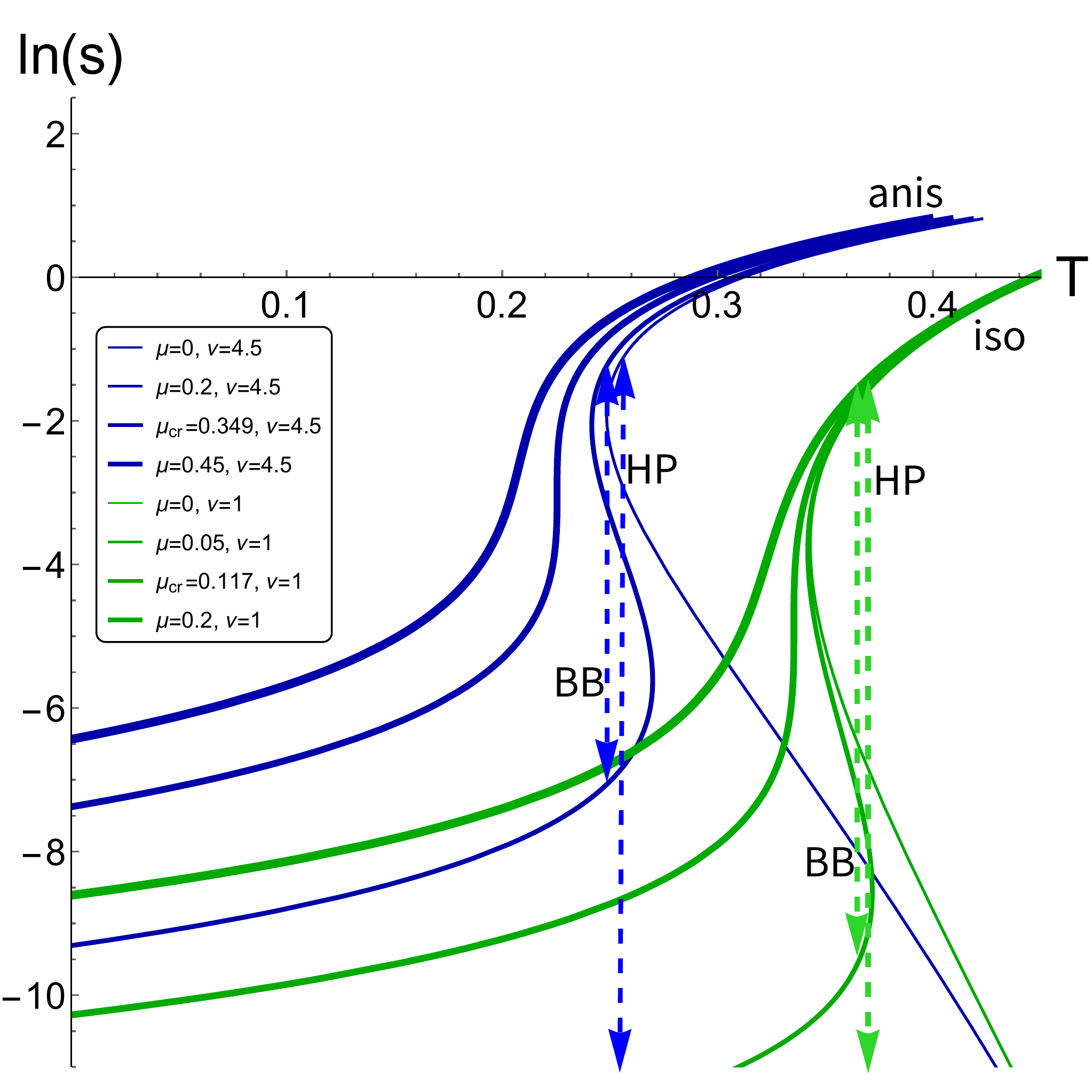}
  \\{\bf A)}\qquad\qquad\qquad\qquad\qquad\qquad\qquad{\bf B)}
 \caption{ {\bf A)} The BB phase transition in the $(\mu,T)$-plane for the isotropic
    background (green curve) and for the anisotropic backgrounds for various $\nu=1.5,2,3,4.5$ (green-gray, khaki, blue-gray and blue 
   curves, respectively). 
$\mu_{cr}(\nu=1)=0.117,T_{cr}(\nu=1)=0.334,$ 
$\mu_{cr}(\nu=1.5)=0.189,T_{cr}(\nu=1.5)=0.294,$ 
$\mu_{cr}(\nu=2)=0.270,T_{cr}(\nu=2)=0.237,$ 
$\mu_{cr}(\nu=3)=0.299, T_{cr}(\nu=3)=0.245,$ 
$\mu_{cr}(\nu=4.5)=0.349,T_{cr}(\nu=4.5)=0.226.$ 
Dots indicate the critical points.
    {\bf B)} The thermal entropy dependence on temperature for various values of chemical potential for $\nu=1$ (green curves) and $\nu=4.5$ (blue curves).
    Dashed lines with arrows show the thermal entropy jumps at the HP transition point (for zero chemical potential) and at the BB phase transition points
    (for non-zero chemical potential).
        }
  \label{Fig:FTmu45}
\end{figure}
 The line of free energy intersects itself at $T = T_{BB}\left(\nu,\mu\right)$, and here  a small black hole transits to a  large one.  Note that for the non-zero values of chemical potential, the Hawking-Page (HP) transition occurs at the temperatures larger than the temperature of black hole to black hole transition (BB) (Fig.\ref{T_zh.pdf}.{\bf B)}). So the black hole solution is always dominant with respect to thermal AdS  for  $\mu\neq0$.
 The position of BB phase transition line determines the phase diagram of the model \cite{AR,1808.05596}. The entropy function $s(T)$ is multivalued for  $\mu<\mu_{cr}$ and  becomes one-to-one for chemical potentials $\mu\geq\mu_{cr}$. The value of the $\mu_{cr}$ depends on the anisotropic parameter $\nu$,
$\mu_{cr}(\nu=1)=0.117$,
$\mu_{cr}(\nu=1.5)=0.189$,
$\mu_{cr}(\nu=2)=0.270$,
$\mu_{cr}(\nu=3)=0.299$, 
$\mu_{cr}(\nu=4.5)=0.349$.

In  Fig.\ref{Fig:FTmu45}.{\bf A)} the
BB phase transitions in the $(\mu,T)$-plane are presented  for the isotropic and anisotropic backgrounds with different $\nu$.  At the temperature values $T=T_{BB}(\nu,\mu)$ the thermal entropy undergoes a significant  jump (see Fig.\ref{Fig:FTmu45}.{\bf B)}), where the entropy is presented in the logarithmic scale. On these plots we see that the jumps disappear 
 for $\mu \geq \mu_{cr }$ ($\mu_{cr }=0.117$ for $\nu =1$ and $\mu_{cr }=0.349$ for $\nu =4.5$).

\section{Entanglement Entropy }\label{Sec: EE}
\subsection{General Framework}\label{Sec: GFW}
The  entanglement entropy is used  to probe correlations in  quantum systems. If the system is divided into two spatially disjoint parts $A$ and $\bar {A}$, the entanglement entropy $S(A)$ gives an estimation of the amount of information loss corresponding to the restriction of an $A$.
It is not simple to calculate the entanglement entropy from the  strongly coupled system side, in particular in QCD.
However, one can compute its holographic dual.
For some boundary region $A$ the HEE is obtained
by extremizing the 3-surface functional
\be\label{calAm}
\mathcal{A} = \int d^{3}\xi   \sqrt{|\det g_{MN}\partial_{\alpha}X^{M}\partial_{\beta}X^{N}|},
\ee
that ends on the boundary surface A. In the dual field theory the entanglement entropy of a subsystem $A$ is given by the formula \cite{RT1,RT2,HRT}
\be S_{EE}=\frac{\mathcal{A} }{4 G_5}. \label{HEE} \ee
In what follows we set $G_{5}=1$.

From  \eqref{calAm} we see that the entanglement entropy depends on the geometry of the area $A$.
 It is difficult to do  calculations  for arbitrary $A$, compare, for example, with \cite{Tonii}.
 We will do the calculations for the areas having the shape of parallelepipeds, two sides of which are long and one short, i.e. for parallelepipeds which look as slabs.

\begin{figure}[h!]
\centering
  \includegraphics[width=8cm]{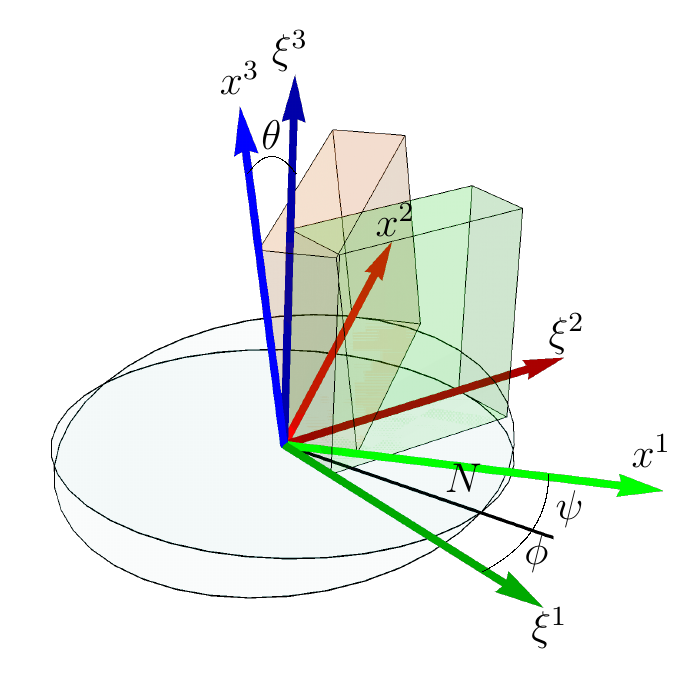}
\caption{The entangling subsystem is presented as a green slab. 
Rotating the green slab  by the Euler angles  $ (\phi, \theta, \psi) $ we get the pink slab that is oriented along the  axes $(x^1,x^2,x^3)$ associated with the HIC geometry and shown in Fig.\ref{fig:orient}.  }
\label{fig:arb}
    \end{figure}

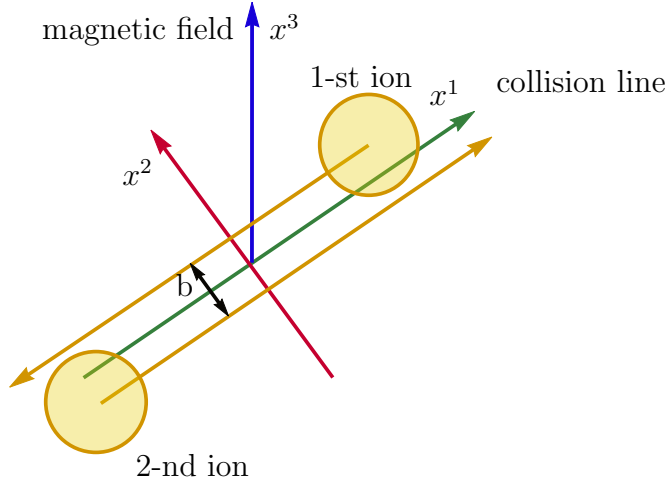
\begin{figure}[h!]
\centering
  \includegraphics[width=7cm]{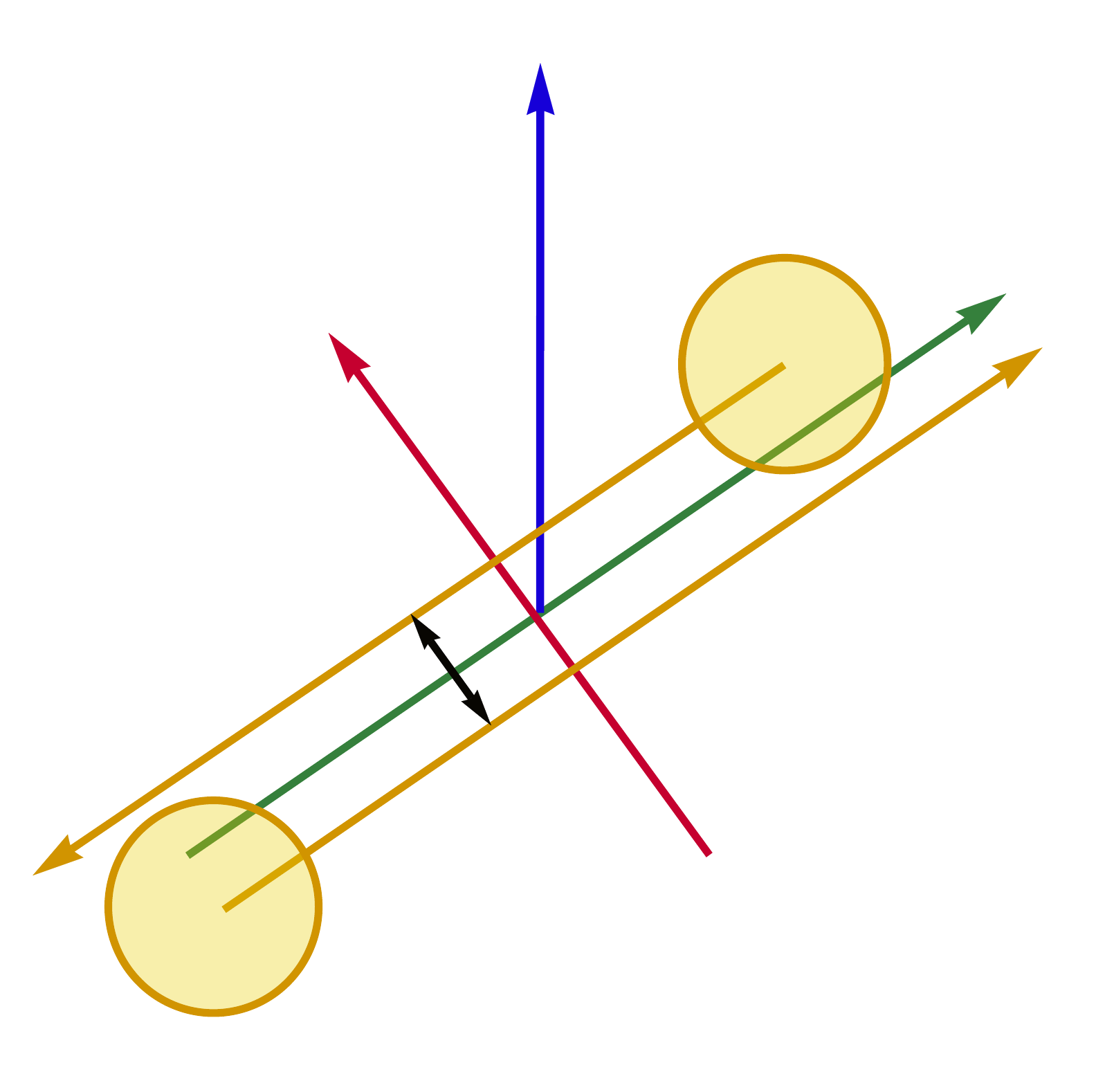}
  \begin{picture}(50,50)\put(-40,145){$x^1$}
  \put(-155,115){$x^2$}
  \put(-100,170){$x^3$}
  \put(-85,152){1-st ion}
   \put(-185,170){magnetic field}
  \put(-150,5){2-nd ion}
  \put(-15,150){collision line }
  \put(-135,73){b}\end{picture}

\caption{
 The orientation of the coordinate system ($x^1,x^2,x^3$) in respect to colliding ions. }
\label{fig:orient}
    \end{figure}

The orientation of these parallelepipeds can be specified by the Euler angles $(\phi, \theta,\psi )$ relative to the axes $x^1,x^2,x^3$, see Fig.\ref{fig:arb}. The axis $x^1$ is chosen along the collision line, the axis $x^2$ is chosen in the transversal direction along the direction of the impact parameter $b$, and the axis $x^3$ is  chosen along the emerging magnetic field, see Fig.\ref{fig:orient}. In Fig.\ref{fig:arb}  the initial slab is oriented along the axes specified by the HIC geometry and is shown in green. The rotated slab shown in pink defines the entanglement area.
 We are going to calculate the HEE for the rotated pink parallelepiped.

\begin{figure}[h!]\centering
 \includegraphics[width=6cm]{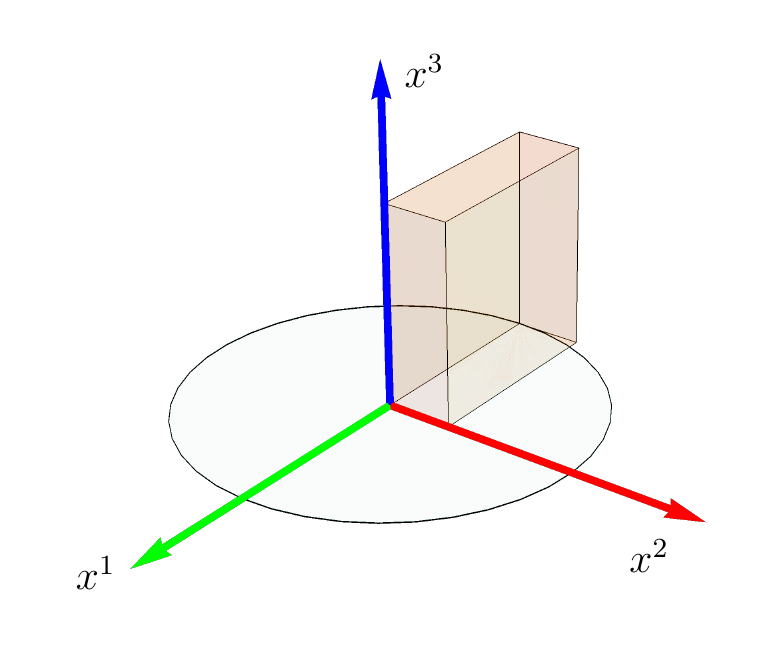}\quad\quad\includegraphics[width=6cm]{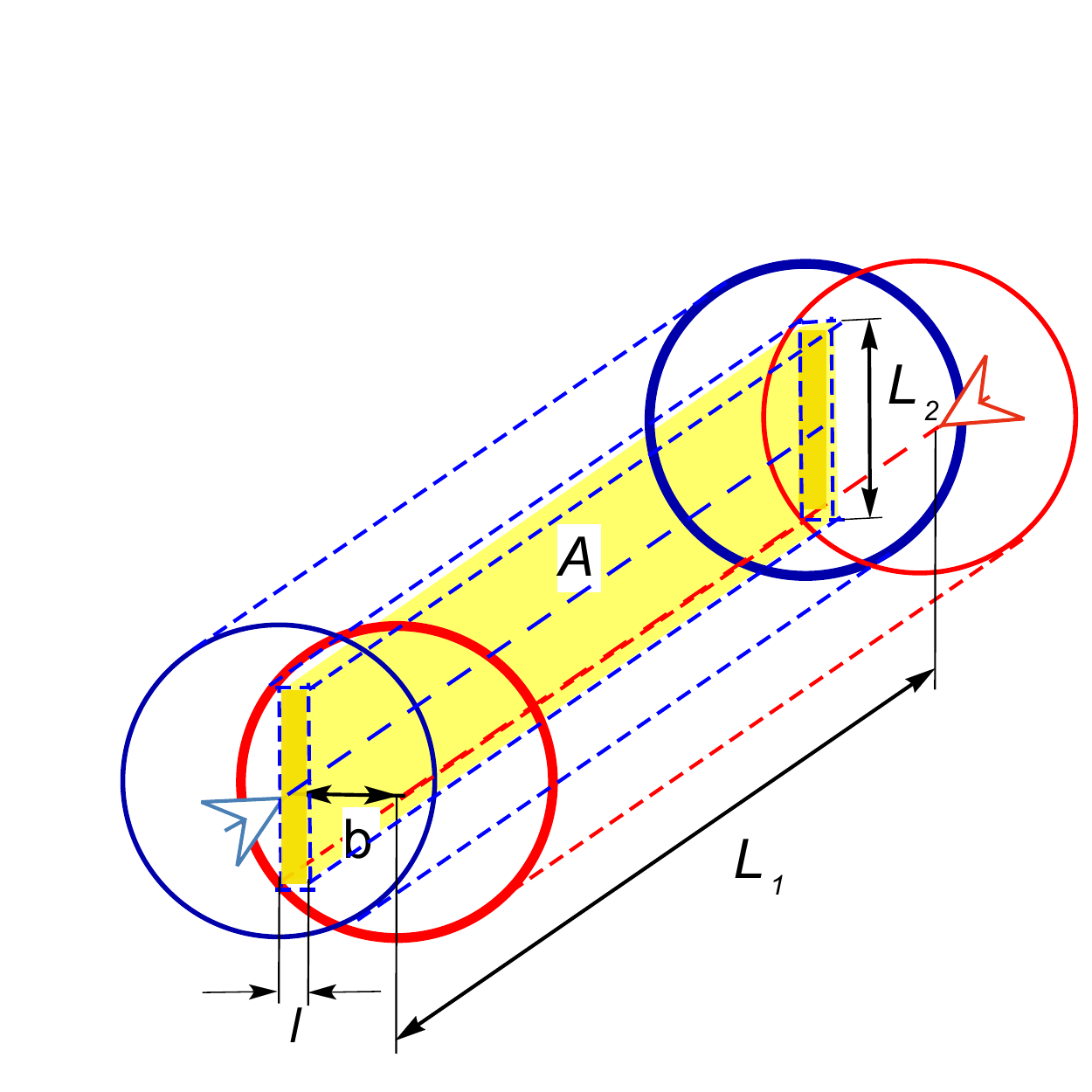}\\{\bf A)}\\
\includegraphics[width=6cm]{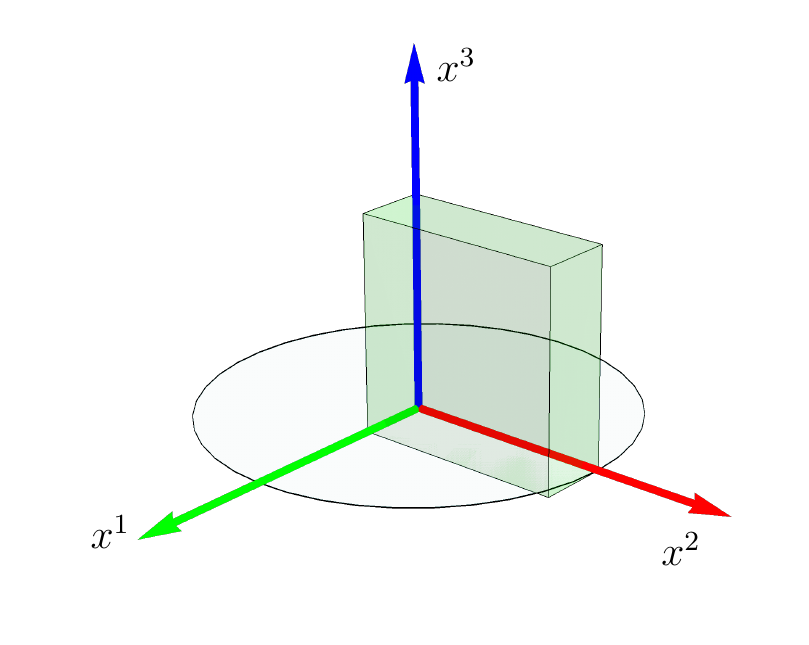}\quad\quad\includegraphics[width=6cm]{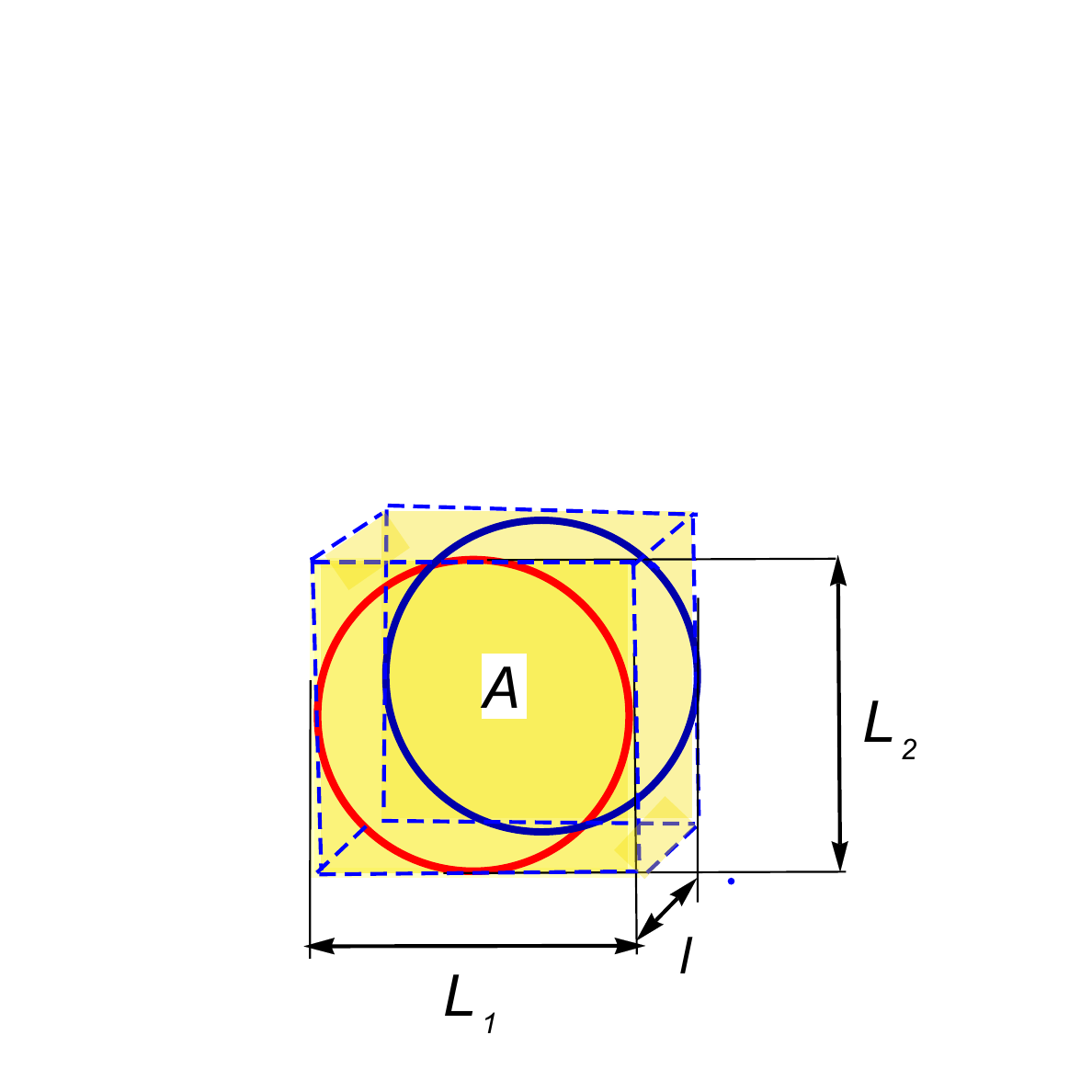}\\{\bf B)}\caption{The schematic picture of two ions collisions. {\bf A)} Each ion is presented as a disk with radius $D/2$ (blue and red disks). The trajectories of centers of two ions depicted by the points are shown by  dashed blue and red lines and the directions of their movement are indicated by thick arrows. The overlapping area of two ions considered as two disks has the shape of a region bounded by two arcs of two circles  (in the cross section perpendicular to the line of the ions collision).  {\bf B)} An almost central collision. On the left side of each graph, we show the orientation of the slabs  that we   are considering.}
    \label{fig:Glauber}
  \end{figure}
   
The entangled slabs  are supposed to be in
the 3-dimensional overlap of two nuclei  regions, so called  Glauber regions, see Fig.\ref{fig:Glauber}. The overlapping region of two nuclei depends on time. At a fixed point of time it is a three-dimensional body  in which the cross section, perpendicular to the axis of collisions, has a shape bounded by arcs of two circles shifted relative to each other according to an impact parameter. The area and the shape  of this area depend on the impact parameter $b$. The overlapping area for a peripheral collision is approximated by the parallelepiped with sizes $L_1,L_2$ and $\ell=D-b$. We assume $L_1,L_2>>\ell$, each ion is presented as a disk with radius $D/2$ and the impact parameter $b$ is essentially less than $D$, $\ell<<D$. The overlapping area of two ions is almost cylindrical and we can consider the case of very short time after collision, $\ell<<D$, $L_1,L_2\approx D$. There are only two specific cases: peripheral collision and central one, see Fig.\ref{fig:Glauber}. 

Let us show that for the spacetimes with
the metric \eqref{Gmetric} the problem of finding the extremal
area functional \eqref{calAm} for the slab with an arbitrary orientation effectively reduces to finding geodesics in some auxiliary 2-dimensional Euclidean space.
To show this we consider the embedding in the static gauge and assume that orientation of the slab in respect to the HIC axes is given by the Euler angles, see Fig.\ref{fig:arb}.

 We have    
\bea \label{GPar}
x^0(\xi)&=&const,\nn\\
x^i(\xi)&=& \sum _{j=1,2,3}a_{ij}(\phi, \theta,\psi )\,\xi^j,  \,\,\,\,\,i=1,2,3,\label{Emb}\\
x^4(\xi)&=&z(\xi^1),\nn 
\eea
$x^i$ are spatial coordinates in \eqref{Gmetric},
$a_{ij}(\phi, \theta,\psi )$ are entries of the rotation matrix
\bea M(\phi, \theta,\psi )={\begin{pmatrix}a_{11}(\phi, \theta,\psi ) & a_{12}(\phi, \theta,\psi ) & a_{13}(\phi, \theta,\psi ) \\a_{21}(\phi, \theta,\psi ) & a_{22}(\phi, \theta,\psi ) & a_{23}(\phi, \theta,\psi )  \\a_{31}(\phi, \theta,\psi ) & a_{32}(\phi, \theta,\psi ) & a_{33}(\phi, \theta,\psi ) \end{pmatrix}}
\label{Eulmat}
\eea
and
\bea
\begin{array}{lll}
&a_{11}(\phi, \theta,\psi )=\cos \phi \cos\psi -\cos \theta \sin \phi \sin\psi,\,\,\,  \\
&a_{12}(\phi, \theta,\psi )=-\cos \psi \sin\phi -\cos\phi \cos \theta \sin\psi,\,\,\,\, \\
&a_{13}(\phi, \theta,\psi )=\sin\theta \sin \psi, \\
&a_{21}(\phi, \theta,\psi )=\cos \theta \cos \psi \sin\phi+\cos \phi \sin\psi, \\
&a_{22}(\phi, \theta,\psi )=\cos \phi \cos \theta \cos\psi-\sin \phi \sin\psi,\\ 
&a_{23}(\phi, \theta,\psi )=-\cos \psi \sin \theta, \\
&a_{31}(\phi, \theta,\psi )=\sin\phi \sin \theta, \\
&a_{32}(\phi, \theta,\psi )=\cos\phi\sin \theta , \\
&a_{33}(\phi, \theta,\psi )=\cos \theta.
 \end{array}\\\ \nn \label{EP}
\eea
Here  $\phi$
 is the angle between the $\xi^1$ axis and the node line  (N),  shown in Fig.\ref{fig:arb}
 in black, $\theta$ is the angle between the $\xi^3$ and $x^3$ axis, $\psi$ is the angle between the node line N and the $x^1$ axis.

We write the line element  for the induced metric as
\bea
ds^2&=&g_{\alpha \beta }\,d\xi^\alpha d\xi^\beta, \,\,\,\,\alpha,\beta=1,2,3
\eea
and substitute the differentials $dx^M$  from the embedding relations \eqref{Emb} in the RHS of \eqref{Gmetric}:
\bea
ds^2&=&\frac{L^2 b_s(z)}{z^2}\Big(\sum _i\fg_i(z)dx^{i\,2}+\frac{dx^{4\,2}}{g}\Big)\nn\\&=&\frac{L^2 b_s(z)}{z^2}\Big( \sum _i\fg_i(z)\,\Big( \sum _ja_{ij}(\phi, \theta,\psi ) d\xi^{j}\Big)^2 
+z'^{2}\frac{d(\xi^1)^2}{g(z)}\Big).
\eea
We have
\bea g_{\alpha\beta}&=&\frac{L^2 b_s(z)}{z^2}\,\bar{g}_{\alpha\beta}
\eea
and
\bea
\bar g_{11}(z,\phi, \theta,\psi )&=&\fg_{1}a_{11}^2+\fg_{2}a_{21}^2+\fg_{3}a_{31}^2+\frac{z'^{2}}{g},\nn
\\
\bar g_{22}(z,\phi, \theta,\psi )&=&\fg_{1}a_{12}^2+\fg_{2}a_{22}^2+\fg_{3}a_{32}^2,\nn\\
\bar g_{33}(z,\phi, \theta,\psi )&=&\fg_{1}a_{13}^2+\fg_{2}a_{23}^2+\fg_{3}a_{33}^2,\nn
\\
\bar g_{12}(z,\phi, \theta,\psi )&=&\fg_{1}a_{11} a_{12}+\fg_{2}a_{21}a_{22}+\fg_{3}a_{13}a_{32},\nn
\\
\bar g_{13}(z,\phi, \theta,\psi )&=&\fg_{1}a_{11} a_{13}+\fg_{2}a_{21}a_{23}+\fg_{3}a_{31}a_{33},\nn
\\
\bar g_{23}(z,\phi, \theta,\psi )&=&\fg_{1}a_{12}a_{13} +\fg_{2}a_{22}a_{23}+\fg_{3}a_{32}a_{33},\nn\\
\bar g_{21}&=&\bar g_{12},\,\,\,\,\,g_{32}=\bar g_{23},\,\,\,\,\,\bar g_{32}=\bar g_{23}.\label{barg}
\eea

The determinant of the induced metric is
\bea
\det g_{\alpha\beta}&=&\left(\frac{L^2 b_s}{z^2}\right)^3\,\left(\fg_{1}\fg_{2}\fg_{3}+\frac{z'^{2}}{g}\,(\bar g_{22}\bar g_{33}-\bar g_{23}^2)\right)
\eea
and the Nambu-Goto action is
\be
{\cal S}=\int _{{\cal P}}\left(\frac{L^2 b_s}{z^2}\right)^{3/2}\sqrt{\left(\fg_{1}\fg_{2}\fg_{3}+\frac{z'^{2}}{g}\,(\bar g_{22}\bar g_{33}-\bar g_{23}^2)\right)}d\xi^{1}d\xi^{2}d\xi^{3},
\label{S_gen}\ee
where $g, \fg_{1}, \fg_{2}, \fg_{3}$ are functions of $z$ and $\bar g_{22},\bar g_{33},\bar g_{23}$ are functions of $z$ and the Euler
angles. Since there is no dependence of the integrand on $\xi^2,\xi^3$ we can perform the integration in these variables that gives the sizes of 
the parallelepiped in second and third directions:
\be
\frac{{\cal S}}{L_1L_2}=\int _{-\ell/2}^{\ell/2}\left(\frac{L^2 b_s}{z^2}\right)^{3/2}\sqrt{\left(\fg_{1}\fg_{2}\fg_{3}+\frac{z'^{2}}{g}\,(\bar g_{22}\bar g_{33}-\bar g_{23}^2)\right)}d\xi^1.\label{GAct}\ee
Here $\ell,L_1,L_2$  are the  lengths of the parallelepiped in the first, second and third directions. $\ell$ can be fixed by boundary conditions, see below \eqref{ell1}. In what follows we assume $L_1=L_2=1$.

The action \eqref{GAct} is a particular case of the BI action
\be
\label{BI}
{\cal S}=\int _{-\ell/2}^{\ell/2} M(z(\xi))\sqrt{{\cal {\cal F}}(z(\xi))+(z^{\prime}(\xi))^ 2}d\xi.
\ee
This action defines the dynamical system with dynamical variable $z=z(\xi)$ and  time $\xi$.
An effective potential is
\be
\label{EfPot}
{\cal V}(z(\xi))\equiv M(z(\xi))\sqrt{{\cal F}(z(\xi))}.
\ee
This system has the  first integral:
\bea
\label{FI}
\frac{M(z(\xi)){\cal F}(z(\xi))}{\sqrt{{\cal F}(z(\xi))+(z'(\xi))^2}}={\cal I}. \eea
From \eqref{FI} we can find the ``top''  point  ${z_{*}}$ (the closest position of the minimal surface to the horizon), where ${z'(\xi)=0}$:

\be
M(z_{*})\sqrt{F(z_{*})}={\cal I}.
\ee

Finding  $z^{\prime}$  from \eqref{FI}  one gets representations for the length $\ell$ and 
the action ${\cal S}$  \eqref{BI}, that defines the HEE  $S$ \eqref{HEE} up to the factor 1/4:

\bea\label{ell1}
\frac\ell2 &=&
\int_0^{z_*}\frac{1}{\sqrt{{\cal F}(z)}}\frac{dz} {\sqrt{\frac{{\cal V}^2(z)}{{\cal V}^2(z_*)} -1 }},\\
\frac{{\cal S}}{2}
&=&
\int_\epsilon^{z_*}\frac{M(z)dz}{\sqrt{1-\frac{{\cal V}^2(z_*)}{{\cal V}^2(z)}}}.
\label{calS1}
\eea

For action \eqref{GAct} 
we have

\bea
M(z)=\left(\frac{L^2 b_s}{z^2}\right)^{3/2}\sqrt{\frac{(\bar g_{22}\bar g_{33}-\bar g_{23}^2)}{g}},
\eea
\bea
{\cal F}(z)=\frac{\fg_{1}\fg_{2}\fg_{3}g}{(\bar g_{22}\bar g_{33}-\bar g_{23}^2)},
\eea
\bea
{\cal V}(z)=\left(\frac{L^2 b_s}{z^2}\right)^{3/2}\sqrt{\fg_{1}\fg_{2}\fg_{3}}. \label{GV}
\eea

Few remarks concerning \eqref{ell1} and \eqref{GAct}  are in order.
\begin{itemize}
\item The form of the effective potential ${\cal V}(z)$ does not depend on the slab orientation.
Therefore, the location of the dynamical wall, defined by location of the minimum of the effective potential ${\cal V}(z)$, is the same for all orientations for fixed $z_h$ and $\nu$.
\item The expressions for $M(z)$ and ${\cal F}(z)$ depend only on the combination
\be
\bar g_{22}\bar g_{33}-\bar g_{23}^2,\ee
that in its turn depends on 3 angles $\phi,\psi$ and $\theta$: 
\bea
\bar g_{22}\bar g_{33}-\bar g_{23}^2&=&\fg_{1}\fg_{2}a^2_{31}(\phi, \theta,\psi) +
\fg_{1}\fg_{3}a_{21}^2(\phi, \theta,\psi) +\fg_{2}\fg_{3}a_{11}^2(\phi, \theta,\psi ) .\eea
 \end{itemize}
 \subsection{Geometric Renormalization}
It is convenient  to perform renormalization of the  entanglement entropy by subtraction the "disconnected" surface contribution from the "connected" one  
\cite{IK,Kol:2014nqa,Dudal:2017max}.  The contribution of the "disconnected" surfaces  is given by the doubled area of the surface hanging from the boundary to the  horizon $z=z_{h}$ or to the dynamical wall $z_{DW}$ and the area of the surface along $z=z_{h}$ or $z=z_{DW}$ (if  the dynamical wall exists for the considered parameters).

 The difference between the "connected" and "disconnected" parts $S_{CD} \equiv S_{conn}-S_{diccon}$
 for  arbitrary oriented  slab is
 
\bea
{\cal S}_{CD}&=&2
\int_\epsilon^{z_*}\frac{L^3 b^{3/2}_s(z)}{z^3} \sqrt{\frac{\fg_{1}\fg_{2}a^2_{31}+
\fg_{1}\fg_{3}a_{21}^2+\fg_{2}\fg_{3}a_{11}^2}{g(z)}}\frac{dz}{\sqrt{1-\frac{{\cal V}^2(z_*)}{{\cal V}^2(z)} }}
\nn\\
&-&2\int_\epsilon^{z_D}\left(\frac{L^2 b_s(z)}{z^2}\right)^{3/2} \sqrt{\frac{\fg_{1}\fg_{2}a^2_{31}+
\fg_{1}\fg_{3}a_{21}^2+\fg_{2}\fg_{3}a_{11}^2}{g(z)}}\,dz\nn\\
   &-&\left(\frac{L^2 b_s(z_D)}{z_D^2}\right)^{3/2} \sqrt{\fg_1 \fg_2
   \fg_3}\Big|_{z=z_D} \ell,\label{calSdc}
\eea

where $z_D$ is the minimum of the  two values $z_{DW}$ and $z_{h}$ ($z_{DW}$ is the position  of the dynamical wall, at this point ${\cal V}^{\prime }(z_{DW})=0$).

Here we used  the fact that the determinant of $3\times 3$ induced  matrix  corresponding to  the 3-dim body hanging along the z-axis is
\bea
\det \bar M&=&\frac{\left(a_{13} a_{22}-a_{12}a_{23}\right){}^2 \fg_1
   \fg_2+\left(a_{13}
   a_{32}-a_{12}
   a_{33}\right){}^2 \fg_1
   \fg_3+\left(a_{23}
   a_{32}-a_{22}
   a_{33}\right){}^2 \fg_2 \fg_3}{g}\nn\\
  &=&\frac{ \fg_{1}\fg_{2}a^2_{31}+
\fg_{1}\fg_{3}a_{21}^2+\fg_{2}\fg_{3}a_{11}^2}{g},\eea
In \eqref{calSdc} $z=z_D$ is the position of the horizon or the dynamical wall. Here we consider  the dynamical wall for HEE, that is defined as the position of the minimum of the  effective potential ${\cal V}(z)$,
\bea
{\cal V}(z)=\left(\frac{L^2 b_s(z)}{z^2}\right)^{3/2}\sqrt{\fg_{1}(z)\fg_{2}(z)\fg_{3}(z)}. \label{GV}
\eea
It is interesting to note that ${\cal V}$ does not depend on the Euler angles.
The HEE dynamical wall does not coincide with  Wilson loop  dynamical walls, which depend
on the orientation of the loop in respect to the collision axes \cite{1808.05596}.

\subsection{HEE for $\fg_1=1$, $\fg_2=\fg_3=\fg$
and arbitrary orientation }

In this section we consider the metric \eqref{ARG}.
From the general formula \eqref{GAct} we get
 
\bea
\frac{\cal S}{2}=\int _{-\ell/2}^{\ell/2}\left(\frac{L^2 b_s(z)}{z^2}\right)^{3/2}\sqrt{s(\fg,\phi,\theta,\psi)}\,dz,
\eea
where
\bea
&&s(\fg,\phi,\theta,\psi)=\fg^2(z)+
\Big[a_{21}^2(\phi,\theta,\psi)+
a_{31}^2(\phi,\theta,\psi)+a_{11}^2(\phi,\theta,\psi)\fg(z) \Big]\,\frac{\fg(z)\,z^{\prime2}}{g(z)}, \nn
\eea
and $a_{ij}$ are given by \eqref{EP} and satisfy the relation
\be
a_{21}^2(\phi,\theta,\psi)+a_{31}^2(\phi,\theta,\psi)+a_{11}^2(\phi,\theta,\psi)=1.\ee
So we can use a new parametrization
\bea
a_{11}(\phi,\theta,\psi)=\cos \varphi,\eea
and in this case 
\be
s(\fg,\phi,\theta,\psi)=\fg^2(z)+
\Big[\sin ^2\varphi+\fg(z) \cos^2 \varphi\Big]\,\frac{\fg(z)\,z^{\prime2}}{g(z)}.\ee
Note that $\varphi$ is nothing but the angle between the $\xi^1$ and $x^1$ axes.
In this parametrization  we have
\bea
{\cal V}(\fg,\phi,\theta,\psi)&=&
\left(\frac{L^2 b_s(z)}{z^2}\right)^{3/2}\fg(z),
\\M(\fg,\phi,\theta,\psi)&=&
\left(\frac{L^2 b_s(z)}{z^2}\right)^{3/2}\sqrt{\frac{\fg(z)}{g(z)}(\sin^2\varphi+\cos^2\varphi\fg(z)) },
\\
{\cal F}(\fg,\phi,\theta,\psi)&=&\frac{\fg(z) g(z)}{\sin^2\varphi+\fg(z)\cos^2\varphi}.
\eea

 \begin{figure}[h!]\centering
  \includegraphics[width=8cm]{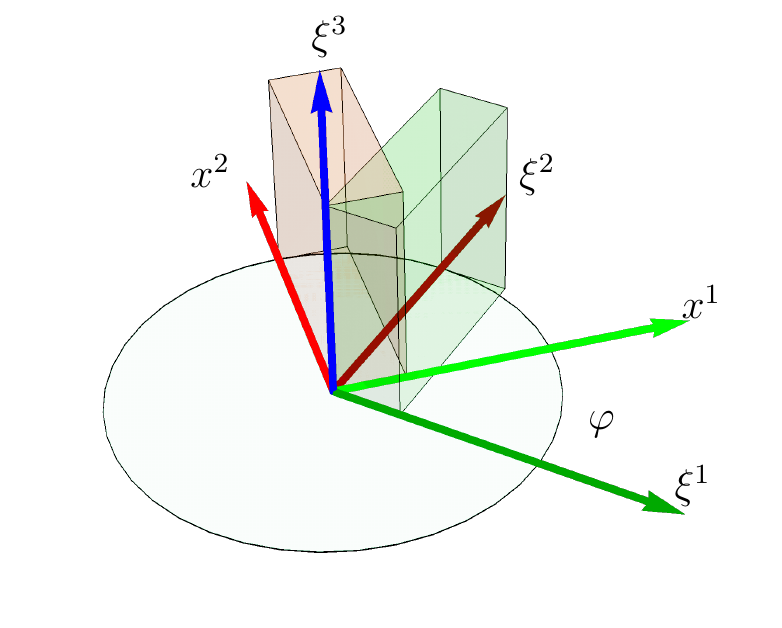}
\caption{
 The entangling subsystem defined by the Euler angles $(\varphi, 0,0)$. The HEE in the geometry  \eqref{ARG} is invariant with respect to rotations of the slab around the $x^1$-axis.}
\label{fig:theta}
    \end{figure}

   This answer is equivalent to the HEE of the slab that is obtained by rotation of the initial slab, oriented along the natural HIC axes, around the $\xi^3$ axis by the angle $\varphi$, see Fig.\ref{fig:theta}.

\subsubsection{Application of Geometric Renormalization}
According to \eqref{calSdc} we have

\bea
{\cal S}_{CD,\varphi}&=&2
\int_\epsilon^{z_*}\left(\frac{L^2 b_s(z)}{z^2}\right)^{3/2} \sqrt{\frac{\fg(z)(\sin^2\varphi+\cos^2\varphi\,\fg(z))}{g(z)}}
\frac{dz}{\sqrt{1-\frac{{\cal V}^2(z_*)}{{\cal V}^2(z)} }}
\nn\\
&-&2\int_\epsilon^{z_D}\left(\frac{L^2 b_s(z)}{z^2}\right)^{3/2}
 \sqrt{\frac{\fg(z)(\sin^2\varphi+\cos^2\varphi\,\fg(z))}{g(z)}}\,dz\nn\\
   &-&\left(\frac{L^2 b_s(z)}{z^2}\right)^{3/2}\fg(z)
   \Big|_{z=z_D}\,
   \ell(\fg,\varphi),
   \label{calSdc-phi}
\eea
where
\bea
\frac{\ell(\fg,\varphi)}2 &=&
\int_0^{z_*}\frac{\sqrt{\sin^2\varphi+\fg(z)\cos^2\varphi}}{\sqrt{\fg(z) g(z)}}\frac{{\cal V}(z_*)}{{\cal V}(z)}\frac{dz} {\sqrt{1 -\frac{{\cal V}^2(z_*)}{{\cal V}^2(z)} }}.\eea

\subsection{The Holographic Entanglement Entropy for $\fg_1=\fg_2=(z/L)^{2-2/\nu}$}\label{Sect:varphi}

    We get the following representations for the character length and the action. For \eqref{ARmetric} we have
\bea \label{elltheta}
\frac{\ell_{\varphi}}2&=&
\int_0^{z_*}\frac{{\cal V}(z_*)}{{\cal V}(z)}
\sqrt{\cfrac{(\frac{z}{L})^{-2+2/\nu} \sin^2\varphi+ \cos^2\varphi} {g(z)( 1-\frac{{\cal V}^2(z_*)}{{\cal V}^2(z)})} }
\,dz,~~~~
{\cal V}(z)=\left(\frac{L}{z}\right)^{1+2/\nu}b^{3/2}_s(z),
\nn\\\,
\label{elltheta}\\\label{Stheta}
\frac{{\cal S_{\varphi}}}{2}
&=&\int _{\epsilon}^{z_{*}}\,{\cal V}(z)\sqrt{\cfrac{(\frac{z}{L})^{-2+2/\nu} \sin^2\varphi+ \cos^2\varphi} {g(z)( 1-\frac{{\cal V}^2(z_*)}{{\cal V}^2(z)})} }
\,dz.
\eea
$M_{\varphi}(z)$ defined as 
\bea
M_{\varphi}(z)&=&
{\cal V}(z)\sqrt{\cfrac{\Big(\frac{z}{L}\Big)^{-2+2/\nu}\sin^2\varphi+\cos^2\varphi } {g(z)} }
\eea
determines the degree of divergence of ${\cal S_{\varphi}}$. Near $z=0$ it has a different behavior for $\varphi=0$ and $\varphi\neq0$. However, we can use the universal renormalization \eqref{calSdc}.

For the particular case of the longitudinal ($\varphi=0$, subscript $xYY$) orientation, the  difference between "connected" and "disconnected" parts according to \eqref{calSdc} is

\bea
 S_{CD,xYY}=\frac{1}{2}\int_\epsilon^{z_*}\,\frac{{\cal V}(z)\,dz}{{\sqrt{g(z)}}}\left[\frac{1}{\sqrt{1-\frac{{\cal V}^2(z_*)}{{\cal V}^2(z)}}}-1\right]
-\frac{1}{2}\int^{z_D}_{z_*}\frac{{\cal V}(z)dz}{\sqrt{g(z)}}-\frac{{\cal V}(z_D)}{4}\,\ell_{xYY},
\nn\\
\label{
diccyXY2}
\eea
where
\bea
\frac{\ell_{xYY}}{2}&=&\int_0^{z_*}\frac{{\cal V}(z_*)}{{\cal V}(z)}\frac{dz} {\sqrt{g(z)( 1-\frac{{\cal V}^2(z_*)}{{\cal V}^2(z)}) }}.\eea
The difference between the "connected" and "disconnected" parts 
for the transversal orientation ($\varphi=\pi/2$, subscript $yXY$) is

\bea
{S}_{CD,yXY}=\frac{1}{2}\int_\epsilon^{z_*}\frac{L^{1-1/\nu}{\cal V}(z)}{z^{1-1/\nu}}\frac{dz}{\sqrt{g(z)}}\left[\frac{1}{\sqrt{1-\frac{{\cal V}^2(z_*)}{{\cal V}^2(z)}}}-1\right] \nn\\
-\frac{1}{2}\int^{z_D}_{z_*}\frac{z^{1-1/\nu}{\cal V}(z)}{z^{1-1/\nu}}\frac{dz}{\sqrt{g(z)}}-\frac{{\cal V}(z_D)}{4}\,\ell_{yXY},
\nn\\
\label{
diccxYY2}
\eea
where
\bea \label{elltheta}
\frac{\ell_{yXY}}2&=&
\int_0^{z_*}\frac{{\cal V}(z_*)}{{\cal V}(z)}\frac{L^{1-1/\nu}}{z^{1-1/\nu} }
\frac{dz} {\sqrt{g(z)( 1-\frac{{\cal V}^2(z_*)}{{\cal V}^2(z)})} }.\eea

The HEE is a UV divergent quantity, so the  HEE needs renormalization for  $z \sim 0$.
Note that we can analyze the behavior of the integral at the upper limit $z_{*}$. If $z_*\neq z_{DW}$ ($z_{DW}$
 is the point, where ${\cal V}'(z_{DW})=0$) we have an integrable singularity (because ${\cal V}'(z)\neq 0$ for $0<z<z_*$). If $z_*=z_{DW}$ we have a logarithmic singularity \cite{1808.05596}.

 Now we can determine the power of the integrand singularity at $z=0$. It is defined by $M$ behavior near $z=0$:
\bea
M(z)\left[\frac{1}{\sqrt{(1-\frac{{\cal V}^2(z_*)}{{\cal V}^2(z)})}}-1\right]\label{diccapp} \underset{z\sim 0}{\sim}
 \frac{M(z)}{({\cal V}(z))^2}\underset{z\sim 0}{\sim}z^{\kappa_{0}}. \,\,\,\,\label{diccapp2}
\eea
Here $\kappa_{0} $  is defined by the asymptotic of $M$ and $\cal V$ at $z=0$ and depends on the orientation, so we use subscripts to specify the different orientations.

Taking into account the dilaton field $\phi(z,z_h,c,\nu)$ the asymptotic is given by (eq. (2.58) in \cite{AR}):
\be
  \phi(z,z_h,c,\nu) \sim - \, k(z_h,\nu,c) + \frac{2 \, \sqrt{\nu -1}}{\nu} \, \log
  \left(\cfrac{z}{z_h}\right),
\ee
 where
$k(z_h,\nu,c)$ does not depend on $z$. Therefore we have the following asymptotic of the functions $b_s(z,\nu)$ and $M(z)$ at $z \to 0$: 
\bea
b_s(z)&\underset{z\sim 0}{\sim} &B_{s}(\nu,c,z_h)\,z^{\frac{\sqrt{\frac83(\nu -1)}}{\nu}},\\
M_{xYY}(z)&\underset{z\sim 0}{\sim}&z^{\kappa_{xYY}(\nu)}B^{3/2}_s(\nu,c,z_h) ,\\
M_{yXY}(z)&\underset{z\sim 0}{\sim}&\,z^{\kappa_{yXY}}B^{3/2}_s(\nu,c,z_h), \label{Mass}
\eea
where
\bea
B_s(\nu,c,z_h)&\equiv &e^{-\sqrt{\frac{2}{3}} \Big(\frac{2 \, \sqrt{\nu -1}}{\nu} \, \log
  z_h
+ \, k(z_h,\nu,c) \Big)}, \label{Bs}\\
 \kappa_{xYY}(\nu)&\equiv&
\frac{\sqrt{6(\nu -1)}}{\nu}-1-\frac{2}{\nu}, \label{kappa}
\eea
\bea
\kappa_{yXY}(\nu)&\equiv &
\frac{\sqrt{6(\nu -1)}}{\nu}-2-1/\nu.\label{MxYY-z0}\eea
The expression for $B_s(\nu,c,z_h)$:
\bea
\nn&&B_{s}(\nu,c,z_h)^{3/2}=2^{-\frac{\sqrt{6} \sqrt{\nu-1}}{\nu}-1} 3^{-\frac{\sqrt{\frac{3}{2}} \sqrt{\nu-1}}{\nu}} \left(\frac{\sqrt{6} \sqrt{\nu-1}}{\nu}-\frac{1}{\nu}-1\right) z_{*}^{-\frac{\sqrt{6} \sqrt{\nu-1}}{\nu}+\frac{1}{\nu}+2} \times 
\\  \nn&&\times\left(\frac{ {\cal N}(\nu,c,z_h)+\sqrt{3} \nu \left(c z_{h}^2-3\right)}{2 \sqrt{2} \sqrt{\nu-1}-3 \sqrt{3} \nu}\right)^{9/4} \left(\frac{{-9 c \nu^2 z_{h}^2+\sqrt{2(\nu-1)} }{\cal N}(\nu,c,z_h)+8 \nu-8}{\nu-1}\right)^{\frac{\sqrt{\frac{3}{2}} \sqrt{\nu-1}}{\nu}}\times 
\\ &&\times\exp \left(\frac{\sqrt{\frac{3}{2}} \sqrt{\nu-1}}{\nu}-\frac{\sqrt{3}{\cal N}(\nu,c,z_h)}{4 \nu}\right) \left(\frac{z}{z_{h}}\right)^{\frac{\sqrt{6} \sqrt{\nu-1}}{\nu}},
\eea
where \be{\cal N}(\nu,c,z_h)= \sqrt{3 c \nu^2 z_{h}^2 \left(c z_{h}^2-6\right)+8 \nu-8}.\ee
The potential $\cal V$ asymptotic at $z\sim 0$:
\bea
{ \cal V}(z)=\frac{b_s(z)^{3/2}}{z^{2+1/\nu}}\underset{z\sim 0}{\sim} M_{xYY}(z).
\eea
Comparing with $\kappa_{xYY}(\nu)$ we obtain
\bea
\kappa_{yXY}(\nu)&=&\kappa_{xYY}(\nu)-1+1/\nu.
\eea
The integrand \eqref{diccapp2} for the $yXY$ case has a larger  degree of divergence than for  $xYY$ case, so
\bea
\kappa_{0}=-\left(2-2/\nu+\frac{\sqrt{6(\nu -1)}}{\nu}-2-1/\nu \right)=
3/\nu-\frac{\sqrt{6(\nu -1)}}{\nu}.
\eea
 Note that  $\kappa_{0}>-1$ for $\nu\geq1$ (see Fig.\ref{kappa}), which means that the HEE is finite after the geometric renormalization for $\nu\geq1$.

\subsubsection{Minimal renormalization for $\varphi=0$}
\label{Sect:longitudinal}
Here we consider \eqref{Stheta} in a particular case $\varphi=0$, i.e.  the smallest size of the slab is oriented along the longitudinal direction.  
In notation \eqref{BI} we have:
\bea
{\cal V}_{xYY}(z)=\frac{b^{3/2}_s(z)}{z^{1+2/\nu}},~~~
M_{xYY}(z)=\frac{b^{3/2}_s(z)}{z^{1+2/\nu}\sqrt{g(z)}},~~~
{\cal F}_{xYY}(z)=g(z).\label{VMFxYY}
\eea

The UV divergencies  are defined by behavior of $M_{xYY}(z)$ at $z \sim 0 $ \eqref{Mass}. We see that $M_{xYY}(z)$ has  an integrable singularity at $z=0$ for $\nu \geq 1.67$.

 For $\nu=1$ we  have to perform a renormalization and for the renormalized HEE we get:
\be
S_{ren}=\frac{1}{2}\int_\epsilon^{z_*}dz\left[\frac{M(z)}{\sqrt{(1-\frac{{\cal V}^2(z_*)}{{\cal V}^2(z)})}}-M_{as}(z)
\right]+\frac{1}{2}\int^{z_*}M_{as}(z)dz,
\label{calSren}\ee
where  the last term means the indefinite integral of $M_{as}(z) $ at $z=z_*$.

For $xYY$ case and $\nu >1.67$ we have an integrable singularity. For $xYY$ case and $1<\nu<1.67$ we  have to perform renormalization:

\bea
S_{xYY,ren}&=&\frac{1}{2}
\int_\epsilon^{z_*}dz\left[ \frac{M_{xYY}(z)}{\sqrt{(1-\frac{{\cal V}_{xYY}^2(z_*)}{{\cal V}_{xYY}^2(z)})}}-M_{xYY,as}(z)
\right]+\frac{1}{2}\int^{z_*}M_{xYY,as}(z)dz,\nn\\
\label{calSrenB}
\eea
where we use the expression for  the indefinite integral  of $M$:
\bea
\int^{z_*}M_{xYY,as}(z)= B_{s}(\nu,c,z_h)^{3/2}\frac{z_{*}^{\kappa_{xYY}(\nu)+1}}{\kappa_{xYY}(\nu)+1}.
\label{calSren2}
\eea

\begin{figure}[h!]\centering
 \includegraphics[width=6cm]{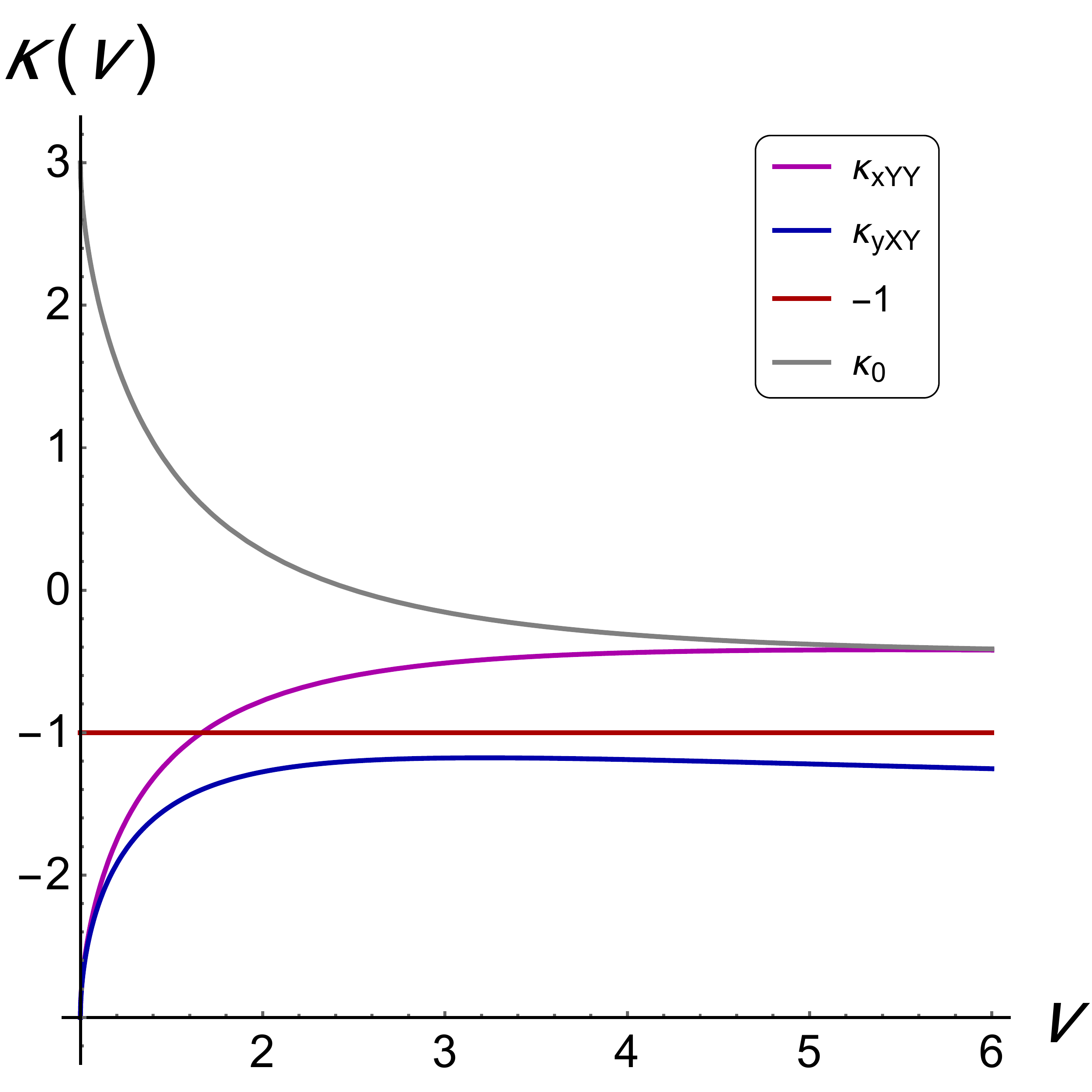}

     \caption{ Functions $\kappa_{yXY}(\nu)$, $\kappa_{xYY}(\nu)$ and $\kappa_0$ are shown by blue, magenta and gray lines. 
     }
    \label{kappa}
 \end{figure}
  
\subsubsection{Minimal renormalization for $\varphi \neq 0$}\label{Sect:transversal}

Here we consider a  configuration for $\varphi \neq 0$. As we can see from \eqref{Stheta}, the type of UV asymptotic  is the same for all $\varphi \neq 0$ and further we consider  the case $\varphi =\pi/2$  for simplicity. Due to the invariance in the transversal directions we can choose $\theta=0$, $\psi=0$. This corresponds to parametrization \eqref{GPar} with  $\phi=\pi/2$,  $\theta=0$, $\psi=0$ and in this configuration the small subsystem orientation is delineated along the transversal direction $y_{1}$. We call this configuration the transversal and denote with the subscript $yXY$.
In notation \eqref{BI} we have:
\bea
{\cal V}_{yXY}(z)=\frac{b^{3/2}_s(z)}{z^{1+2/\nu}},~~~
M_{yXY}(z)=\frac{b^{3/2}_s(z)}{z^{2+1/\nu}\sqrt{g(z)}},~~~
{\cal F}_{yXY}(z)=g(z)\,z^{2-2/\nu},\label{VMFyXY}
\eea

UV divergences are now defined by the asymptotic of
$M_{yXY}(z)$ at $z\sim 0$ \eqref{Mass}.

The plots of functions $\kappa_{yXY}(\nu)$ and $\kappa_{xYY}(\nu)$
are presented in Fig.\ref{kappa}.
We see that $M_{yXY}(z)$ has a
nonintegrable singularity for all $\nu$ values at $z=0$.

For $yXY$ case and $\nu>1$ there is  a nonintegrable singularity and we have to perform a renormalization:

\bea
S_{yXY,ren}= \frac{1}{2}\int_\epsilon^{z_*}dz\left[\frac{M_{yXY}(z)}{\sqrt{(1-\frac{{\cal V}_{yXY}^2(z_*)}{{\cal V}_{yXY}^2(z)})}}- M_{yXY,as}(z)
\right]+\frac{1}{2}B_{s}(\nu,c,z_h)^{3/2}\frac{z_{*}^{\kappa_{yXY}(\nu)+1}}{\kappa_{yXY}(\nu)+1},\nn\\
\label{calSrenC}\eea
where
\bea
\nn
\int M_{yXY,as}\,dz= B_{s}(\nu,c,z_h)^{3/2}\frac{z^{\kappa_{yXY}(\nu)+1}}{\kappa_{yXY}(\nu)+1}.
\label{calSren2B}
\eea

\newpage
\section{Entanglement Entropy Density and c-functions}\label{Sect:density1}

It is also instructive to consider the entanglement entropy density that is defined as  \cite{NNT,BHRT} 
\be
  \eta=\frac{d S(\ell)}{d\ell}
  \ee
 (compare with \cite{1708.09376,1709.07016}).   The advantage of dealing with the HEE density is that it has no  divergences. For fixed temperature $T$ (i.e. fixed $z_h$) the entanglement entropy density  can be obtained from equation \eqref{calS1}. We see that it can be expressed in terms of the value of the effective potential  
${\cal V}$,
  \be
  \eta(z_{*})=\frac{d S(z_{*})}{d\ell(z_{*})}=\frac{\frac{d S(z_*)}{dz_*}}{\frac{d \ell(z_*)}{dz_*}}=\frac{{\cal V}(z_*)}{4}.\label{eta} \ee

  The potential ${\cal V}(z)$ in the  Born-Infield  action \eqref{GAct}  is the same for different orientations of the entangling domain. From \eqref{GV} we get
\be
\eta(z_*)=\frac{L^3 b^{3/2}_s(z_*)}{4 z_*^3} (\fg_{1}(z_*)\fg_{2}(z_*)\fg_{3}(z_*))^{1/2}. \label{GVeta}
\ee
However, the angular dependence of the density exists because the length $\ell$, given by \eqref{elltheta}, depends on the orientation. 

Definition \eqref{GVeta} requires a comment.  Despite the fact that the density of entropy does not contain divergences, its definition may depend on a finite renormalization. In particular, adopting the geometric renormalization \eqref{calSdc} we have 

\bea
\eta_{CD}&=&\frac{1}{4}\Big( {\cal V}(z_*)-{\cal V}(z_{D})\Big)\label{etaR}\\
&=&\frac{L^3}{4}\Big(\frac{ b^{3/2}_s(z_*)}{z_*^3} \, (\fg_{1}(z_*)\fg_{2}(z_*)\fg_{3}(z_*))^{1/2} -\frac{ b^{3/2}_s(z_{D})}{z_D^3}(\fg_{1}(z_D)\fg_{2}(z_D)\fg_{3}(z_D))^{1/2}\Big)\nn,
\eea
where $z_D=z_{DW}$ or $z_D=z_{h}$.

The anisotropic c-function can be defined as
\be
c_\varphi=\ell _\varphi ^{m_\varphi}\frac{d S(\ell_\varphi)}{d\ell_\varphi},
\ee
 Here $m$ is the scaling power. It depends on the model and we  make few comments about $m$ in the next subsection \ref{Sect:c-function}. Generally speaking, unlike conformal theories 
 \cite{Zamolodchikov:1986gt, Cardy:1988cwa, Komargodski:2011xv} and especially the holographic theories with conformal invariance   \cite{Girardello:1998pd,Freedman:1999gp,
 Myers:2010tj, Myers:2010xs, Myers:2012ed,Ryu:2006ef}, $m_\varphi$ 
 is different in UV and  IR, can depend on the orientation of the slab as well (see  considerations of candidates for c-functions in theories with Lorentz violation \cite{Liu:2012wf,Swingle:2013zla,Cremonini:2013ipa,Bea:2015fja}, and especially, \cite{Chu:2019uoh,Ghasemi:2019xrl,Hoyos:2020zeg}). 

 We can also  consider
\be
c_{CD,\varphi}=\frac{\ell^m_{\varphi}L^3}4\,\Big(\frac{ b^{3/2}_s(z_*)}{z_*^3} (\fg_{1}(z_*)\fg_{2}(z_*)\fg_{3}(z_*))^{1/2} -\frac{b^{3/2}_s(z_{D})}{z_D^3} (\fg_{1}(z_D)\fg_{2}(z_D)\fg_{3}(z_D))^{1/2}\Big).\label{GA}
\ee

$$\,$$

\section{Numerical Results}\label{Sect:NR}
In this section we display and discuss our main results using numerical calculations. In what follows  we set $L=1$.
We present all the plots according to the  color scheme of the lines indicated in Table 1.
\begin{table}[h!]
\centering
\begin{tabular}{|l|c|c|c|l|c|l|c|l|}
\hline
 & \multicolumn{4}{c|}{SF}                                                                        & \multicolumn{4}{c|}{EF}                                \\ \cline{2-9} 
                  & 
                  {ISO}       & \multicolumn{3}{c|}{ANIS}                                        
                   & 
                   {ISO}       & \multicolumn{3}{c|}{ANIS} \\ \cline{3-5} \cline{7-9} 
                  &                            & TRAN                      & $\varphi$ & \multicolumn{1}{c|}{LONG} &                            & TRAN  & $\varphi$  & LONG \\ \hline
COLOR             & \multicolumn{1}{l|}{Green} & \multicolumn{1}{l|}{Blue} & various   & Magenta                   & \multicolumn{1}{l|}{Brown} & Cyan  & various    & Gray \\ \hline
\end{tabular}
\caption{The color scheme of the lines used in the plots in this article.}
\end{table}
\label{Tab:color}
\subsection{Entanglement Entropy
near the Background Phase Transition}\label{Sect:HEE-TL}

 In this section we present plots of the entanglement entropy dependence on the geometric characteristics of the  entangling region (orientation and thickness of the slab) and the thermodynamic characteristics of the medium (temperature and chemical potential)
for the model \eqref{ARmetric}.
We find the slab  HEE dependence on the smallest length $\ell$ numerically performing  integration of \eqref{elltheta} and \eqref{Stheta}, and then excluding the dependence on $z_*$ by solving equation \eqref{elltheta} for a given $\ell$. \\

\begin{figure}[h!]\centering
 \includegraphics[width=6.5cm]{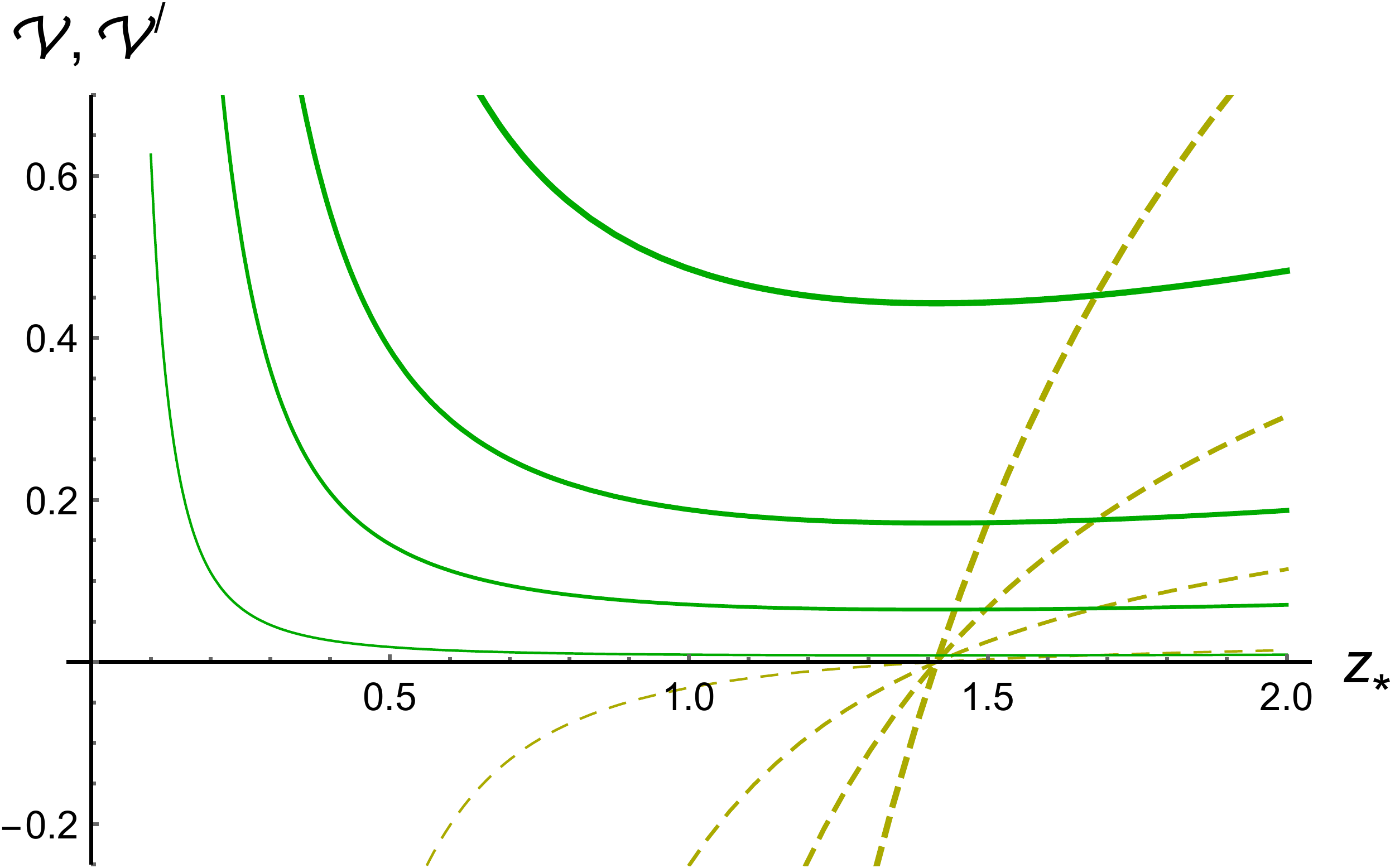}\quad \quad
  \includegraphics[width=6.5cm]{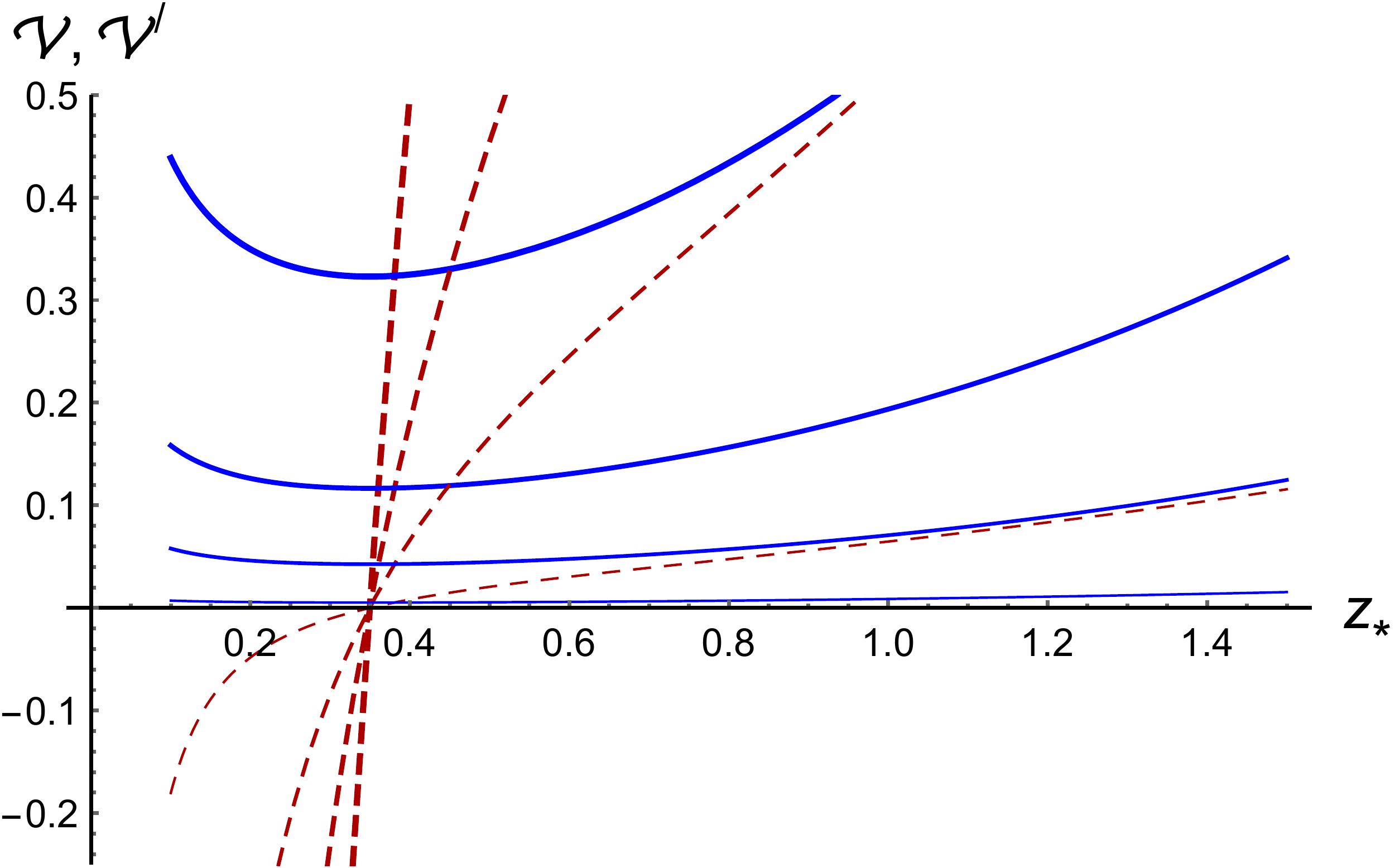}
 \\
 {\bf A)}\quad \quad \quad \quad \quad \quad \quad \quad \quad \quad \quad \quad \quad {\bf B)}\\
  \includegraphics[width=6.5cm]{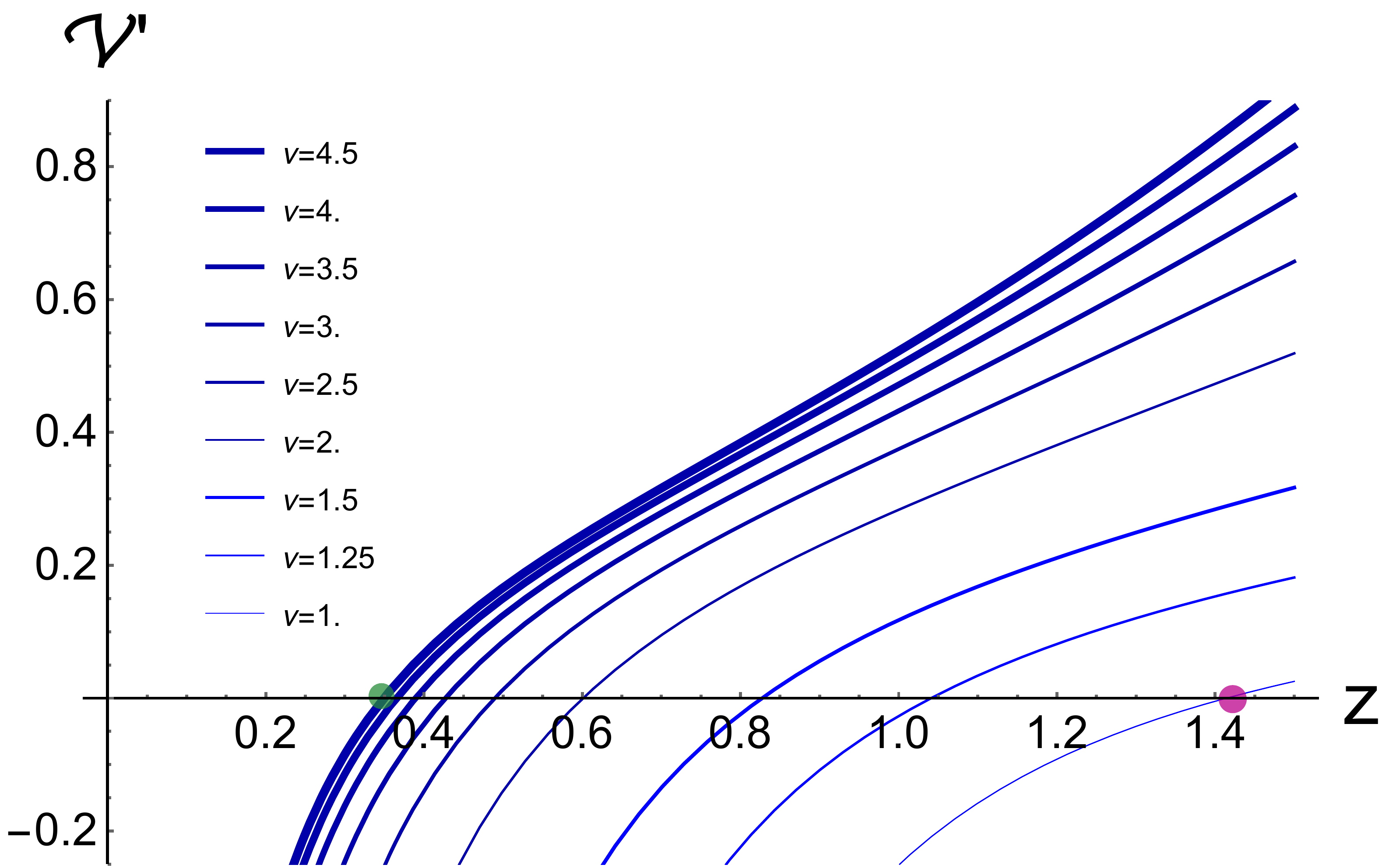}\quad \quad
  \includegraphics[width=6.5cm]{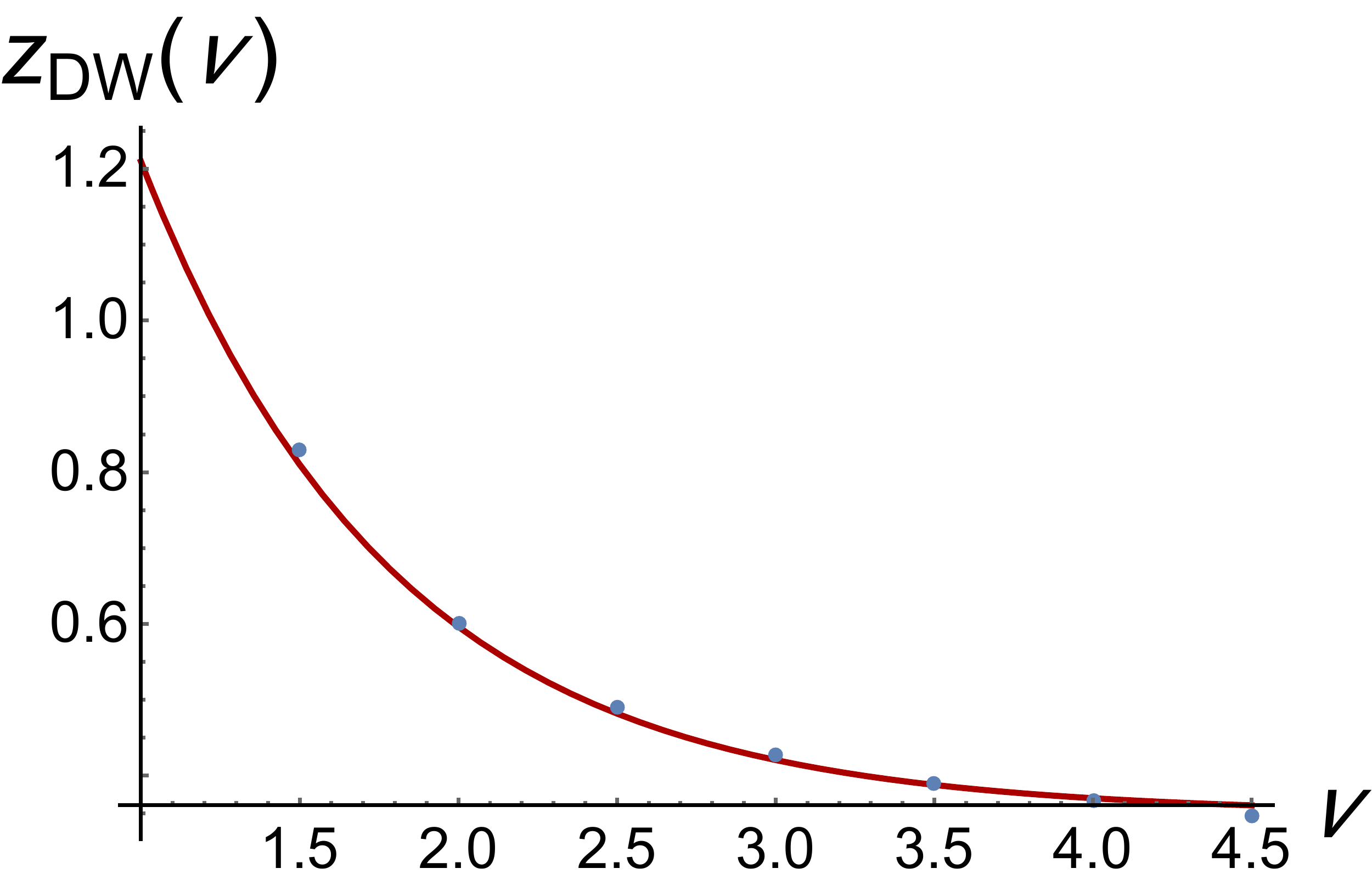}\\
   {\bf C)}\quad \quad \quad \quad \quad \quad \quad \quad \quad \quad \quad \quad \quad {\bf D)}\\
\caption{ Locations of  the dynamical wall for HEE for {\bf A)}  $\nu=1$ and {\bf B)} $\nu=4.5$.
     Solid lines present $0.15\cdot {\cal V}$, dashed lines present ${\cal V}'$. We see that the location of the dynamical wall does not depend on $z_h$.  Note that only for $z_h<z_{HP}(\nu)$ 
    we obtain the stable solutions, $z_{HP}(1)= 1.505$, $z_{HP}(4.5)=1.138$.
}
\label{fig:3D-DW}
  \end{figure}

In this construction the location of the dynamical walls and the horizon plays a special role. In Fig.\ref{fig:3D-DW} we show the appearance of the dynamical wall. 
The dynamical wall appears when the effective potential 
${\cal V}$ gets a saddle point.
  
  In Fig.\ref{fig:3D-DW} we also illustrate the location of   $z_{DW}$  for various choices of the horizon size $z_h$  for  $\nu= 1$ {\bf A)} and
$\nu =4.5$ {\bf B)}. Solid lines present ${\cal V}$ and dashed lines  present ${\cal V}'$. We see that the location of the dynamical wall does not depend on the horizon. In Fig.\ref{fig:3D-DW}.{\bf C)} we show the location of the dynamical wall for arbitrary $1\leq\nu\leq4.5$.
\\

\begin{figure}[h!]
\centering
 \includegraphics[width=7 cm]{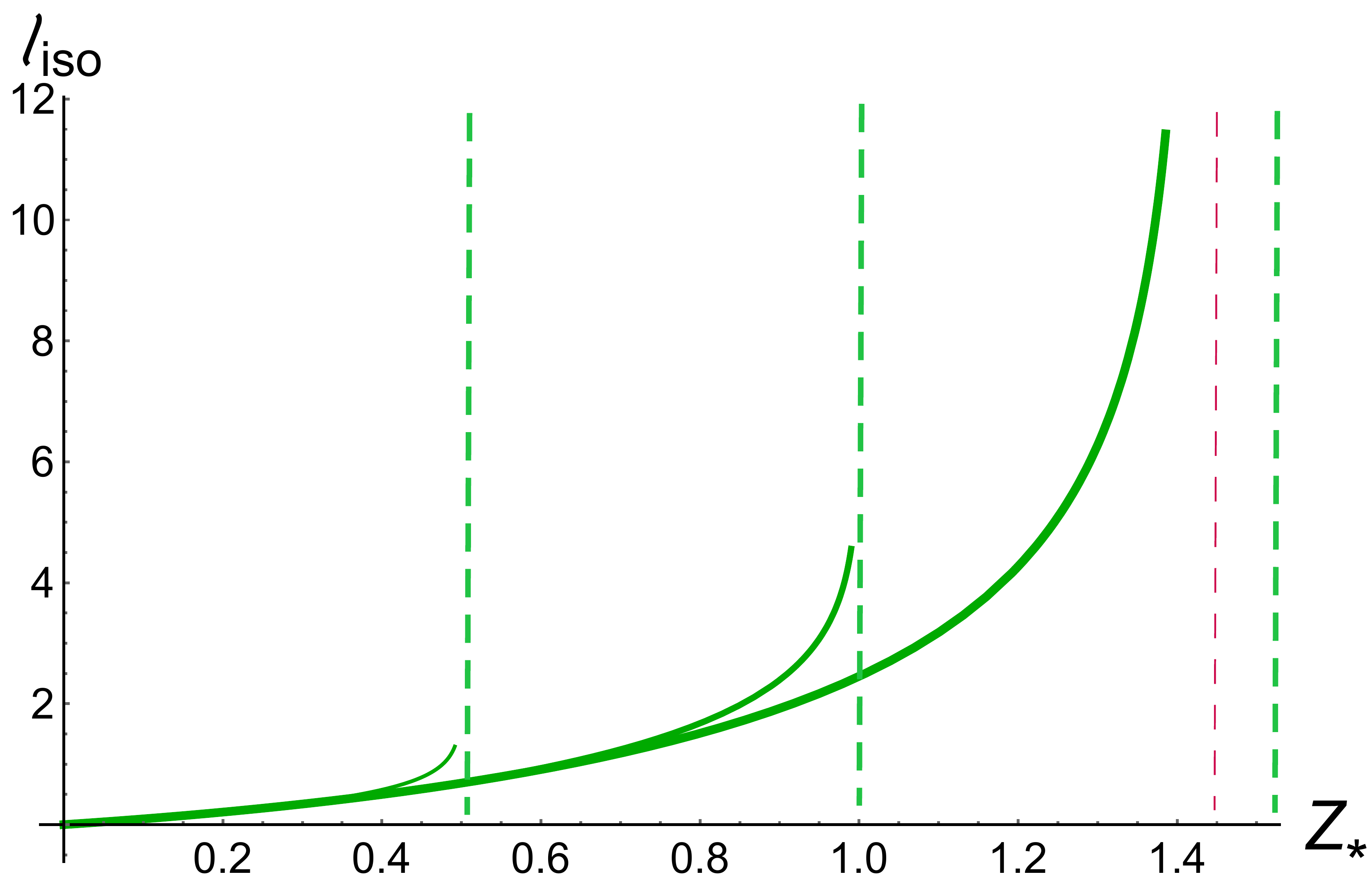}\qquad
 \includegraphics[width=6 cm]{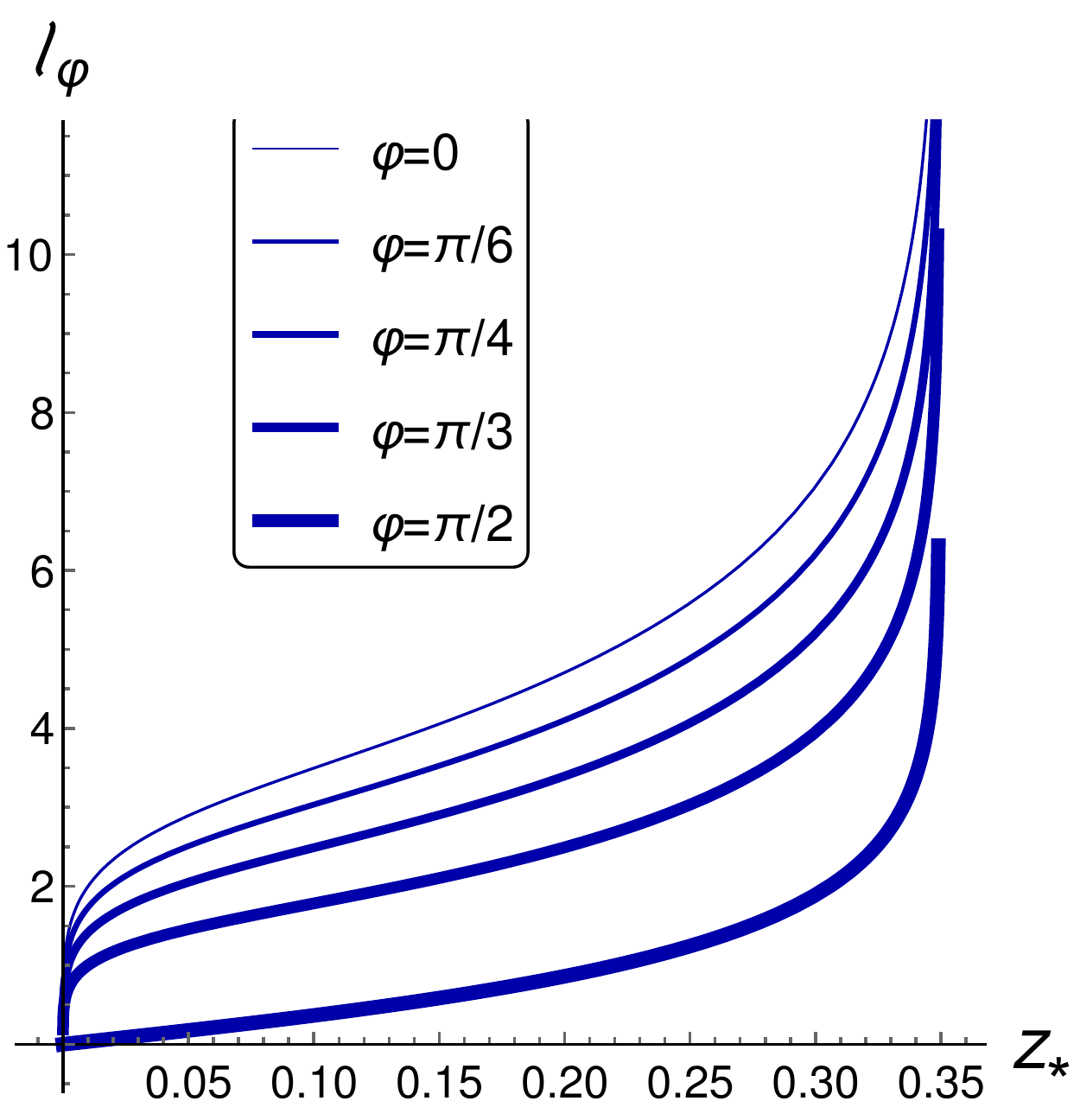}\quad 
 \\
 {\bf A)}\quad \quad \quad \quad \quad \quad \quad \quad \quad \quad \quad \quad \quad {\bf B)}
    \caption{{\bf A)} The $\ell$ dependence on $z_*$  for $\nu=1$,  $z_h=0.5,1,1.5$ and $\mu=0.$
The red dashed line show the position of the dynamical wall, $z_{DW}=1.41$. {\bf B)} The plot shows the
$\ell$ dependence on $z_*$  for $\nu=4.5$,  $T=0.25$ and $\mu=0.2$ for different angles $\varphi=0, \pi/6,\pi/4, \pi/2$.  
Note that $ z _* $ does not exceed $ z_ {DW} (T, \mu) $, the position of the dynamical wall for given $ T $ and $ \mu $.
Here $z_{DW}=0.349$. 
}
\label{fig:toDW}\end{figure}

In Fig.\ref{fig:toDW} we see that $\ell$ increases when $z_* $ approaches  $z_{DW}$ or the horizon. The location of the dynamical wall does not depend on the orientation of slab and  also does not depend on the horizon, see Fig.\ref{fig:3D-DW} below. In Fig.\ref{fig:3D-DW} the plots are presented for $\nu=4.5$,  $T=0.25$ and $\mu=0.2$.  

We present in Fig.\ref{fig:toDW}.{\bf A)} the $\ell$ dependence on $z_*$  for $\nu=1$,  $z_h=0.5,1,1.5$ and $\mu=0.$
The red dashed line shows the position of the dynamical wall.   The plots in Fig.\ref{fig:toDW}.{\bf B)} show the
$\ell$ dependence on $z_*$  for $\nu=4.5$,  $T=0.25$ and $\mu=0.2$ for different angles $\varphi=0, \pi/6,\pi/4, \pi/2$. The value of  
$ z _* $ does not exceed $ z_ {DW} (\mu,T) $, the position of the dynamical wall for given $ \mu $ and $ T$. Here $z_{DW}=0.349$. 

We tried to avoid  divergences in \eqref{Stheta} and used the minimum subtraction of the UV asymptotic  in accordance with the formulas above in \ref{Sect:longitudinal} and Sect. \ref{Sect:transversal}.

\subsubsection{Entanglement entropy dependence on $\ell$}
\begin{figure}[h!]
 \centering\includegraphics[width=8cm]{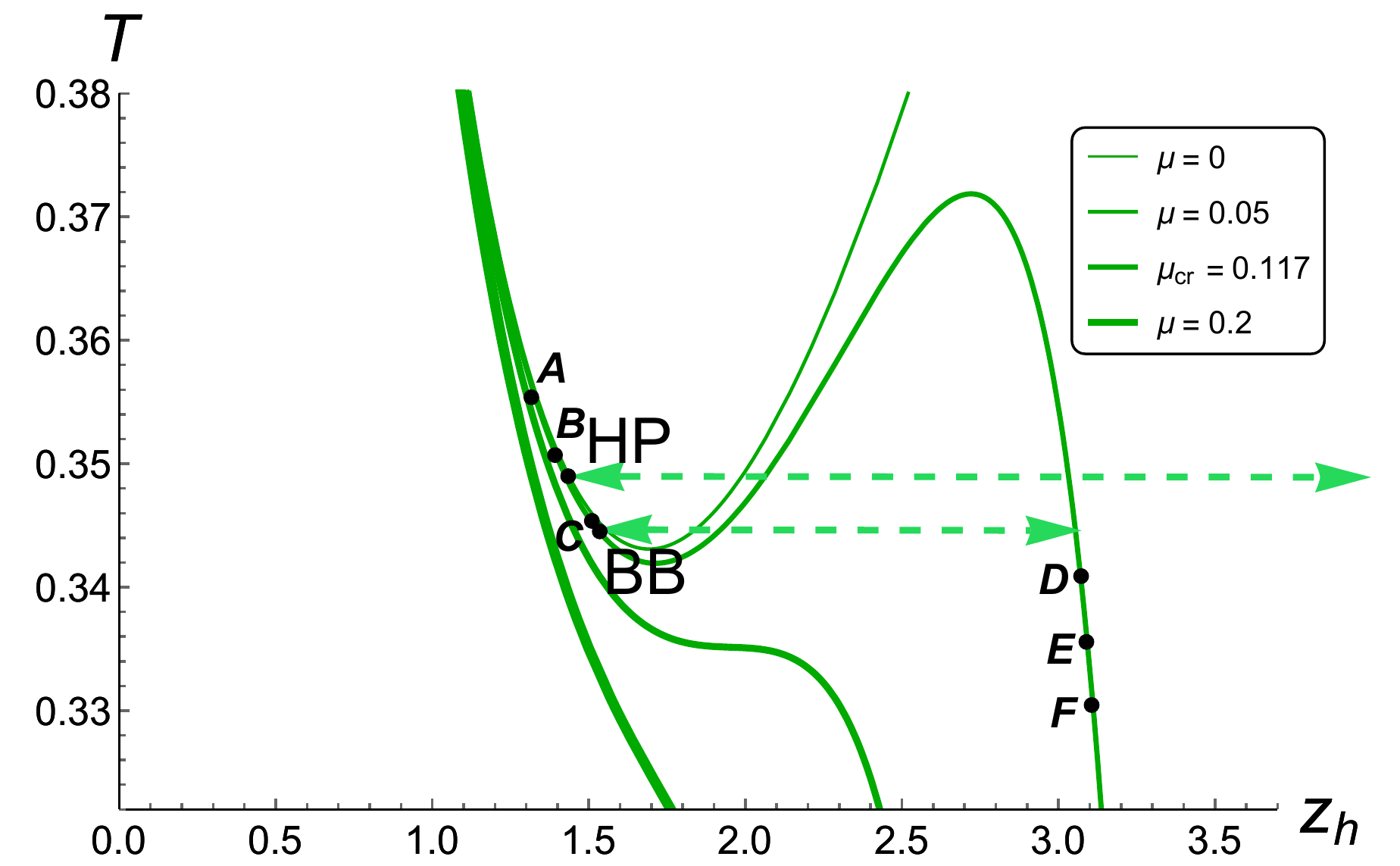}\\{\bf A)}\\
 \includegraphics[width=6cm]{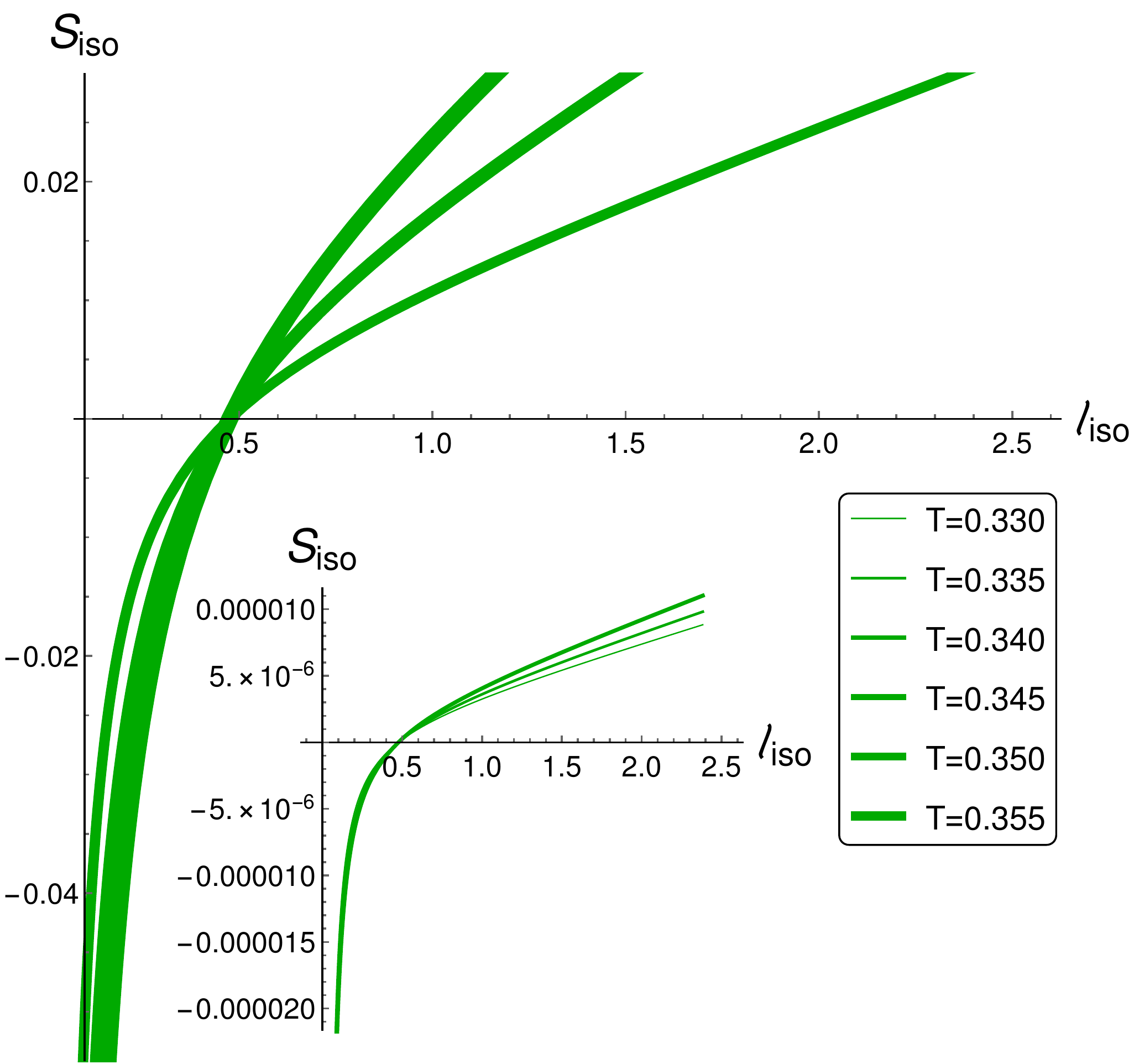}\qquad
\includegraphics[width=6cm]{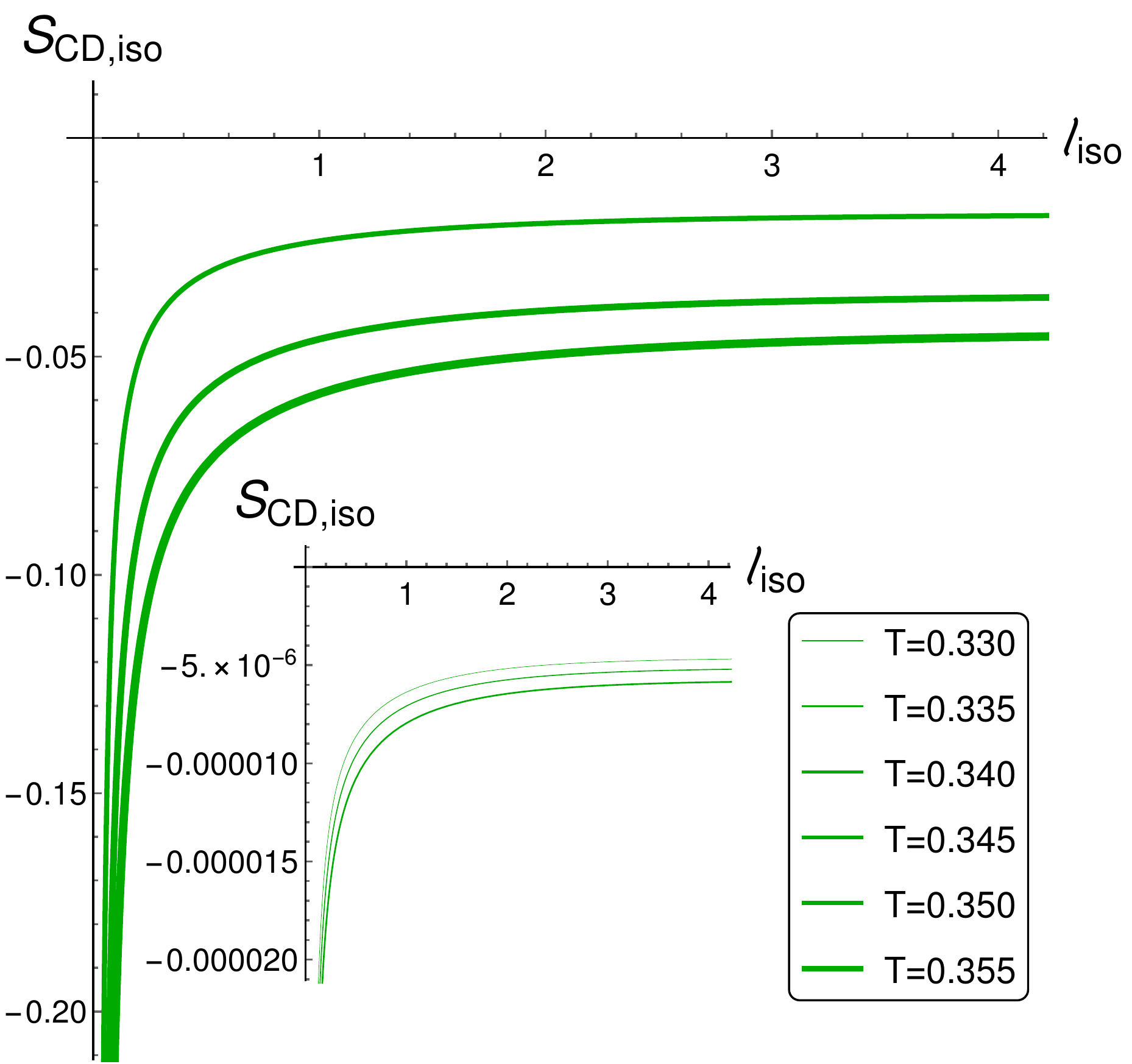}
\\\qquad\qquad\qquad\qquad
 \,{\bf B)}\,\qquad\qquad\qquad\qquad\qquad\qquad\qquad\qquad{\bf C)}
\caption{{\bf A)} $T$ as a function of $z_h$ for various values
of the chemical potential $\mu$. Here  curves with different thicknesses correspond to  $\mu=0, 0.05, 0.117$ and $0.2$. Plots {\bf B)} and  {\bf C)} show the dependence of HEE for a slab on its smallest  length $\ell $ for various  temperatures near the temperature of the BB phase transition, 
$T_{BB}=0.3445$  $(z_h=1.532)$ corresponding to $ \mu = 0.05 $ and $ \nu = 1 $. In plot {\bf B)} we use the minimal renormalization scheme, in {\bf C)} we use the geometric renormalization.}
    \label{fig:S-iso}
  \end{figure}

In Fig.\ref{fig:S-iso} we present the dependence of the slab HEE on its smallest length $\ell$ at  various  values of the temperature. These temperatures are chosen  equidistantly around the temperature  of the BB phase transition at 
 $\mu=0.05$ 
in the  isotropic case. These temperatures are indicated on the $(z_h,T)$-curve  by the points (A,\,B,\,C) above the  $T_{BB}$
 and the points (D,\,E,\,F) below $T_{BB}$ for $\mu=0.05$ (see Fig.\ref{fig:S-iso}.{\bf A)}). For these temperatures and $\mu=0.05$ we depict the dependence of the HEE on $\ell$. Here $\ell $ is chosen to be less than $\ell _c$, the length at which the turning point of the connected entangling surface moves closer towards  the dynamical wall $z_{DW}$ (see Fig.\ref{fig:toDW}). 
We see that the HEE undergoes a jump when  the temperature crosses the  phase transition line at the point $T_{BB}(\mu)$. 
We also see that these jumps depend smoothly on $\ell$ at least  for $\ell _{UV}<\ell<\ell_c$. 

\begin{figure}[h!]
 \centering
 \includegraphics[width=7cm]{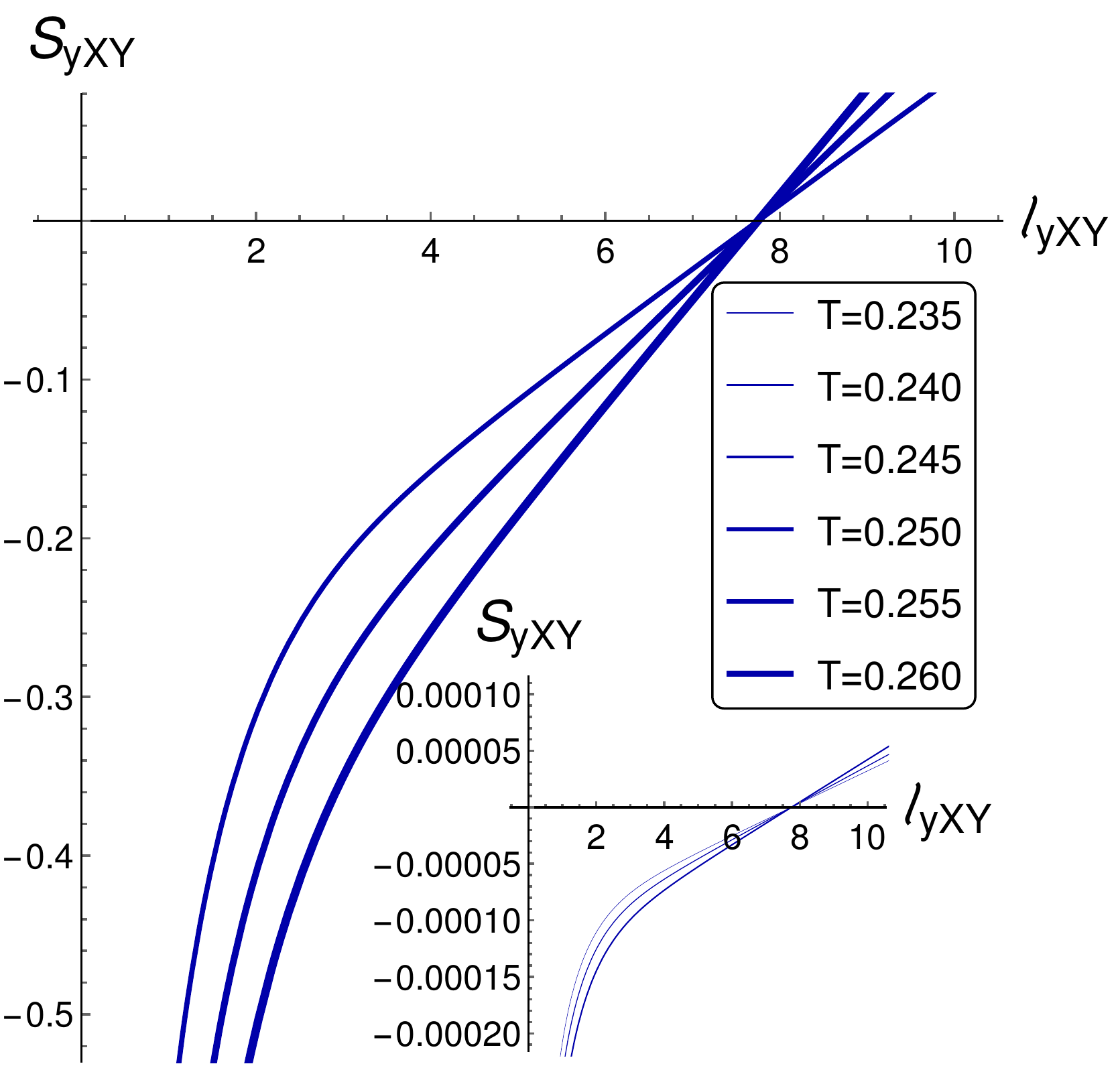}
 \includegraphics[width=7cm]{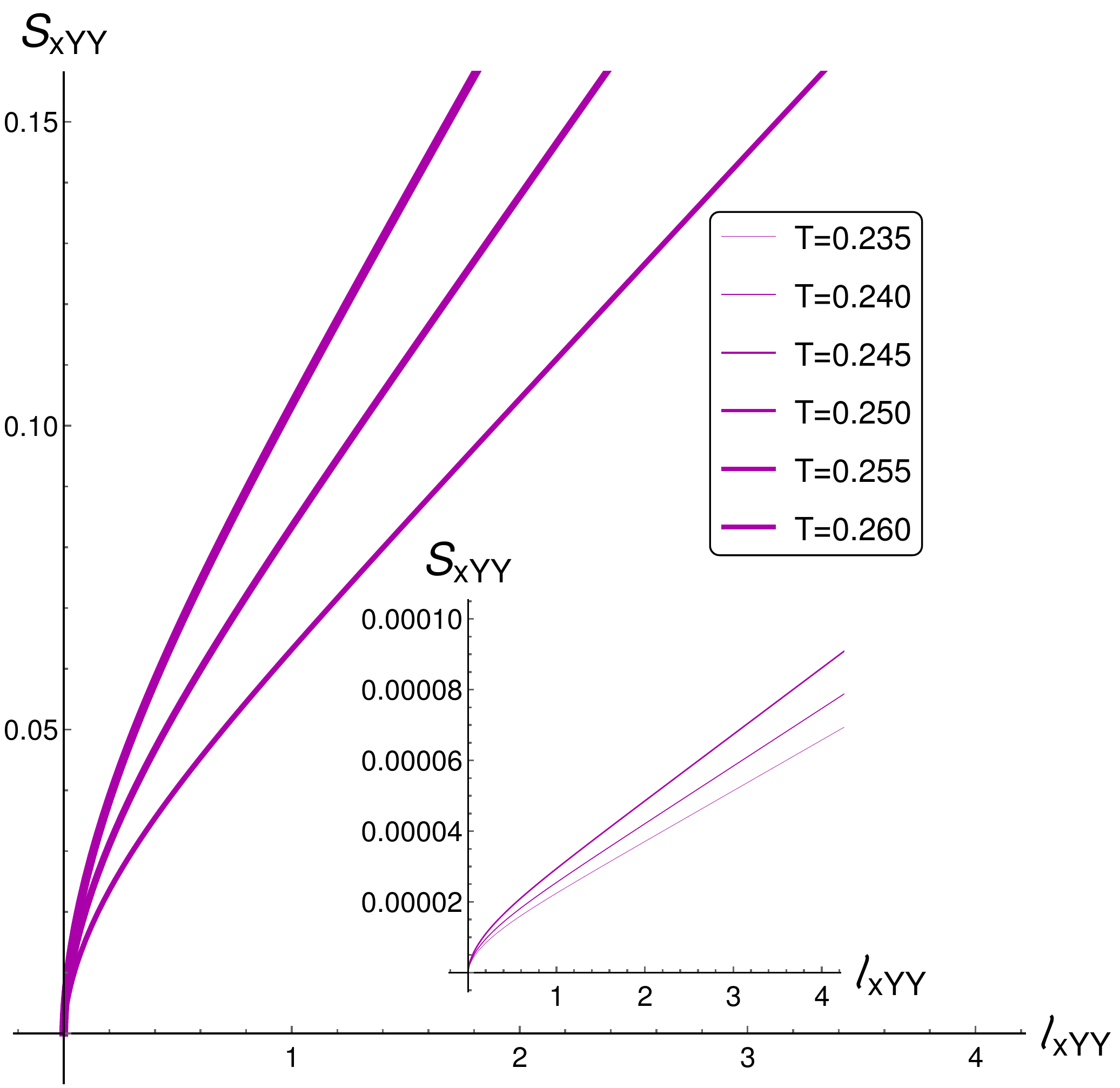}\\
 \qquad\qquad\qquad\qquad
 \,{\bf A)}\,\qquad\qquad\qquad\qquad\qquad\qquad\qquad\qquad{\bf B)}\\
  \includegraphics[width=7cm]{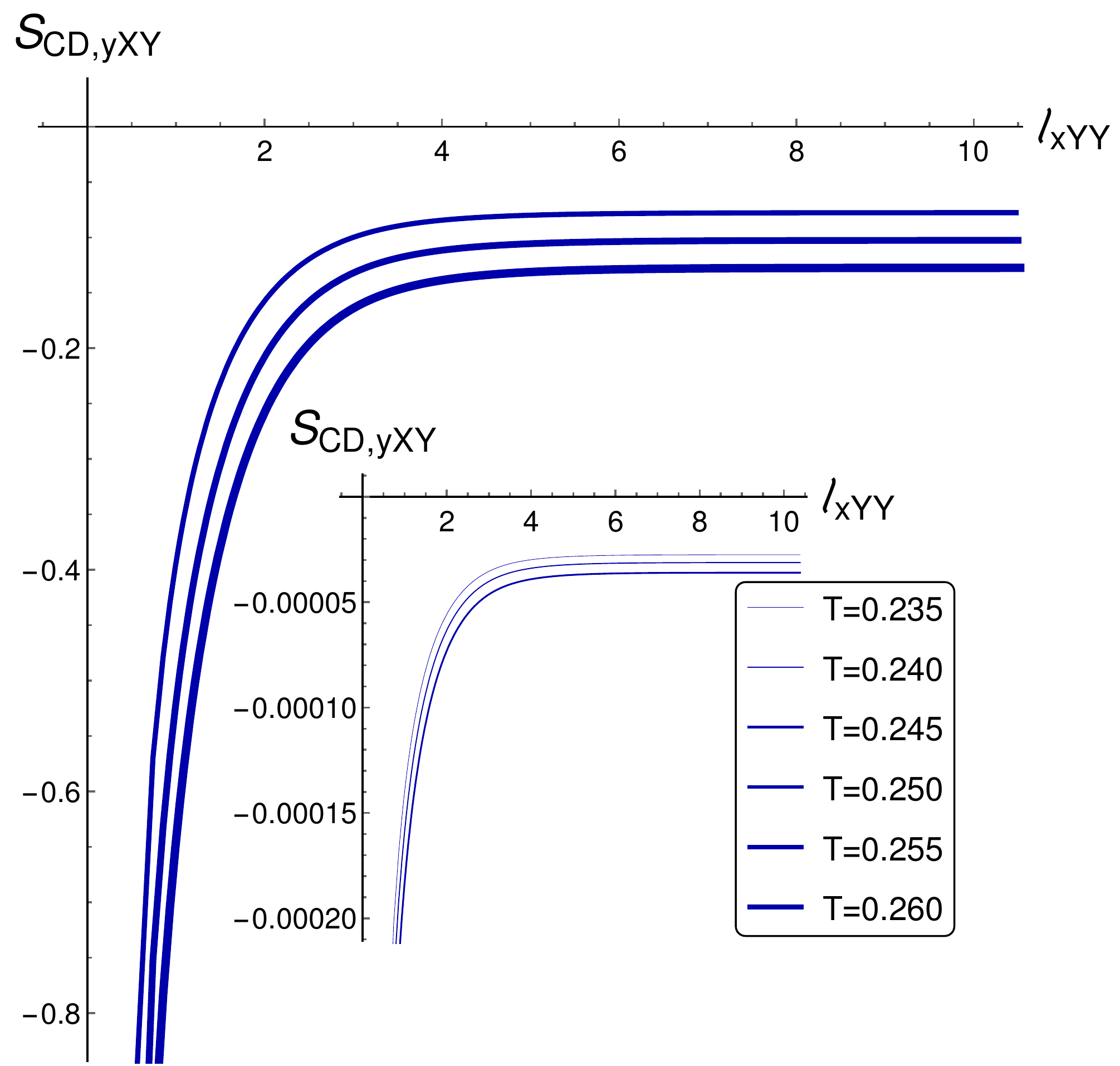}
 \includegraphics[width=7cm]{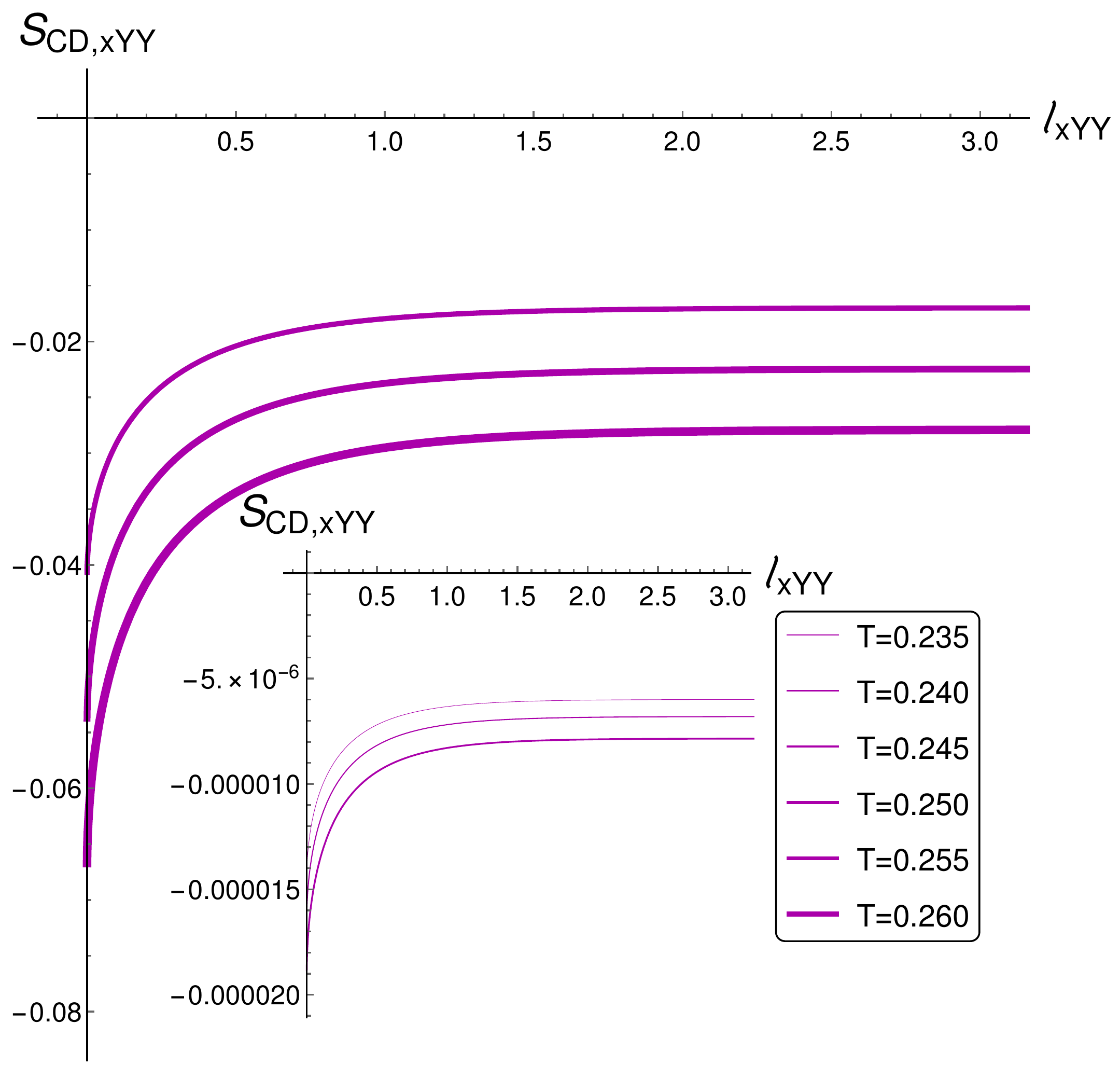}\\
 \qquad\qquad\qquad\qquad
 \,{\bf C)}\,\qquad\qquad\qquad\qquad\qquad\qquad\qquad\qquad{\bf D)}
 \caption{
    The HEE dependencies on the smallest length of the entangling slab $\ell$ for $\varphi =\pi/2$, the transversal orientation, and $\varphi =0$, the longitudinal orientation, at  equidistant values of  the temperature around the BB phase transition  for $\mu=0.2$ and $\nu=4.5$ are presented in {\bf A)} and {\bf B)},  respectively.  Here the renormalizations of the HEE are performed according to \eqref{calSrenB} and \eqref{calSrenC}. In {\bf C)} and {\bf D)} we show the  HEE for renormalizations performed according to
    \eqref{diccxYY2} and  \eqref{diccyXY2},   {\bf C)} for $\varphi =\pi/2$ and {\bf D)} for $\varphi =0$ 
    }
    \label{fig:SanisoL}
  \end{figure}

  The similar picture takes place for the anisotropic background. The HEE dependencies on  $\ell$ for  the transversal and  longitudinal   orientations at  equidistant values of  the temperature around the BB phase transition  for $\mu=0.2$ and $\nu=4.5$ are presented in Fig.\ref{fig:SanisoL}.{\bf A)} and Fig.\ref{fig:SanisoL}.{\bf B)},  respectively. We see that with decreasing of the angle $\varphi$ from $\varphi=\pi/2$ to $\varphi=0$ we increase the HEE at fixed $\ell$ and fixed thermodynamic parameters (with our adopted method to eliminate divergences). 
$$\,$$

  \newpage
$$\,$$
\newpage
 Therefore, we have demonstrated that the HEE depends on the regularization scheme.
\begin{itemize}
\item For the minimal regularisation scheme both the longitudinal  and the transversal HEE
\begin{itemize}
\item increase linearly at large $\ell$
\be
S_{yXY}\underset{\ell \to \infty}{\sim } C_{yXY}(T,\mu)\ell,\,\,\,\,\,S_{xYY}\underset{\ell \to \infty}{\sim } C_{xYY}
(T,\mu)\ell,\,\,\,\,\ee
and $C_{yXY}(T,\mu)>C_{xYY}(T,\mu)$. The linearity region for longitudinal entanglement entropy begins at lower $ \ell $ compared to the transverse entanglement entropy.
\item they   have different behavior for $\ell \to 0$
\be
S_{xYY}\underset{\ell \to 0}=0,\,\,\,\,\,\,\,S_{yXY}\underset{\ell \to 0}{\sim } \ell^{-\xi _{yXY}(T,\mu)},\,\,\,\,\xi _{yXY}(T,\mu)>0,
\ee
(compared with results of \cite{AGG}).\end{itemize}
\item For the CD regularisation scheme both the longitudinal  and the transversal  HEE 
\begin{itemize}
\item tend to constants  at large $\ell$
\be
S_{yXY}\underset{\ell \to \infty}{\sim } a_{yXY}(T,\mu),\,\,\,\,\,S_{xYY}\underset{\ell \to \infty}{\sim } a_{xYY}
(T,\mu)\,\,\,\,\ee
and $a_{yXY}(T,\mu)>a_{xYY}(T,\mu)$;
\item at small $\ell$ they have different behavior for $\ell \to 0$
\be
S_{xYY,CD}\underset{\ell \to 0}{\sim }\ell^{-\xi _{CD,xYY}(T,\mu)},\,\,\,\,\,\,\,
S_{yXY,CD}\underset{\ell \to 0}{\sim } \ell^{-\xi _{CD,yXY}(T,\mu)}\,.
\ee
\end{itemize}

\end{itemize}

$$\,$$
\subsubsection{Entanglement entropy dependence on temperature}
\begin{figure}[h!]
\centering
 \includegraphics[width=6 cm]{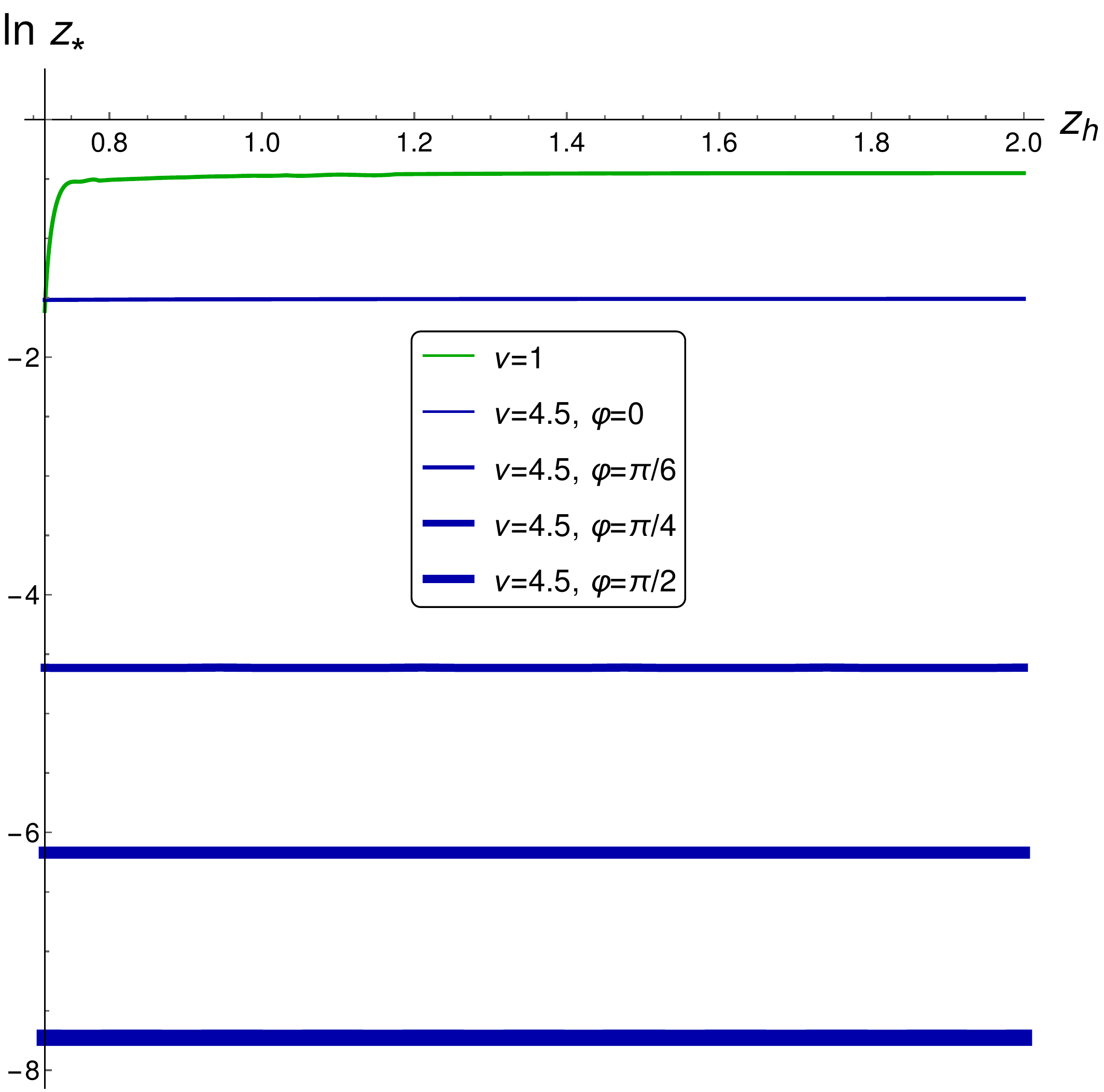}
     \caption{
The dependence of  $z_*$ on $z_h$ at fixed $\ell=1$ and different  angles for the anisotropic case 
(blue lines for angles $\varphi=0,\pi/4,\pi/3,\pi/2$ from bottom to top) and for isotropic case (green lines) at $\mu=0$.
 } 
    \label{Lengthzszh}
  \end{figure}

 It is instructive to depict the HEE dependence on the temperature for a fixed value of $\ell$  near the phase transition. For this purpose we first find the dependence  of $z_*$ on the length $\ell$ (at  fixed horizon) and then calculate the HEE from \eqref{calSren} and \eqref{calSrenB}. The dependence of $z_*$ on $z_h$ for the fixed value of $\ell=1$ and different values of the angle $\varphi$ at the chemical potentials $\mu=0$ are shown by blue lines in Fig.\ref{Lengthzszh}. For comparison,  in this plot the dependence of $z_*$ on $z_h$ is shown for the isotropic case by green lines also for $\mu=0$. We notice a significant dependence on $z_h$ only for small values of $z_h$ (large BHs).

\begin{figure}[h!]
\centering

 \includegraphics[width=6 cm]{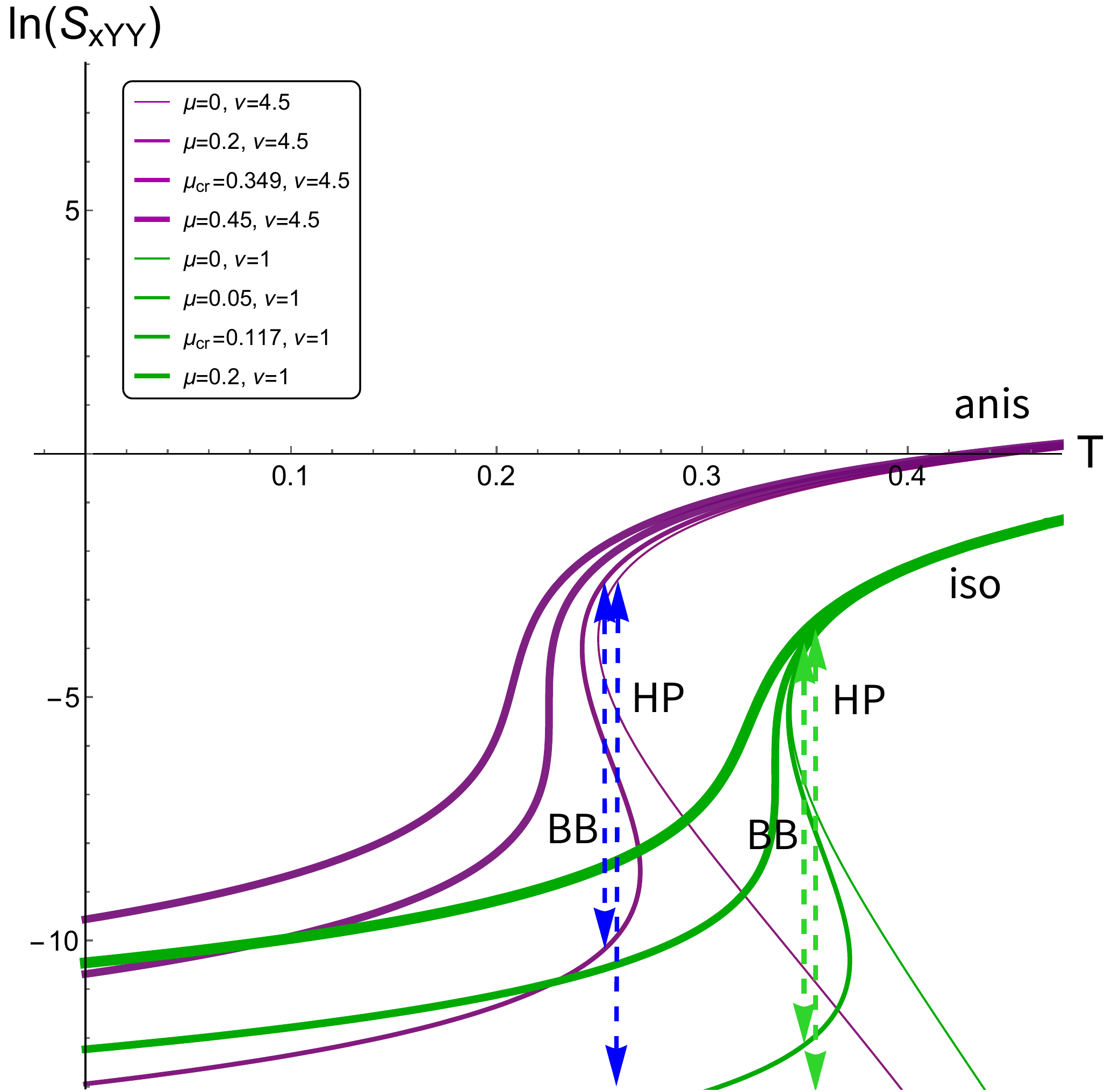}\quad 
 \includegraphics[width=6cm]{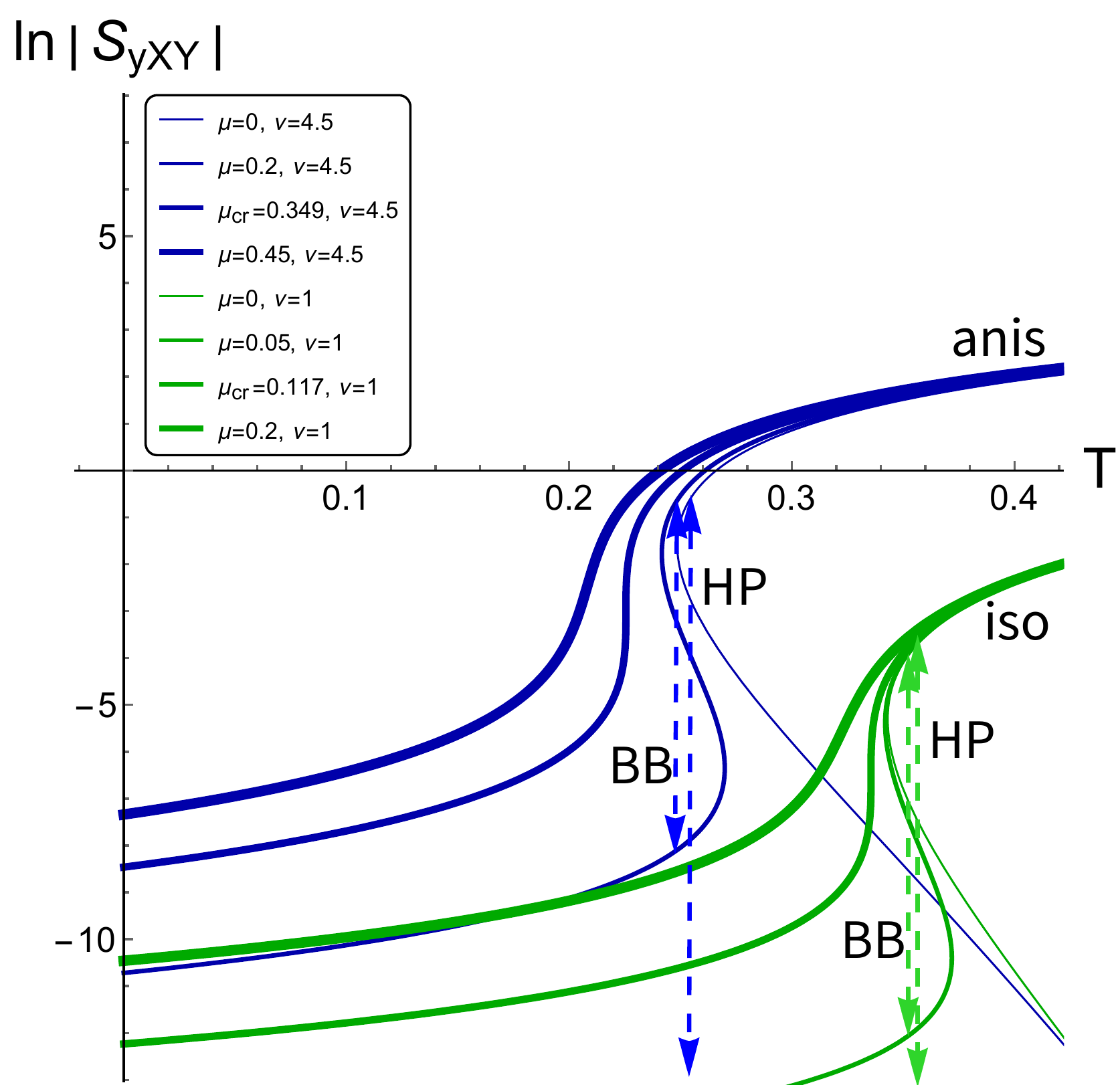}
\qquad\qquad\qquad\qquad
 \,{\bf A)}\,\qquad\qquad\qquad\qquad\qquad\qquad\qquad\qquad{\bf B)}
    \caption{
The HEE of the slab with the fixed value of $\ell=1$ dependences on the temperature for {\bf A)} longitudinal  and 
{\bf B)} transversal  cases in the anisotropic ($\nu=4.5$) case for  chemical potential below and above the critical chemical potential
$\mu_{cr}=0.349$ are shown by the blue lines with different thickness.  For comparison, the dependences of HEE of the same slab in isotropic background are shown   by green lines in  both plots . 
 }
    \label{fig:ST_SL}
  \end{figure}

 Substituting the dependence of $z_{*}$ on $z_h$ to equations \eqref{calSren} and \eqref{calSrenB} we get 
the HEE dependence on the temperature at the fixed value of $\ell=1$ for different chemical potentials $\mu$
for {\bf A)} longitudinal and {\bf B)} transversal  cases. 
These dependences are shown by blue lines in  Fig.\ref{fig:ST_SL}
{\bf A)} and Fig.\ref{fig:ST_SL}
{\bf B)}  for  
various  chemical potentials $\mu$ near  the critical value $\mu_{cr}=0.349$ for the longitudinal and transversal  cases, respectively.
For comparison, the dependences of  the HEE on temperature for the same slab are shown by green lines for
different chemical potentials $ \mu $ near $ \mu_ {cr} = 0.117 $ for the isotropic case, $ \nu = 1 $ in both graphs.

  \begin{figure}[h!]
\centering

 \includegraphics[width=6 cm]{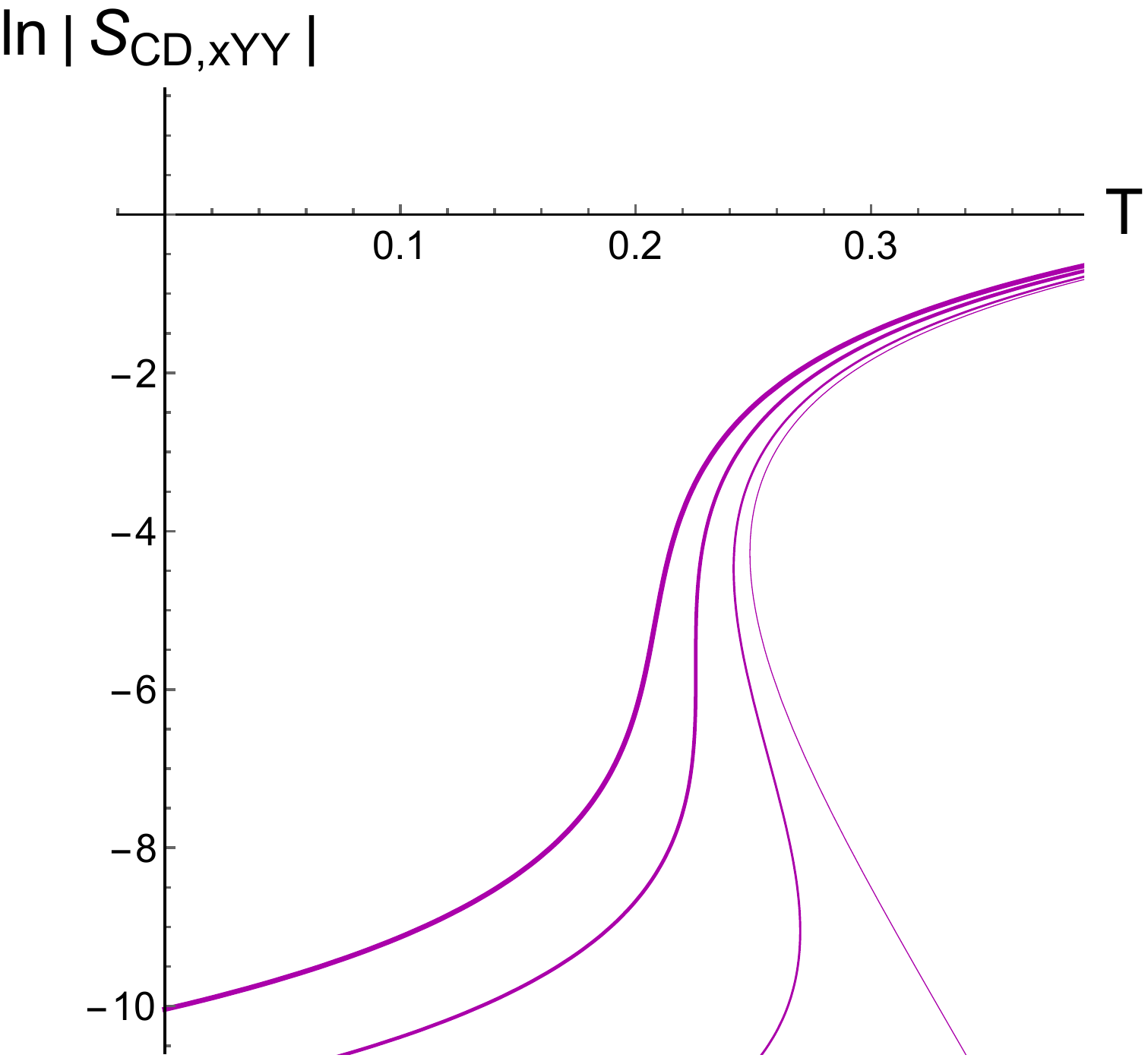}\quad\quad\quad
 \includegraphics[width=6cm]{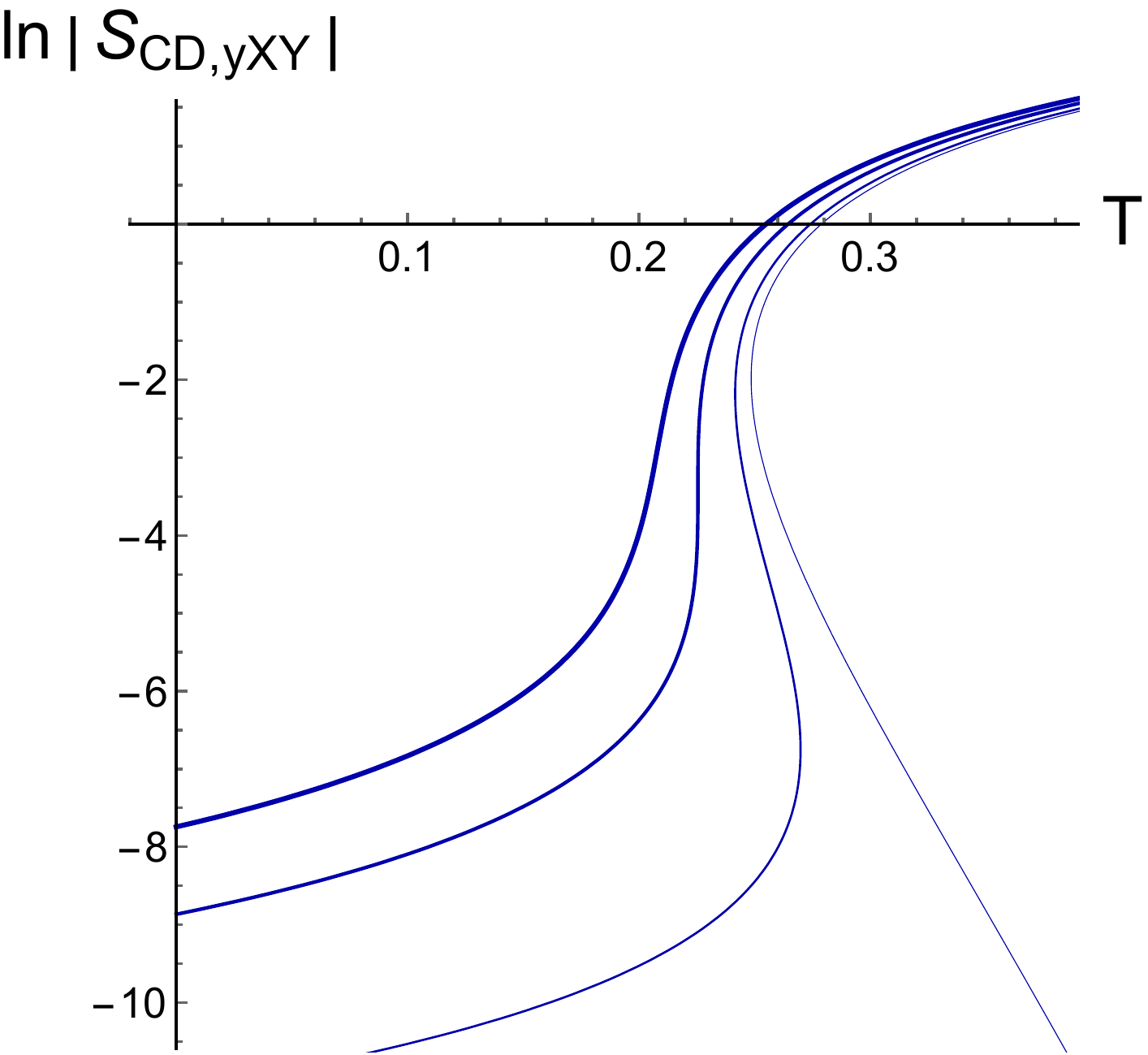}
 \\{\bf A)}\quad\quad\quad\quad\quad\quad\quad\quad\quad\quad{\bf B)}\
    \caption{
The HEE of the slab with the fixed value of $\ell=1$ dependences on the temperature for {\bf A)}  longitudinal and 
{\bf B}) transversal  cases in the anisotropic ($\nu=4.5$) case for  chemical potential below and above the critical chemical potential
$\mu_{cr}=0.349$ are shown by the blue and magenta lines with different thicknesses.   
 }
  \end{figure}

  \begin{figure}[h!]
\centering
\includegraphics[width=6 cm]{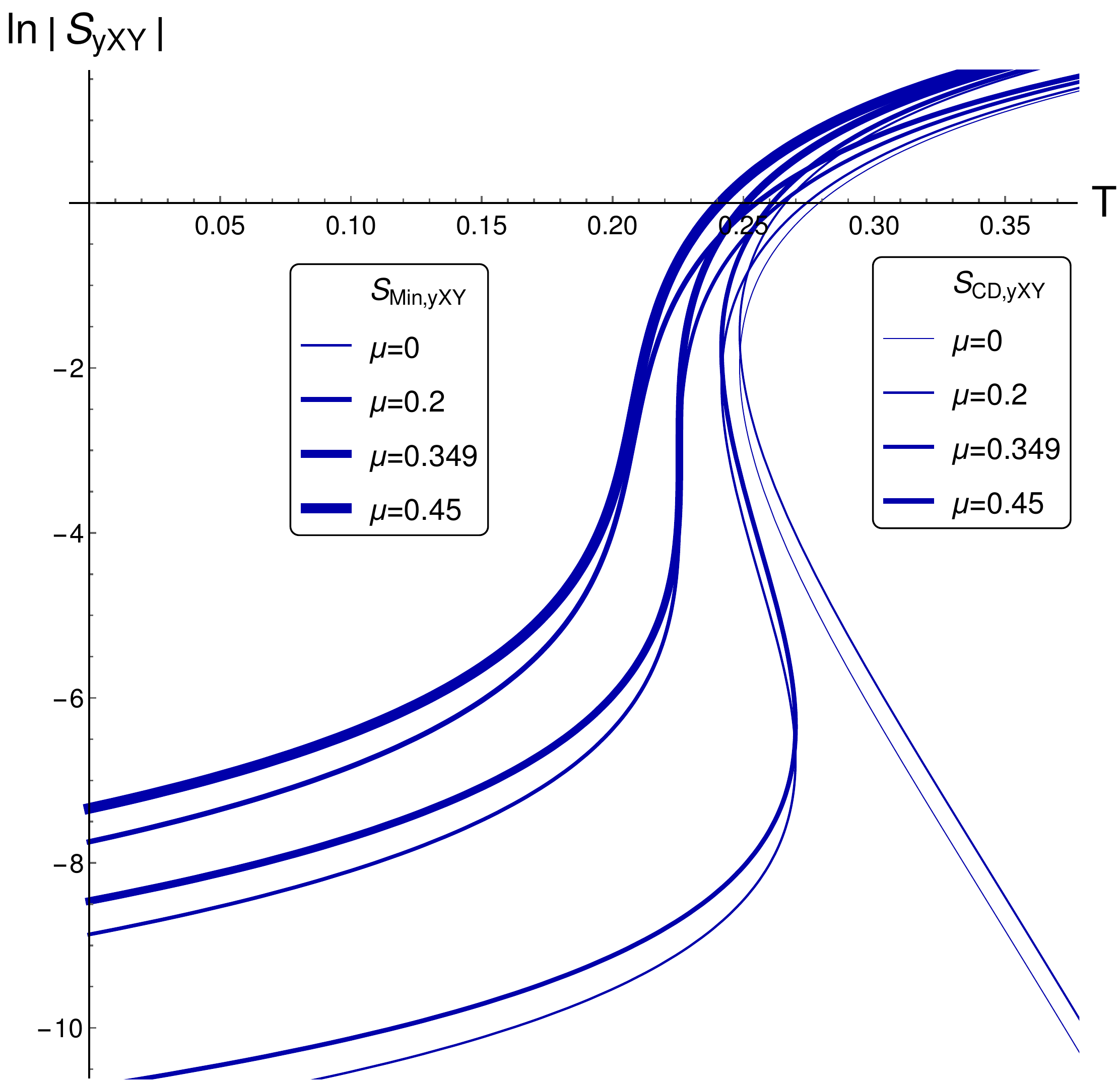}    \caption{ Comparison of 
the HEE of the slab with the fixed value of $\ell=1$ calculated 
according to two different renormalization schemes near the background phase transition line.
 }\label{fig:comHEEM-CD}
  \end{figure}

These plots   Fig.\ref{fig:ST_SL} show that  in the vicinity of the phase transition temperature  $T=T_{BB}(\nu,\mu)$ the HEE, like the thermal entropy (see Fig.\ref{Fig:FTmu45}.{\bf B)}) undergoes  significant  jumps. From these plots we also see that 
 for $\mu >\mu_{cr }$ ($\mu_{cr }=0.117$ for $\nu =1$ and $\mu_{cr }=0.349$ for $\nu =4.5$)  the jumps disappear,
 similar to the thermal entropy  behavior. 
$$\,$$

To summarise this subsection, we note that,
\begin{itemize}

\item the HEE of the slab with constant thickness $\ell$ increases with increasing temperature and fixed chemical potential;

\item the HEE  jumps near the phase transition do not  depend significantly on the type of renormalization (geometric and minimal renormalization), see Fig.\ref{fig:comHEEM-CD}.

\end{itemize}

\subsection{Entanglement Entropy Density}\label{Sect:density}
The entanglement entropy density, defined in the previous Sect. \ref{Sect:density} as \eqref{GVeta}, in our case takes the form  
\be
\label{eta}
\eta(z_*)=\frac14\frac{b^{3/2}_s(z_*)}{z_*^{1+2/\nu}},
\ee
but if defined  according to \ref{etaR}, it  takes the form
\be
\label{etaCD}
\eta_{CD}(z_*)=\frac14\left(\frac{b^{3/2}_s(z_*)}{z_*^{1+2/\nu}}-\frac{b^{3/2}_s(z_D)}{z_D^{1+2/\nu}}\right).
\ee

In this section we study the dependence of the entanglement entropy density 
 \eqref{eta} on temperature and chemical potential near background phase transition. 
Also we compare the results in minimal and geometric renormalizations, \eqref{eta} and \eqref{etaCD}.

\subsubsection{Entanglement entropy density dependence on $\ell$}
\begin{figure}[h!]
\centering
 \includegraphics[width=6.5 cm]{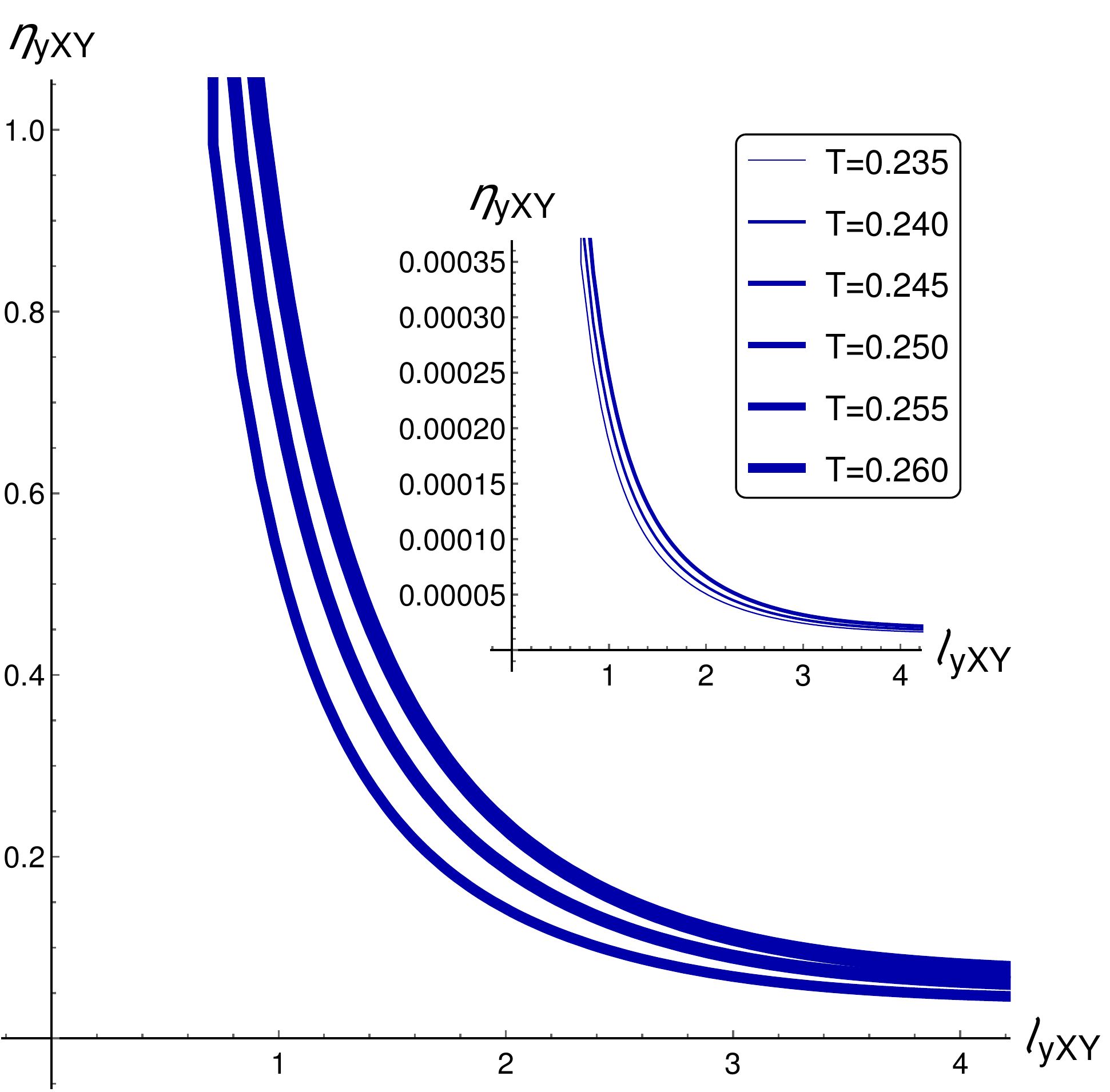}\quad\quad
 \includegraphics[width=6.5 cm]{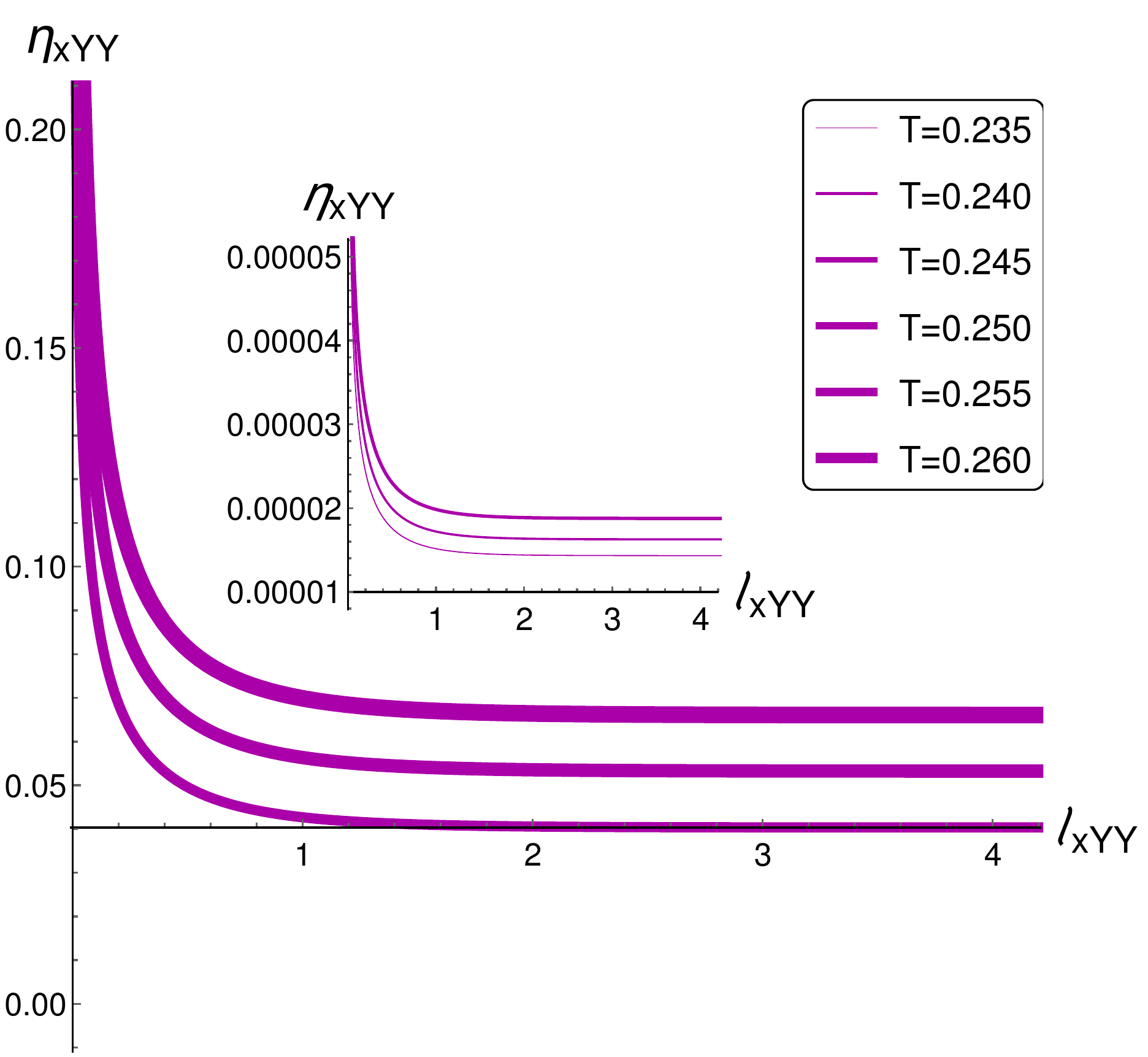}
\qquad\qquad\qquad\qquad
 \,{\bf A)}\,\qquad\qquad\qquad\qquad\qquad\qquad\qquad\qquad{\bf B)}
    \caption{The dependence of the
slab entropy density on $\ell$ for various temperatures around the critical temperature at $ \mu = 0.2$
for {\bf A)} transversal  and {\bf B)} longitudinal  orientations.
The density exhibits a jump near the BB phase transition ($T_{BB}=0.2457$ for $\mu=0.2$, we can compare $T=0.235,0.240,0.245,0.250,0.255,0.260$).
} \label{fig:eta-ell}
\end{figure}

 In Fig.\ref{fig:eta-ell} we show the dependence of the HEE density \eqref{eta}
on $\ell$ at fixed values of the temperature and chemical potential near the background phase transition line for the model \eqref{ARmetric}. We see that the density function decreases monotonically depending on the length. 
Here, similarly  to the considerations in the previous Sect.\ref{Sect:HEE-TL},  we consider the density of the entanglement entropy for different T and $\mu$ near  the  phase transition point  ($T_{BB}=0.2457,\mu_{BB}=0.2)$, namely for the points $A,B,C$ and $D,E,F$ indicated in Fig.\ref{T_zh.pdf}.A.
We see that the densities change significantly when we cross the  phase transition point ($T_{BB},\mu_{BB})$. 

\begin{figure}[h!]
\centering
 \includegraphics[width=6.5 cm]{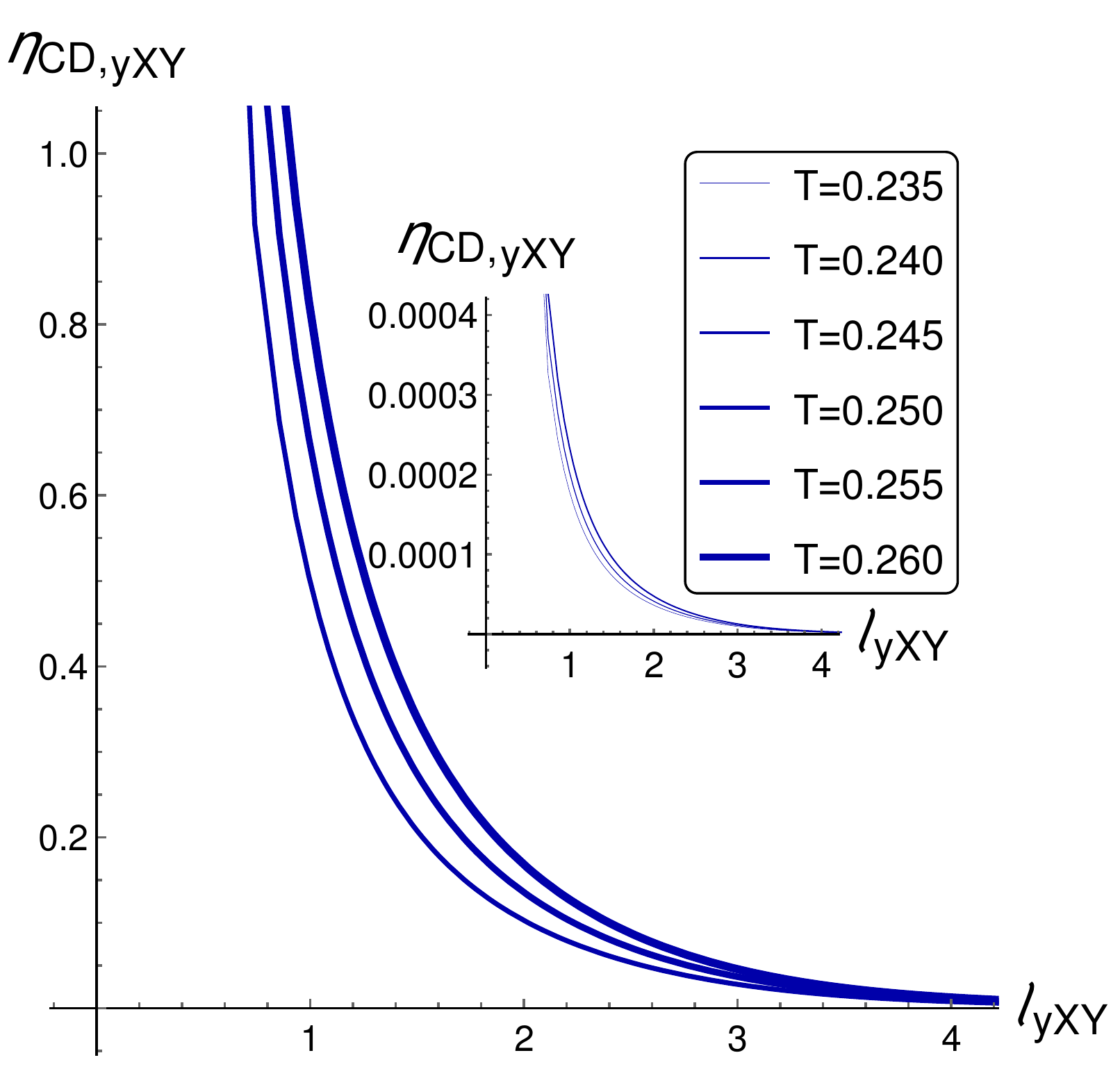}\quad\quad
 \includegraphics[width=6.5 cm]{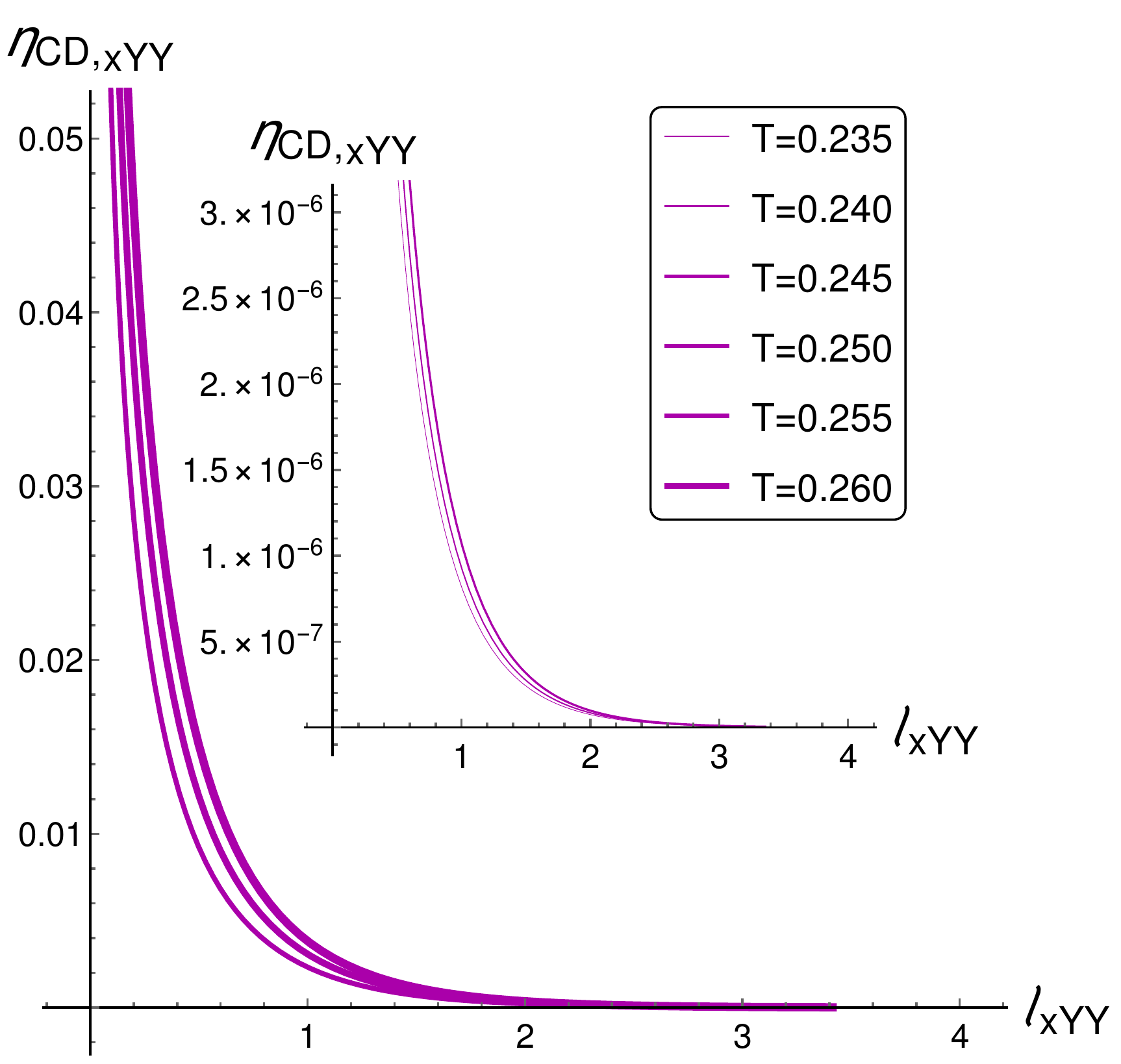}
\qquad\qquad\qquad\qquad
 \,{\bf A)}\,\qquad\qquad\qquad\qquad\qquad\qquad\qquad\qquad{\bf B)}
    \caption{The dependence of the
slab entropy density defined by \eqref{etaCD} on $\ell $ for various temperatures around the critical temperature at $ \mu = 0.2 $
for {\bf A)} transversal  and {\bf B)} longitudinal orientations.  Considered temperatures correspond to the point indicated in Fig.~\ref{T_zh.pdf}. The density exhibits a jump near the BB phase transition 
point ($T_{BB}=0.2457,\mu=0.2)$.
} \label{fig:eta-ell-CD}
\end{figure}

\begin{figure}[h!]
\centering
  \includegraphics[width=6.5 cm]{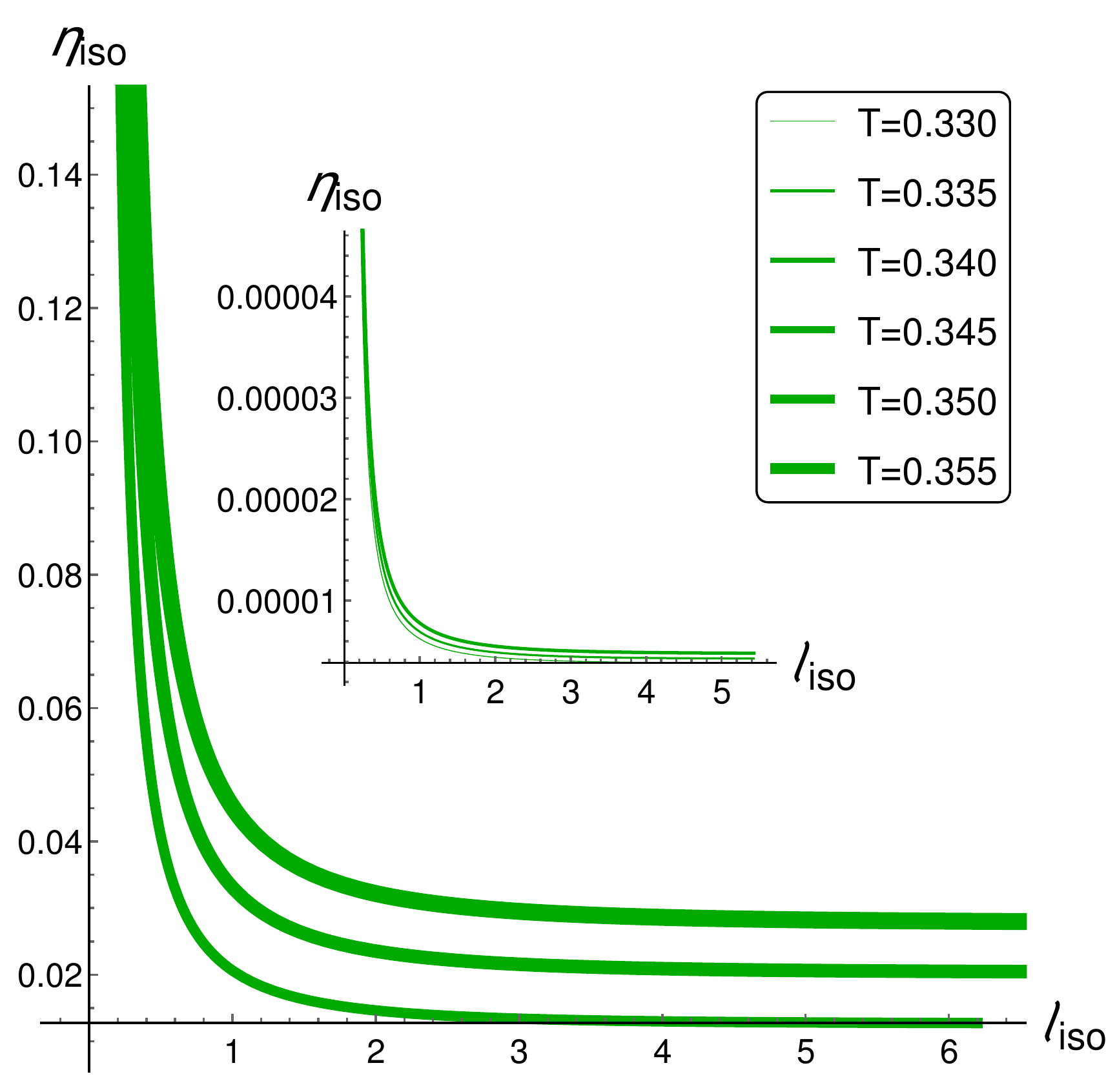}\includegraphics[width=6.5 cm]{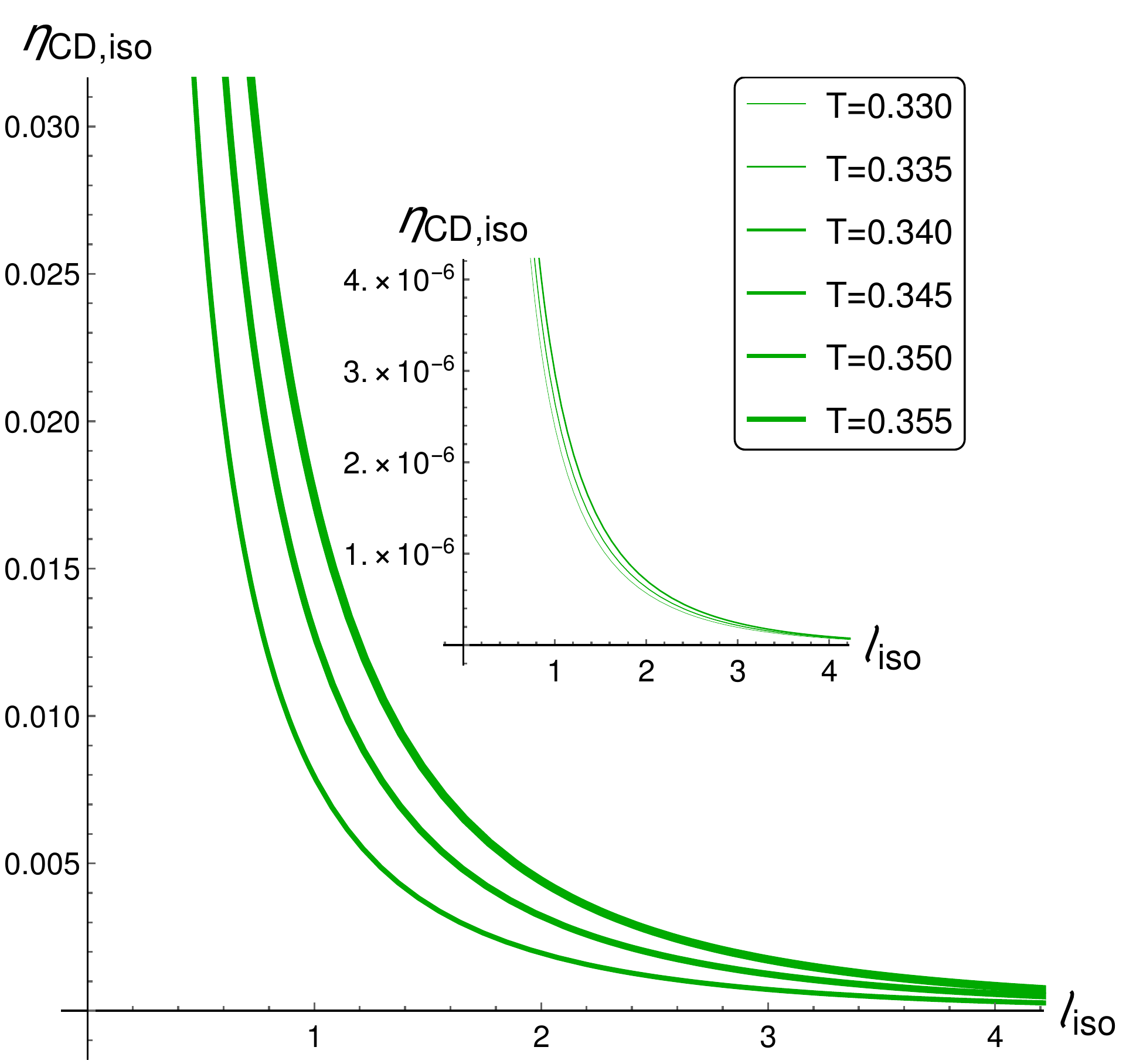}
    \caption{The dependence of the 
slab entropy density \eqref{etaCD} on $\ell $ for various temperatures around the critical temperature at $ \mu = 0.05 $
for isotropic case. Considered temperatures correspond to the point indicated in Fig.~\ref{fig:S-iso}.{\bf A)}. The density exhibits a jump near the BB phase transition $(T_{BB}=0.3445, \mu=0.05)$.
} \label{fig:eta-ell-iso}
\end{figure}

In Fig.~\ref{fig:eta-ell-CD} we show the dependence of the HEE density \eqref{etaCD} on $\ell$ at  the same temperatures and chemical potential as in Fig.~\ref{fig:eta-ell}. We get the similar behavior in both cases. For comparison, we present  the dependence of the densities \eqref{eta} and \eqref{etaCD} on $\ell $ in the isotropic case in Fig.~\ref{fig:eta-ell-iso}.

$\,$
 
\newpage
$$\,$$
\newpage
\subsubsection{Entanglement entropy density dependence on temperature}

\begin{figure}[h!]\centering
 \includegraphics[width=6 cm]{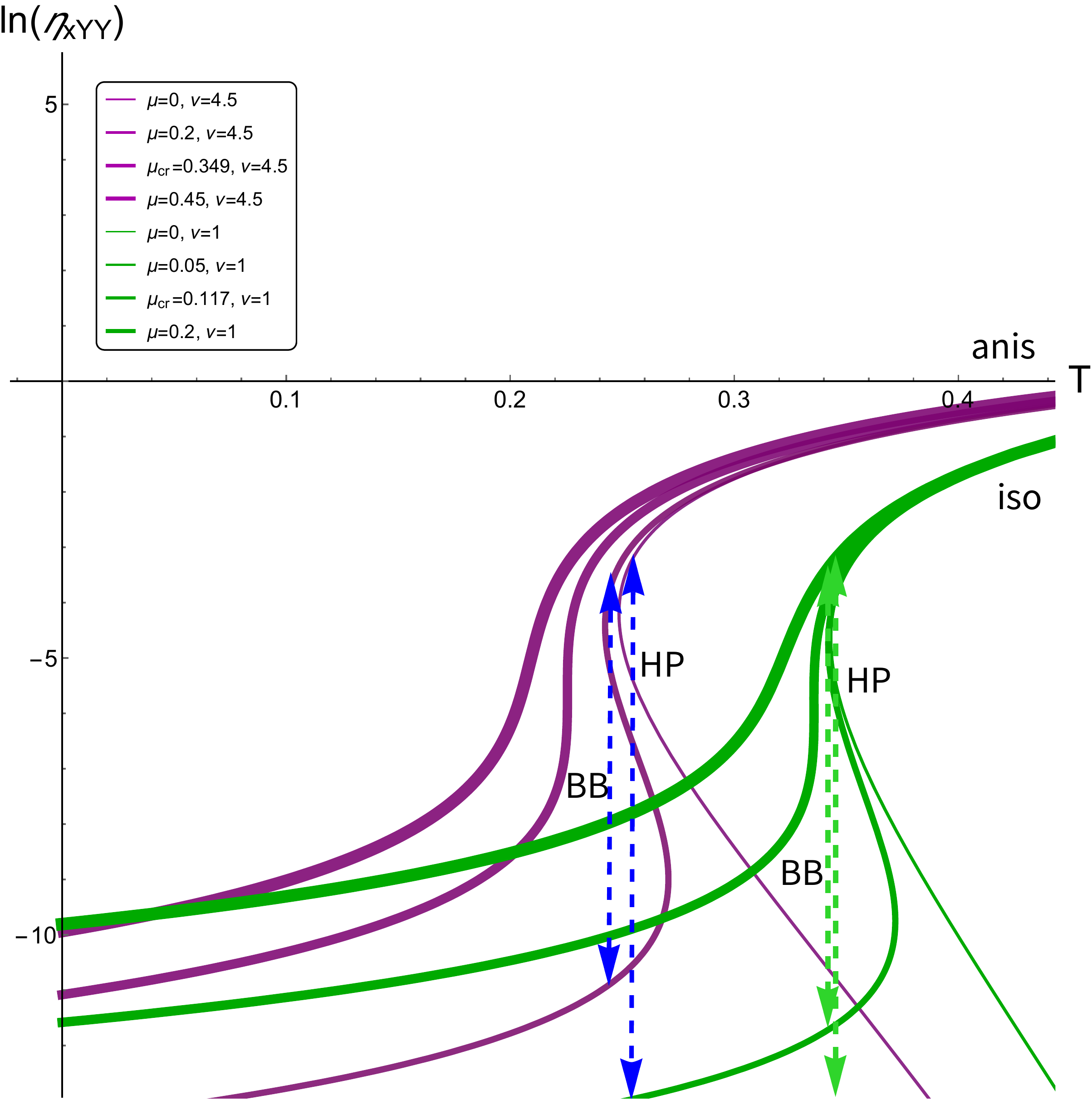}\qquad
  \includegraphics[width=6 cm]{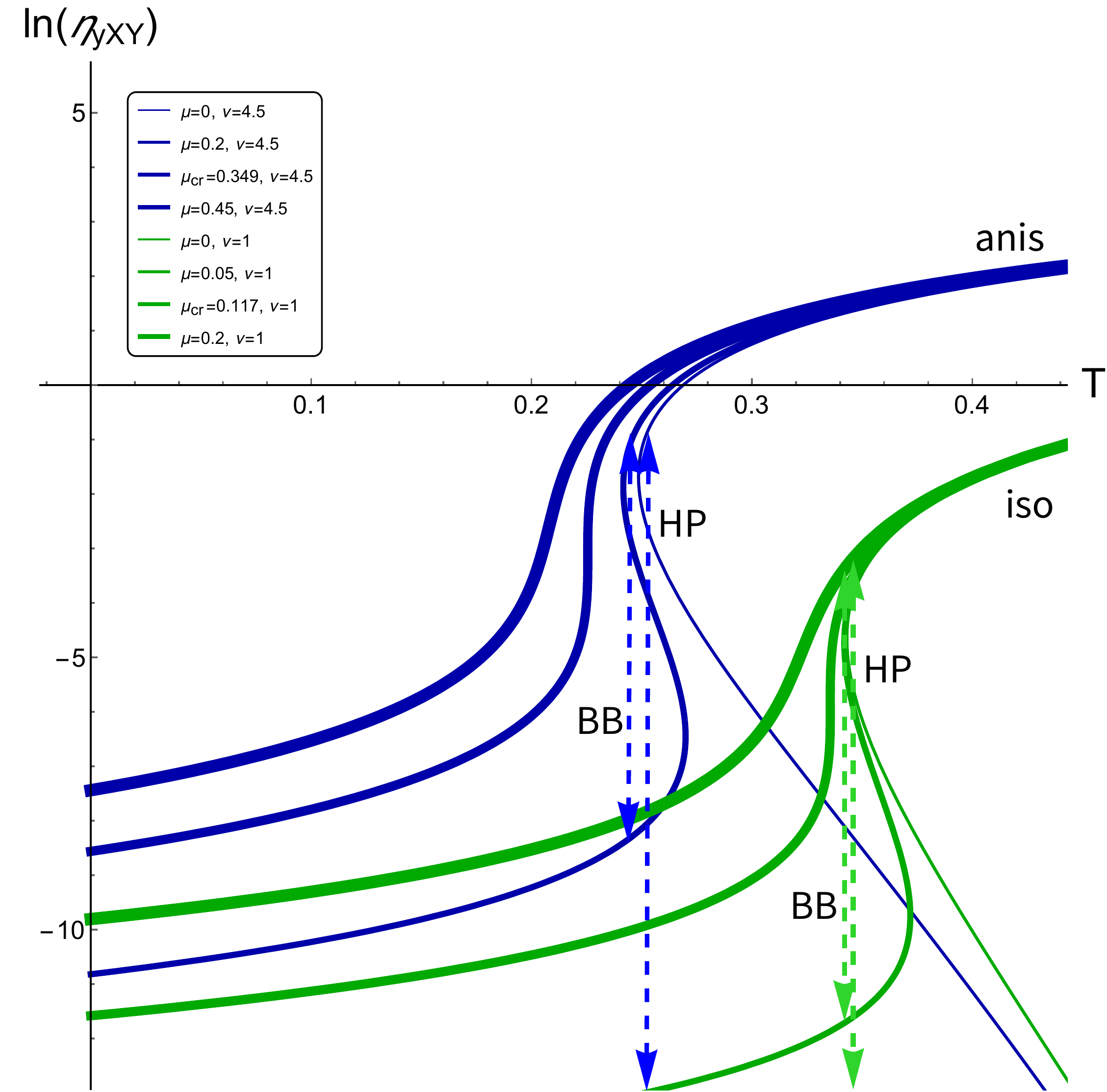}
\qquad\qquad\qquad\qquad
 \,{\bf A)}\,\qquad\qquad\qquad\qquad\qquad\qquad\qquad\qquad{\bf B)}
    \caption{The 
 HEE density dependence on the temperature for fixed value of $\ell=1$.  Blue lines on {\bf A)}  and  {\bf B)} show the 
 HEE density dependence   for different values of the chemical potential ($\mu=0,\,0.2, \,0.349,\, 0.45$) for  longitudinal  and   
 transversal   orientations, respectively,  $\nu=4.5$. Green lines show the same dependencies in the isotropic case for the chemical potential ($\mu=0,\,0.05, \,0.117,\, 0.2$). Arrows indicate the HEE density jumps at the points of HP and BB phase transitions  ($T_{HP} (\mu=0)$ and  $T_{BB} (\mu>0)$).
}
  \label{fig:den}
    \end{figure}

Now, we fix the length $l=1$ and present in Fig.\ref{fig:den} the  slab
HEE density dependence on the temperature for different angles, $\varphi=0$ and $\varphi=\pi/2$,
and different chemical potentials $\mu$,  below and above the critical values $\mu^{(anis)}_{cr}=0.349$ corresponding to 
$\nu =4.5$.  For comparison, we present here the similar plots for isotropic cases where $\mu^{(iso)}_{cr}=0.117$.
We see a dependence on orientation.

\begin{figure}[h!]\centering
\includegraphics[width=6 cm]{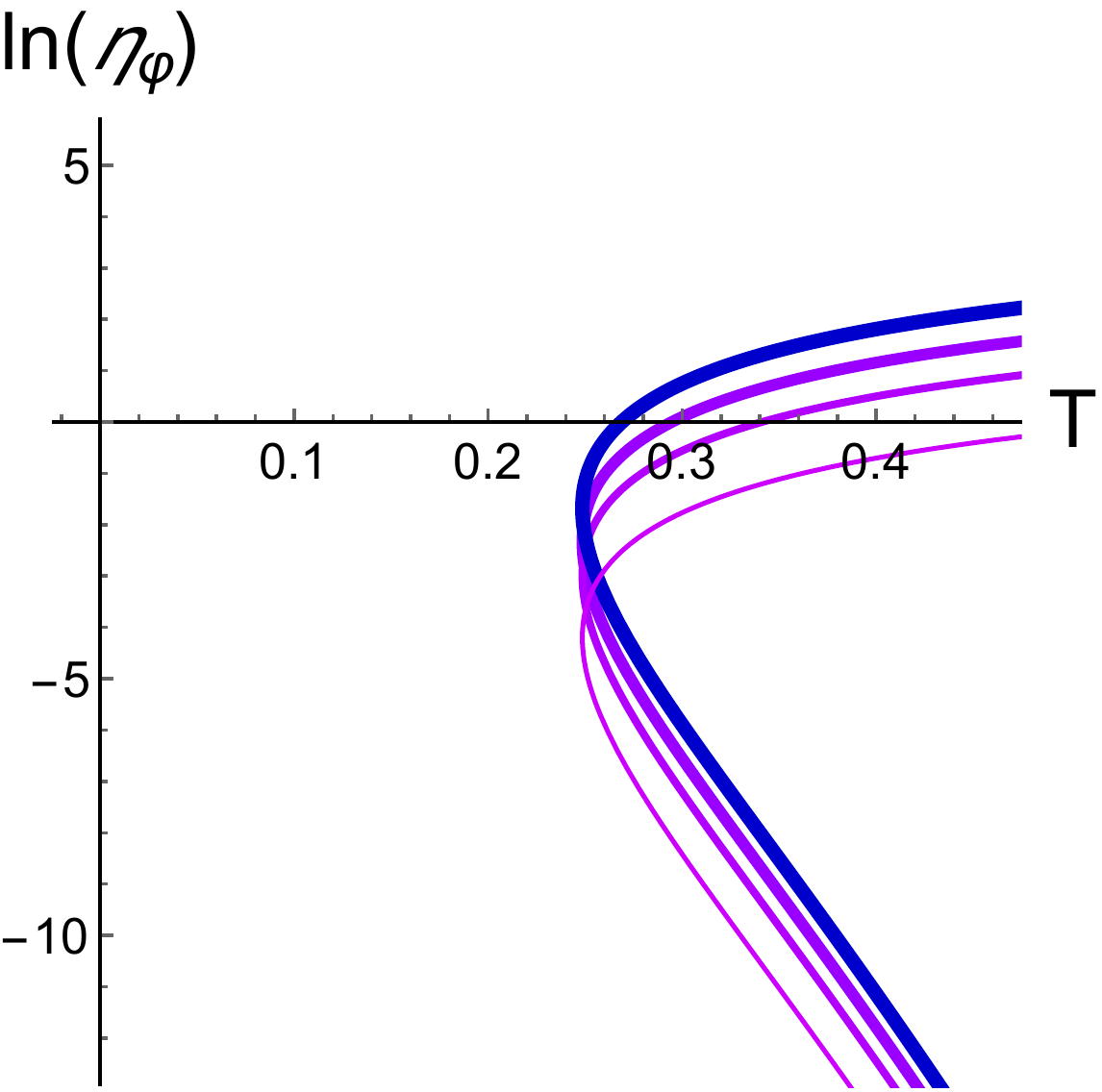}\quad\quad\quad
\includegraphics[width=6 cm]{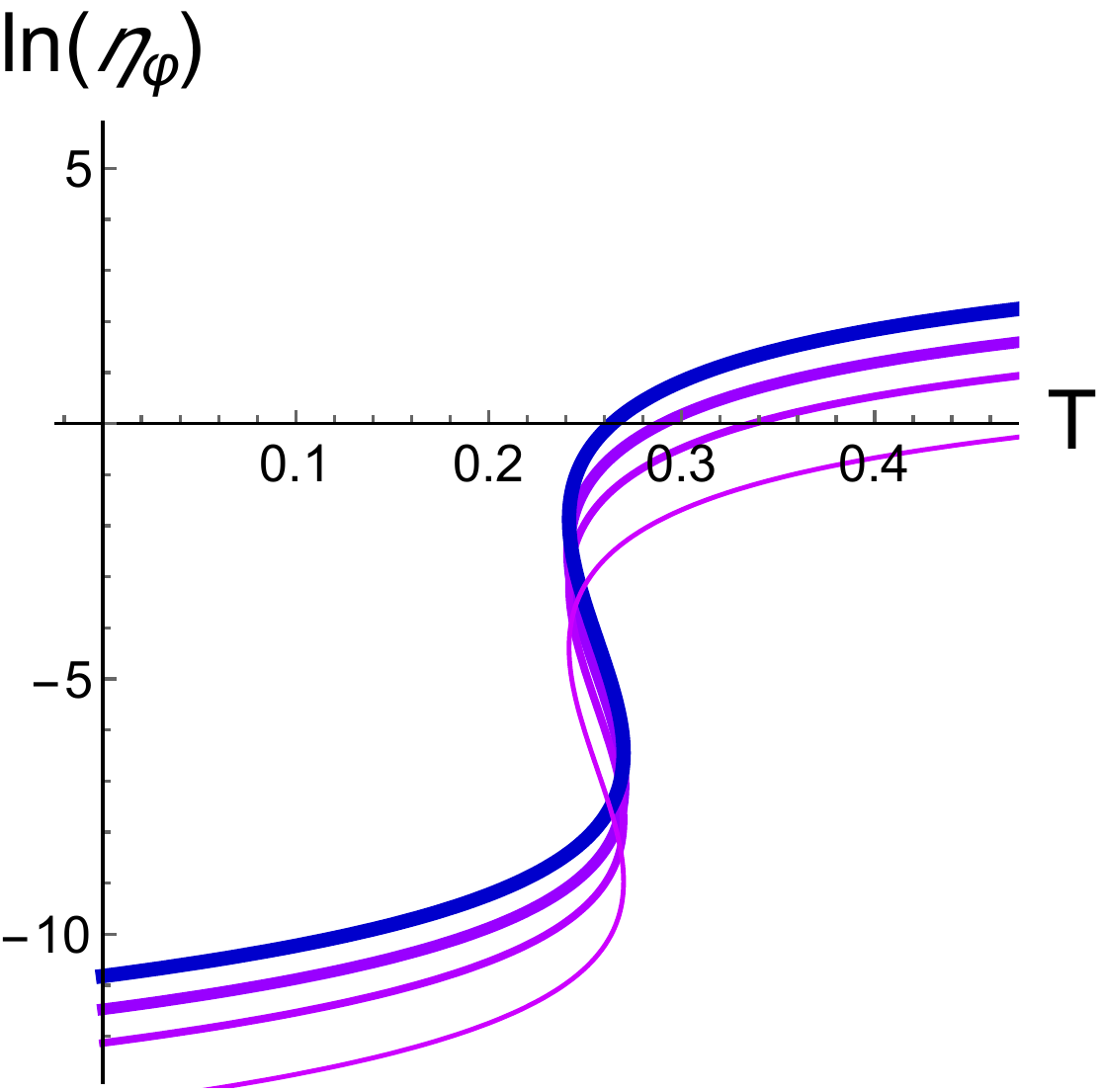}\\
  {\bf A)} \qquad\qquad\qquad\qquad\qquad\qquad\qquad\qquad\qquad{\bf B)} 
 \caption{The temperature dependence of the logarithm  of the entanglement entropy density \eqref{eta} for  angle values $\varphi=0$ (magenta), $ \pi/6$(magenta blue),  $\pi/4$(blue magenta) and $\pi/2$(blue) at the chemical potential {\bf A)} $\mu=0$  and {\bf B)} $\mu=0.2$. 
}
\label{fig:angularden}
    \end{figure}

In Fig.\ref{fig:angularden} the angular dependence of the logarithm of entanglement entropy density for  angle values $\varphi=0,\pi/6,\pi/4,\pi/2$ (thickness increases with increasing angle) at the chemical potential $\mu=0$ and $\mu=0.2$  are shown in {\bf A)} and  {\bf B)}, respectively.
We see that the shapes of the curves shown in {\bf A)} are the same for different orientation angles, and the curves only shift with increasing angle $ \varphi $. The same applies to the curves shown in {\bf B)}.

    \begin{figure}[h!]\centering
    $$\,$$
 \includegraphics[width=6 cm]{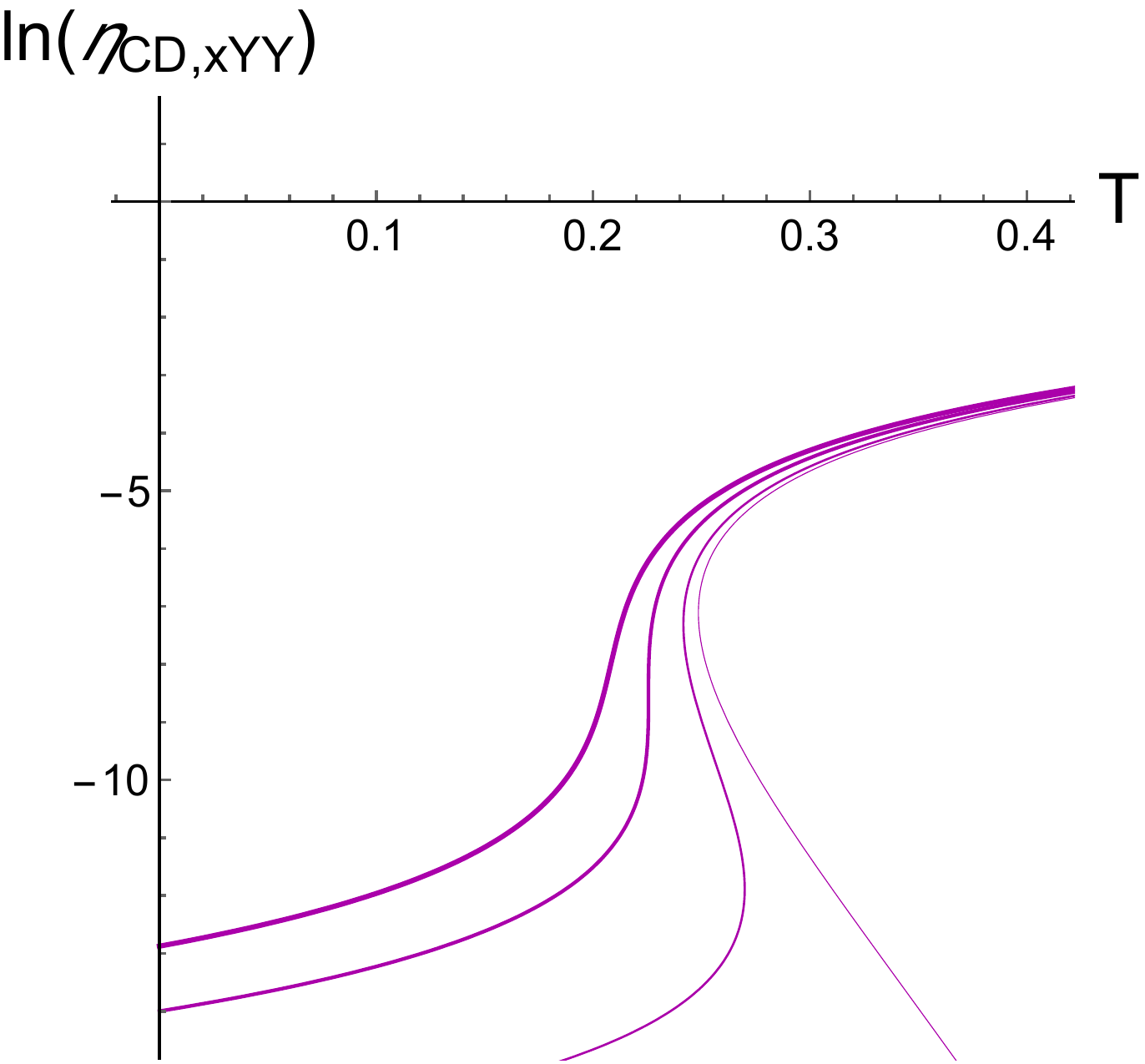} \qquad
  \includegraphics[width=6 cm]{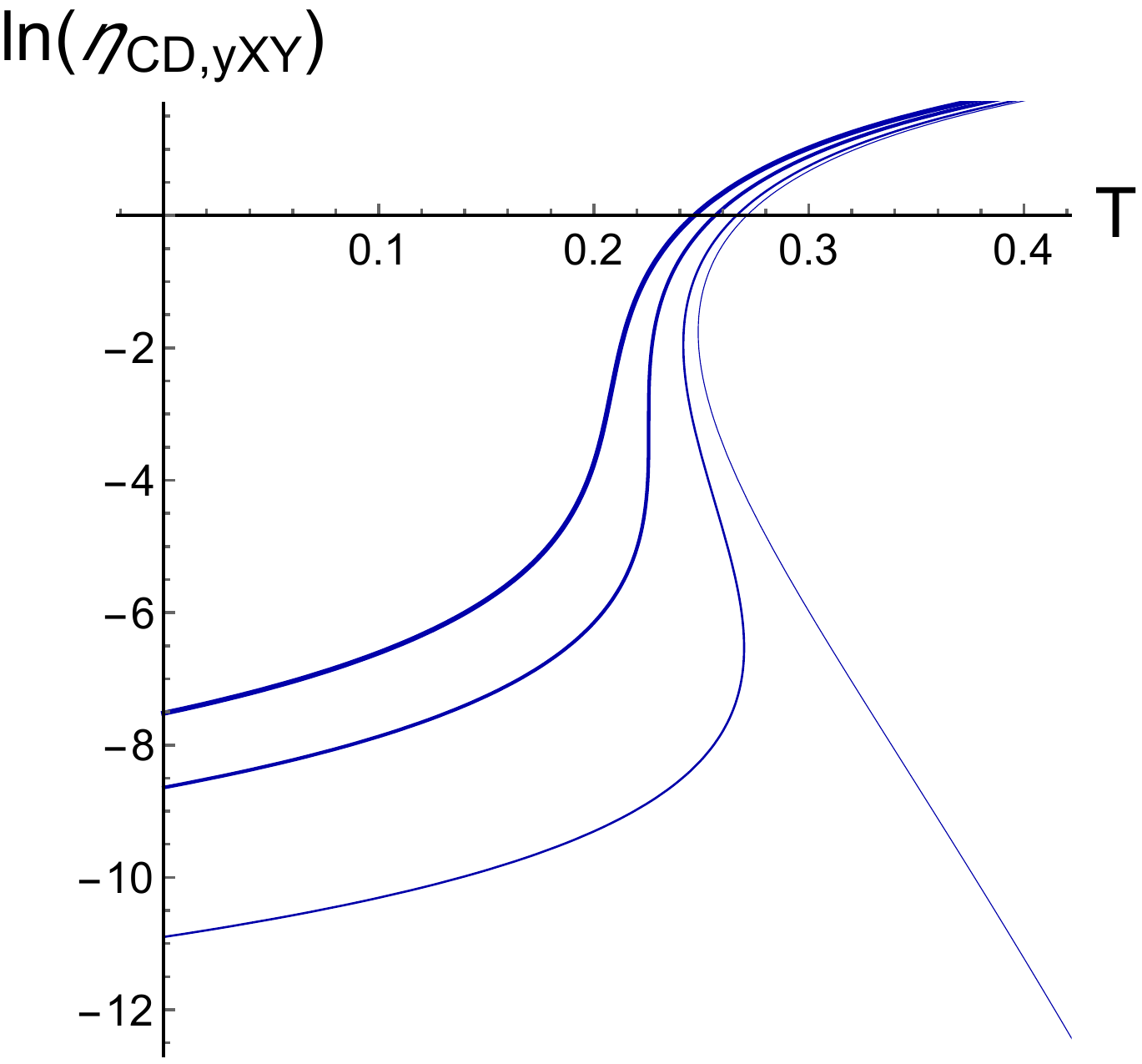}\\
  \qquad\qquad\qquad\qquad
 \,{\bf A)}\,\qquad\qquad\qquad\qquad\qquad\qquad\qquad\qquad{\bf B)}
    \caption{The 
 HEE density dependence on the temperature for fixed value of $\ell=1$.   The lines in {\bf A)}  and  {\bf B)} show the 
 HEE density dependence   for different values of the chemical potential, $\mu=0,\,0.2, \,0.349,\, 0.45$ (thickness increases with increasing $\mu$) for  longitudinal  and    transversal   orientations, respectively,  $\nu=4.5$. 
 }
    \end{figure}

\begin{figure}[h!]
$$\,$$
\centering
\includegraphics[width=6.5 cm]{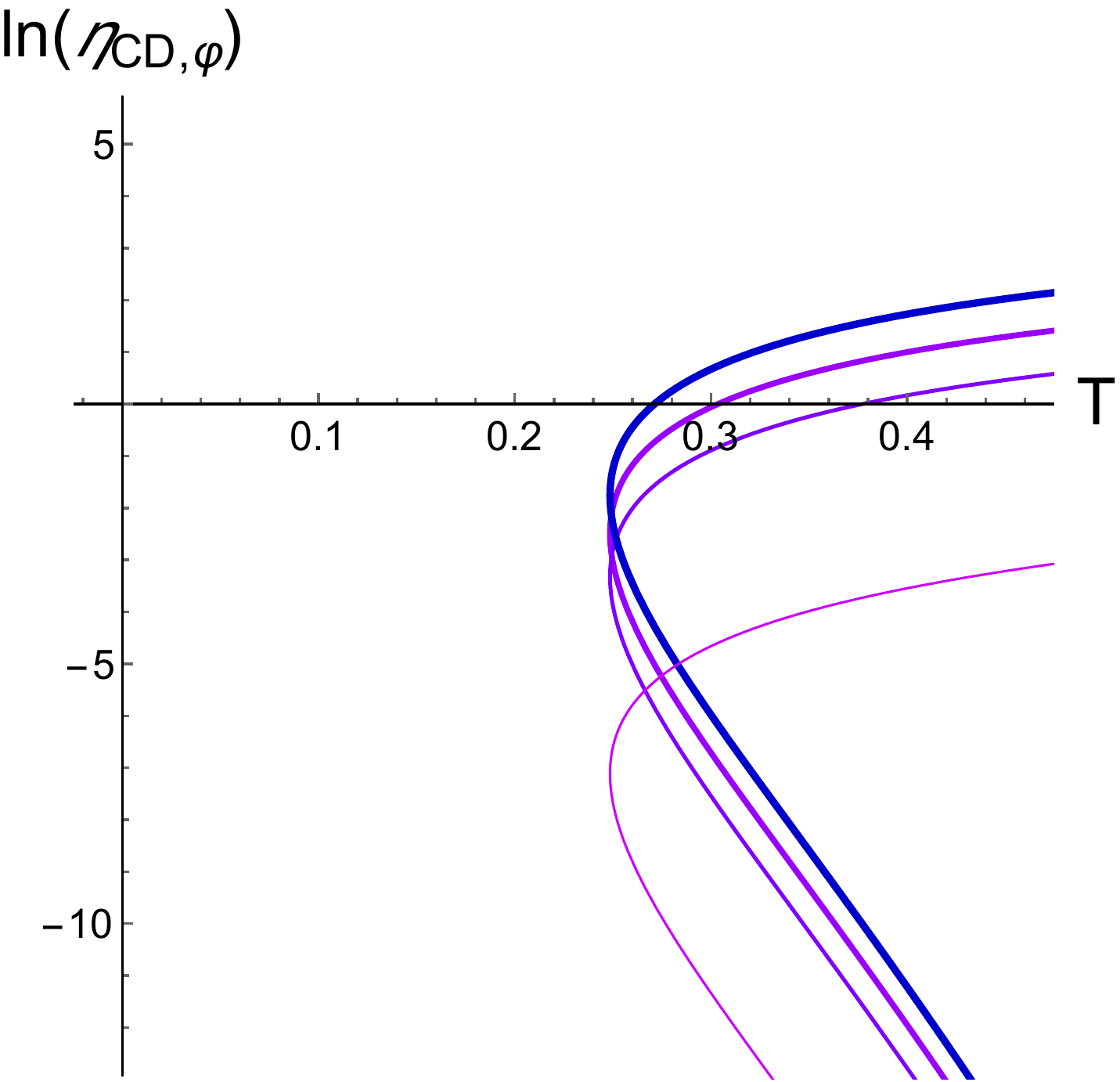}\quad\quad\quad
\includegraphics[width=6.5 cm]{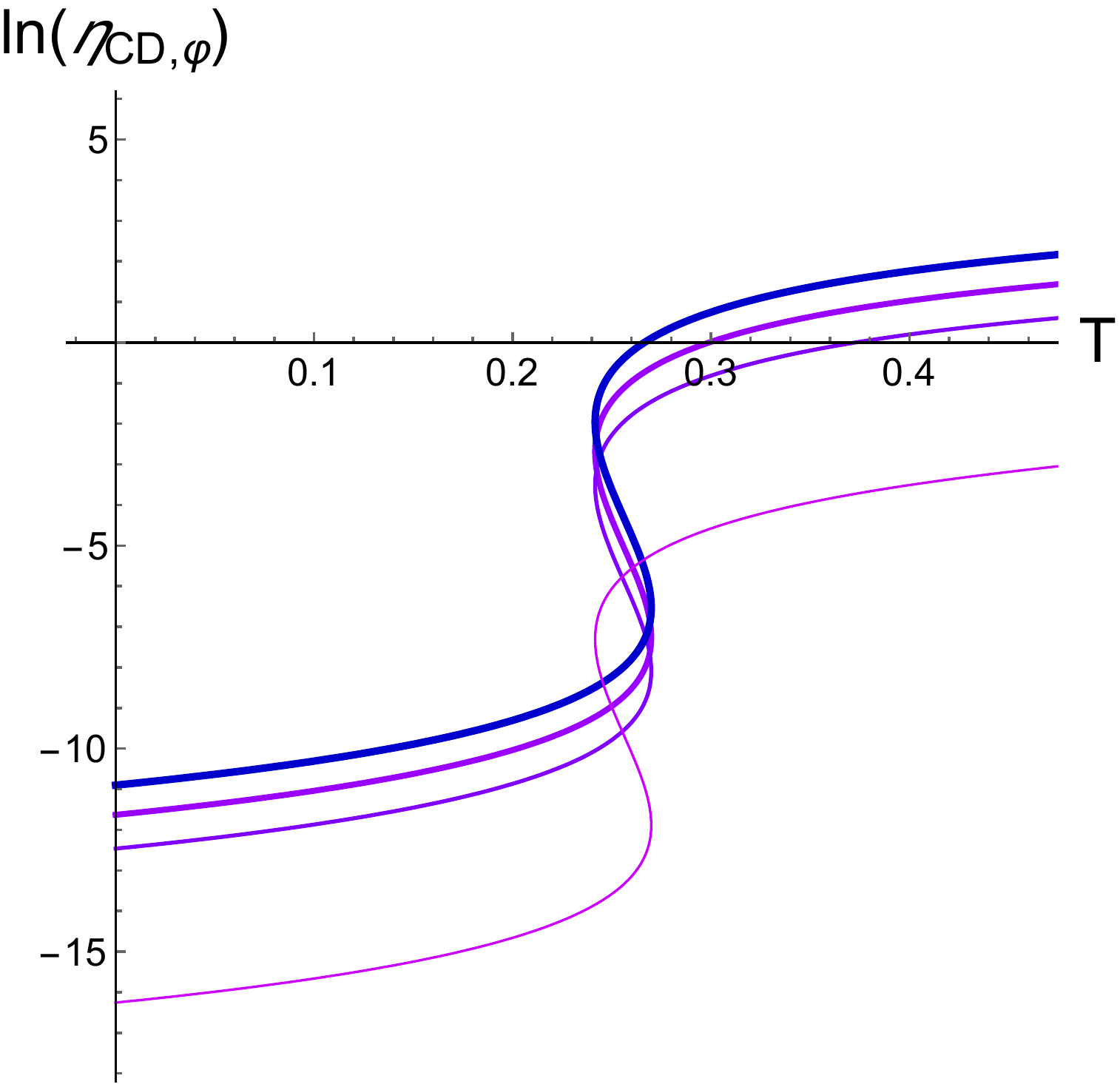}\\
{\bf A)} \qquad\qquad\qquad\qquad\qquad\qquad\qquad\qquad\qquad{\bf B)} 
 \caption{The temperature dependence of the logarithm of entanglement entropy density \eqref{etaCD}  for angle values $\varphi=0 $  (magenta) $,\pi/6, \pi/4, \pi/2$ (blue) at the chemical potential {\bf A)} $\mu=0$ and {\bf B)} $\mu=0.2$. 
}
\label{fig:angularden1CD}
    \end{figure}
    
\begin{figure}[h!]\centering
$$\,$$
\includegraphics[width=5 cm]{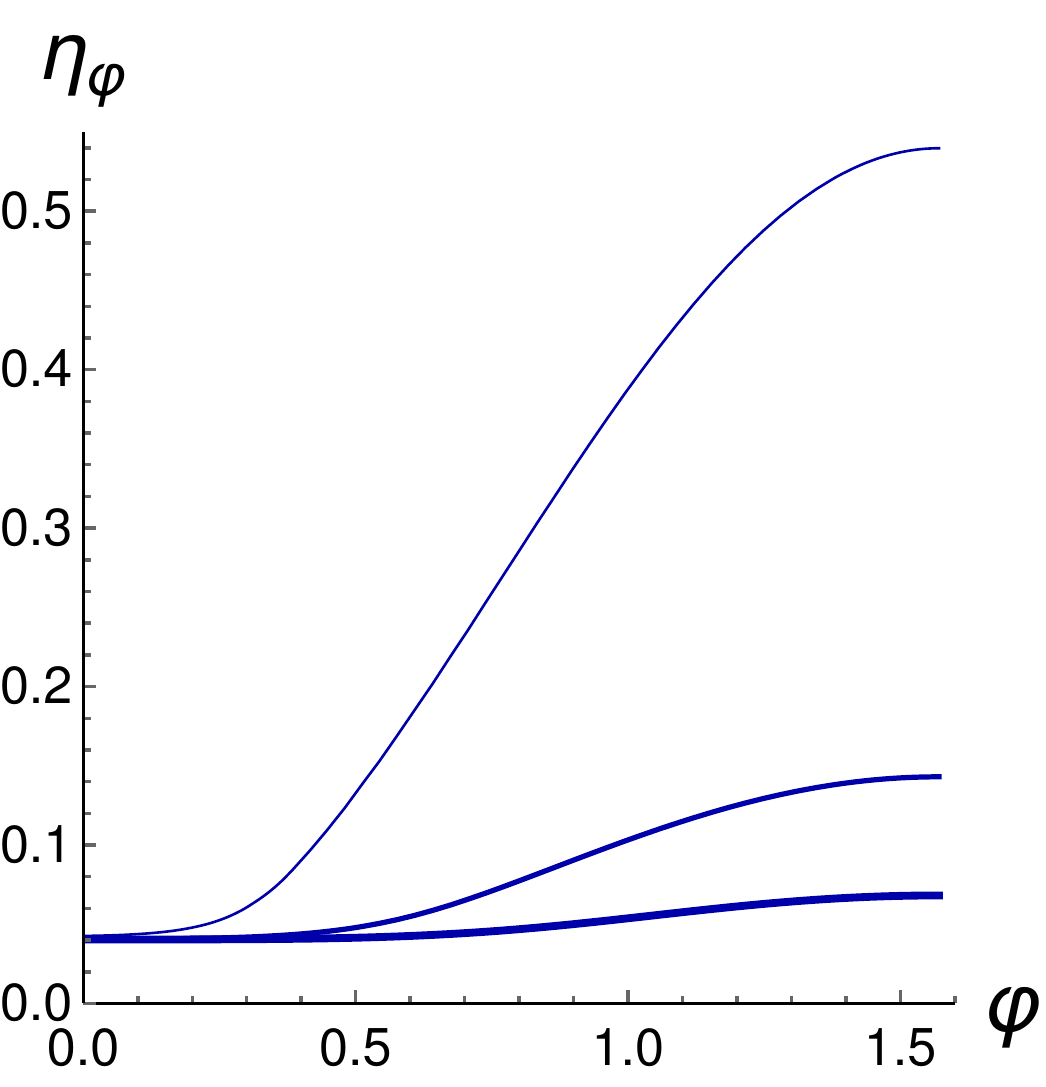}\quad\quad\quad\quad
\includegraphics[width=5 cm]{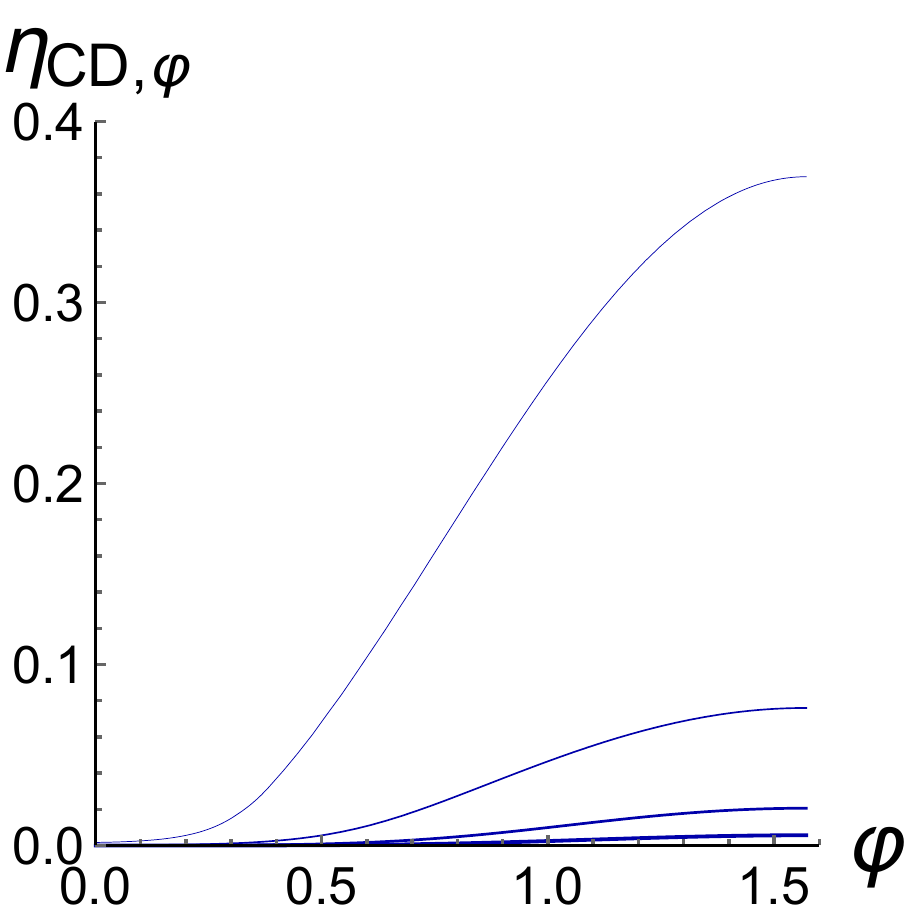}\qquad\qquad\qquad\qquad
 \,{\bf A)}\,\qquad\qquad\qquad\qquad\qquad\qquad\qquad\qquad{\bf B)}
 \caption{
The angular dependence of entanglement entropy density {\bf A)} $\eta_\varphi$  and {\bf B)} $\eta_{CD,\varphi}$  for $T=0.25$, $\mu=0.2$, $\nu=4.5$, $\ell =1,2,3$ (thickness increases with increasing length).}
\label{fig:angularden2}
    \end{figure}

\begin{figure}[h!]
$$\,$$\centering
\includegraphics[width=4.5 cm]{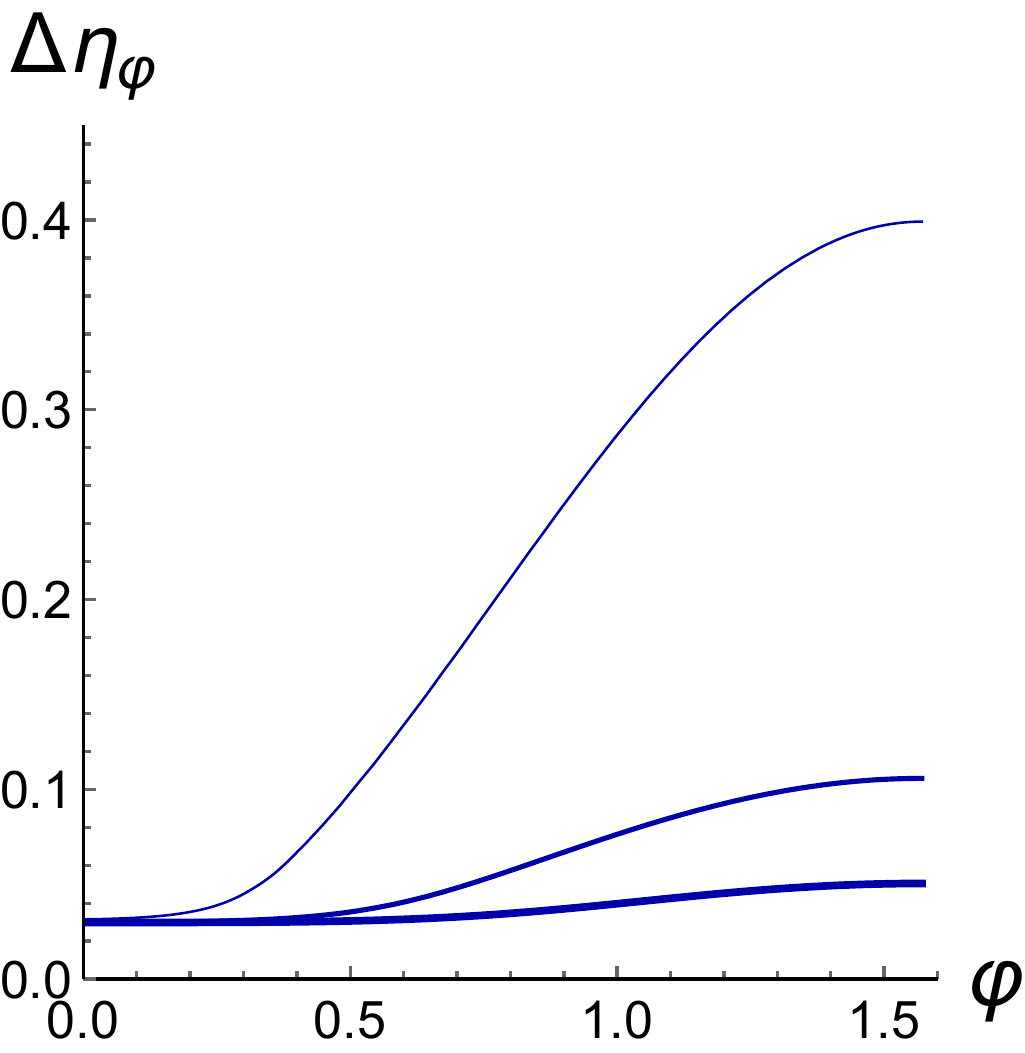}\qquad\qquad\qquad
\includegraphics[width=4.55 cm]{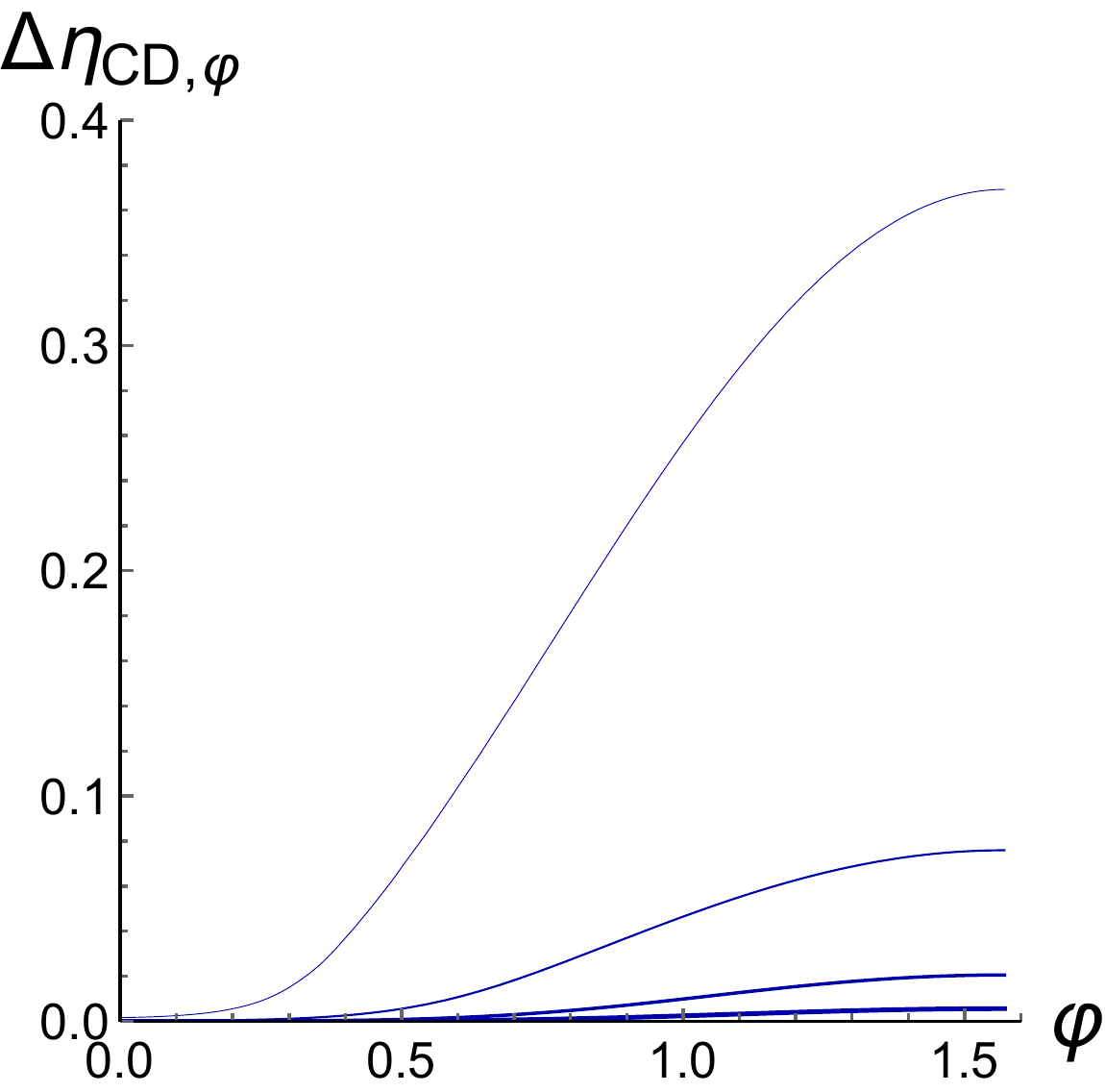}\\
 \,{\bf A)}\,\qquad\qquad\qquad\qquad\qquad\qquad\qquad\qquad{\bf B)}
 \caption{The angular dependence of entanglement entropy density jumps (thickness increases with increasing length, $\ell =1,2,3$),   {\bf A)} $\Delta\eta_{\varphi}$  
 and  {\bf B)} $\Delta\eta_{CD,\varphi}$, at the chemical potential $\mu=0.2$. }
\label{fig:angulardeltaden}
    \end{figure}

In Fig.\ref{fig:angularden1CD} we show the angular dependence of the entanglement entropy density for different values of $\ell$ (thickness increases with slab width increase) at the chemical potential $\mu=0.2$ and $T=0.25$  in {\bf A)}  minimal  and  {\bf B)} geometrical regularizations.

Comparing the plots in Fig.\ref{fig:angularden} and Fig.\ref{fig:angularden1CD} we see that the transversal HEE density  depends weakly on the regularization type while there is  a substantial difference for the longitudinal  density in different regularization schemes. 

In Fig.\ref{fig:angularden2} we show the angular dependence of the entanglement entropy density for different values of $\ell$ (thickness increases with slab width increase) at the chemical potential $\mu=0.2$ and $T=0.25$  in minimal  {\bf A)} and  {\bf B)} geometrical regularization. Since the magnitude of the HEE density above the critical temperatures is substantially larger than the values of HEE density  below the phase transition, the plot of the jumps of HEE density Fig.\ref{fig:angulardeltaden}  is similar to the  Fig.\ref{fig:angularden2}.  

We see that the HEE density undergoes jumps near the BB phase transition and these jumps depend on the anisotropy, slab orientation and chemical potential.\\

\newpage
$$\,$$

  To summarize this subsection, note that  similarly to the plots presented for the entaglement entropy in the previous subsection \ref{Sect:HEE-TL},  in Fig.\ref{fig:eta-ell} - Fig.\ref{fig:angularden2}, we present dependence of the HEE densities \eqref{eta} and  \eqref{etaCD} on the slab width $l$ for equidistant  values of temperature near background phase transition.
  We observe that the HEE density undergoes  a  jump  when  the  temperature  crosses  the  phase  transition  line.
  We also observe a non-substantial dependence on the regularization scheme.

 \subsection{c-functions}\label{Sect:c-function}
Now we consider   the behavior of c-function numerically
\be
c_{\nu,\varphi,F}=\frac{\ell_\varphi^{m_{\varphi,F}}}4\,\Big(\frac{b^{3/2}_F(z_*)}{z_*^{1+2/\nu}}-\frac{b^{3/2}_F(z_D)}{z_D^{1+2/\nu}}\Big),\label{GAg}
\ee
where $m_{\varphi,F}$ is a scaling power. It depends on the model, in our particular case on the anisotropy parameter $\nu$,
on the orientation (index $\varphi$) and frame. The frame is denoted by index $F$, $F=EF$ (Einstein frame),  $F=SF$ (string frame). In isotropic conformal invariant case in 5 dimensions,   $m_{AdS_5}=3$.

 In our model in the EF we have

\be ds^{EF\,2}_{UV} \sim \frac{1}{z^2}\left(-f( z)dt^{2} + dx^{2} + z^{2-2/\nu}(dy^{2}_{1} + dy^{2}_{2})  + \frac{d z^{2}}{ f(z)}\right),\ee
This metric is invariant under the rescaling
\be
t\to \Lambda t\ , \ x_i\to \Lambda x_i \ , \ y_j\to \Lambda^{1/\nu} y_j \ , \ i=1, \, j=1,2 \ ,z\to \Lambda z.
\ee

Using the suggestion of  \cite{Chu:2019uoh}, see also \cite{Ghasemi:2019xrl,Hoyos:2020zeg},
we get
\bea
 m_{x,EF}&=&d_1+d_2\frac{n_2}{n_1}=1+\frac{2}{\nu},\label{mx}\\
 m_{y,EF}&=&d_2-1+d_1\frac{n_1}{n_2}=2+\nu,\label{my}\eea
where $n_1=1, d_1=1$ and $n_2=1/\nu,\,d_2=2$.
 Taking into account the behavior of the dilaton in UV \cite{AR} we get  in the SF
 
 \be\label{bm}
ds^{SF\,2}_{UV} \sim \frac{1}{z^{K_{UV}}}\left(-f( z)dt^{2} + dx^{2} + z^{2-2/\nu}(dy^{2}_{1} + dy^{2}_{2})  + \frac{d z^{2}}{ f(z)}\right),
\ee
where \be
K_{UV}(\nu)=2-\frac{\sqrt{\frac{8}{3}(\nu -1)}}{\nu}\ee
therefore,
\bea
t&\to& \Lambda^{K_{UV}/2} t\, ,
\, x_i\to \Lambda ^{K_{UV}/2}x_i\,,\,\,z\to \Lambda ^{K_{UV}/2}z \\   y_j&\to&=
\Lambda^{K_{UV}/2-1+1/\nu} y_j
 \ , \ i=1, \,\ j=1,2\,
\eea
and \bea
n_1&=&K_{UV}/2,\\
n_2&=&K_{UV}/2-1+1/\nu.\eea

In the SF we take
\bea
 m^{SF}_{x,UV}&=&1+2\frac{K_{UV}/2-1+1/\nu}{K_{UV}/2},\label{mSFx}\\
m^{SF}_{y,UV}&=&2+\frac{K_{UV}/2}{K_{UV}/2-1+1/\nu}.\label{mSFy}
\eea
In the particular case, $\nu=4.5$, we have
\bea
m^{EF}_x\Big|_{\nu=4.5}&=&1.444,\,\,\,\,\,\,\,\,\,\,m^{SF}_x\Big|_{\nu=4.5}=0.645,\label{mSFx-45}\\
m^{EF}_y\Big|_{\nu=4.5}&=&6.5,\,\,\,\,\,\,\,\,\,\,m^{SF}_y\Big|_{\nu=4.5}=-3.635.\label{mSFy-45}\eea

In \eqref{GAg} in the SF $z_D$ means    the minimum of two values, the position of the horizon
and the position of the corresponding dynamical wall. In the EF there is no dynamical wall and $D$
means the position of the horizon. Notice that position of the dynamical wall depends on the anisotropy parameter and 
does not depend on the horizon position, see Fig.\ref{fig:3D-DW}.

\begin{figure}[h!]\centering
\includegraphics[width=7cm]{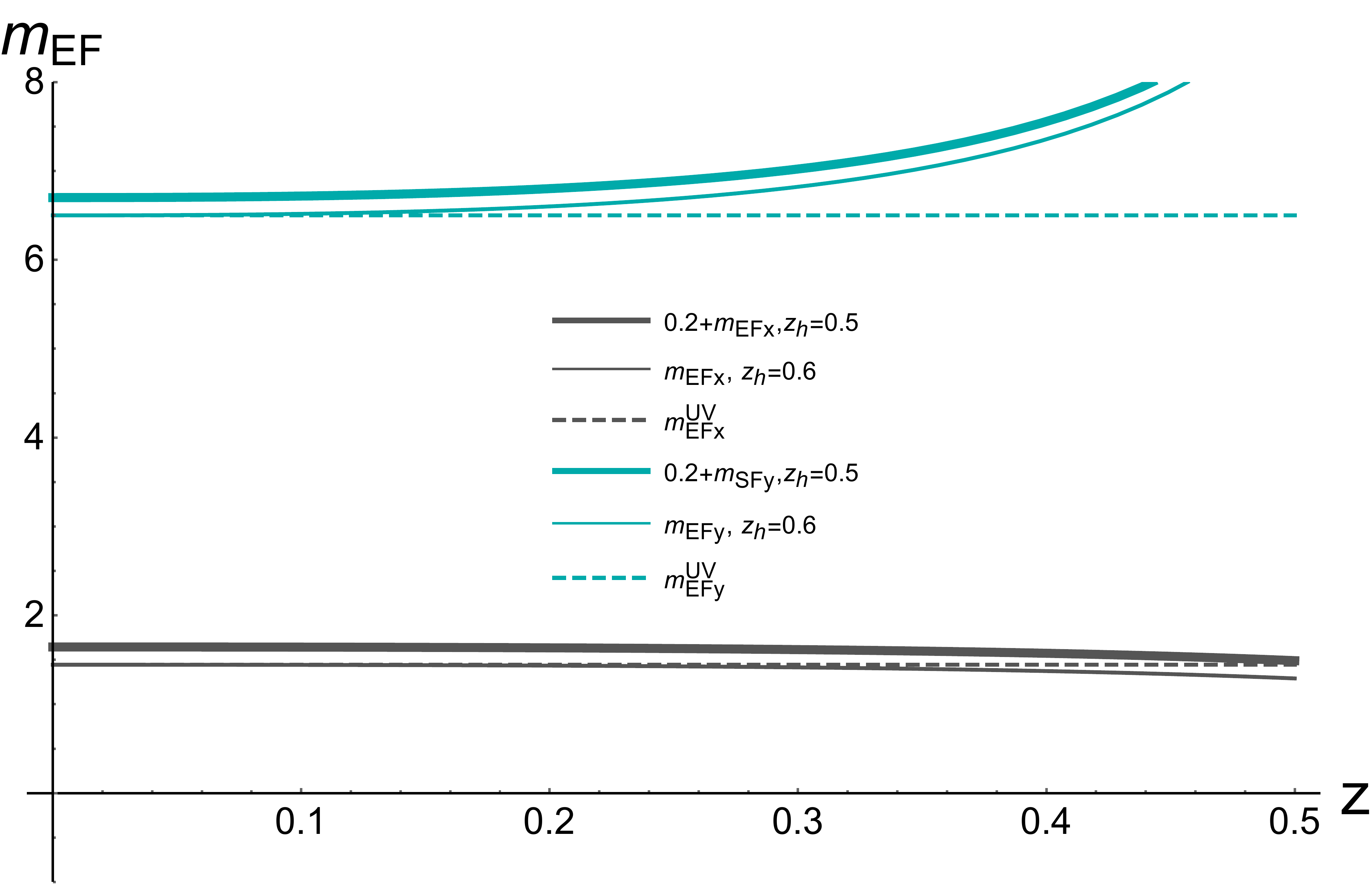} \qquad
\includegraphics[width=7cm]{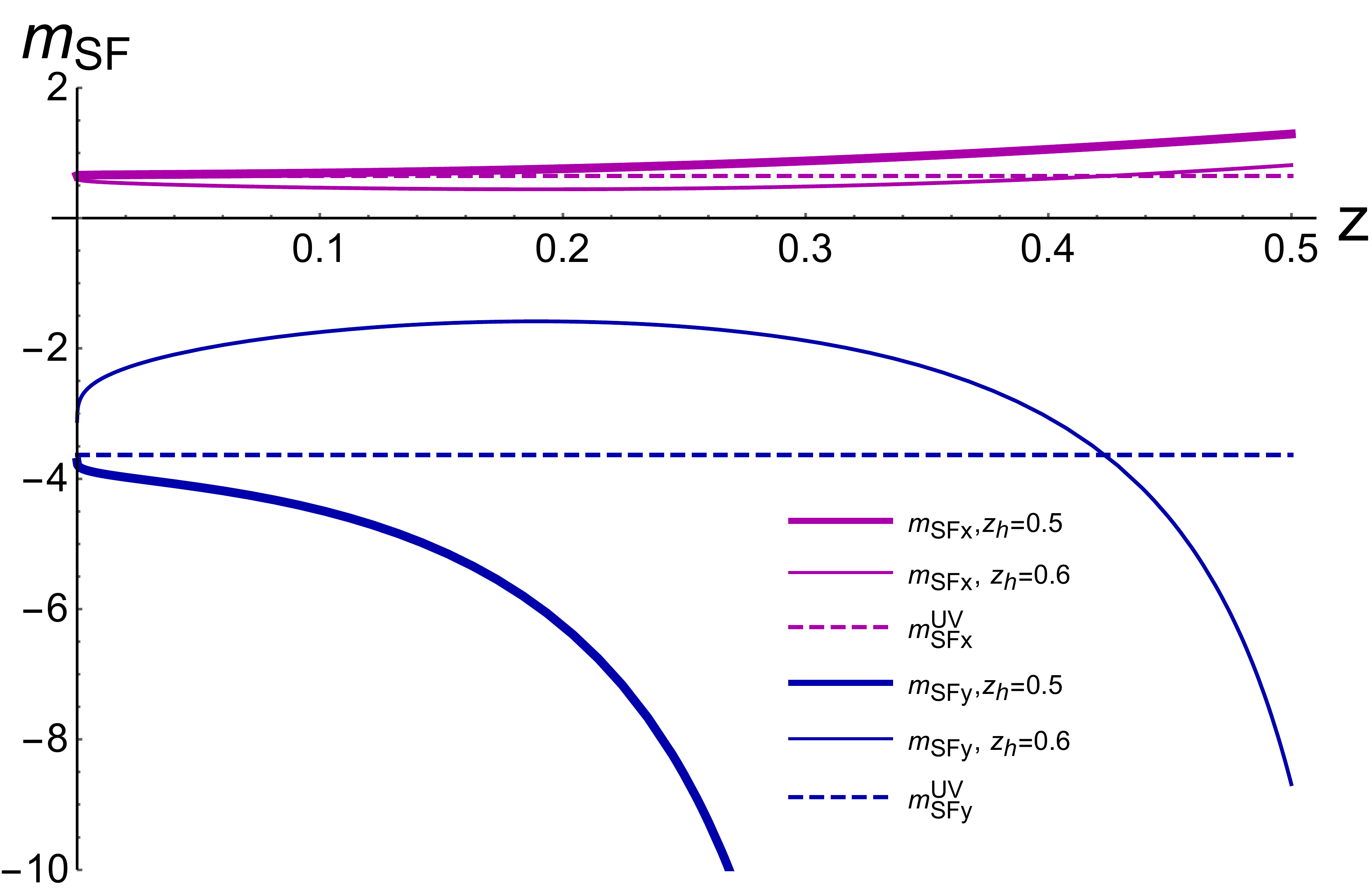} 
 \caption{ 
The "scaling factors" $m^{SF}_{x,y}(z,z_h,\nu)$ and $m^{EF}_{x,y}(z,z_h,\nu)$ defined by \eqref{KmSFx} and \eqref{KmSFy} and their UV approximations 
\eqref{mSFx-45} and \eqref{mSFy-45} shown by the dashed lines. }
    \label{v-F}
  \end{figure}

We have to mention,  that here we use UV asymptotics of the warp factors $b(z)$ and $b_s(z)$. The scaling exponents are changed 
already in the intermittent region. The regions of the validity of the scaling \eqref{mSFx-45} and \eqref{mSFy-45} depend on the size of the horizon, see Fig. \ref{v-F}.
  
 The scaling functions $m^{SF}_{x,y}$ and $m^{EF}_{x,y}$ shown  in \eqref{v-F} are defined by the formula similar to \eqref{mSFx} and \eqref{mSFy}
\bea
 m^{F}_x(z,z_h,\nu)&=&1+2\frac{K^{F}(z,z_h,\nu)/2-1+1/\nu}{K^{F}(z,z_h,\nu)/2},\label{KmSFx}\\
m^{F}_y(z,z_h,\nu)&=&2+\frac{K^{F}(z,z_h,\nu)/2}{K^{F}(z,z_h,\nu)/2-1+1/\nu},\label{KmSFy}
\eea
where $F$ is the index indicated the frame, $F=EF,SF$ and 
\be\label{bm}
ds^{SF\,2} = \frac{1}{z^{K^{SF}(z,z_h,\nu)}}\left(-f( z)dt^{2} + dx^{2} + z^{2-2/\nu}(dy^{2}_{1} + dy^{2}_{2})  + \frac{d z^{2}}{ f(z)}\right),
\ee
\bea
K^{SF}(z,z_h,\nu)=\frac{\log \left(\frac{b_s\left(z,c,\nu ,z_h,\mu
   \right)}{z^2}\right)}{\log
   \left(\frac{1}{z}\right)}\\
   K^{EF}(z,z_h,\nu)=\frac{\log \left(\frac{b\left(z,c,\nu ,z_h,\mu
   \right)}{z^2}\right)}{\log
   \left(\frac{1}{z}\right)}.\eea
We see that the UV scaling factor  $m^{EF}_{UV,x,y}$ approximates the function  $m^{EF}_{x,y}(z,z_h,\nu)$ relatively well up to holographic coordinates about a half of the horizon position. The same concerns also the approximation  $m^{SF}_{UV,y}$ to $m^{SF}_{y}(z,z_h,\nu)$, meanwhile the UV scaling factor $m^{SF}_{UV,x}$ cannot be used as an  approximation for $m^{EF}_{y}(z,z_h,\nu)$.
Moreover, all $m^{F}_{x,y}(z,z_h,\nu)$ exhibit the singular behavior at large  $\ell$.

\begin{figure}[h!]\centering
\includegraphics[width=5cm]{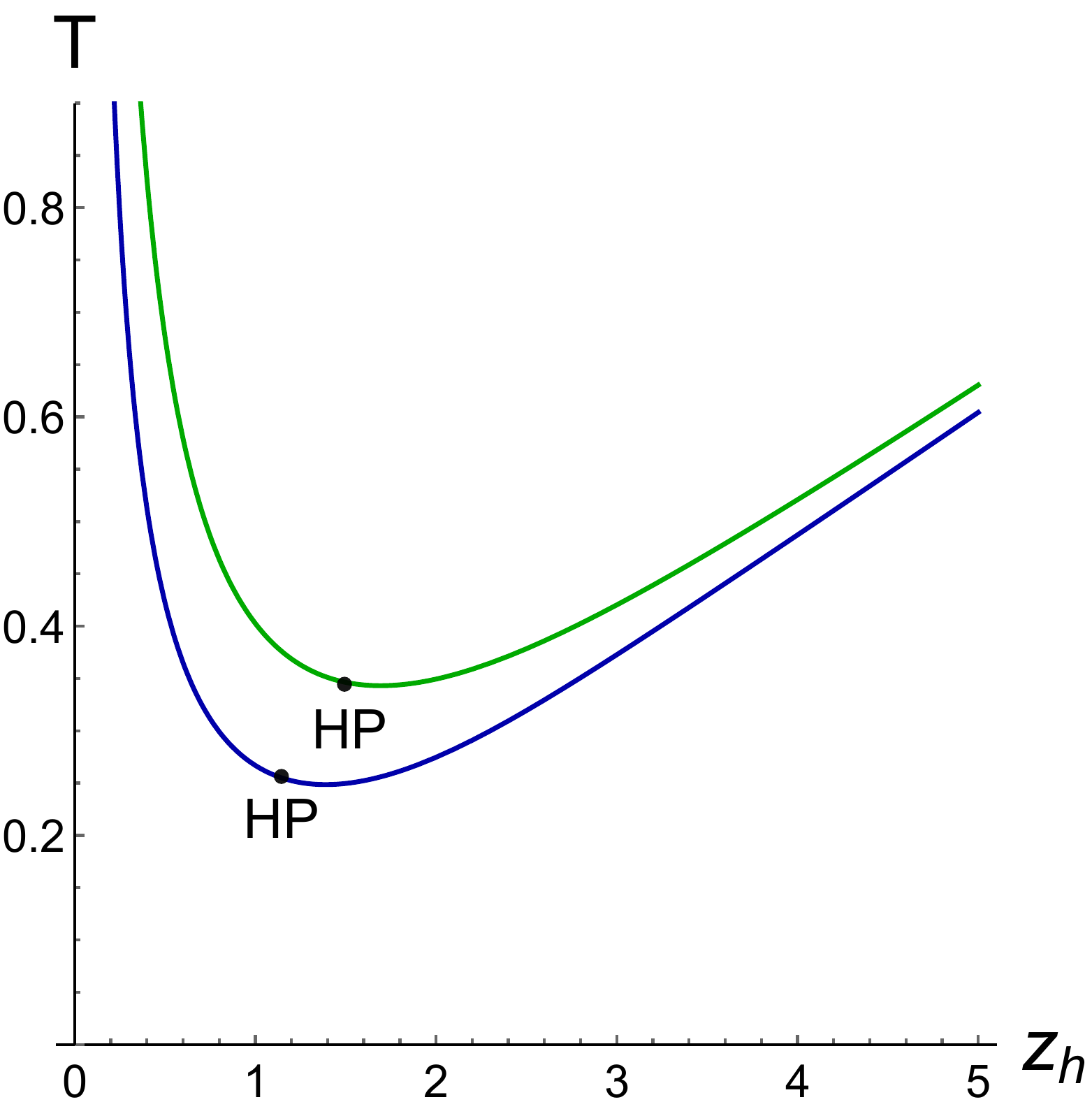} 
 \caption{ 
 The dependence of $T$ on $z_h$ for $\nu=1$ (green line) and  $\nu=4.5$ (blue line) for $\mu =0$. 
  The dots show location of $z_{HP}$ points,  $z_{HP}(1)=1.505$,  $z_{HP}(4.5)=1.138$. }
    \label{T-zh-mu0}
  \end{figure}

Let us first consider the case of zero chemical potential. The dependence of temperature on the horizon for $\mu=0$ is presented in Fig.\ref{T-zh-mu0}. For large BH's the temperature increases when $z_h \to 0$, and for small BH's the temperature increases also for $z_h\to \infty$. At $z_h=z_{HP}$ the HP phase transition takes place and all small BH's are unstable.
  
\subsubsection{The c-function in the isotropic case}
\begin{figure}[h!]\centering
 \includegraphics[width=7cm]{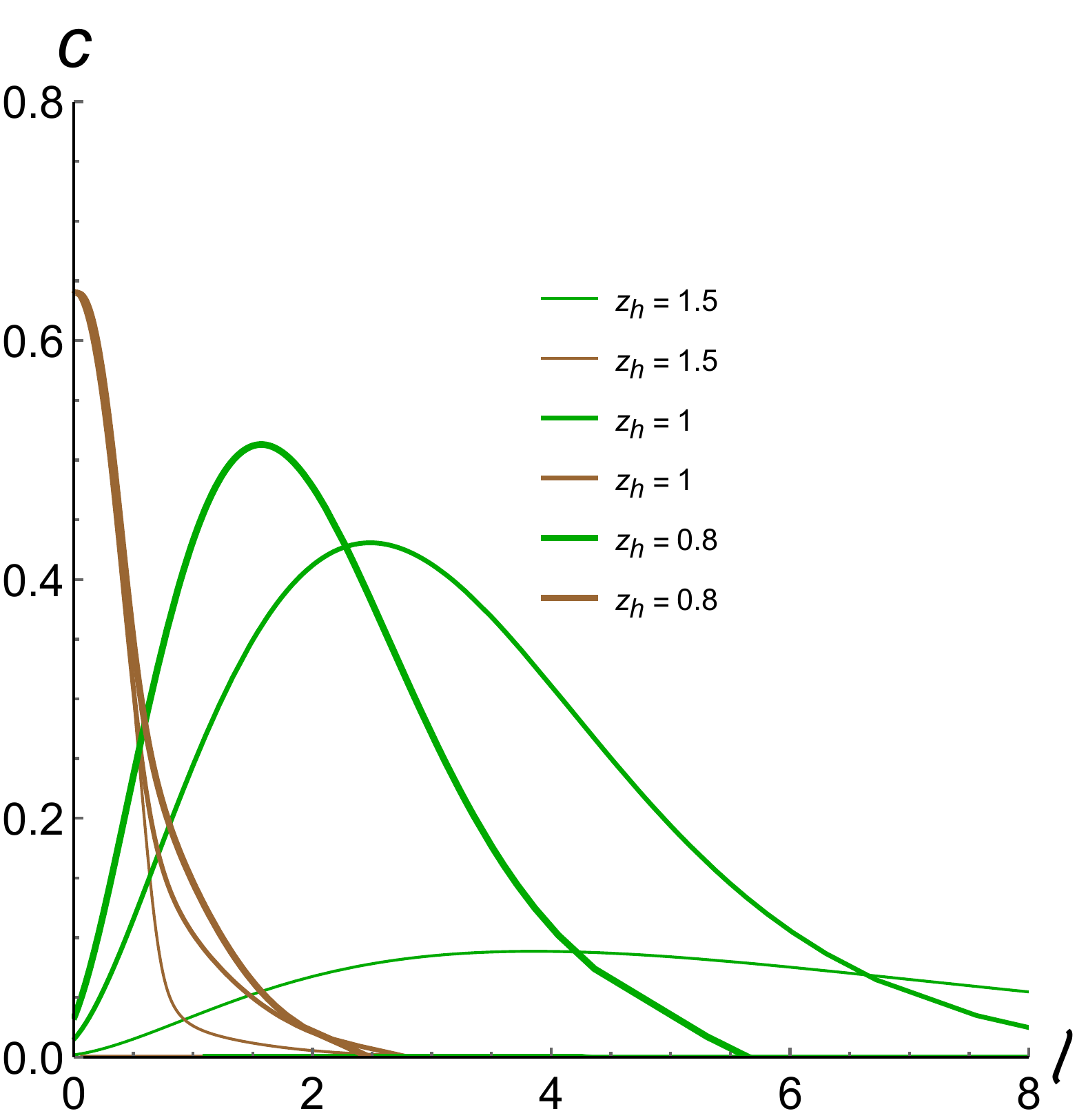}
     \caption{  Comparison of  dependences of $c$ on $\ell$ for  $z_h=0.8,1,1.5$ in  the isotropic case (line thickness increases with decreasing horizon) for calculations that are done in the EF (brown lines) and SF (green lines). The dynamical wall in the SF is at $z_{DW}=1.41$ and there is no dynamical wall in the EF.
    }
    \label{fig:c-b-ISO}
  \end{figure}

For the isotropic case in the EF, i.e. with $b$-factor without a dilaton,  the c-function is  defined as in the conformal invariant case 
\be
c_{\nu=1,EF}=\frac{\ell_{\nu=1}^{3}}4\,\Big(\frac{b^{3/2}(z_*)}{z_*^{1+2/\nu}}-\frac{b^{3/2}(z_D)}{z_D^{1+2/\nu}}\Big).\label{c-ISO}
\ee
In Fig.\ref{fig:c-b-ISO} we depict the $c_{\nu=1,F}$ vs $\ell_{iso}$  by  brown lines
for   $z_h=0.8,1,1.5$ (thickness increases with decreasing $z_h$).
 In all these cases     the disconnected parts hang
  up to the horizon (there is no dynamic wall in the EF) therefore in \eqref{c-ISO} $z_D=z_{h}$.

  For comparison, in Fig.\ref{fig:c-b-ISO} the  dependences of $c_{\nu=1,SF}$ on $\ell$, i.e. for calculations 
     that are done in the SF with 
     $b_S(z)$, 
     \bea
c_{\nu=1,SF}=\frac{\ell_{\nu=1}^{m_{\nu=1,SF}}}4\,\Big(\frac{b_s^{3/2}(z_*)}{z_*^{1+2/\nu}}-\frac{b_s^{3/2}(z_D)}{z_D^{1+2/\nu}}\Big)\label{C-iso-SF}
\eea
 are shown by green  lines. 
In this case there is the dymamical wall  and if it is not covered by the horizon, $z_{DW}=1.413<z_h$, the disconnected parts hang up to the dynamical wall. For the cases depicted in Fig.\ref{fig:c-b-ISO} this corresponds to 
 the cases $z_h=0.8$ and $z_h= 1$. Note that $z_h=1.3$ 
     corresponders to the thermodynamically unstable phase. \\
     
    We observe a completely different behavior of green and brown curves. There are saddle points on  the green lines, but not on the brown ones. These saddle points are related with existence of solution of the equation
    \be
   \label{saddle-point eq} 
c'_{\nu,\varphi,F}(z_{0})=0=\ell \,{\cal V}'+m_F\,\ell'\Big({\cal V}(z_0)-{\cal V}(z_D)\Big).
\ee

Notice, that by definition the DW corresponds to ${\cal V}'\Big|_{z=z_{DW}}=0$, but $\ell\to \infty$ at $z\to z_{DW}$.
In the second term the difference is zero at $z= z_{DW}$, but $\ell'\to \infty$ and we have uncertainty at $z\to z_{DW}$.
Calculations show that $c_{SF}'\to 0$ for $z\to z_{DW}$.

 \subsubsection{The c-function in the anisotropic case}
  In  anisotropic cases the definition of  the c-function  is modified \cite{Chu:2019uoh,Ghasemi:2019xrl,Hoyos:2020zeg}
\be
c_{\varphi}=\frac{\ell_{\varphi}^{m_\varphi}}4\,\Big(\frac{b^{3/2}_s(z_*)}{z_*^{1+2/\nu}}-\frac{b^{3/2}_s(z_D)}{z_D^{1+2/\nu}}\Big).\label{c-aniz}
\ee
  Here the power $m_\varphi$ depends on the orientation.
    For the particular cases of the transversal and longitudinal orientations these powers, 
    $m_y$ and $m_x$, are defined according to different scaling in  transversal and longitudinal directions.
    We use here  the scaling factors $m_\varphi$ defined in accordance with the behavior of the metric in the UV region. Note that since the 
    behavior of  $b_s(z)$ and $b(z)$ are different in the UV region, we get different factors $m_\varphi$ at different frames, and we put a subscript  $m_{\varphi,F}$ to indicate this; $F=EF$ for the Einstein frame and $F=SF$ for the string frame (we have  the same for  the isotropic case).
    The factor $m_{\varphi,F}$ is defined according to the formula  \eqref{mSFx} and \eqref{mSFy} and for $\nu=4.5$ we have  \eqref{mSFx-45} and \eqref{mSFy-45}.\\

    \begin{figure}[h!]\centering
    \includegraphics[width=6cm]{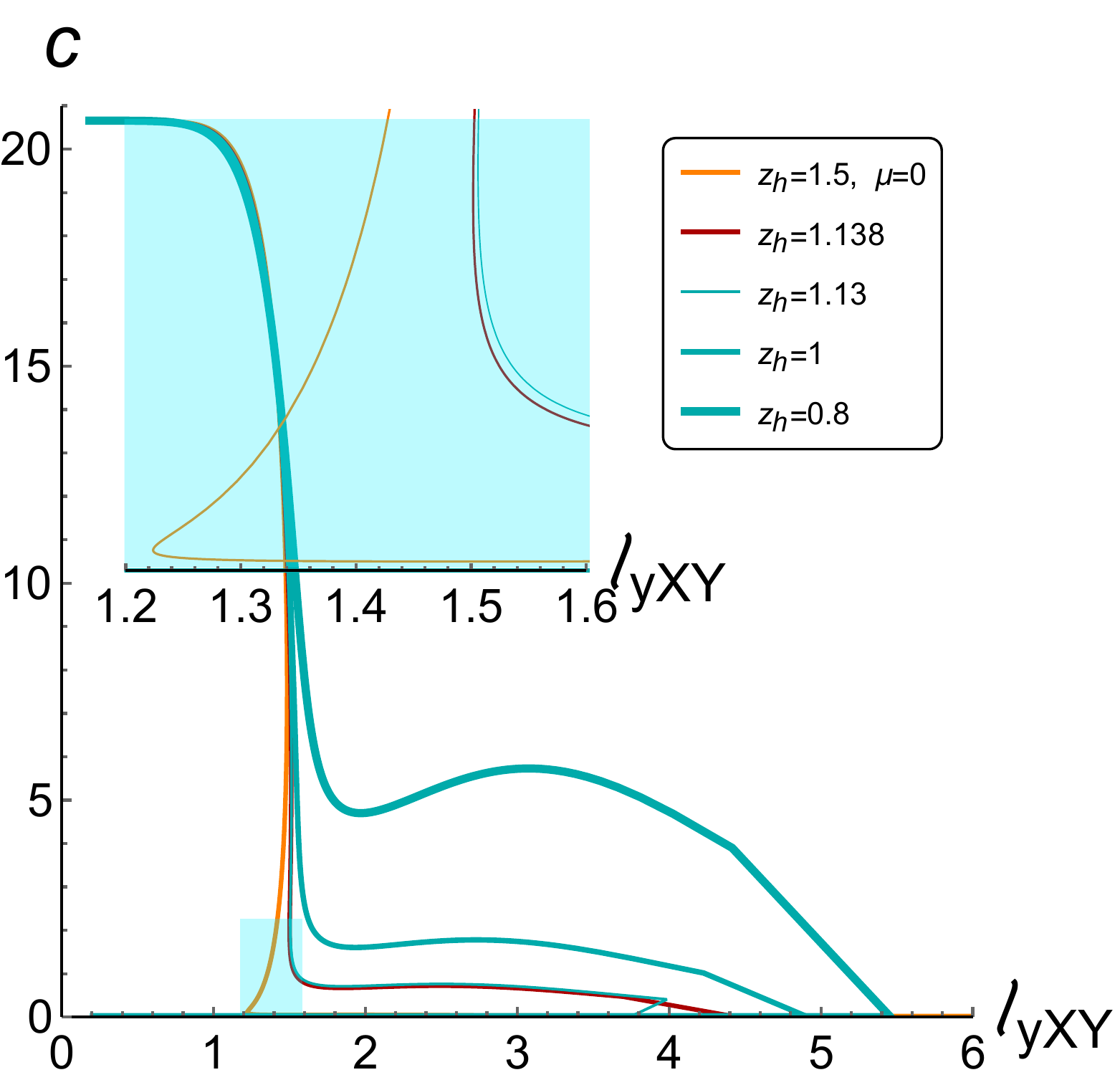}\qquad\qquad\includegraphics[width=6cm]{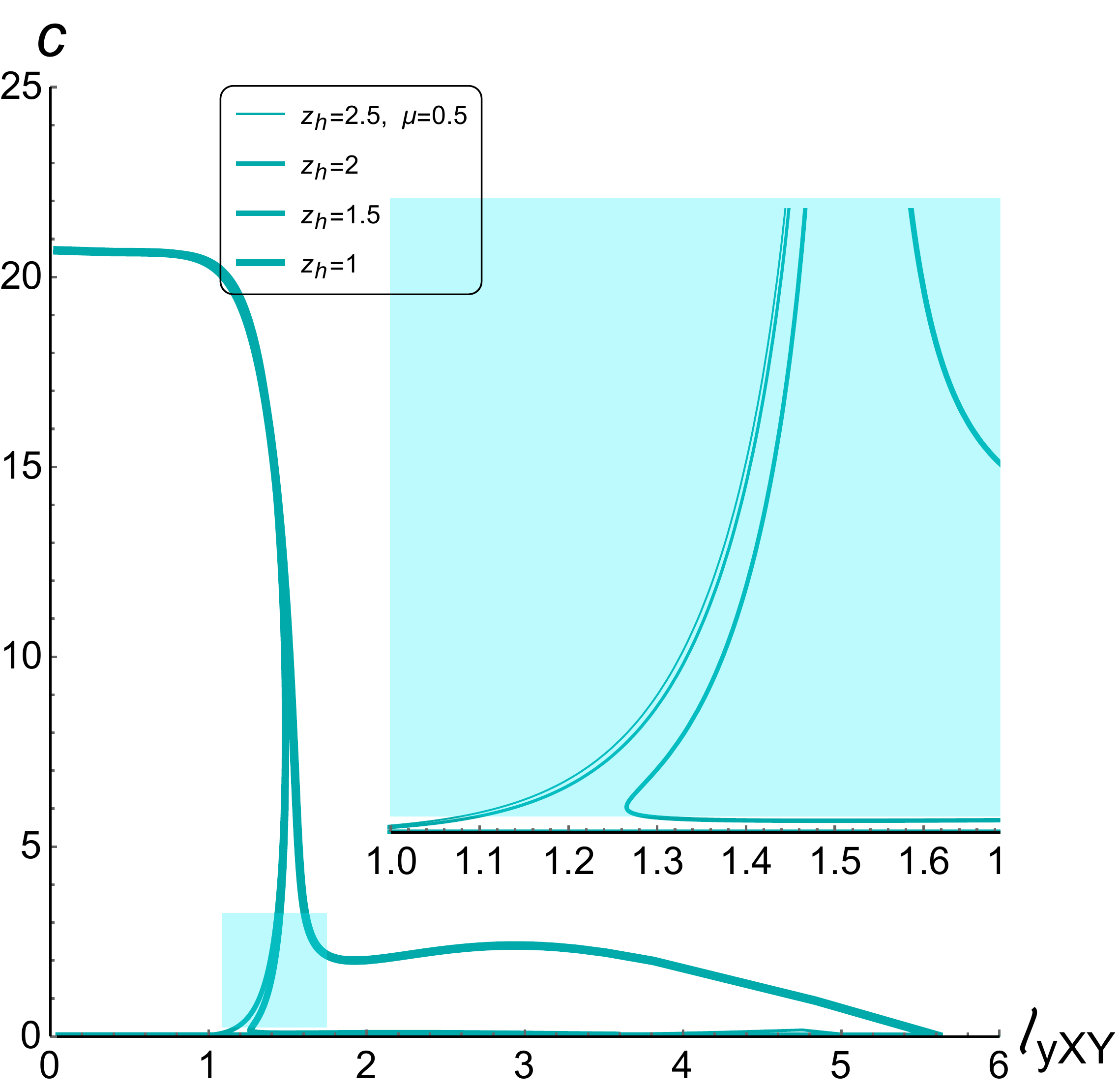}\\
    {\bf A)} \quad\quad\quad\quad\quad\qquad\qquad\qquad{\bf B)} \\
 \includegraphics[width=6cm]{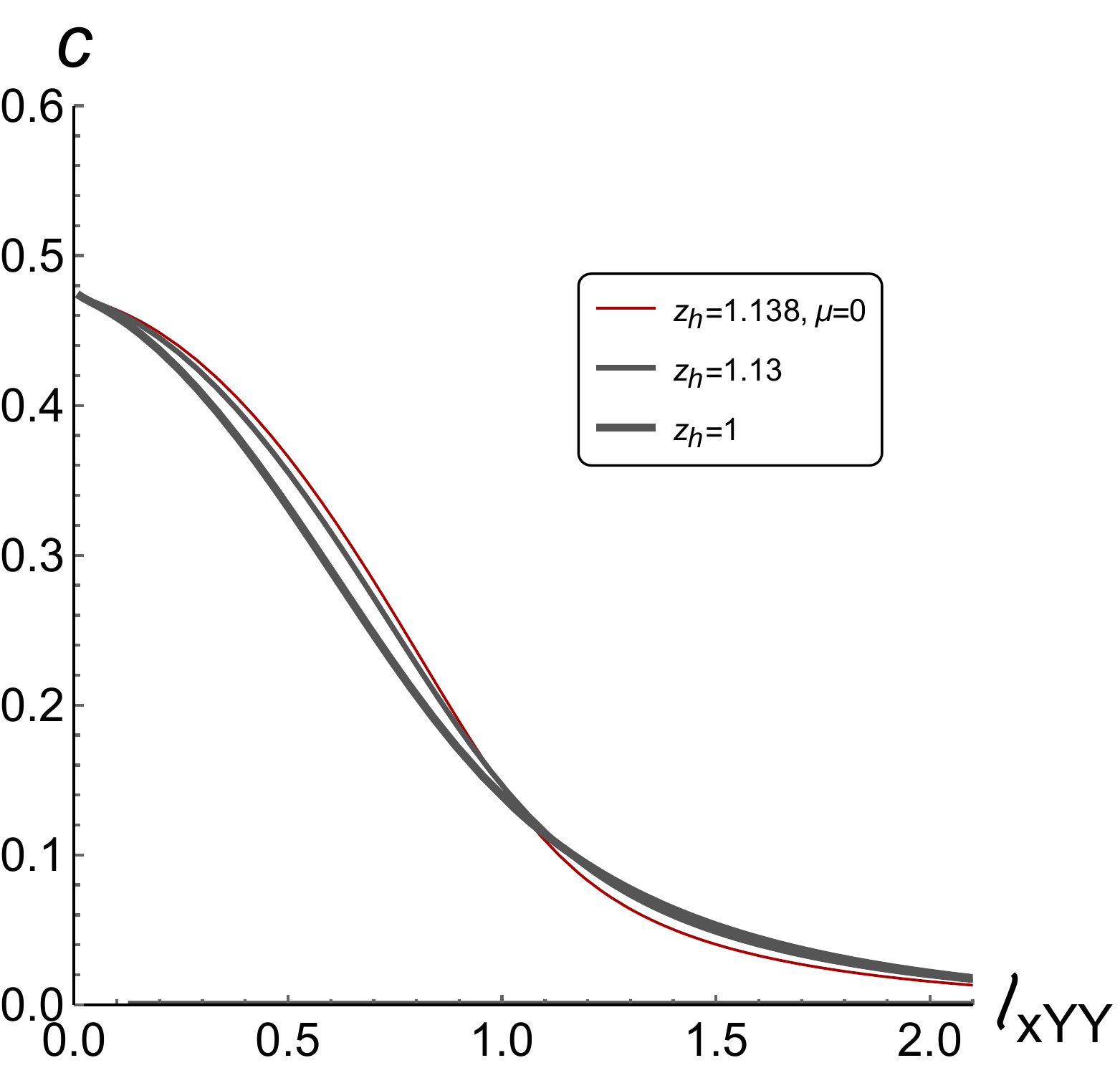}\qquad\qquad
  \includegraphics[width=6cm]{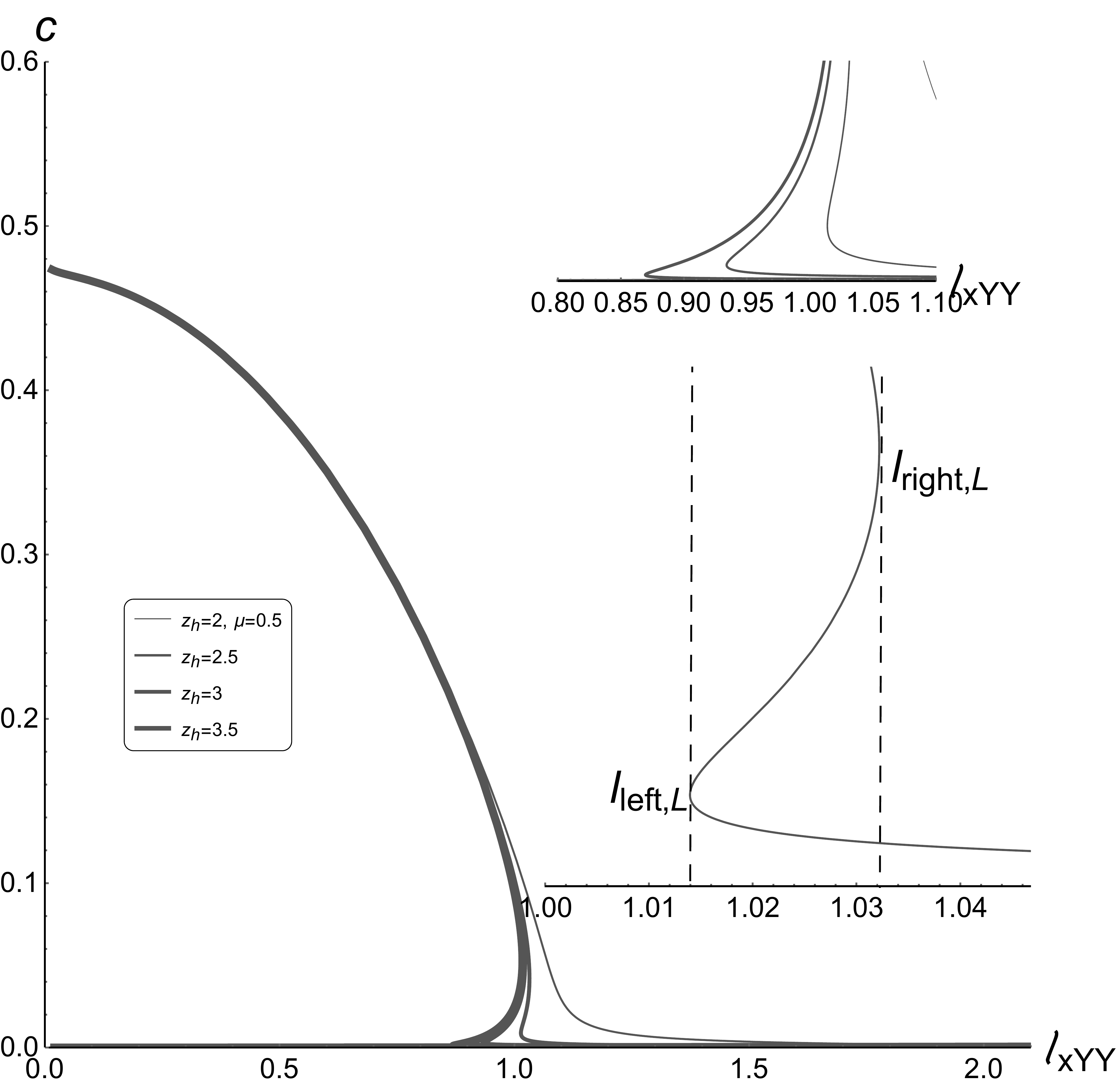}\\{\bf C)} \quad\quad\quad\quad\quad\qquad\qquad\qquad{\bf D)} \\
     \caption{The dependences of $c$, defined by \eqref{c-aniz}, on $\ell$ for $\nu=4.5$, and  various values of $z_h$  for calculations  done in the EF.  {\it Top line}: {\bf A)} transversal orientation and EF, $z_h=1,1.13$ and $z_h=z_{HP}=1.138$ (red lines), here we also insert the lines with an unstable value of $z_h=1.5$ (unstable point, orange line) and a stable one $z_h=0.8$ (the inset shows the multi-valued behavior of $ c = c (\ell) $ in the region near $ z = z_ {HP} $, indicated by a light blue rectangle);  $\mu=0$. {\bf B)} the same  for $\mu=0.5$. 
     {\it Bottom line}: longitudinal orientation and EF for {\bf C)} $\mu=0$ and  {\bf D)} $\mu=0.5$ (the insets show the multi-valued behavior of $ c = c (\ell) $; here $\ell_{left,L}$  and $\ell_{right,R}$ are indicated).
}
    \label{fig:c-b-LT-EF}
  \end{figure}

  \begin{figure}[h!]\centering
    \includegraphics[width=6cm]{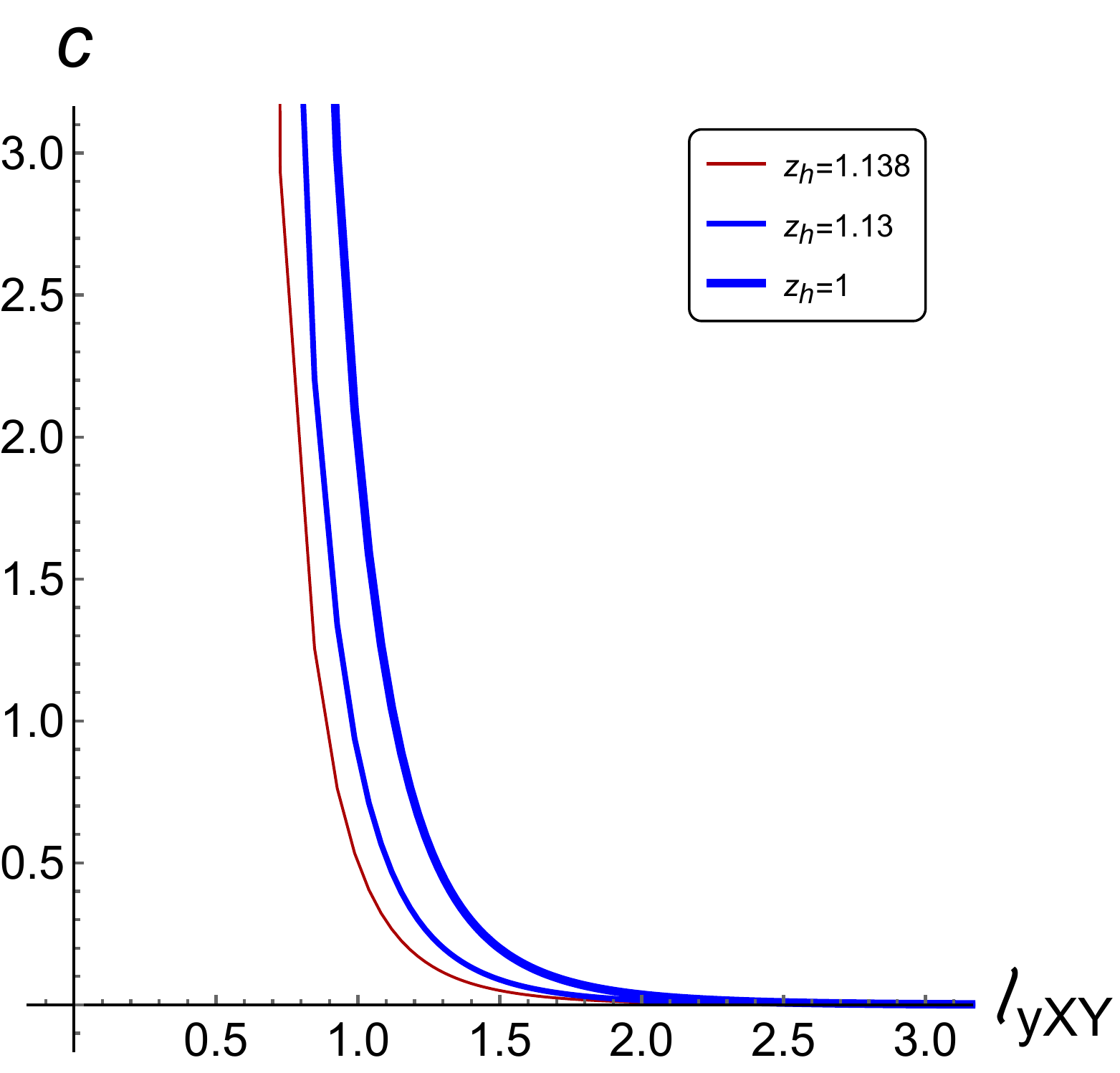}\qquad\qquad
  \includegraphics[width=6cm]{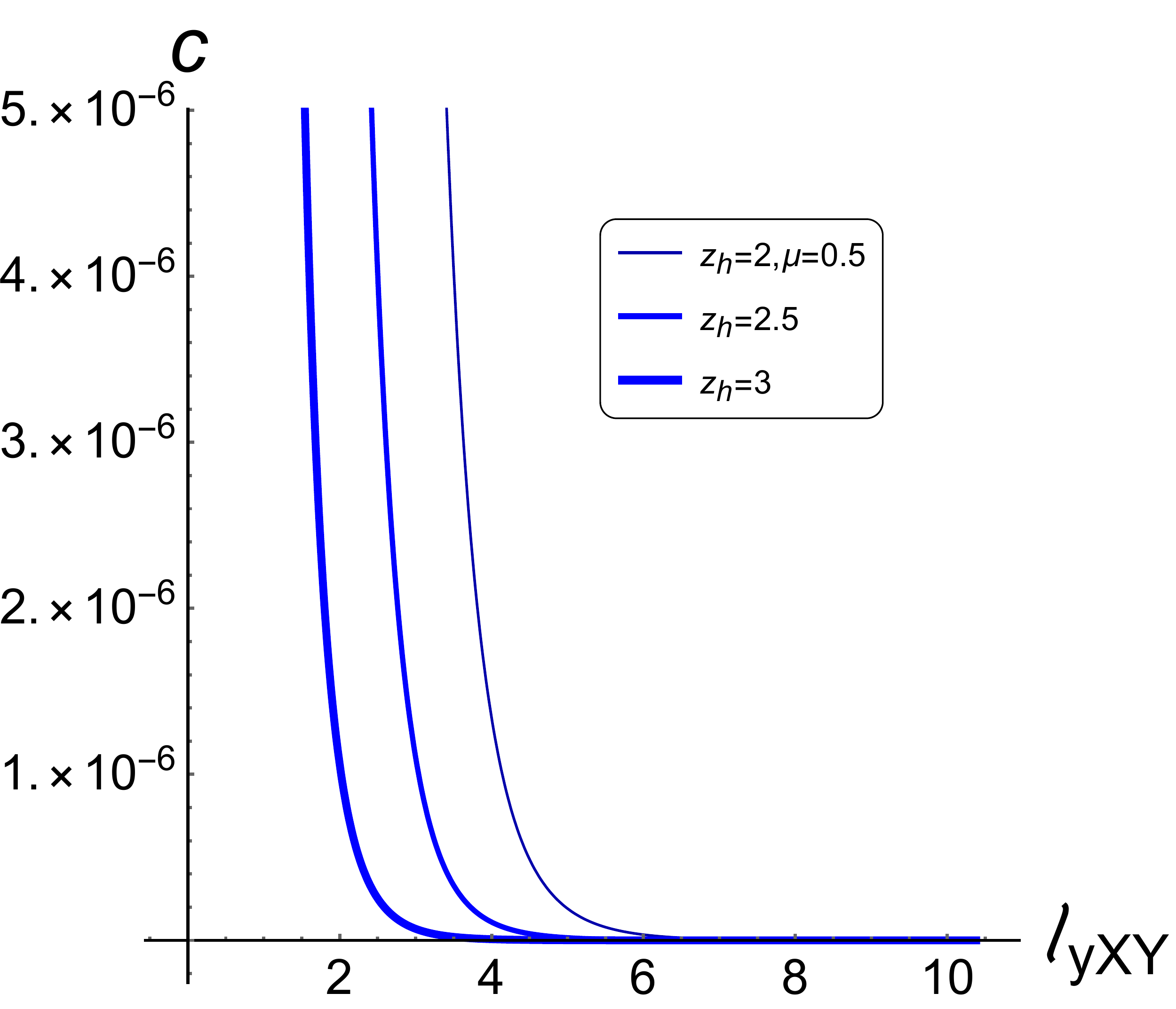}\qquad\qquad\qquad
  \\
 {\bf A)}\qquad\qquad\qquad\qquad\qquad\qquad\qquad\qquad{\bf B)}\\
 \includegraphics[width=6cm]{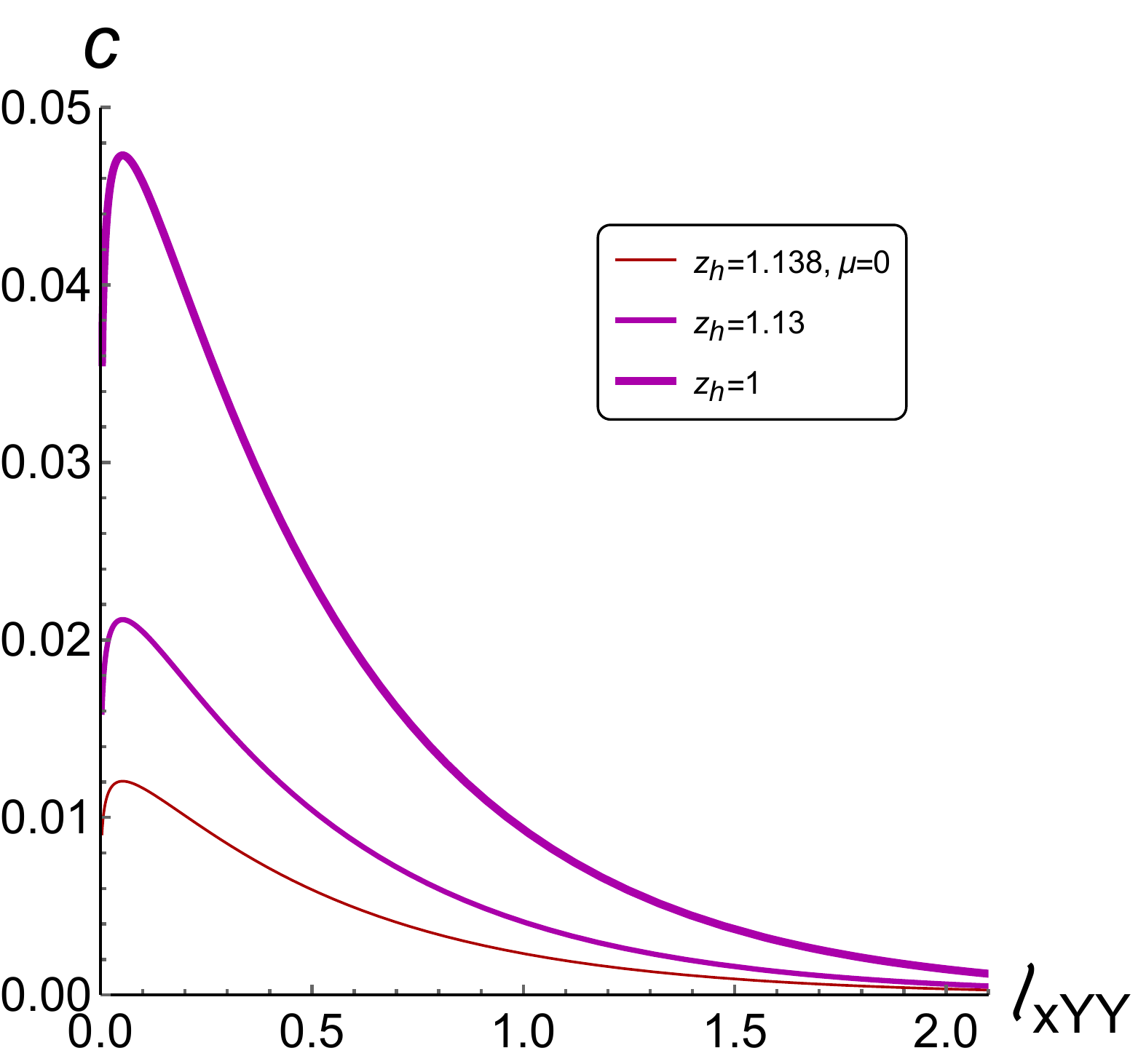}\qquad\qquad
  \includegraphics[width=6cm]{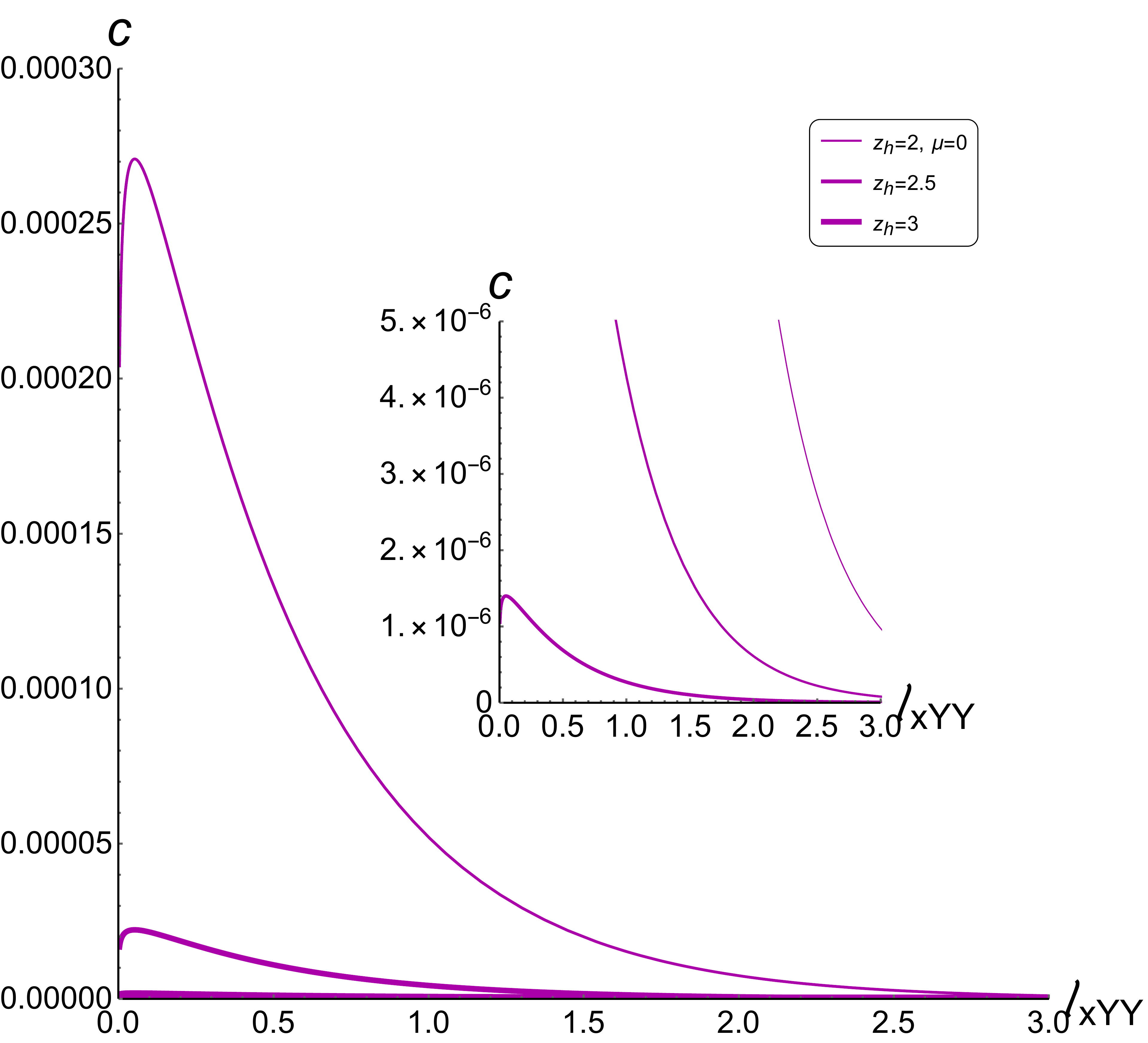}\\
 \,{\bf C)}\,\qquad\qquad\qquad\qquad\qquad\qquad\qquad\qquad{\bf D)}
     \caption{The dependences of $c$, defined by \eqref{c-aniz}, on $\ell$ for $\nu=4.5$, and  various values of $z_h$  for calculations  done in the  SF. 
      {\it Top line}: transversal orientation for  {\bf A}) $\mu=0$ and  {\bf B}) $\mu=0.5$.
       {\it Bottom line}:  longitudinal orientation for  {\bf C}) $\mu=0$ and {\bf D}) $\mu=0.5$.  In the SF the dynamical wall is at $z_{DW}=0.34$. 
 }
    \label{fig:c-b-LT-SF}
  \end{figure}
  
  In Fig.\ref{fig:c-b-LT-EF}  and Fig.\ref{fig:c-b-LT-SF}  we present dependences of $c$ on  $\ell$ for $\nu=4.5$, various $z_h$ 
and $\mu=0,0.5$ calculated  in the EF and SF for transversal and  longitudinal orientations.  
  We see that we get rather different behavior in $EF$ and $SF$. 
\begin{itemize}
\item We see in Fig.\ref{fig:c-b-LT-EF}.{\bf A)} that in the transversal case the c-function has rather nontrivial behavior in the EF for $\mu=0$  already. Namely we see that there is a region of $z_h$, here we see multivalued c-function as function of $\ell$. This  takes place not only at nonstable points (one of them is shown by orange), but also at $z_{h,HP}=1.138$, and this multivalued phenomena  disappears  at $z_h=1.13$.
We also see that c-function has local minimum and maximum at the points $\ell_{min}$ and $\ell_{max}$ correspondingly, and there is a region of $\ell_{min}<\ell<\ell_{max}$ where $c$ increases when 
$\ell$ increases.
\item We see in Fig.\ref{fig:c-b-LT-EF}.{\bf B)}  that this multi-validity  is preserved at $\mu=0.5$ and the region of $z_h$
where this multi-validity  takes place,
 is  wider in comparison with $\mu=0$ case.
\item The c-function for the longitudinal orientation for $\mu=0 $  monotonically decreases.
\item While at  $\mu=0.5 $ and rather large $z_h$ (small BH), the  c-function exhibits a non-monotonic behavior around some $\ell_{cr}$ (see the plot in the inset of Fig.\ref{fig:c-b-LT-SF}.{\bf D)}).
\item In the SF  the c-function for the transversal orientation decreases monotonically.
\item For the longitudinal orientation there is a saddle point $\ell_{0}$ (which depends on $z_h$ and $\mu$) and  the c-function starts to  decrease
only for  $\ell>\ell_{0}$.
\end{itemize}

\subsection{Origin of non-monotonic behavior of  c-functions}\label{sect:origine}
Let us make a few comments about the origin of the behavior of the c-functions presented in Fig.\ref{fig:c-b-LT-EF} and 
Fig.\ref{fig:c-b-LT-SF}.
  To understand the origin of such an untypical behavior we first consider the dependence of c-function on $z_*$ for the cases presented here, and then incorporate the dependence of $\ell $ on $z_*$.

\subsubsection{The c-function as a function of $z_*$}

   \begin{figure}[h!]\centering
\includegraphics[width=4cm]{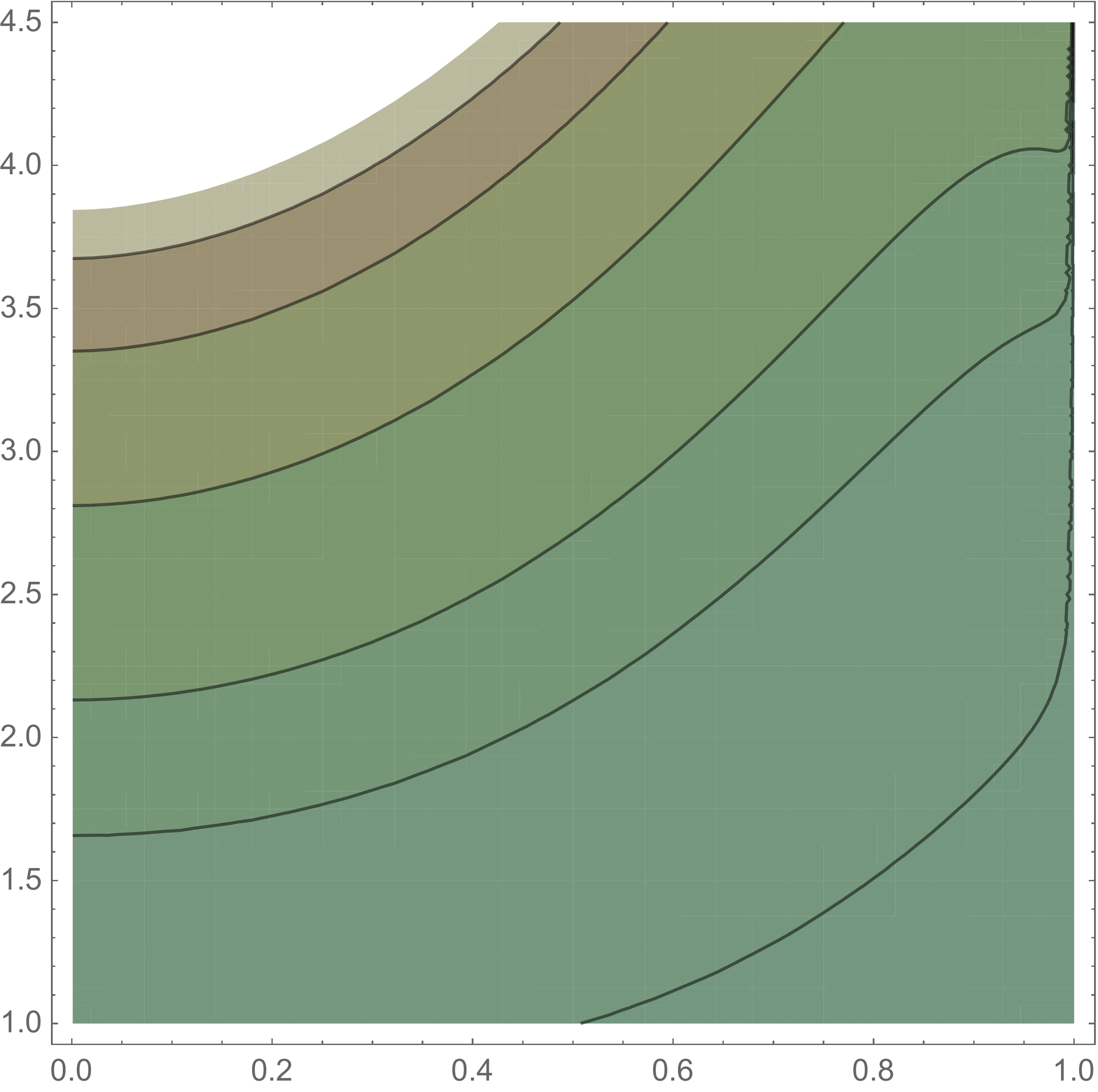}
\includegraphics[width=0.5cm]{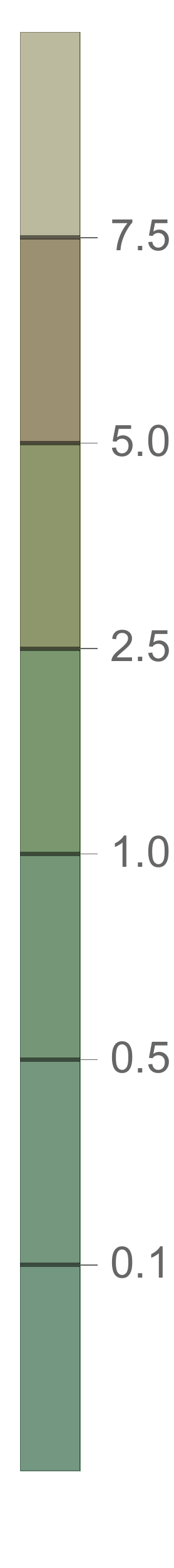}
\includegraphics[width=4cm]{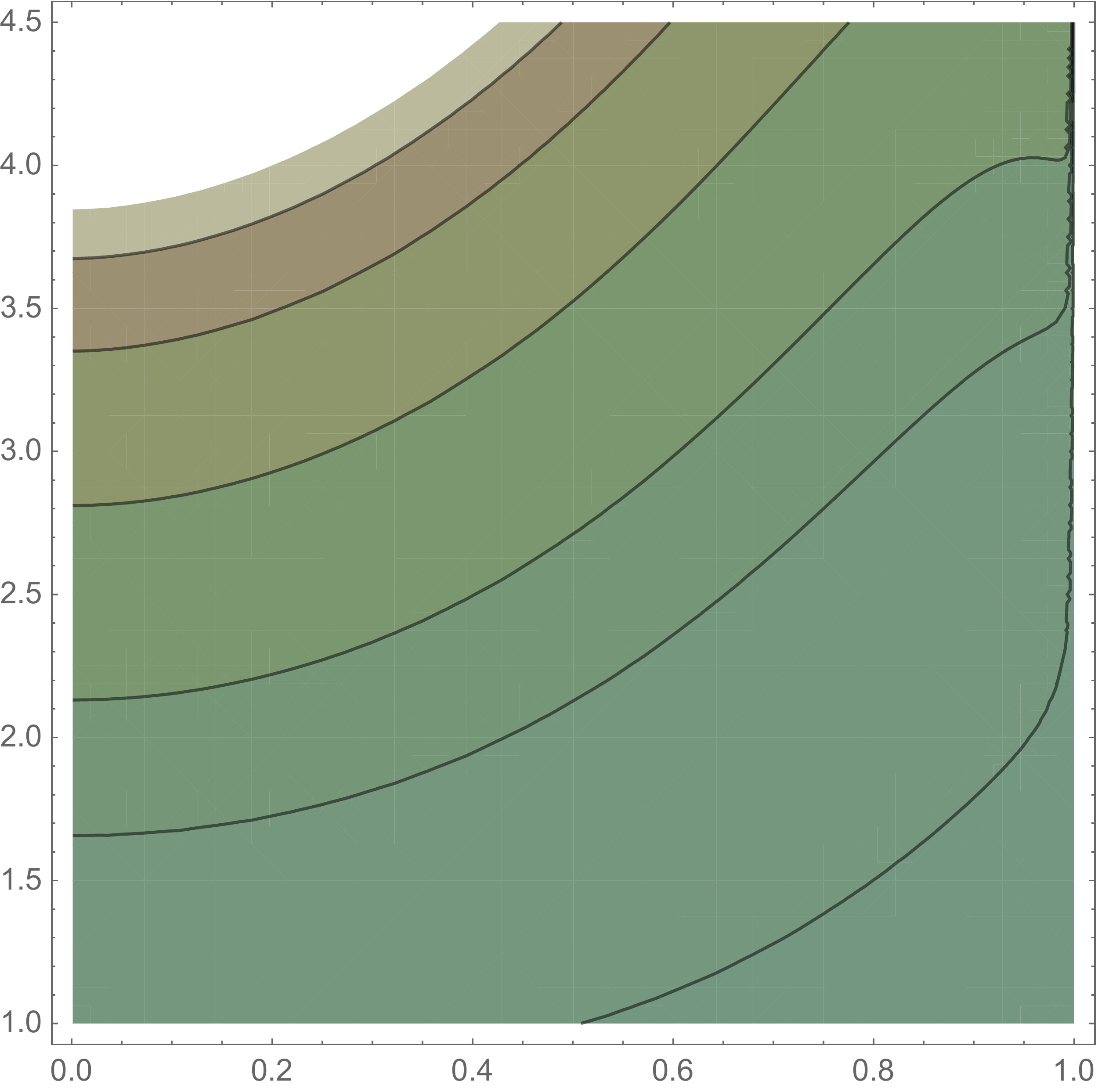}\quad\quad
\includegraphics[width=4cm]{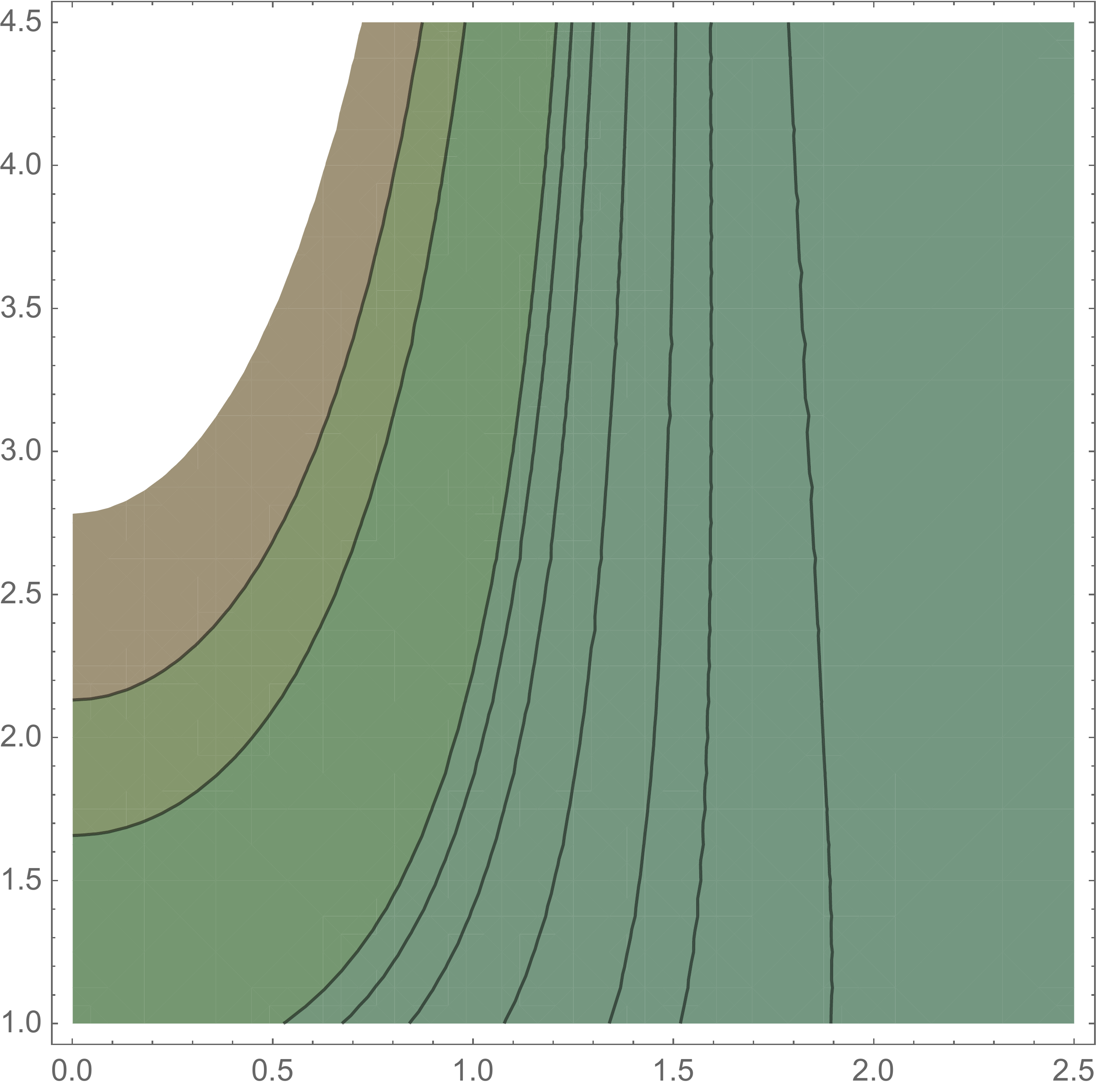}
\includegraphics[width=0.7cm]{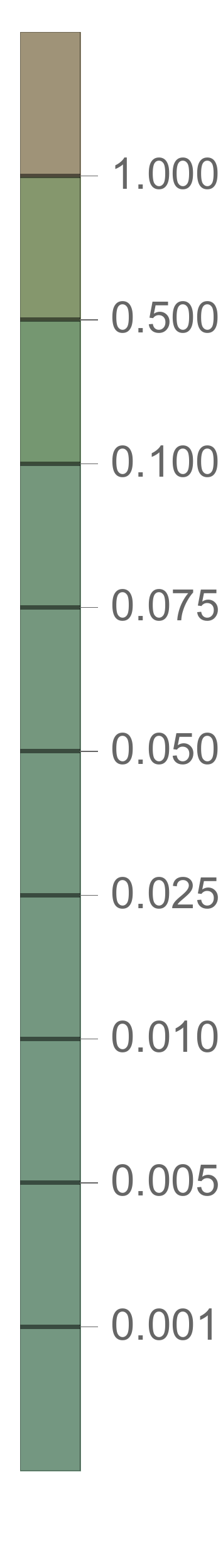}\\
\,{\bf A)}\,\qquad\qquad\qquad\qquad{\bf B)}\qquad\qquad\qquad\qquad{\bf C)}\\
$\,$\\
\includegraphics[width=4cm]{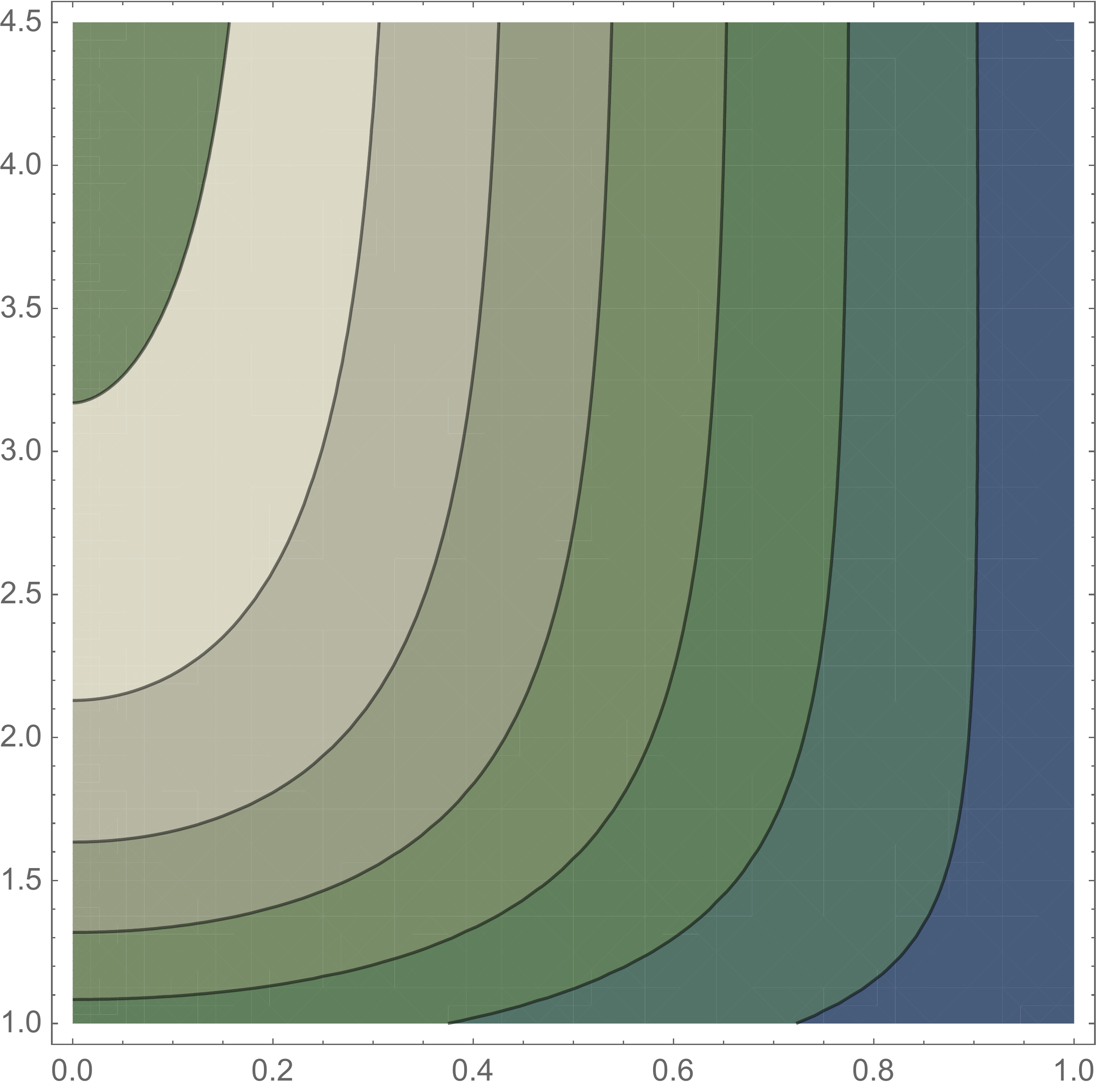}
\includegraphics[width=.6cm]{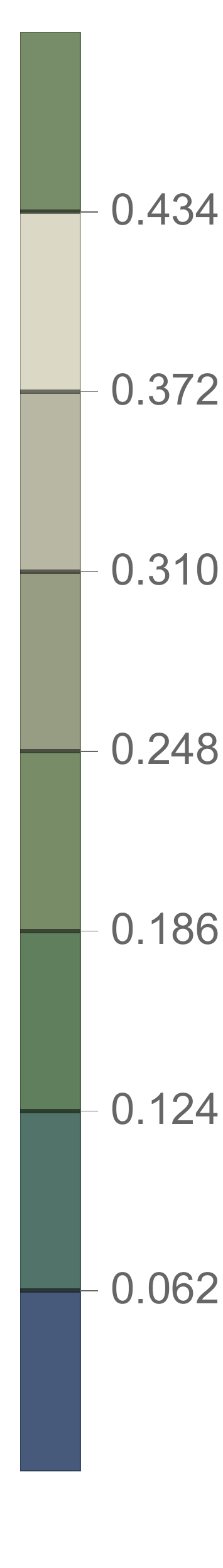}
\includegraphics[width=4cm]{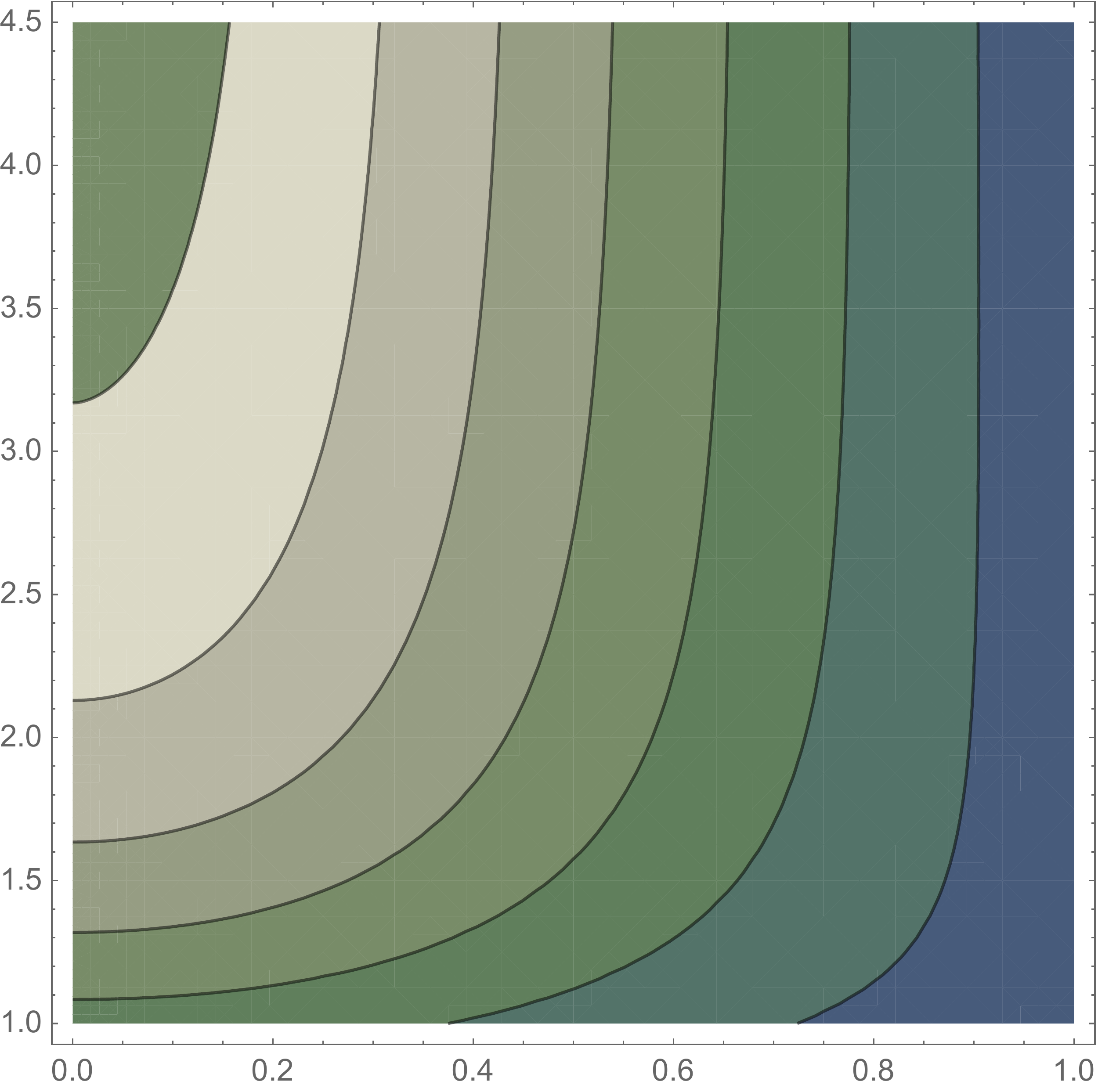}\quad\quad
\includegraphics[width=4cm]{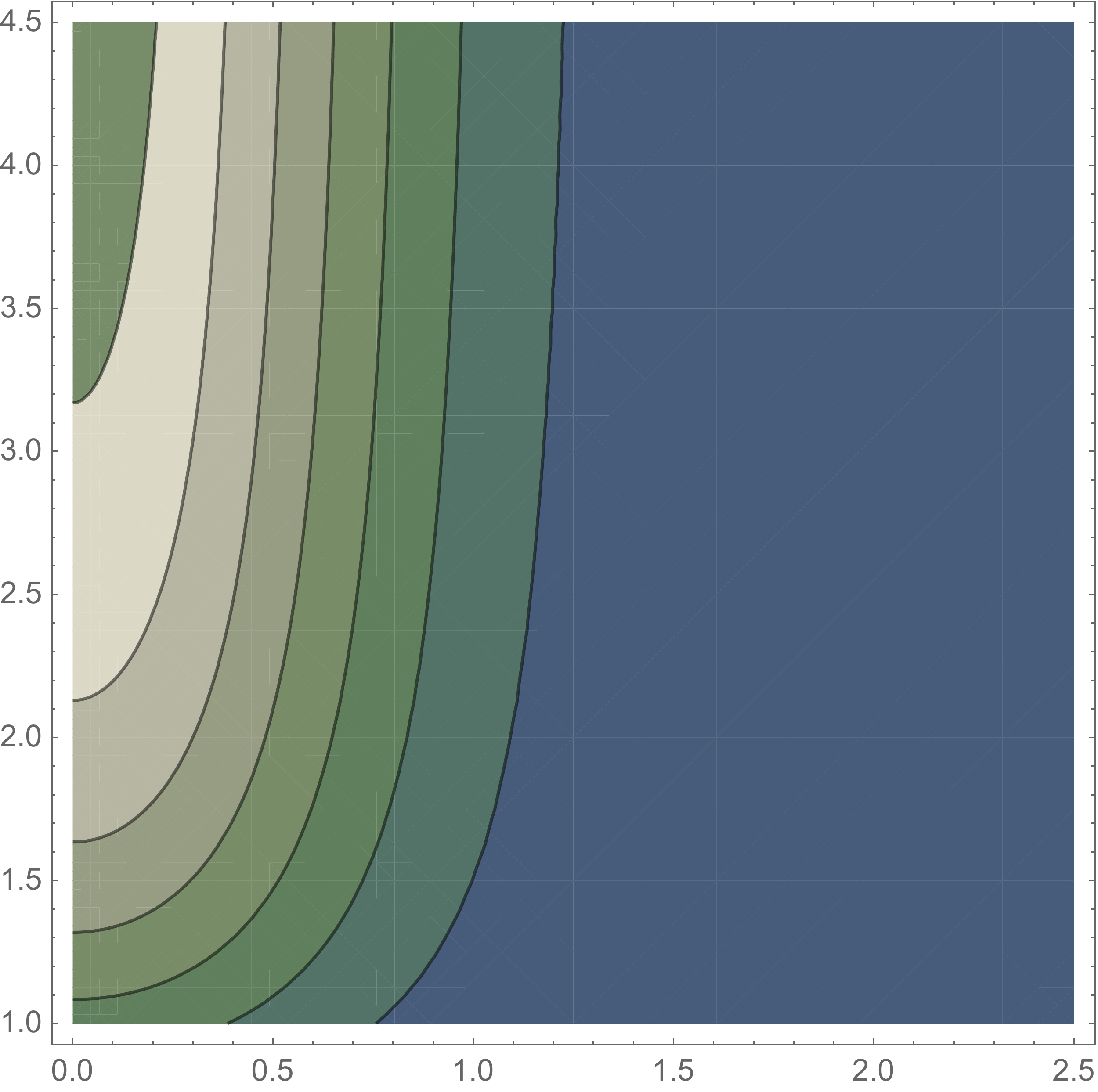}
\includegraphics[width=.6cm]{APSfig/c-nu-EF-l-leg.pdf}
\\ \,{\bf D)}\,\qquad\qquad\qquad\qquad{\bf E)}\qquad\qquad\qquad\qquad{\bf F)}\\$\,$\\
 \caption{{\it Top line}: $\ell_\phi ^{m_F(\nu)}\eta_{CD}(\nu)$ vs $z_*$ for F=EF  and  $\varphi=\pi/2$ (transversal case) at {\bf A)} $z_h=1$, $\mu=0$, {\bf B)} $z_h=1$, $\mu=0.2$   and 
 {\bf C)} $z_h=2.5$, $\mu=0.2$ . $\,$ 
 {\it Bottom line}:  $\ell_\phi ^{m_F(\nu)}\eta_{CD}(\nu)$ vs $z_*$ for F=EF  and 
$\varphi=0$ (longitudinal case)  at {\bf D)} $z_h=1$, $\mu=0$,  {\bf E)} $z_h=1$, $\mu=0.2$ and  {\bf F)}   $z_h=2.5$, $\mu=0.2$. }
    \label{fig:eta-zs-EF}
  \end{figure} 

Fig.\ref{fig:eta-zs-EF}   shows  contour plots of $\ell_\phi ^{m_F(\nu)}\eta_{_{CD}}(\nu)$ over the ($z_*$,$\nu$) plane  for  the EF  at different $z_h$ and  $\mu$, and different orientations ({\it top line} for the transversal case and {\it bottom line} for the longitudinal one).
All these plots demonstrate that the c-function as a function of $ z _* $ monotonically decreases while increasing $ z_* $.
One can also see that the c-function essentially depends on the orientation,  c-function is larger for the transversal case for the same thermodynamical parameters
and the same $z_*$. Also   the  c-function falls with temperature  faster  in the transversal case. 
Note that the $c(z_*)$ does not change substantially when we change the chemical potential.
This is due to the fact that the c-function depends mainly on the effective potential, which does not depend on the blackening function.
There is a weak dependence of the form of the function $ c = c (z _ *) $ on $ z_h $ and $ \mu $, compare the graphs on the same lines in Fig.\ref{fig:eta-zs-EF}.\\

\begin{figure}[h!]\centering
$$\,$$\\
\includegraphics[width=6cm]{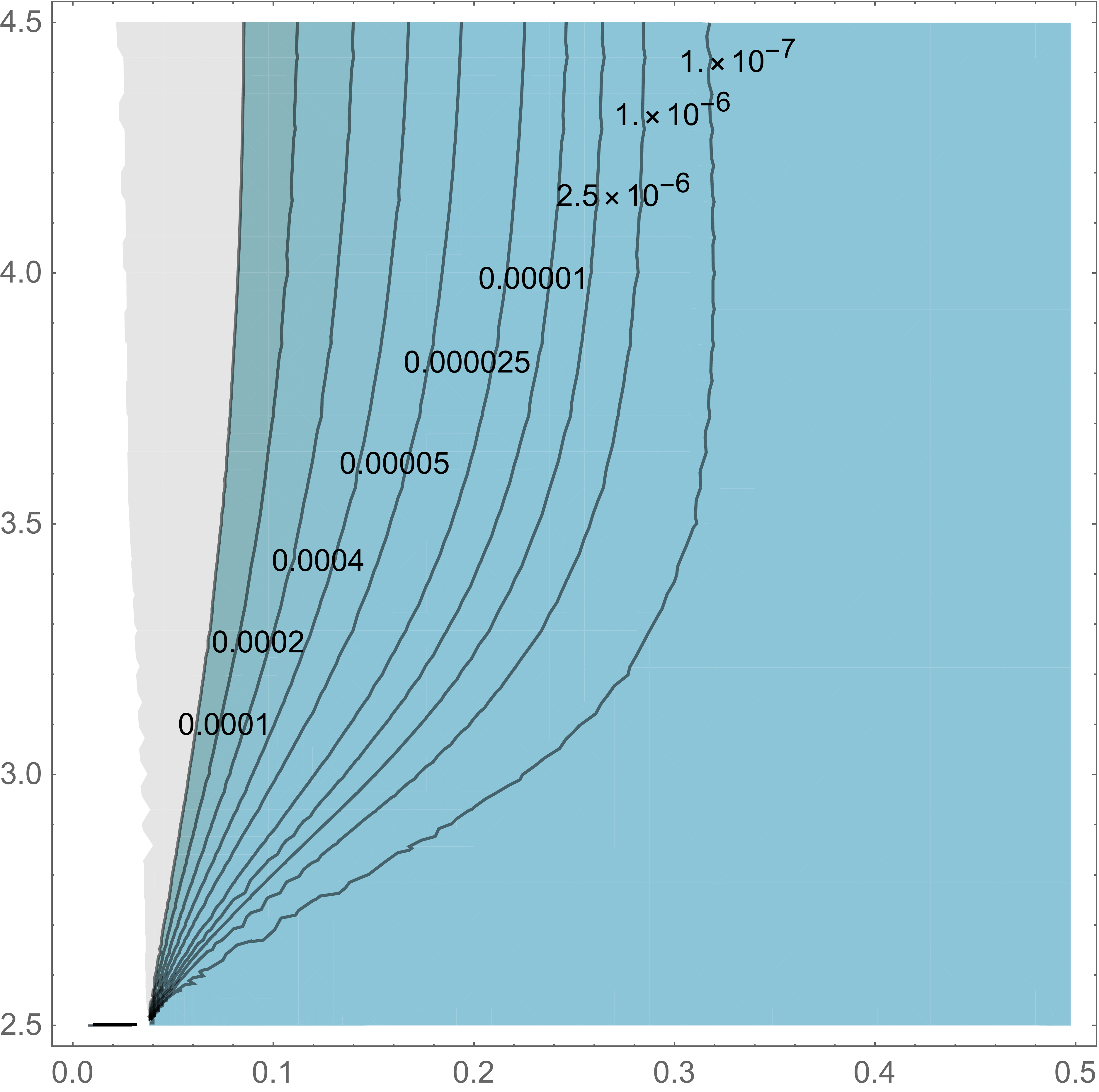}\qquad
\qquad\qquad
\includegraphics[width=6cm]{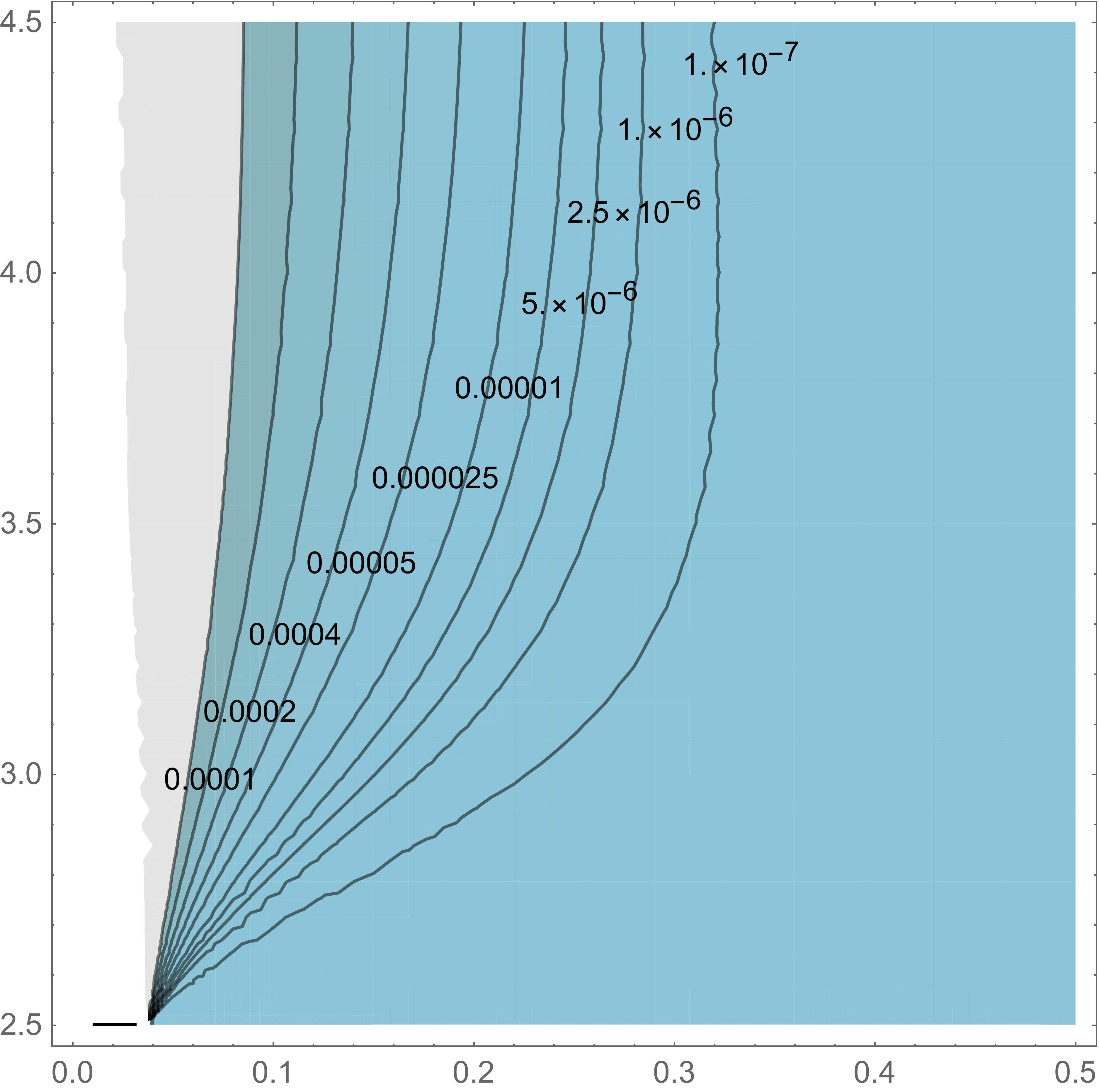}
\\
{\bf A)}\qquad\qquad\qquad\qquad\qquad\qquad\qquad{\bf B)}\\$$\,$$\\
\includegraphics[width=6cm]{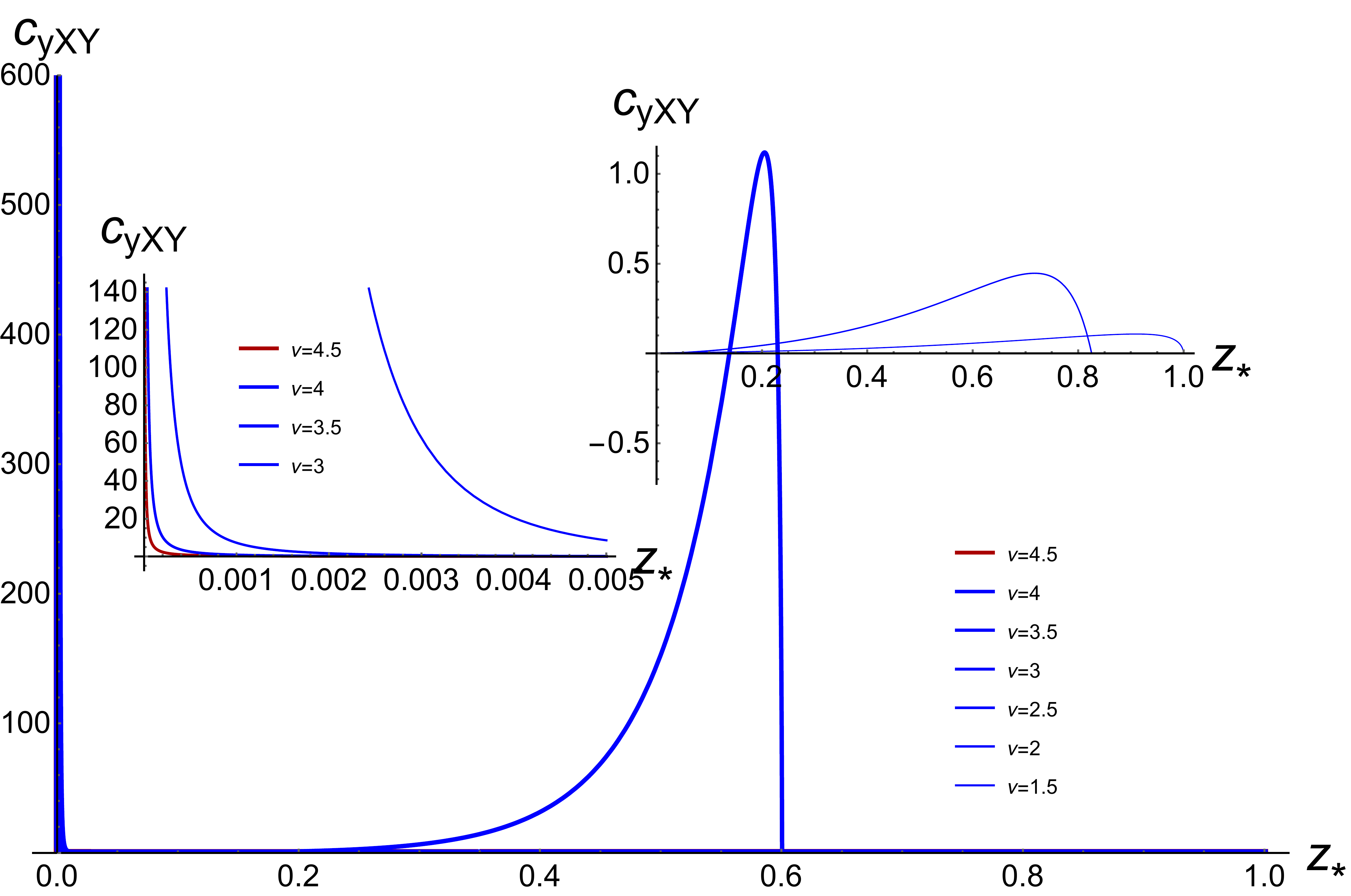}\qquad\qquad\qquad
\includegraphics[width=6cm]{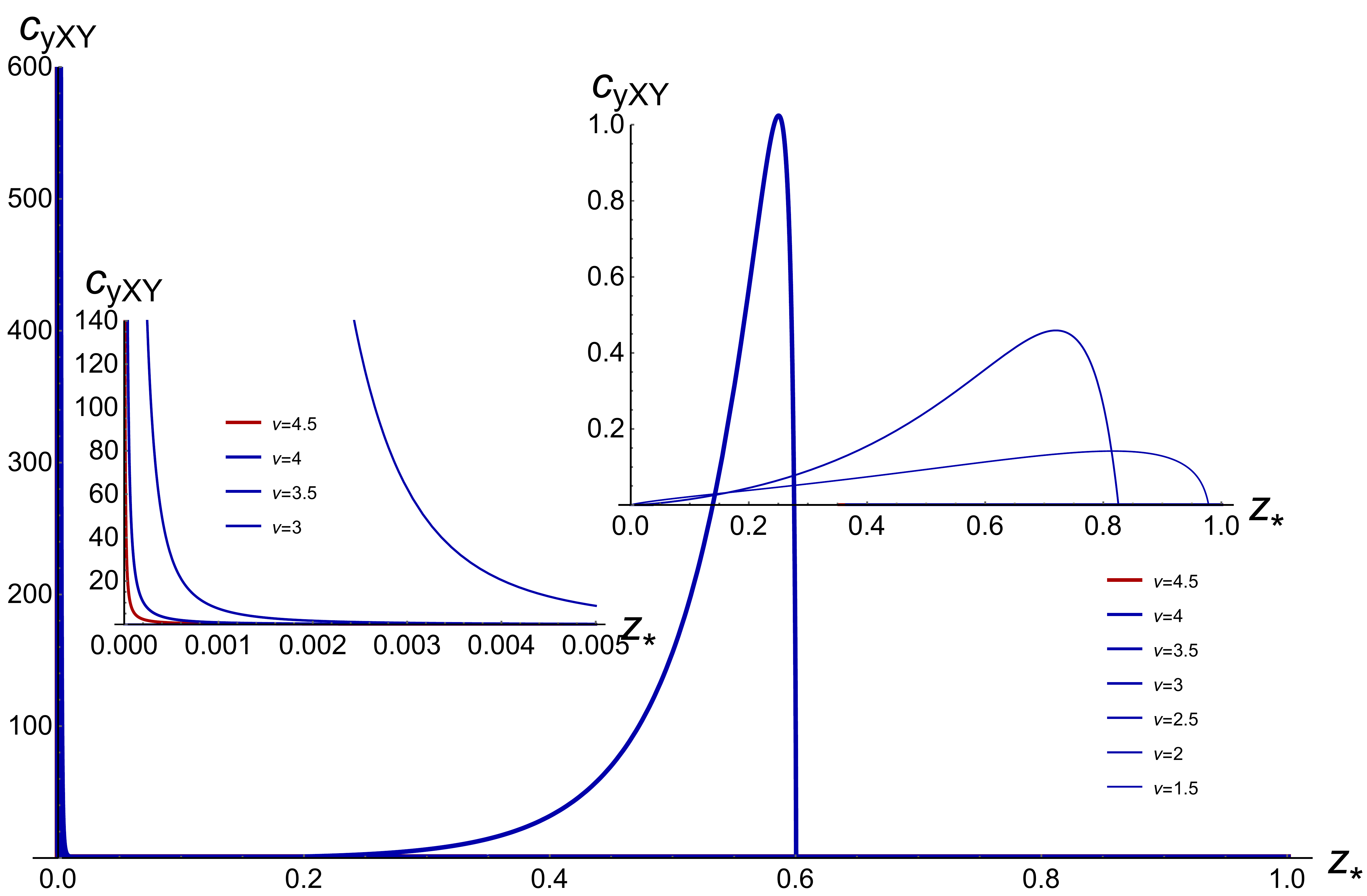}\\
 {\bf C)}\qquad\qquad\qquad\qquad\qquad\qquad\qquad{\bf D)}
 \caption{{\it Top line}: contour plots for $\ell_\phi ^{m_F(\nu)}\eta_{_{CD}}(\nu)$ vs $z_*$ (horizontal axis) and $\nu$ (vertical axis) for F=SF at $z_h=1$ and  $\varphi=\pi/2$ (transversal case), and for different values of the chemical potential:  {\bf A)} $\mu =0.2$,     and   {\bf B}) $\mu =0.5$. We see that two contour plots are almost identical.
 {\it Bottom line}: {\bf C)} the same as in {\bf A)}   for 
 various discrete values of $\nu$; {\bf D)} the same as in {\bf B)}  for 
 various discrete values for $\nu$. We see that there is a small quantitive difference in the right inserts of both plots: 
 the coordinates of the saddle points for {\bf A)} $\mu=0.2$   are $z_*|_{\nu=1.5}=0.7146, \,\,c|_{\nu=1.5}=0.4590$
 and $z_*|_{\nu=1}=0.9361, \,\,c|_{\nu=1}=0.1170$, and for {\bf B)}
 $\mu=0.5$ are $z_*|_{\nu=1.5}=0.71748, \,\,c|_{\nu=1.5}=0.46135$
 and $z_*|_{\nu=1}=0.8489, \,\,c|_{\nu=1}=0.14450$.
}
    \label{fig:eta-zs-SFT}
  \end{figure}

  In Fig.\ref{fig:eta-zs-SFT} contour plots for $\ell_\varphi ^{m_F(\nu)}\eta_{_{CD}}(\nu)$ over  $(z_*,\nu)$-plane for the SF at $z_h=1$ and  $\varphi=\pi/2$ (transversal case), and for different values of the chemical potential are shown:  Fig.\ref{fig:eta-zs-SFT}.{\bf A}) $\mu =0.2$     and   Fig.\ref{fig:eta-zs-SFT}.{\bf B}) $\mu =0.5$. We see that two contour plots are almost identical.
 On the {\it bottom line} of   Fig.\ref{fig:eta-zs-SFT}.{\bf C)}  the plots similar to plots  in Fig.\ref{fig:eta-zs-SFT}.{\bf A})   for 
 various discrete values of $\nu$ are presented; also the similar  plots are presented in Fig.\ref{fig:eta-zs-SFT}.{\bf D)}    for 
 various discrete values of $\nu$. We see that there is a small quantitive difference in the right inserts of both plots: 
 the coordinates of the saddle points for Fig.\ref{fig:eta-zs-SFT}.{\bf A)} $\mu=0.2$ are $z_*|_{\nu=1.5}=0.7146, \,\,c|_{\nu=1.5}=0.4590$
 and $z_*|_{\nu=1}=0.9361, \,\,c|_{\nu=1}=0.1170$, and for Fig.\ref{fig:eta-zs-SFT}.{\bf B)}
 $\mu=0.5$ are $z_*|_{\nu=1.5}=0.71748, \,\,c|_{\nu=1.5}=0.46135$
 and $z_*|_{\nu=1}=0.8489, \,\,c|_{\nu=1}=0.14450$.

       \begin{figure}[h!]\centering
$$\,$$\\
\includegraphics[width=6cm]{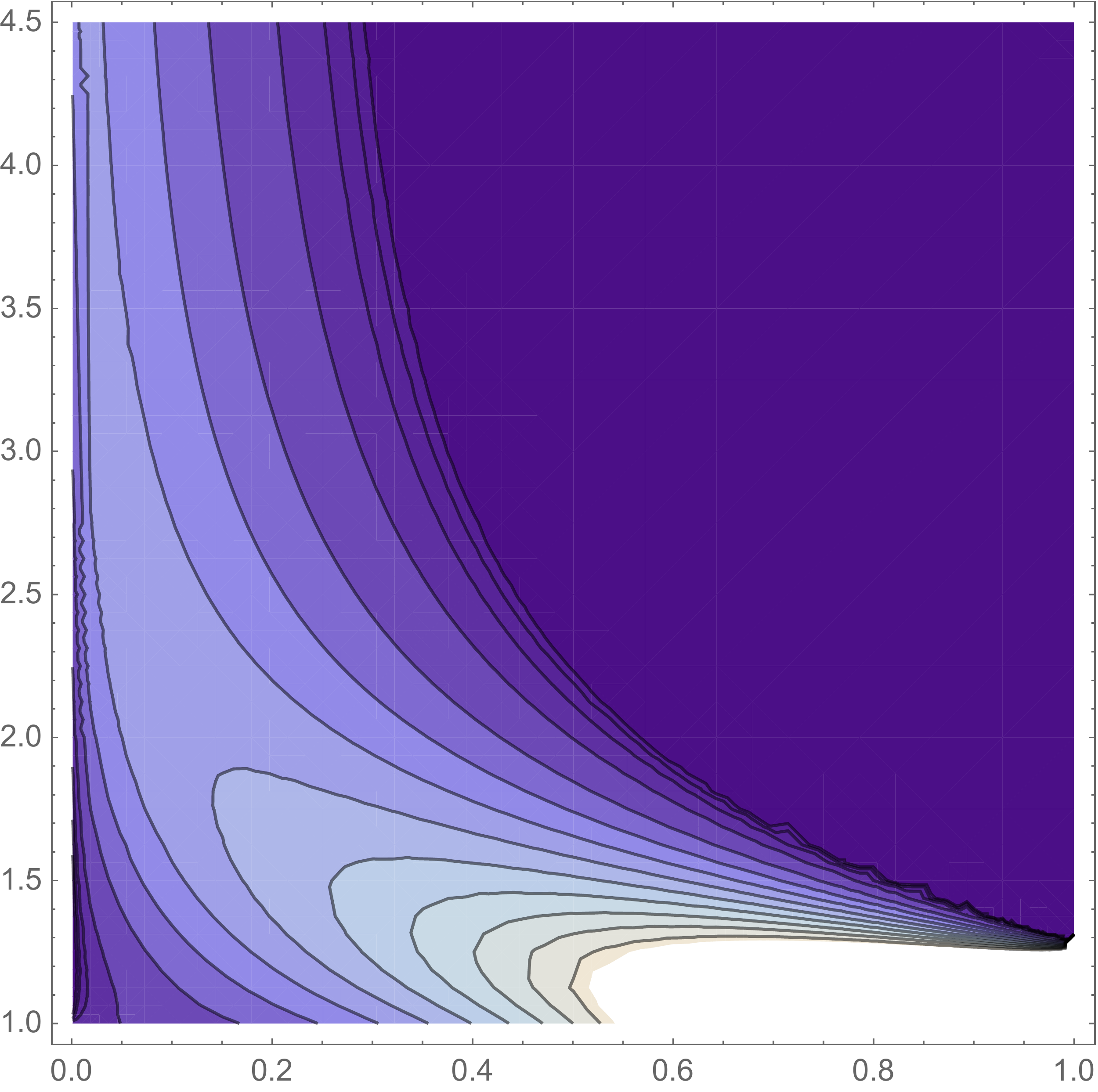}
\quad
\includegraphics[width=6cm]{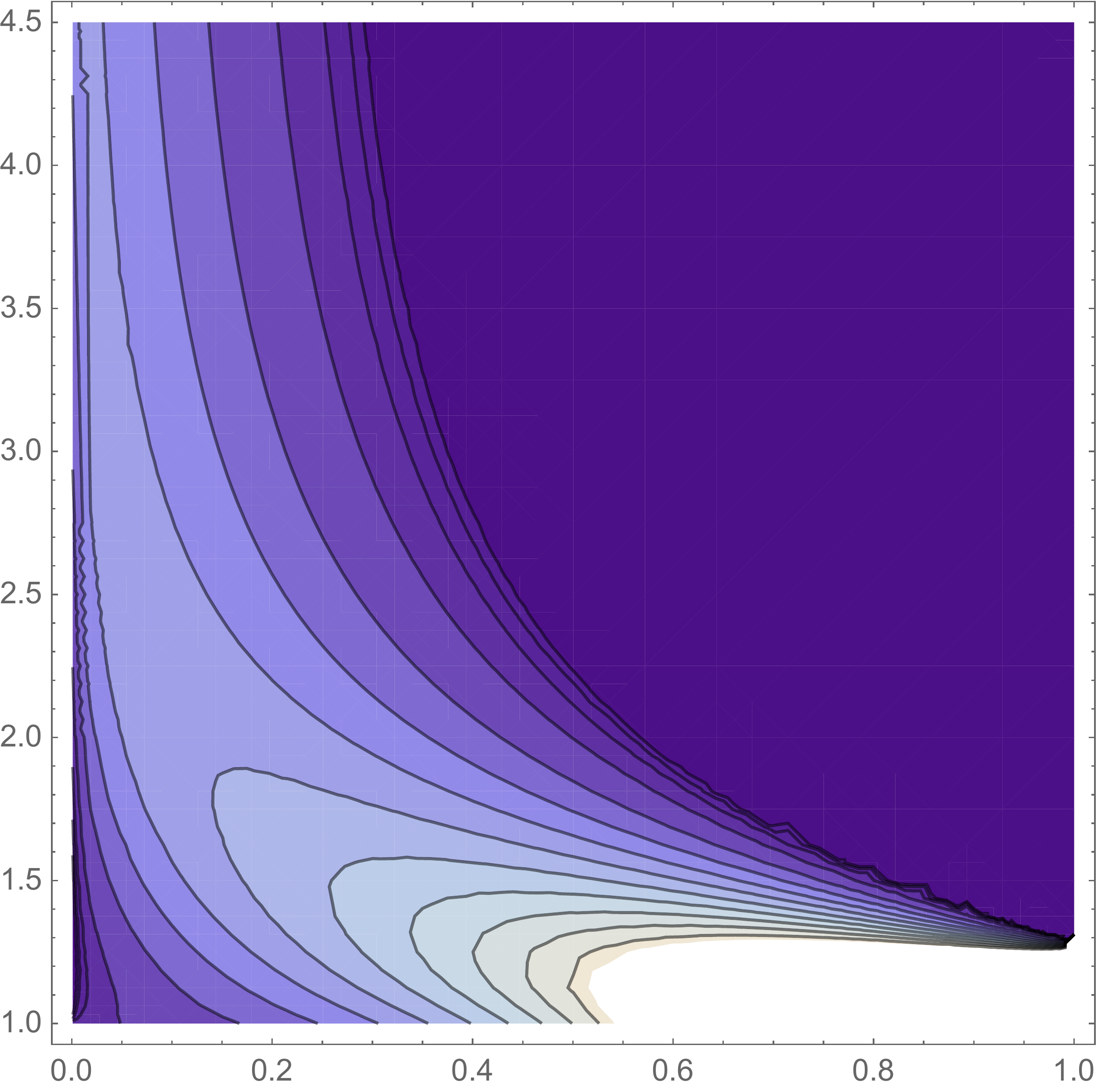}\quad\quad
\includegraphics[width=0.8cm]{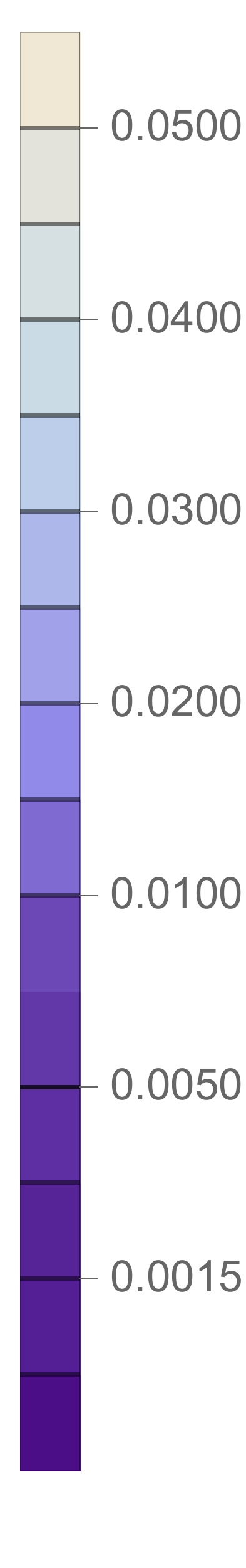}
\\ \,{\bf A)}\,\qquad\qquad\qquad\qquad\qquad\qquad\qquad{\bf B)}\\
\includegraphics[width=6cm]{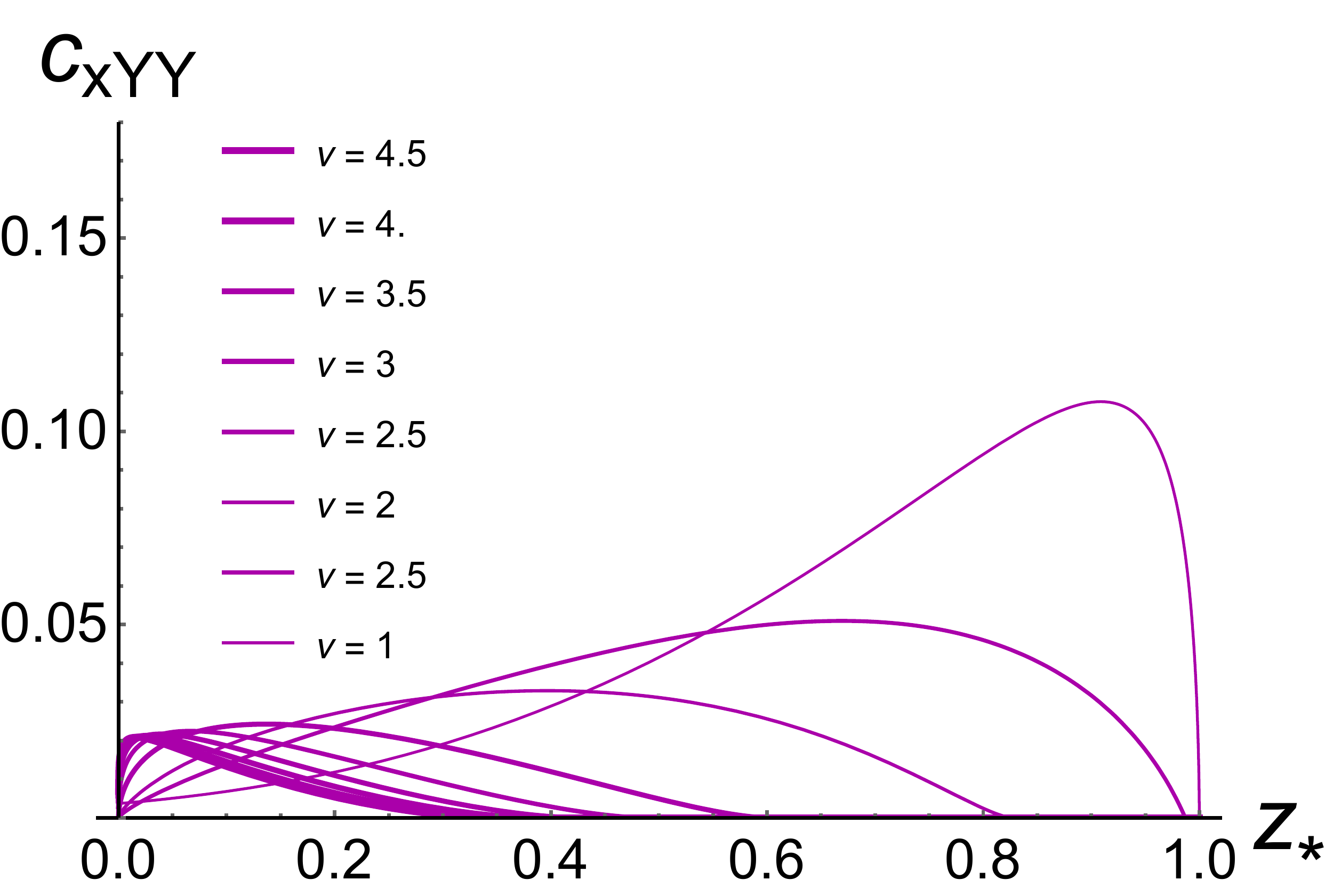}\quad\quad\quad
\includegraphics[width=6cm]{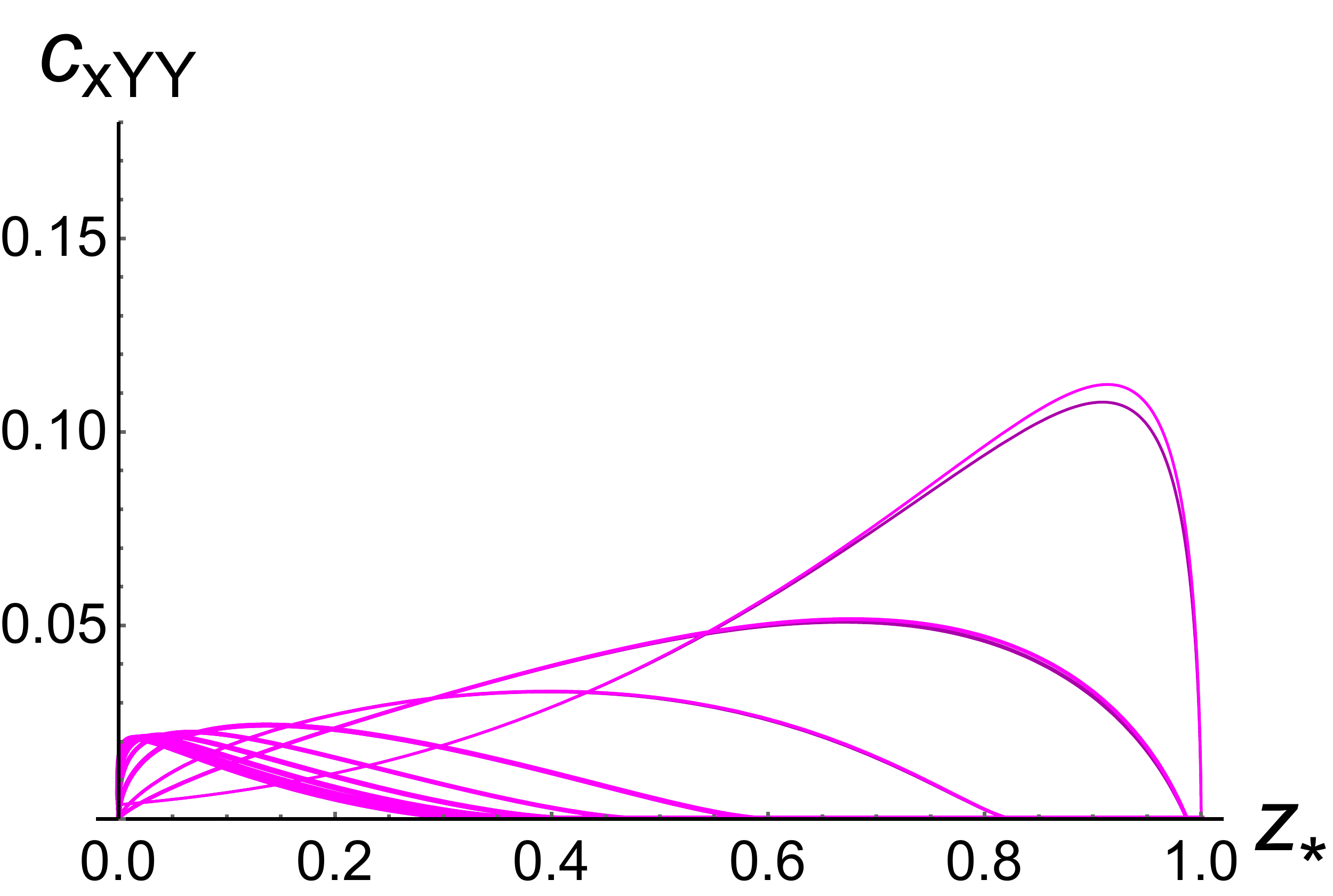}\\{\bf C)}\quad\quad\quad\qquad\qquad\qquad\qquad\qquad\qquad\qquad{\bf D)}
 \caption{{\it Top line}: contour plots for $\ell_\phi ^{m_F(\nu)}\eta_{_{CD}}(\nu)$ vs $z_*$ (horizontal axis) and $\nu$ (vertical axis) for F=SF at $z_h=1$ and  $\varphi=0$ (longitudinal case), and for different values of the chemical potential:  {\bf A)} $\mu =0$  and   {\bf B)} $\mu =0.5$. We see that two contour plots are practically identical.
 {\it Bottom line}: {\bf C)} and  {\bf D)} are the same as  {\bf A)} and {\bf B)}, but for 
 discrete values of $\nu$; comparison of c-functions for  {\bf C)} $\mu=0$ (darker magenta lines) and  {\bf D)} $\mu =0.5$ (magenta lines) for  discrete values of $\nu$. We see that two sets of lines almost coincide. 
}
    \label{fig:eta-zs-SFL}
  \end{figure} 

In the {\it top line} of Fig.\ref{fig:eta-zs-SFL}  contour plots of $\ell_\phi ^{m_F(\nu)}\eta_{_{CD}}(\nu)$ over $(z_*,\nu)$-plane for
the SF at $z_h=1$ and  $\varphi=0$ (longitudinal case), and for different values of the chemical potential are presented. Here in:  Fig.\ref{fig:eta-zs-SFL}.{\bf A)} $\mu =0$  and in Fig.\ref{fig:eta-zs-SFL}.{\bf B)} $\mu =0.5$. We see that two contour plots are practically identical.
In the {\it bottom line} of Fig.\ref{fig:eta-zs-SFL}  $\ell_\phi ^{m_F(\nu)}\eta_{_{CD}}(\nu)$  as  functions of $z_*$ for 
 discrete values of $\nu$   are presented.  For comparison, c-functions for $\mu=0$ (magenta lines) and $\mu =0.5$ (darker magenta lines) for  discrete values of $\nu$ are presented on the same plot in Fig.\ref{fig:eta-zs-SFL}.{\bf D)}. We see that two sets of lines almost coincide.

\newpage
$$\,$$
\newpage
\subsubsection{$\ell$ as a function of $z_*$}
\begin{figure}[h!]\centering
   $$\,$$
\includegraphics[width=6cm]{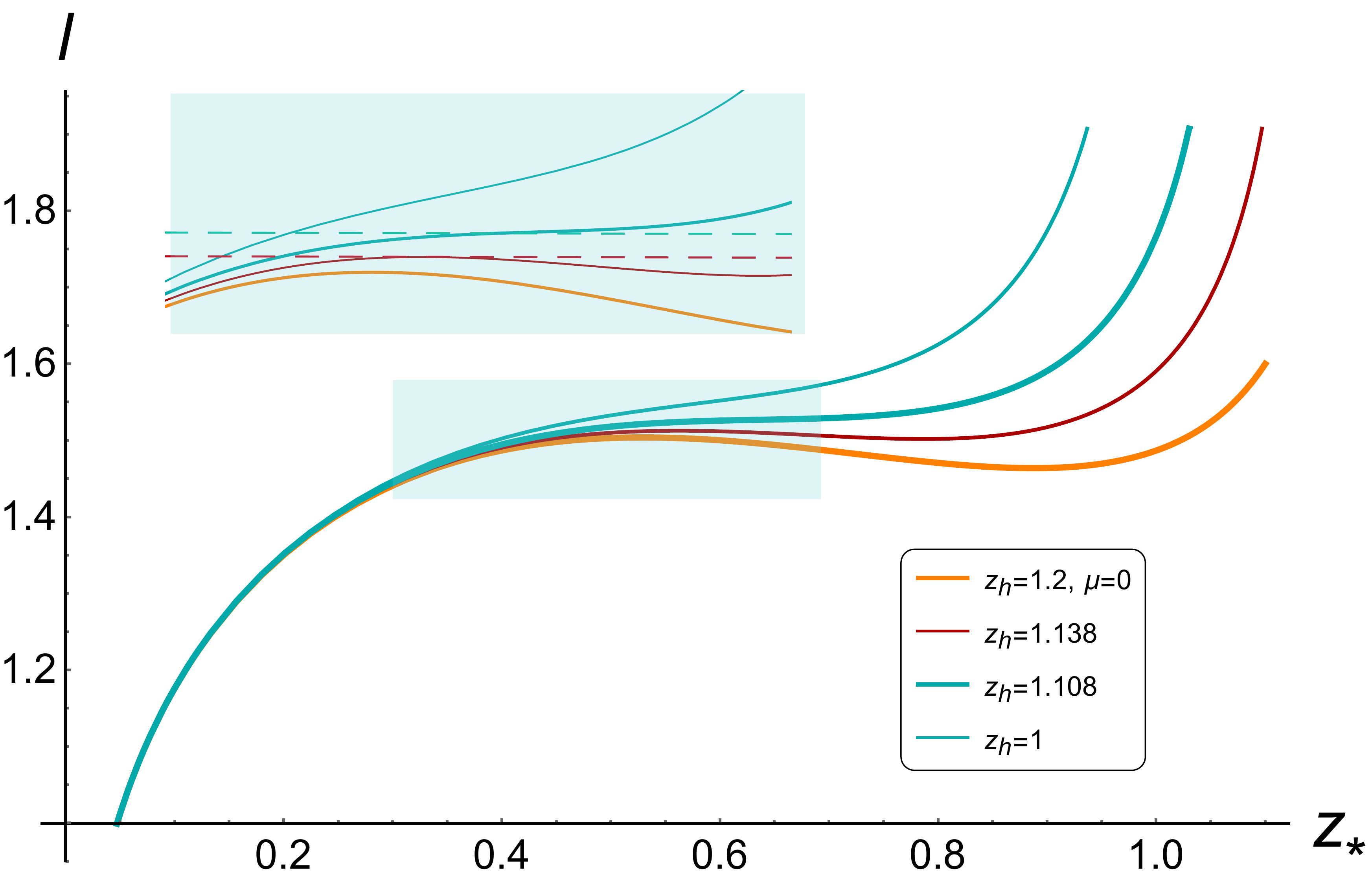}\qquad\qquad\qquad
\includegraphics[width=6cm]{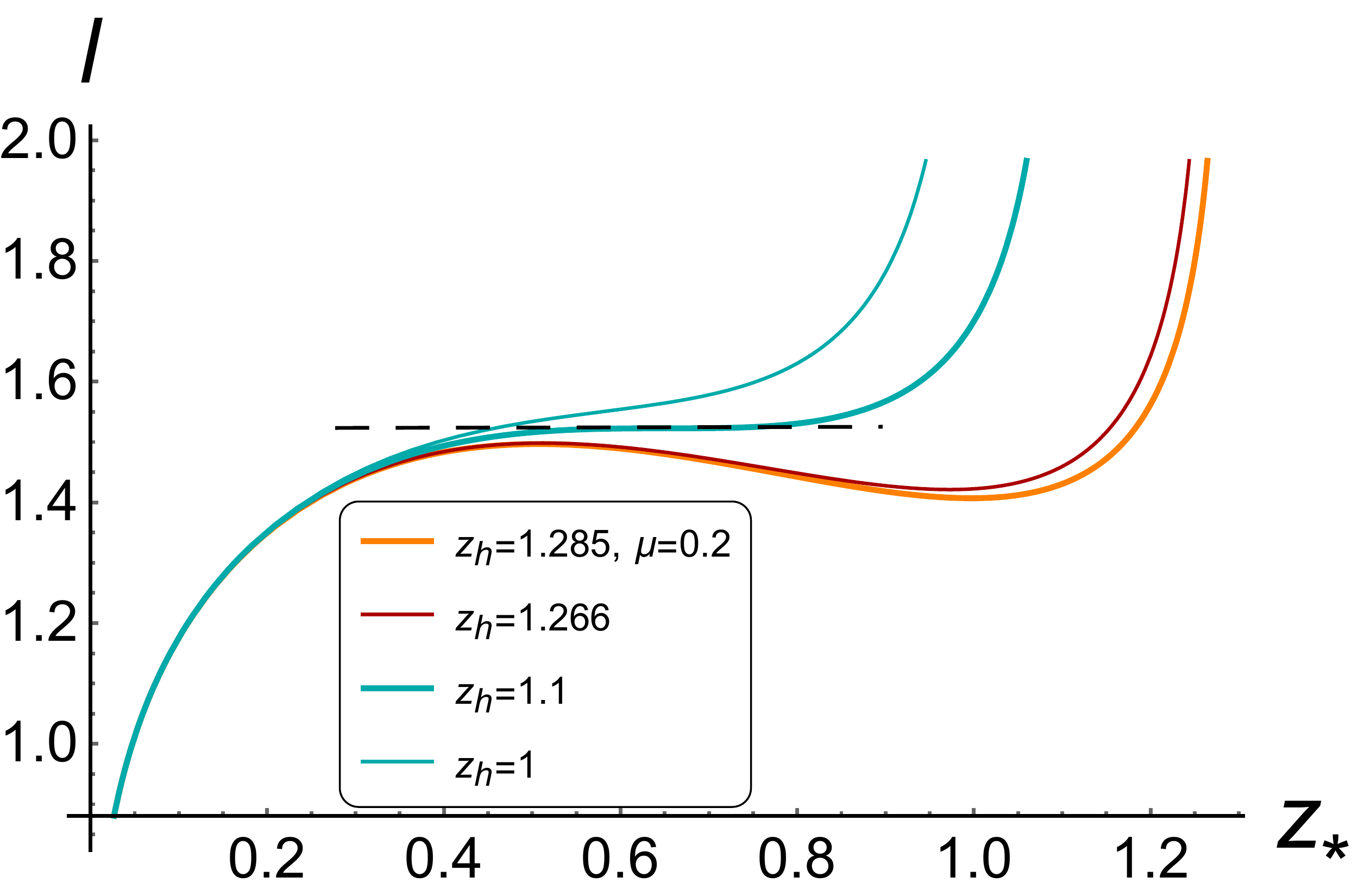}\\
 \,{\bf A)}\,\qquad\qquad\qquad\qquad\qquad\qquad\qquad\qquad \qquad\qquad {\bf B)}\\
\includegraphics[width=6cm]{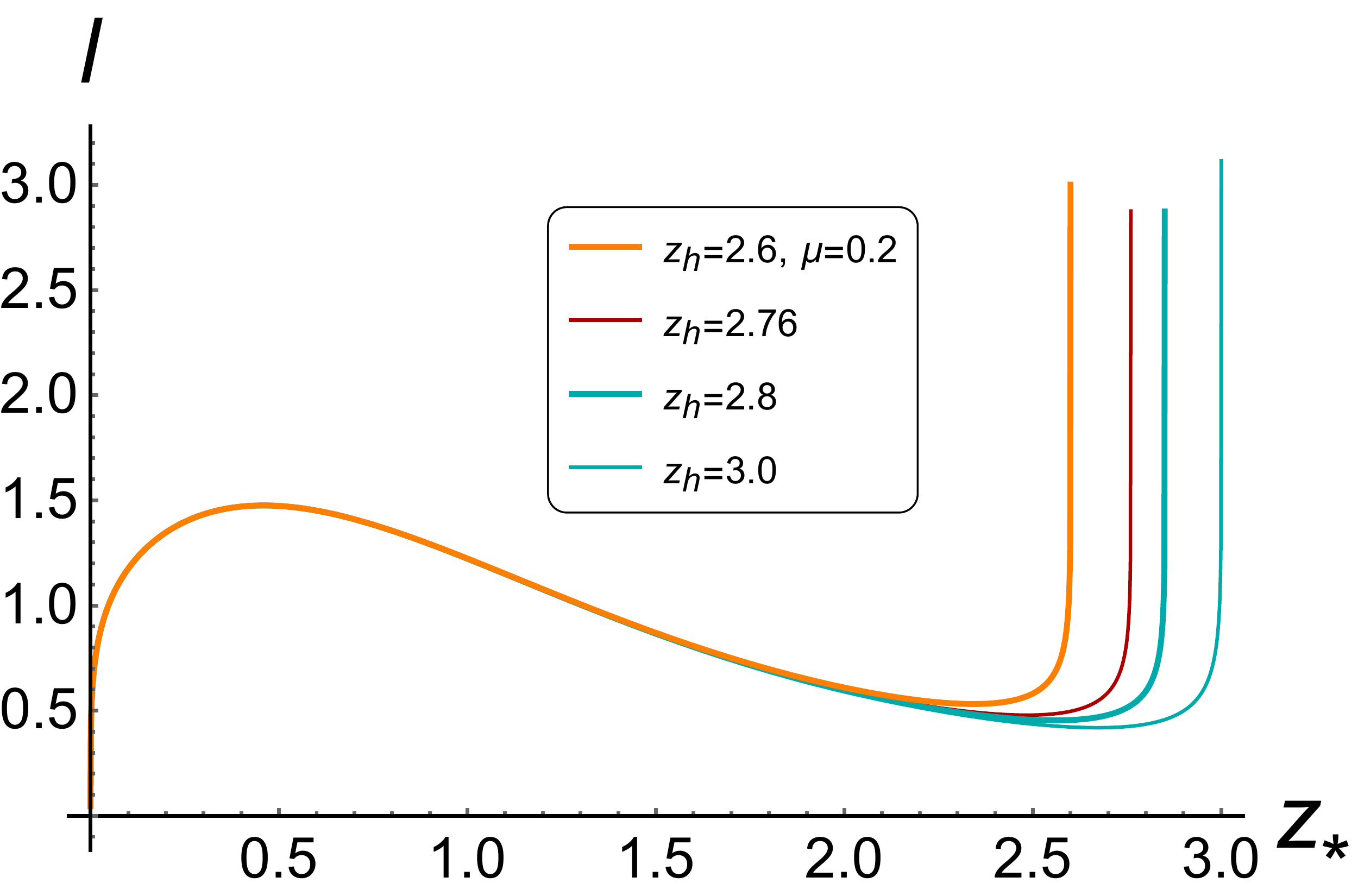}\qquad\qquad\qquad
\includegraphics[width=6cm]{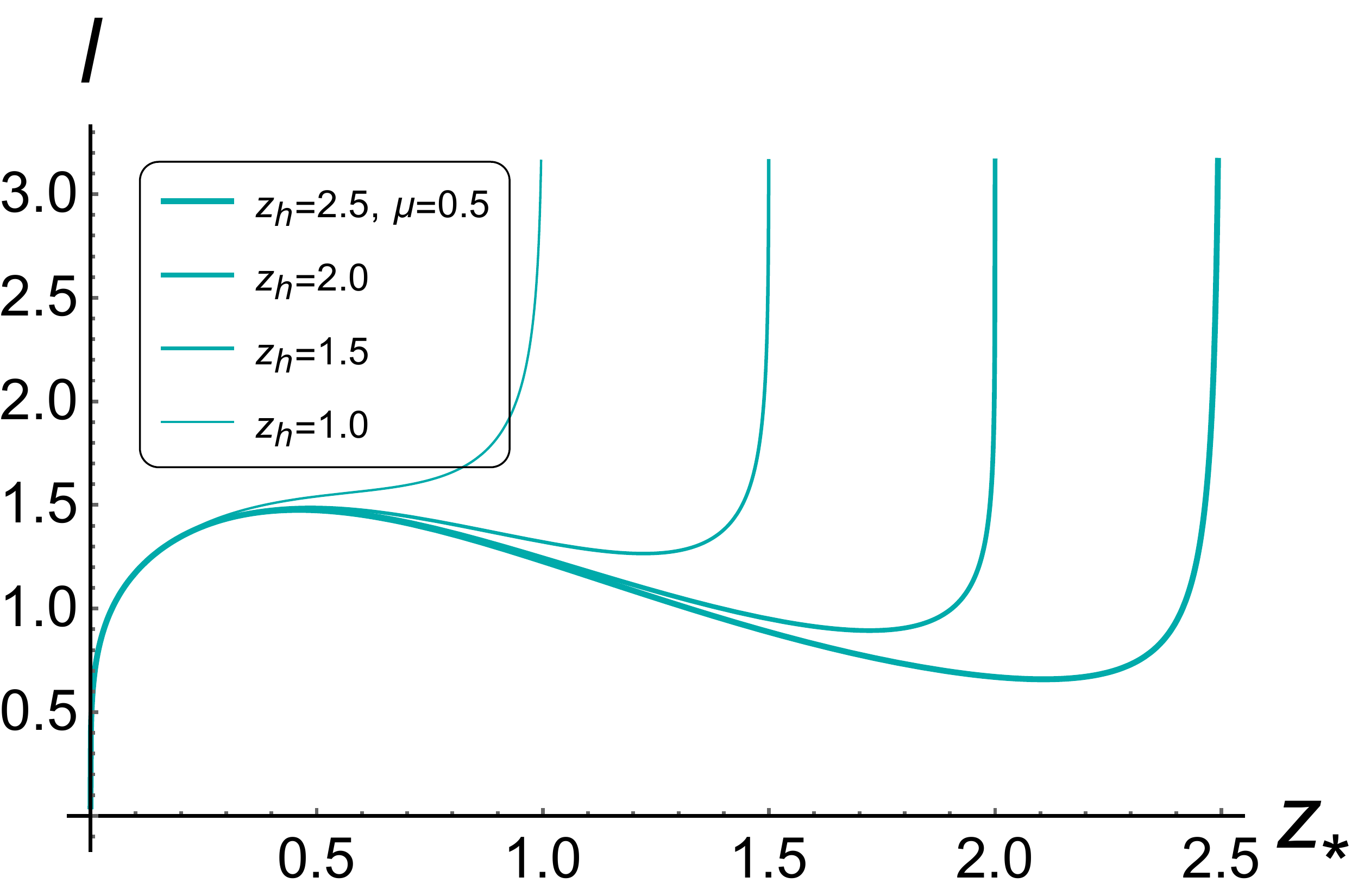}\\
{\bf C)}\,\qquad\qquad\qquad\qquad\qquad{\bf D)}\caption{$\ell$ vs $z_*$ for F=EF and  $\varphi=\pi/2$ (transversal case). {\bf A)} Here  $z_h=1.2,\,1.138,\,1.108,\, 1$ and   $\mu=0$ (in vicinity of the HP phase transition).  Here $z_{HP}=1.138$ is displayed in red  and $z_{h}=1.12$, corresponding to the unstable background, in orange. (Inset: the lines with $z_{HP}=1.138$ and $z_h=1.2$ have local maxima and the line with $z_h=1.108$ has an inflection point).
 {\bf B)}  Here $\mu = 0.2 $ and $z_h=1.285$ (an unstable point, the corresponding curves are displayed in orange), $z_h=1.1266$
 (the point of the BB PT, depicted in red),  $z_h=1.1$ (the corresponding curve has an inflection  point) and $z_h=1$.
   {\bf C)} Here $\mu = 0.2 $ and points with $z_h>2.76$ correspond to the stable backgrounds.   An unstable point  with $z_h>2.6$ is in orange. {\bf D)} Here $\mu=0.5$
   and all curves correspond to stable background.}    \label{fig:ell-zs-EF-T}
  \end{figure}

The length $\ell $ as a function of $ z _* $ has a rather nontrivial form for some values of thermodynamic parameters, especially in the longitudinal case.
In Fig.\ref{fig:ell-zs-EF-T}.{\bf A)}  the functions $\ell=\ell(z_*,z_h,\mu)$ for F=EF for transversal case at $z_h=1,1.06,1.138,1.2$ and $\mu=0$ are presented. It is interesting to note that even on stable backgrounds, i.e.  
 $z_{h,ip}=1.108<z_h<z_{HP}=1.138$,
the functions $\ell=\ell(z_*,z_h,\mu)$ are non-monotonic. Here $z_{h,ip}=1.108$ corresponds to the line, that has  an inflection point
(see the inset of Fig.\ref{fig:ell-zs-EF-T}.{\bf A)}).
This means that the same $\ell$ can be realized at different $z_*$, or  3 different values of entanglement entropy, entropy density and  c-function  corresponds to the same length. 

Fig.\ref{fig:ell-zs-EF-T}.{\bf B)}-Fig.\ref{fig:ell-zs-EF-T}.{\bf D)} show what happens when we change the chemical potential.
In Fig.\ref{fig:ell-zs-EF-T}.{\bf B)}  $\mu = 0.2 $ and $z_h=1.285$ (an unstable point, the corresponding curves are displayed in orange), $z_h=1.1266$
 (the point of the BB phase transition depicted in red),  $z_h=1.1$ (the corresponding curve has an inflection  point) and $z_h=1$.
  In Fig.\ref{fig:ell-zs-EF-T}.{\bf C)}  $\mu = 0.2$ and points with $z_h>2.76$ correspond to the stable backgrounds.   An unstable point  with $z_h>2.6$ is in orange.  In Fig.\ref{fig:ell-zs-EF-T}.{\bf D)} $\mu=0.5$
   and all curves correspond to the stable backgrounds.

     \begin{figure}[h!]\centering
   $$\,$$
\includegraphics[width=6cm]{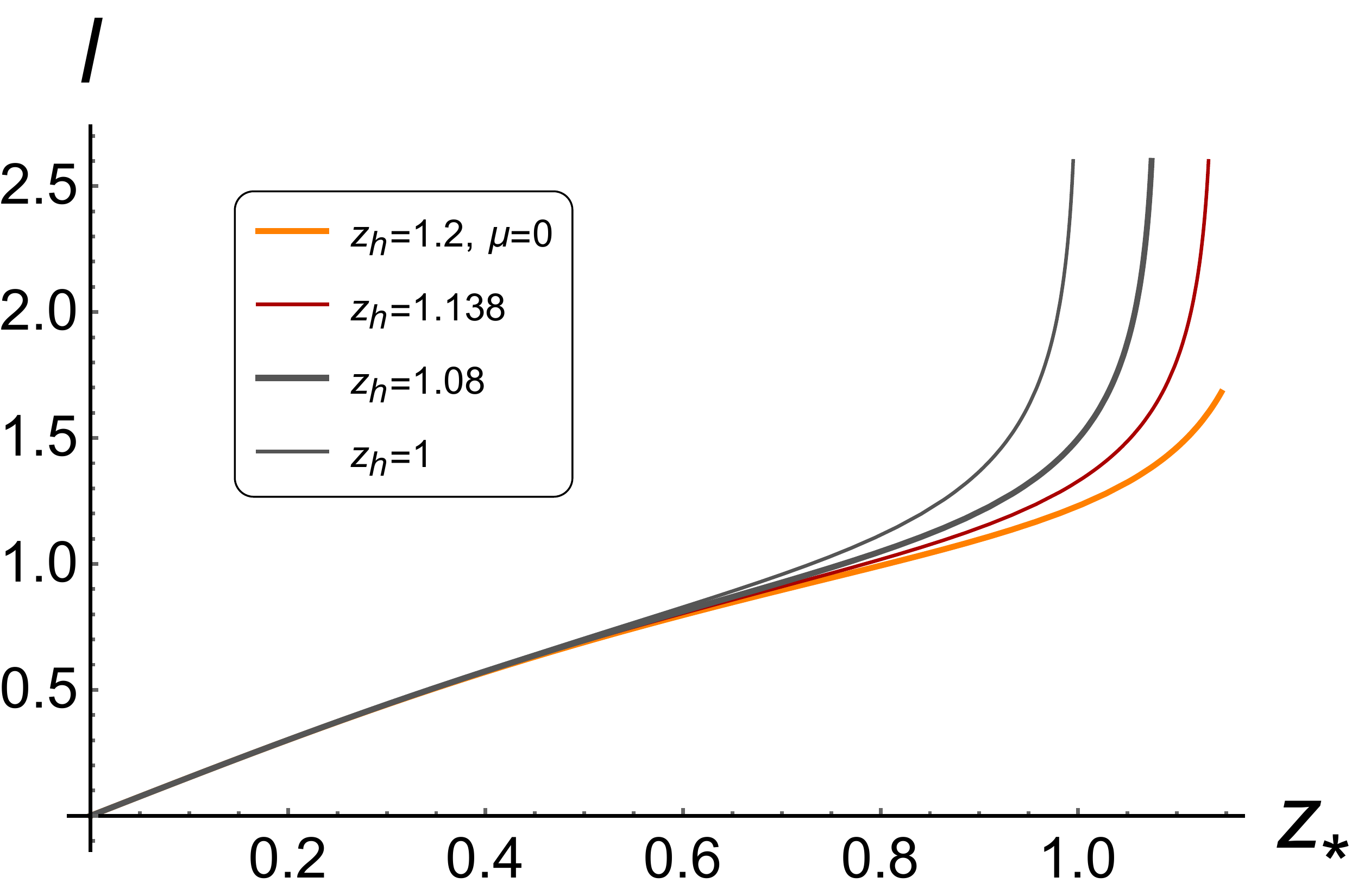} \qquad\qquad 
\includegraphics[width=6cm]{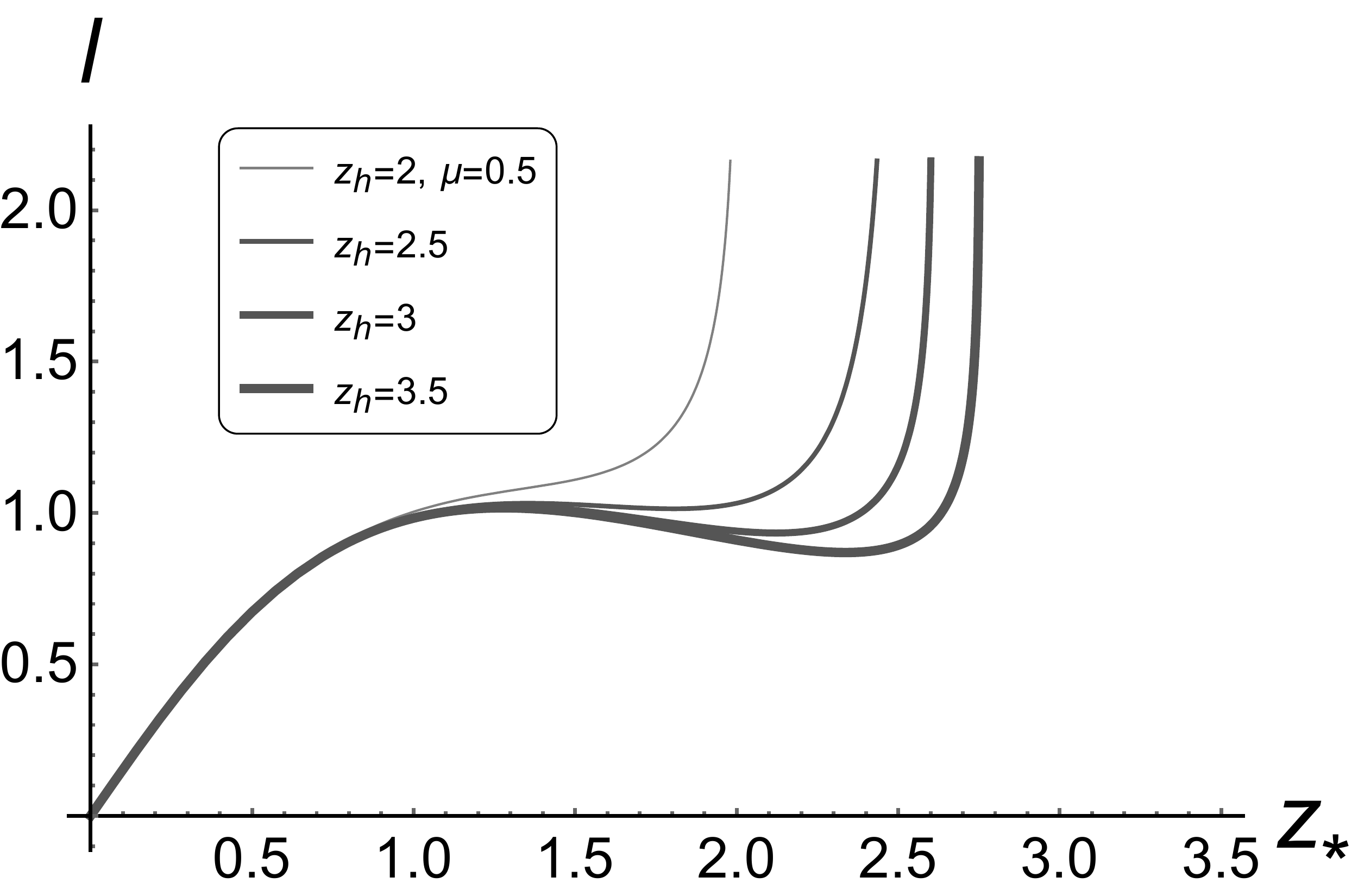}\\{\bf A)}\,\qquad\qquad\qquad\qquad\qquad{\bf B)}\\
\includegraphics[width=6cm]{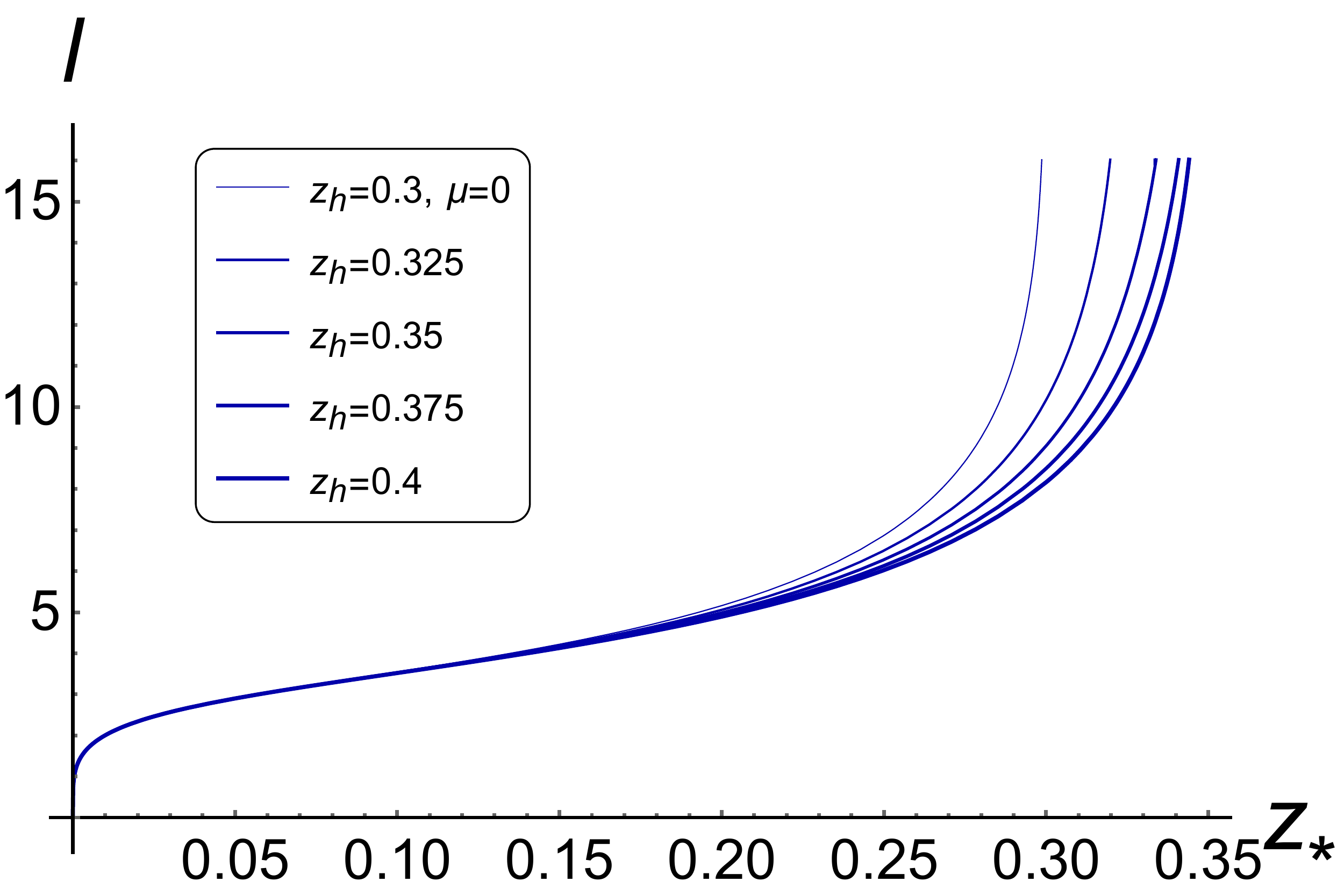} \qquad\qquad 
\includegraphics[width=6cm]{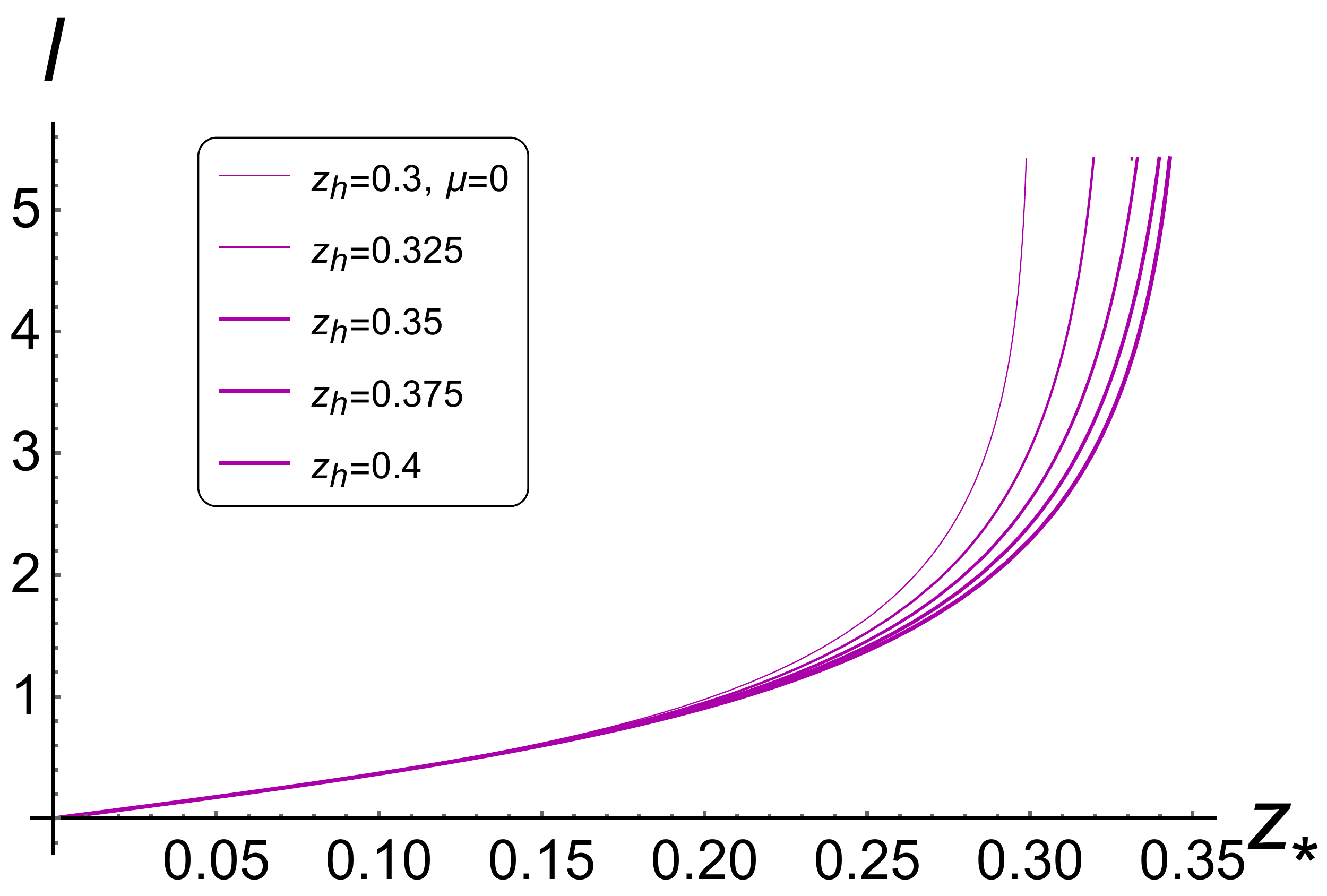}\\
{\bf C)} \qquad\qquad\qquad \qquad\qquad {\bf D)}
\\
\caption{{\it Top line}: $\ell$ vs $z_*$ for F=EF for $\varphi=0$ (longitudinal case)
  at {\bf A)} $z_h=1.2,\,1.138,\,1.108,\, 1$ and   $\mu=0$, at {\bf B}) $z_h=2,\,2.5,\,3.0,\, 3.5$ and   $\mu=0.5$.
{\it Bottom line}:   $\ell$ vs $z_*$ for F=SF  for {\bf C)} $\varphi=\pi/2$ (transversal case) at $z_h=0.3,0.325,0.35,0.375$ and 0.4.  
 The right curve  corresponds to surfaces touching a dynamic wall located at $ z_ {DW}  = 0.354 $
and for the rest, $ z _ * $ is located near the horizons.  The plot {\bf D)} is the same as  {\bf C)} 
 for $\varphi=0$ (longitudinal case). 
}
    \label{fig:ell-zs-EF}
  \end{figure} 
  
  In Fig.\ref{fig:ell-zs-EF} on the {\it top line} $\ell$ vs $z_*$ for the EF and $\varphi=0$ (longitudinal case) is shown: 
  Fig.\ref{fig:ell-zs-EF}.{\bf A)} at $z_h=1.2,\,1.138,\,1.108,\, 1$ and   $\mu=0$; Fig.\ref{fig:ell-zs-EF}.{\bf B)} at $z_h=2,\,2.5,\,3.0,\, 3.5$ and   $\mu=0.5$.
On the {\it bottom line}   $\ell$ vs $z_*$ for  the SF is shown:   Fig.\ref{fig:ell-zs-EF}.{\bf C)} for $\varphi=\pi/2$ (transversal case) at $z_h=0.3,0.325,0.35,0.375$ and 0.4.  
 The right curve  corresponds to surfaces touching a dynamic wall located at $ z_ {DW}  = 0.354 $
and for the rest, $ z _ * $ is located near the horizons. In  Fig.\ref{fig:ell-zs-EF}.{\bf D)} the  plot  for $\varphi=0$ (longitudinal case) is shown.
 
   \begin{figure}[h!]\centering
\includegraphics[width=4.5cm]{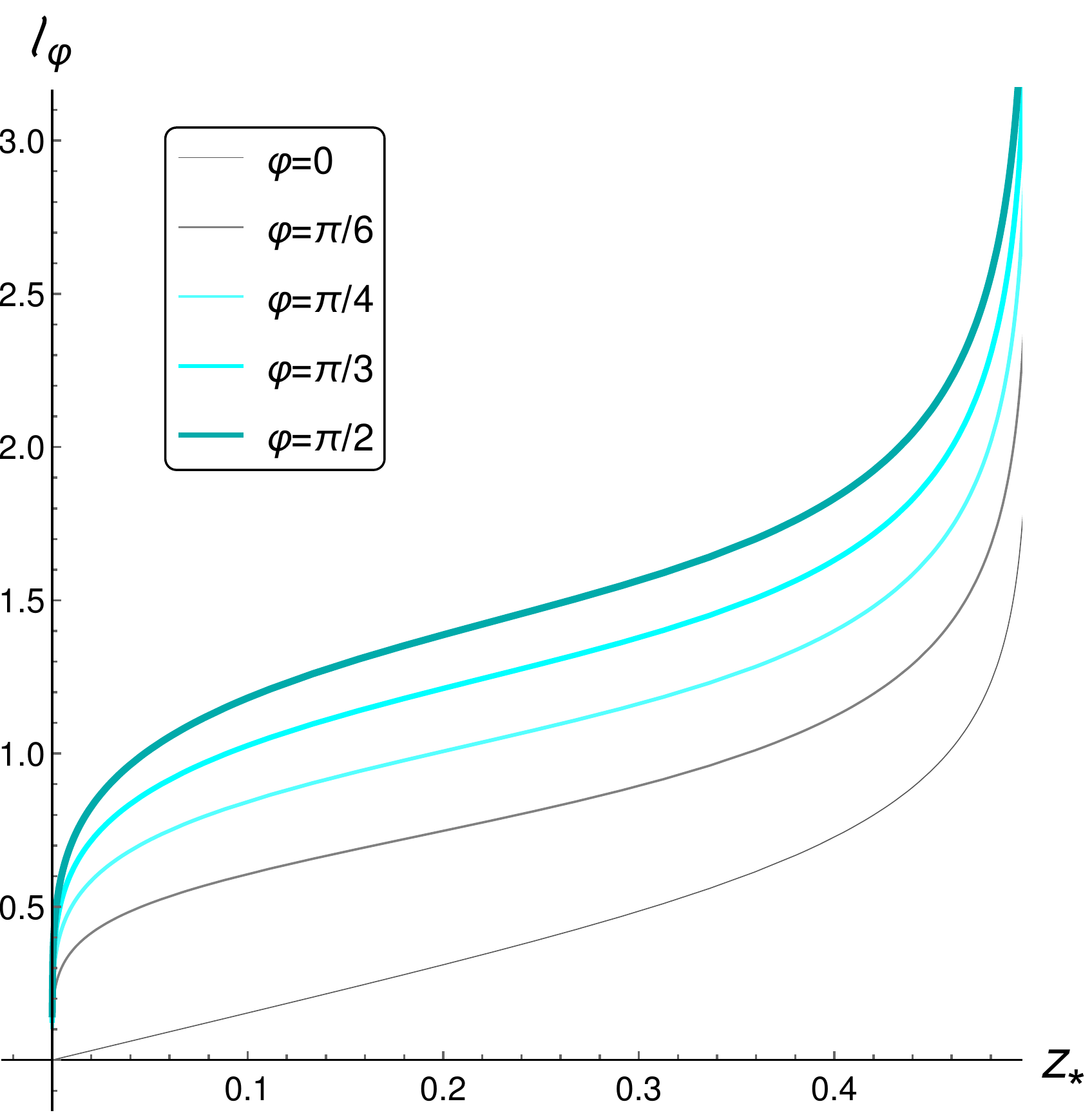}
\includegraphics[width=4.5cm]{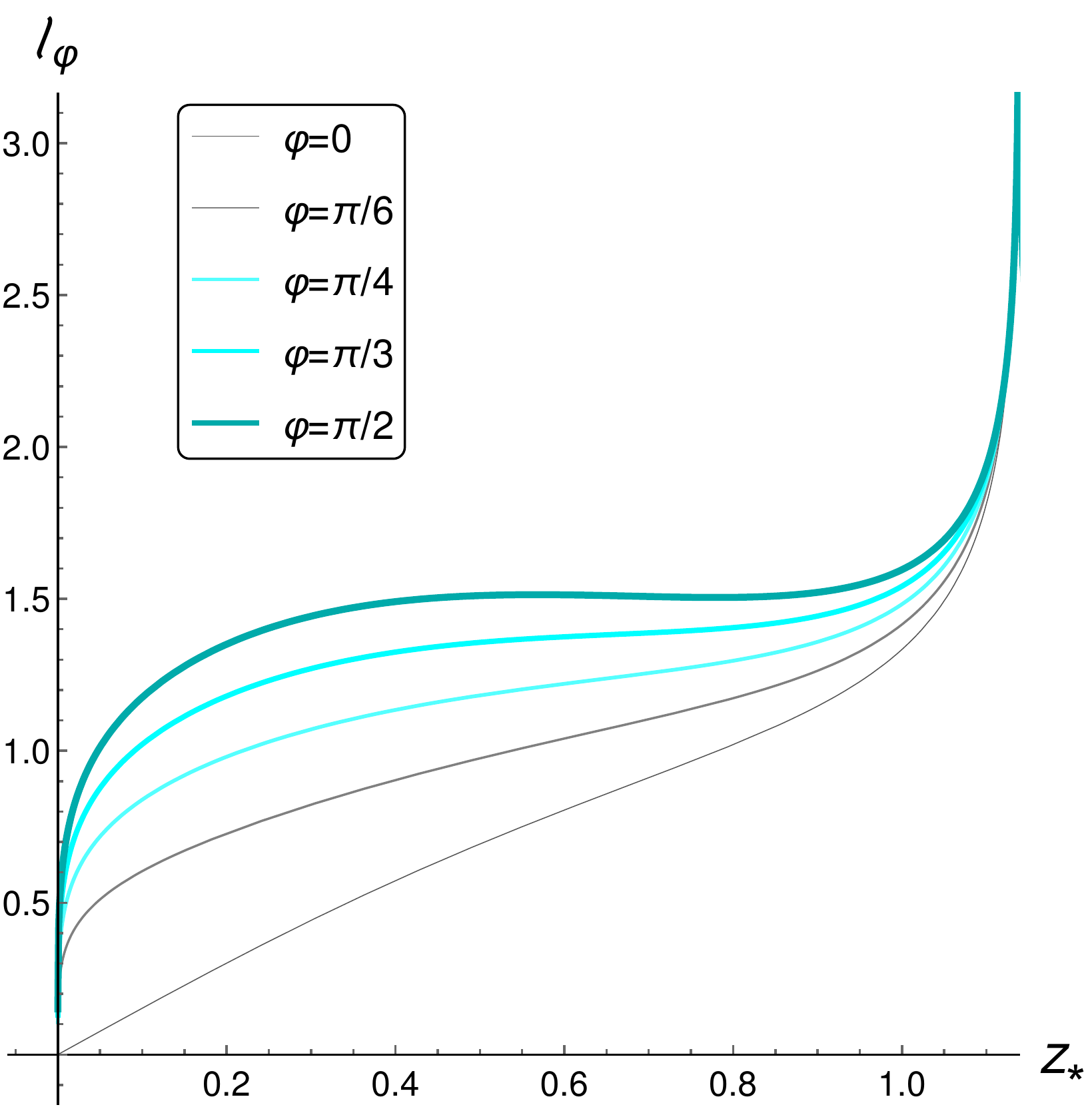}
\includegraphics[width=4.5cm]{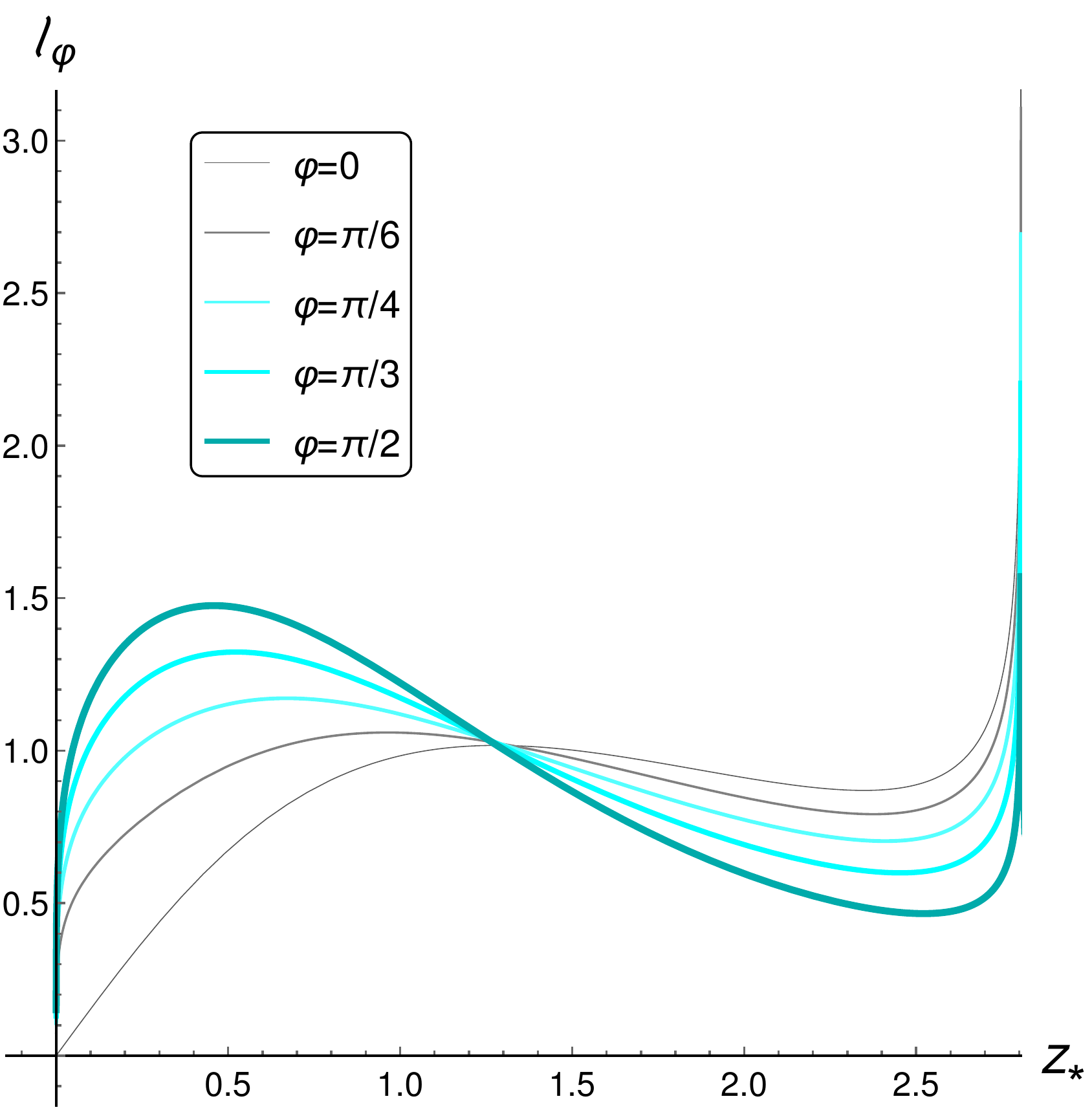}
\\ \,{\bf A)}\,\qquad\qquad\qquad\qquad\qquad{\bf B)}\qquad\qquad\qquad \qquad\qquad {\bf C)}\\
 \caption{ $\ell$ vs $z_*$ for F=EF at $\mu=0.2$, different orientations $\varphi=0,\pi/6,\pi/4,
 \pi/3,\pi/2$,  $\mu=0.2$,   and  $z_h=0.5, \,1.1388, \,2.80463$ (corresponding temperatures are $T=0.418,\,0.25,\,0.235$),  are presented in  
  {\bf A)}, {\bf B)} and  {\bf C)}, respectively.  }
    \label{fig:ell-zs-EFangle}
  \end{figure}

\begin{figure}[h!]\centering
  \includegraphics[width=6cm]{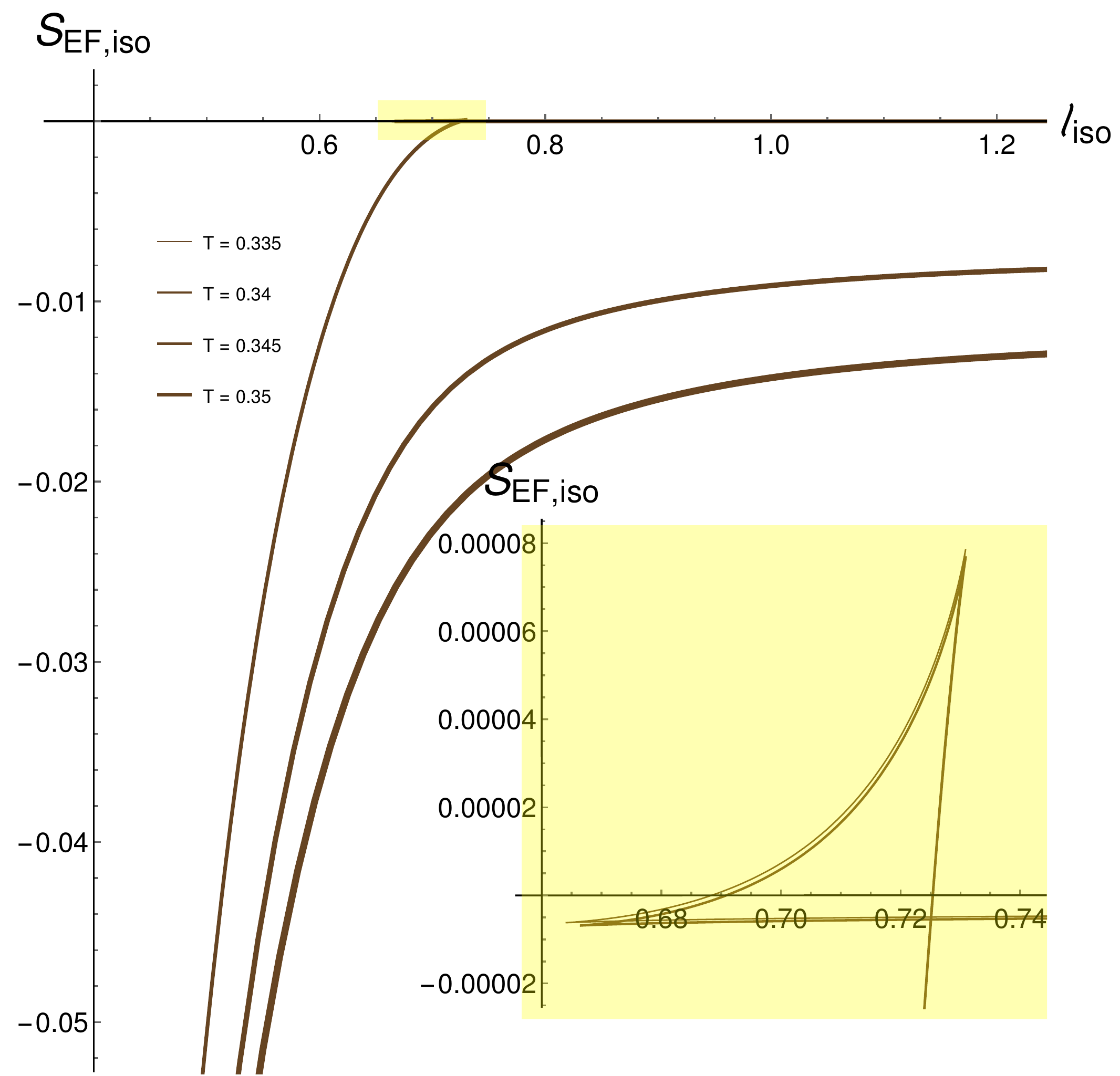} 
  \caption{ The entanglement entropy  $S_{HEE}$ vs $\ell$ in the EF for $\nu=1$ and  various values of the temperature below and above the phase transition point $(T=0.3445,\,\mu=0.05)$:   $T=0.335,0.34$ (below)  and $T=0.345,0.35$ (above) (thickness increases with increasing temperature) and $\mu=0.05$.
The inset shows a zoom of the yellow domains.}
 \label{fig:SEF(l)}
  \end{figure}

\begin{figure}[h!]\centering

\includegraphics[width=6cm]{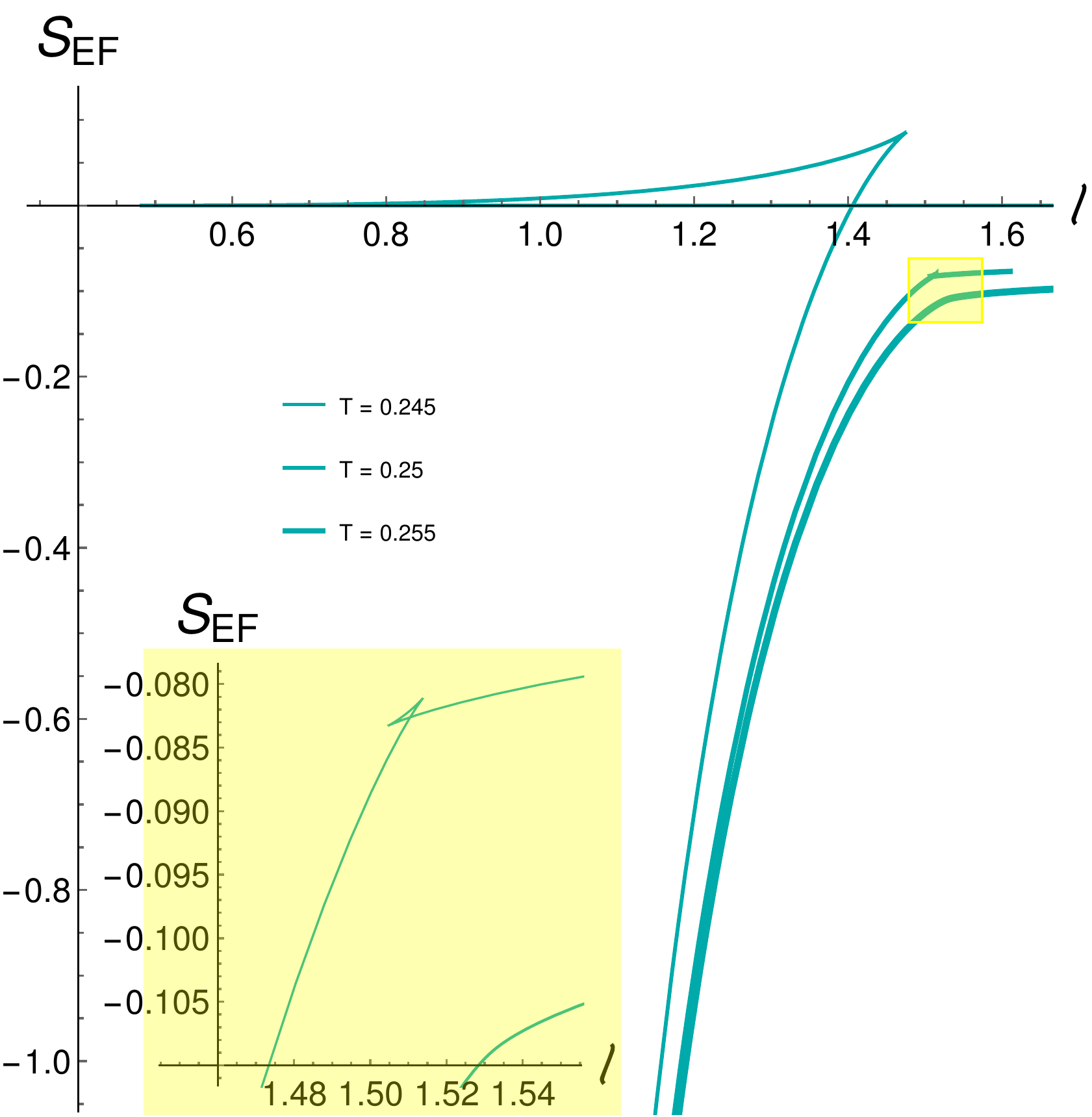}\qquad\qquad
\includegraphics[width=6cm]{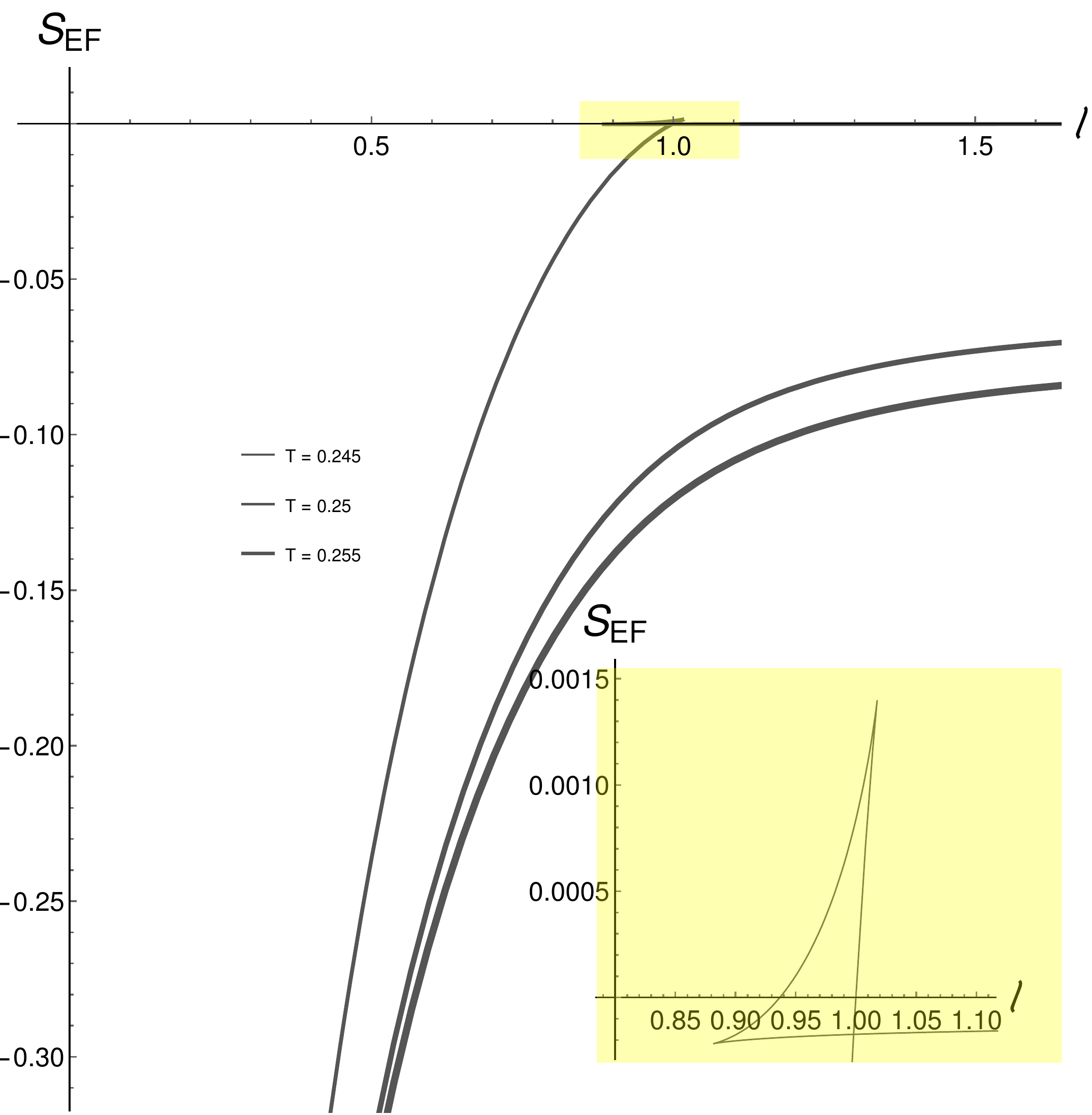}\\
  {\bf A)}\,\qquad\qquad\qquad\qquad\qquad\qquad\qquad\qquad{\bf B)}
  \caption{ The entanglement entropy  $S_{HEE}$ vs $\ell$ in the EF for $\nu=4.5$, transversal  and longitudinal  cases are presented in 
  {\bf A)} and {\bf B)}, respectively.  Here   values of temperature are below and above  the phase transition point $(T=0.2457,\,\mu=0.2)$.
In all panels the insets show zooms of the yellow domains.}
 \label{fig:SEFA(l)}
  \end{figure}

The multi-valued dependency of  $\ell$ on $z_*$ in holographic models was previously observed in \cite{1805.02938}. The authors established a new type of the phase transition associated with the swallow-tail like structure for $S_{HEE}$ as the function of $\ell$. For completeness, we present the dependence of the entanglement entropy in the EF on $\ell$ in Fig.\ref{fig:SEF(l)} and Fig.\ref{fig:SEFA(l)}.
For the isotropic case the Van-der-Waals behavior of the entanglement entropy depends on a slab width  takes place only  for small black holes. For the anisotropic there is a small region of temperatures where $S_{HEE}$ is multi-valued also for large black holes. The c-function undergoes a jump at $\ell_{crit}$ obtained from the self-intersection of $(S_{HEE},\ell)$ diagram. The lengths
$\ell_{left,L}$, $\ell_{right,L}$ in Fig.\ref{fig:c-b-LT-EF}.{\bf D)} correspond to the turning points. 

\subsection{The c-function near the Background Phase Transition}\label{nearBB}
\begin{figure}[h!]\centering
 \includegraphics[width=7cm]{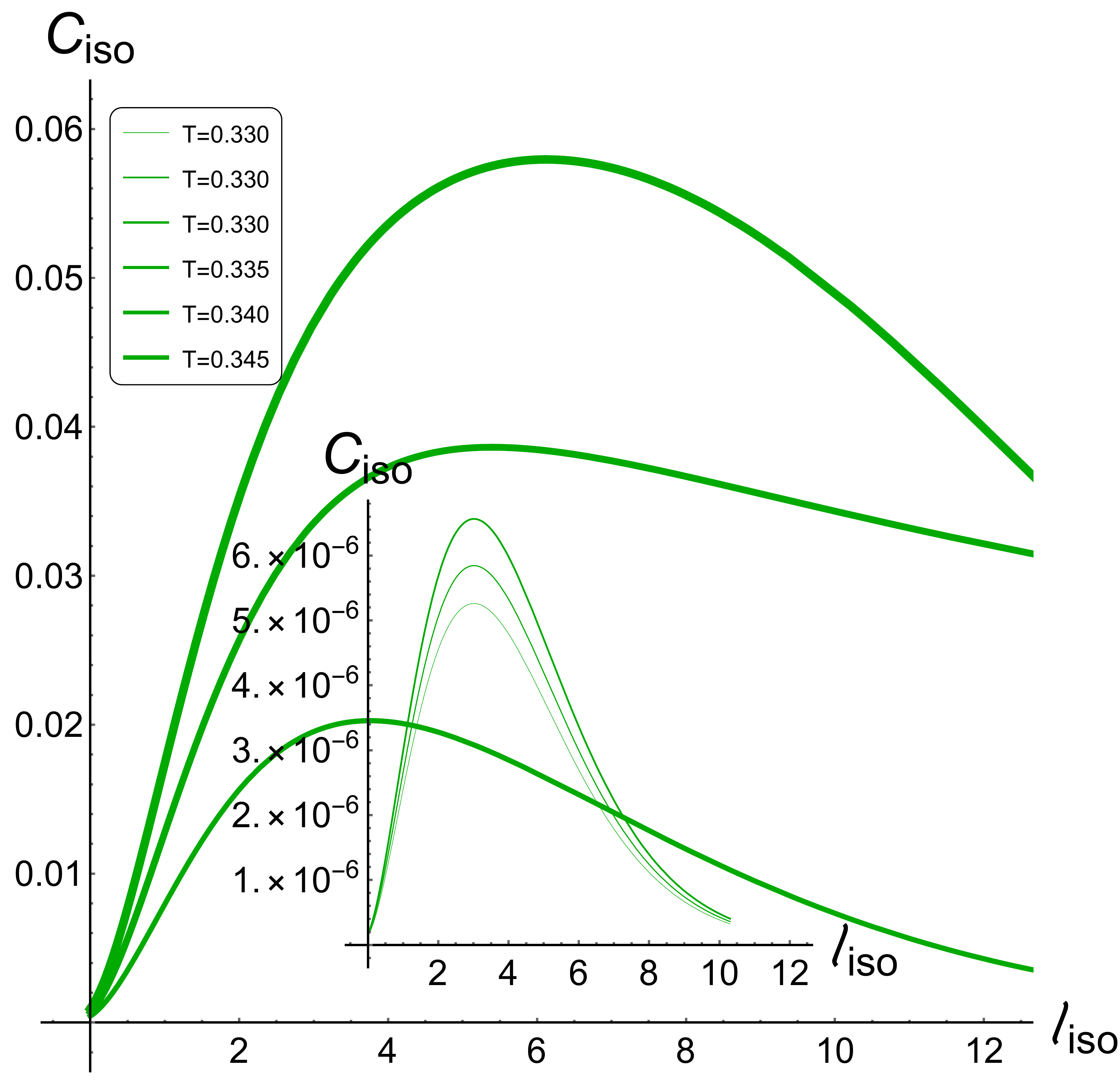} 
 \\{\bf A)}\\$$\,$$\\
 \includegraphics[width=7cm]{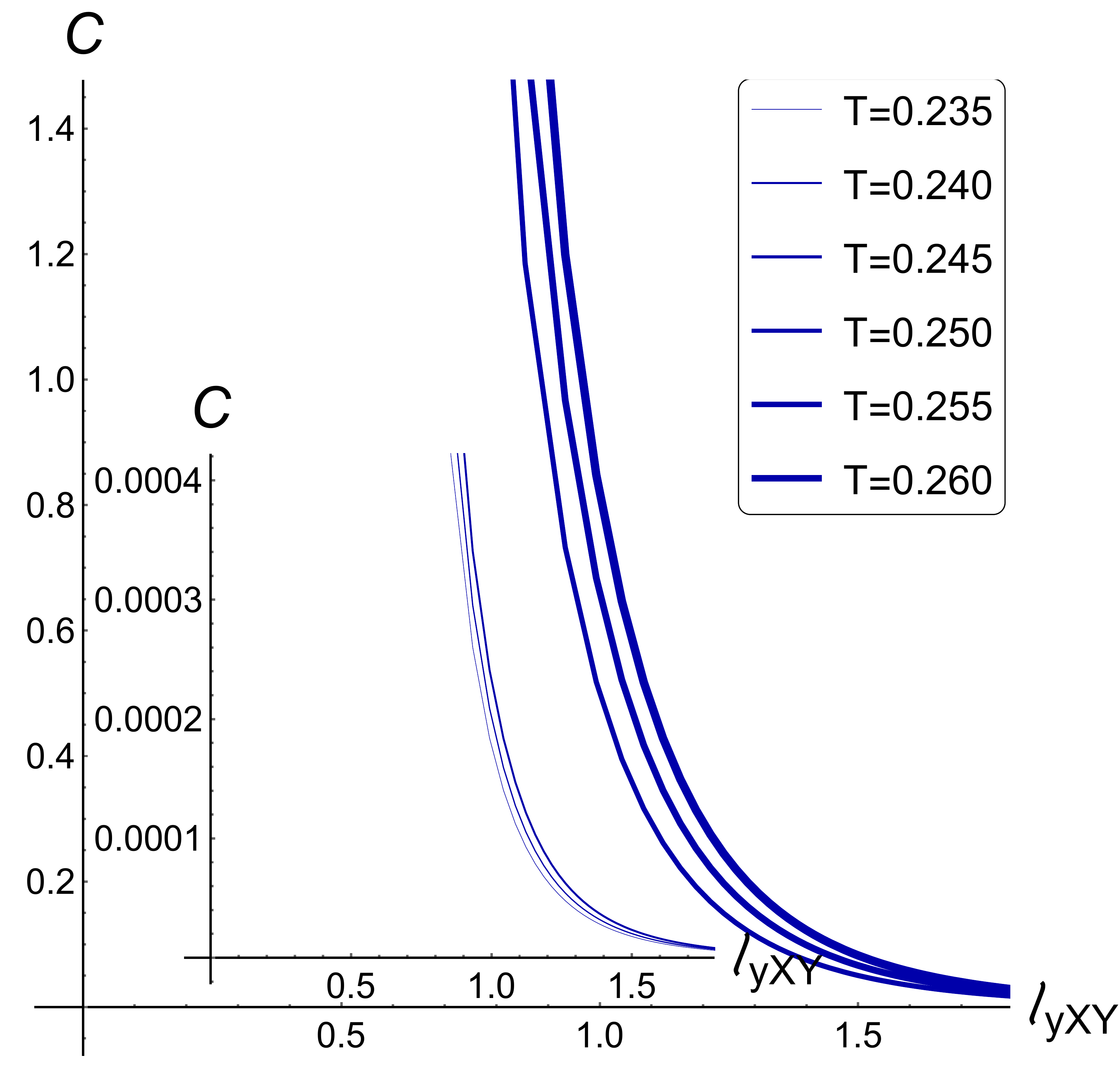}
 \qquad\includegraphics[width=7cm]{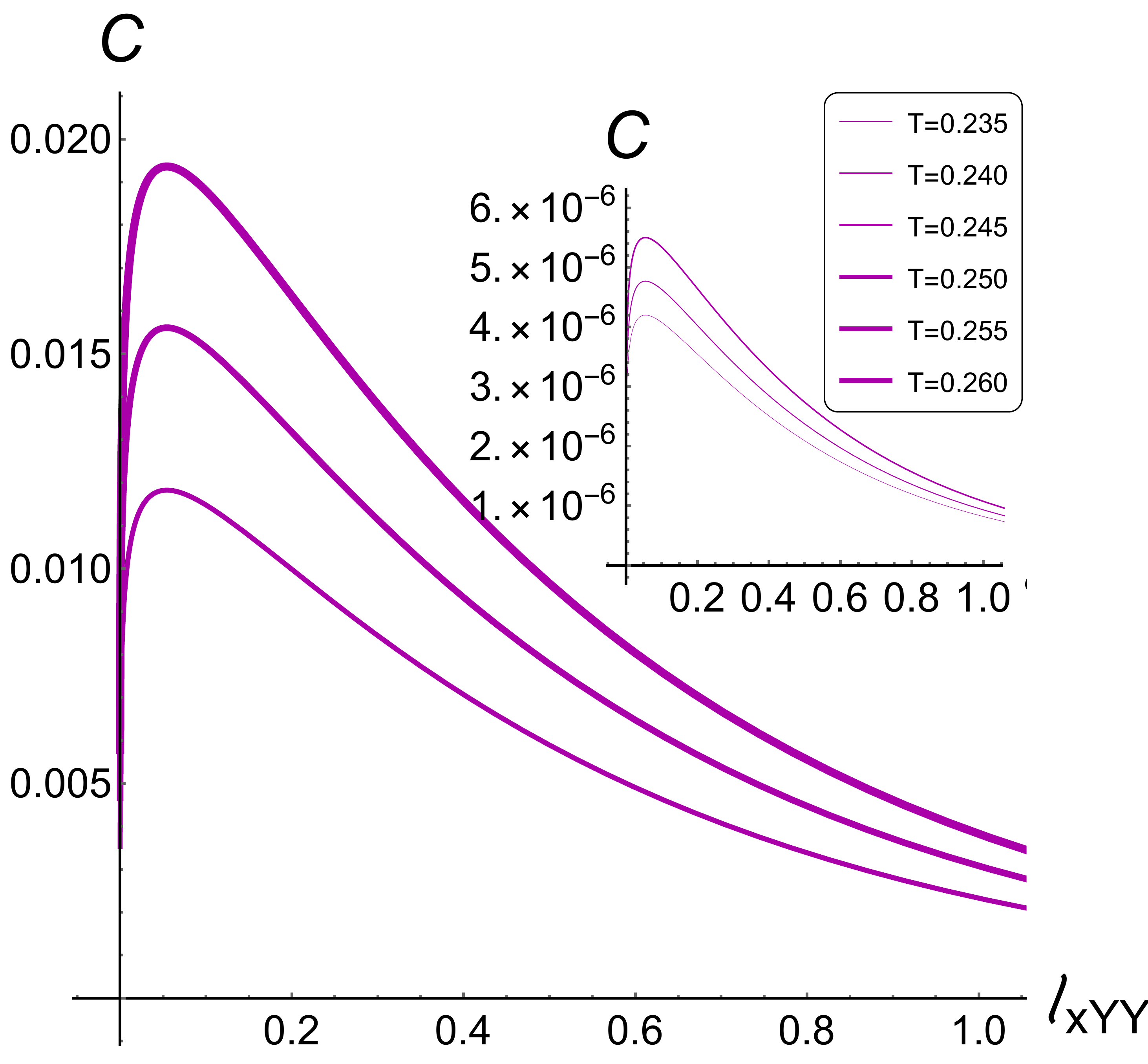}
\qquad\qquad\qquad\qquad
 {\bf B)}\qquad\qquad\qquad\qquad\qquad\qquad\qquad\qquad{\bf C)}
 \quad\quad\quad\quad
     \caption{The dependencies of $c$, defined by \eqref{c-aniz}, on $\ell$  in proximity of  the phase transitions (background shown in Fig.\ref{fig:phaseHEE}) in SF: {\bf A}) $\nu=1$; {\bf B}) $\nu=4.5$ and transversal orientation;  
     {\bf C}) $\nu=4.5$ and transversal orientation. {\it Insets}: $c(\ell)$ above the corresponding phase transitions.   }
    \label{fig:c-b-LT-PT}
  \end{figure}

In this subsection we present  dependences of $c$ on $\ell$  in proximity of  the background phase transitions. First we 
consider c-functions defined by \eqref{c-aniz} in the SF. In Fig.\ref{fig:c-b-LT-PT} we present this c-function for   $\nu=1$ and $\nu=4.5$ (see Fig.\ref{fig:c-b-LT-PT}.{\bf A)}) for transversal (see  Fig.\ref{fig:c-b-LT-PT}.{\bf B)}) and longitudinal orientations (see  Fig.\ref{fig:c-b-LT-PT}.{\bf C)}). In the main panels plots for the temperature below the phase transitions are shown, while the insets show graphs at temperatures above phase transitions. 
We see that the c-functions below the phase transition line are negligibly small as compared with c-functions above this line. Note that for isotropic case c-functions have  saddle points, but there are no saddle points in the anisotropic case for transversal configuration.
     
\begin{figure}[h!]\centering
 \includegraphics[width=7cm]{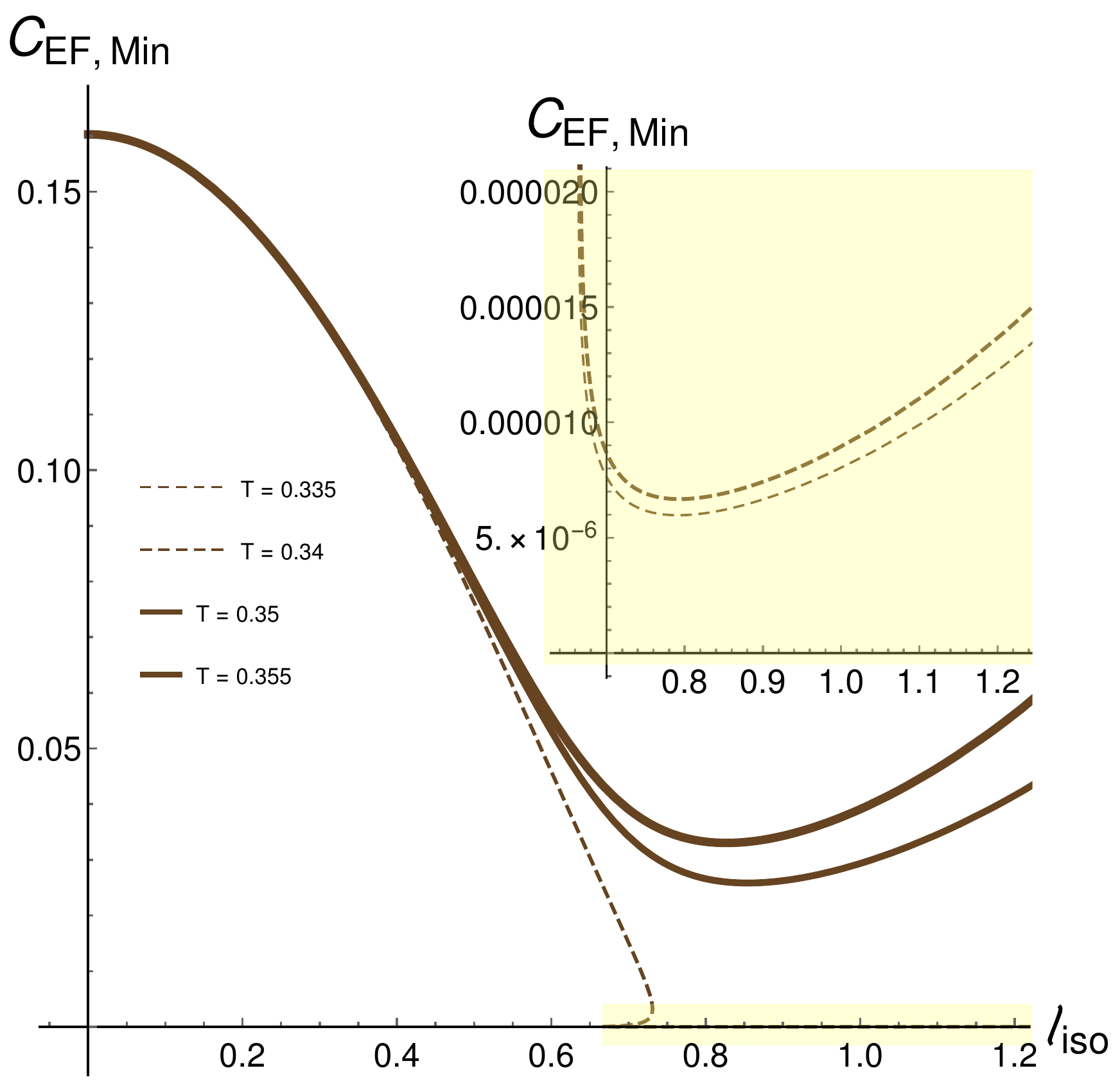}\qquad
  \includegraphics[width=7cm]{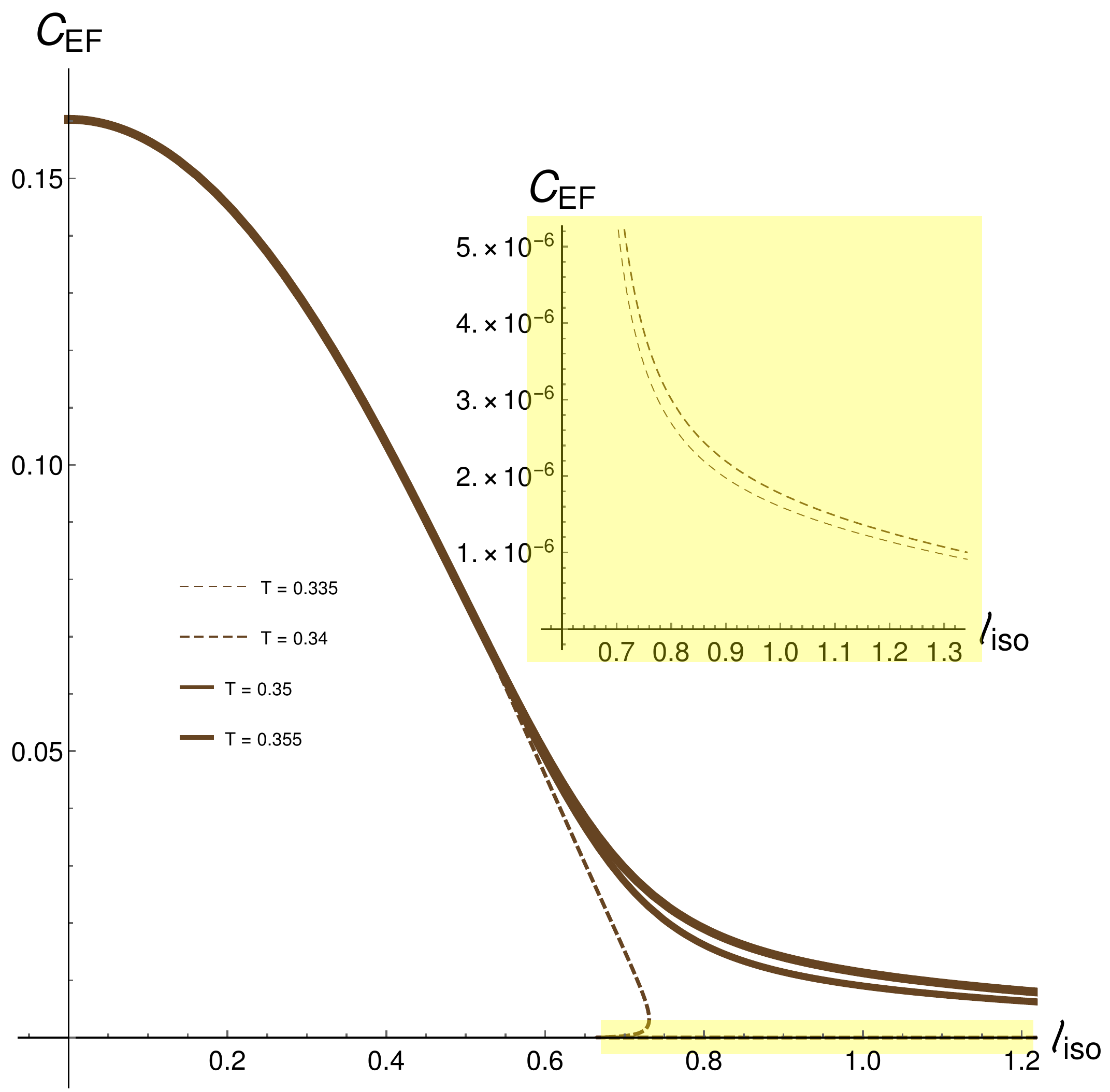} \\
 {\bf A)}\qquad\qquad\qquad\qquad\qquad\qquad\qquad\qquad  {\bf B)}\\$$\,$$\\
\includegraphics[width=7cm]{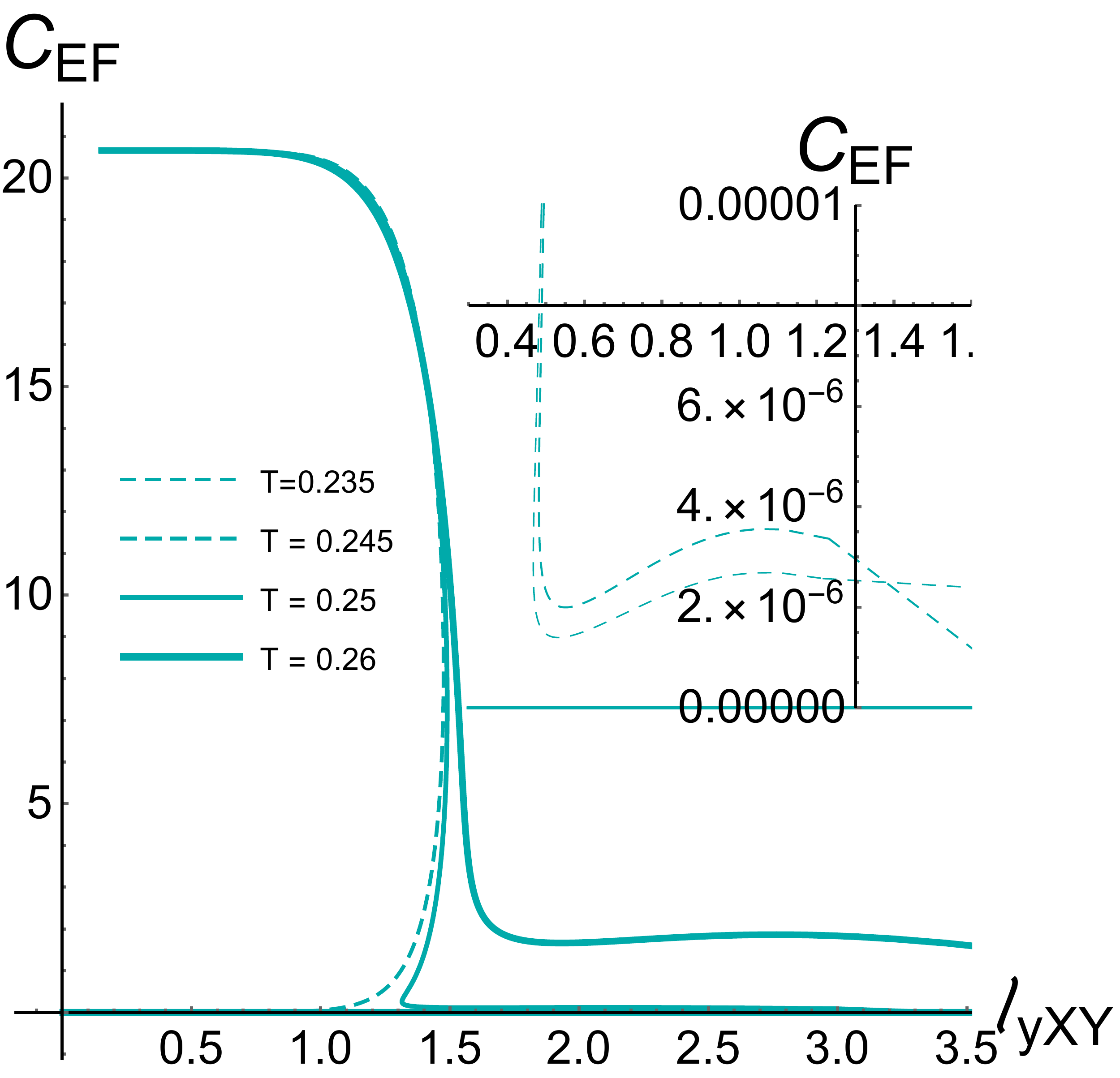}\qquad
\includegraphics[width=7cm]{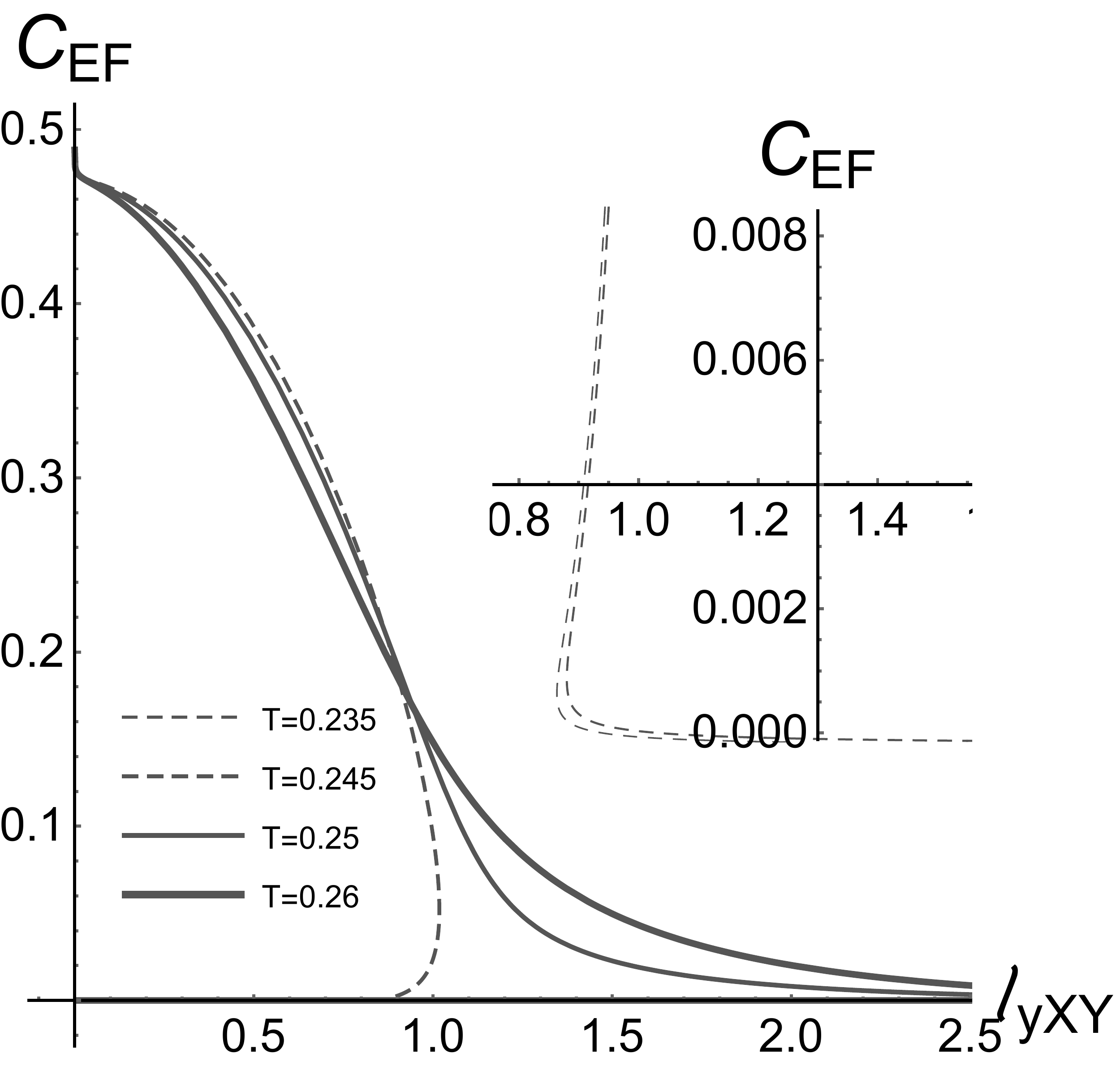}\\
  {\bf C)}\,\qquad\qquad\qquad\qquad\qquad\qquad\qquad\qquad{\bf D)}
  \caption{  Dependence of $c$ on $\ell$ in EF. {\it Top line}: $\nu=1$ and the c-function calculated with $\eta$ given   {\bf A)} by  \eqref{eta}    and  {\bf B)} by \eqref{etaCD}.  Various values of the temperature  above and below the     phase transition point $T=0.3445,\,\mu=0.05$; $T=0.33,0.335,0.34$ (below)  and $T=0.345,0.35, 0.355$ (above)  (thickness increases with increasing temperature) and $\mu=0.05$.
  {\it Bottom line}: {\bf C)} $\nu=4.5$, transversal orientation and values of temperature  above and below  the phase transition;  
     {\bf D)} $\nu=4.5$, longitudinal orientation and values of temperature below and above the phase transition. }
    \label{fig:c-b-SLT-PT}
  \end{figure}

 In Fig.\ref{fig:c-b-SLT-PT}    c-fuctions versus $\ell$ in the  EF  are shown. For  $\nu=1$ we calculate  the c-function  with $\eta$ given by  \eqref{eta} (see  Fig.\ref{fig:c-b-SLT-PT}.{\bf A)}) and with \eqref{etaCD} (see  Fig.\ref{fig:c-b-SLT-PT}.{\bf B)}).  Here one could see curves corresponding to various values of the temperature below and above the     phase transition point $T=0.3445,\,\mu=0.05$:   $T=0.33,0.335,0.34$ (below)  and $T=0.345,0.35, 0.355$ (above) (with increasing thickness with increasing temperature) and $\mu=0.05$.
 We also present the plots of c-functions
 for $\nu=4.5$ and  transversal and longitudinal orientation and values of temperature below and above  the phase transition (see Fig.\ref{fig:c-b-SLT-PT}.{\bf C)} and  Fig.\ref{fig:c-b-SLT-PT}.{\bf D)}).  Here we see the prints of the length phase transition similar to \cite{Dudal:2016joz}. This phase transition is observed particularly clearly for transversal configurations and temperatures above the phase transition line.

$$\,$$
\newpage
$$\,$$
\newpage

\subsection{Various c-functions as functions of $\ell$}\label{sect: summary}
Let us summarize and comment on the results obtained in the previous Sect.\ref{Sect:c-function} and Sect.\ref{nearBB}. The summary of these results is presented also in Table 2.

\begin{table}[t!]
\begin{tabular}{|c|c|c|c|c|c|c|}
\hline
                           & \multicolumn{3}{c|}{EF} & \multicolumn{3}{c|}{SF} \\ \hline
            $\mu$               & ISO    & T      & L     & ISO    & T     & L     \\ \hline
$0$
 & \includegraphics[width=0.15\textwidth, height=20mm]{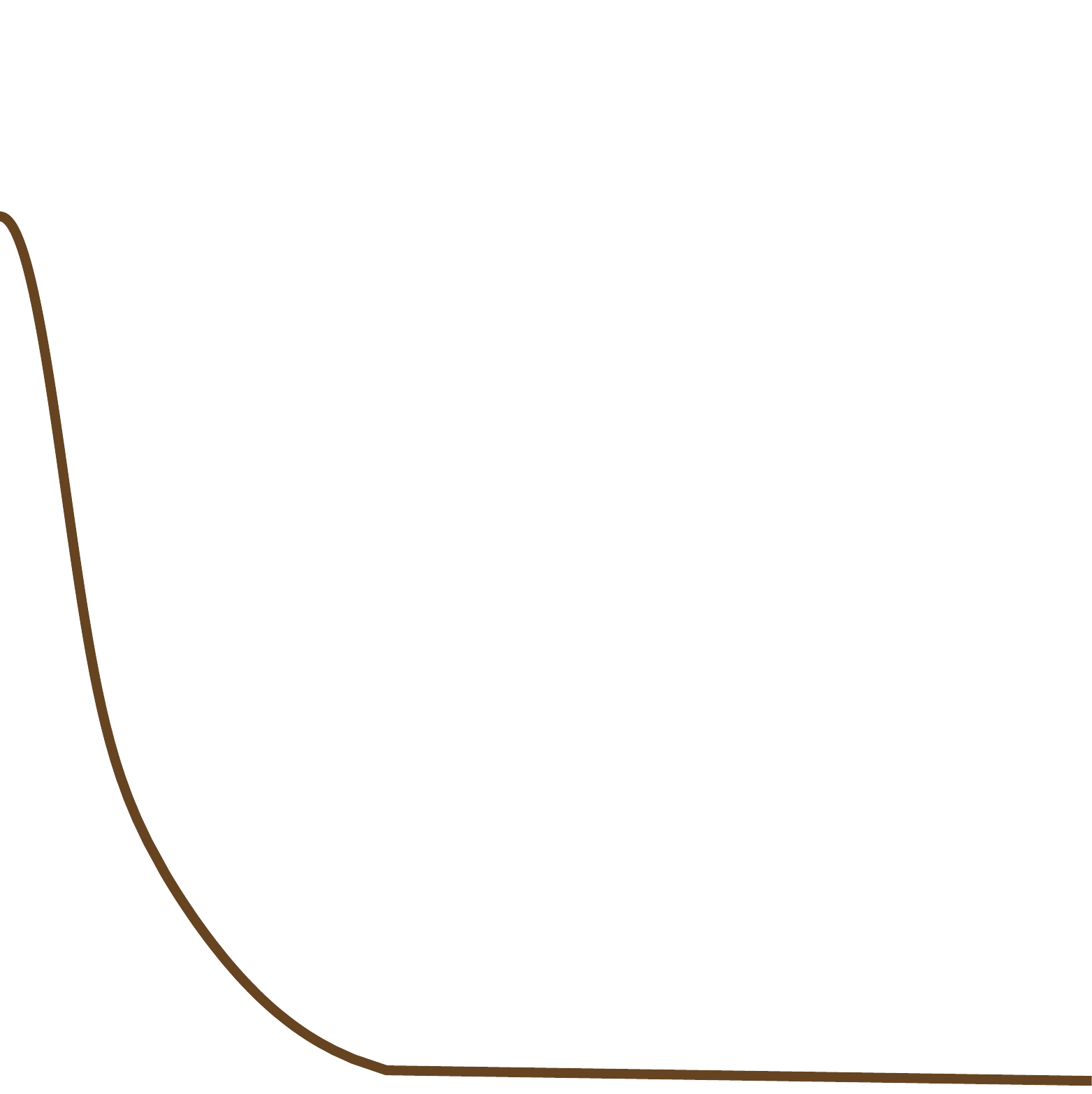}     & \includegraphics[width=0.15\textwidth, height=20mm]{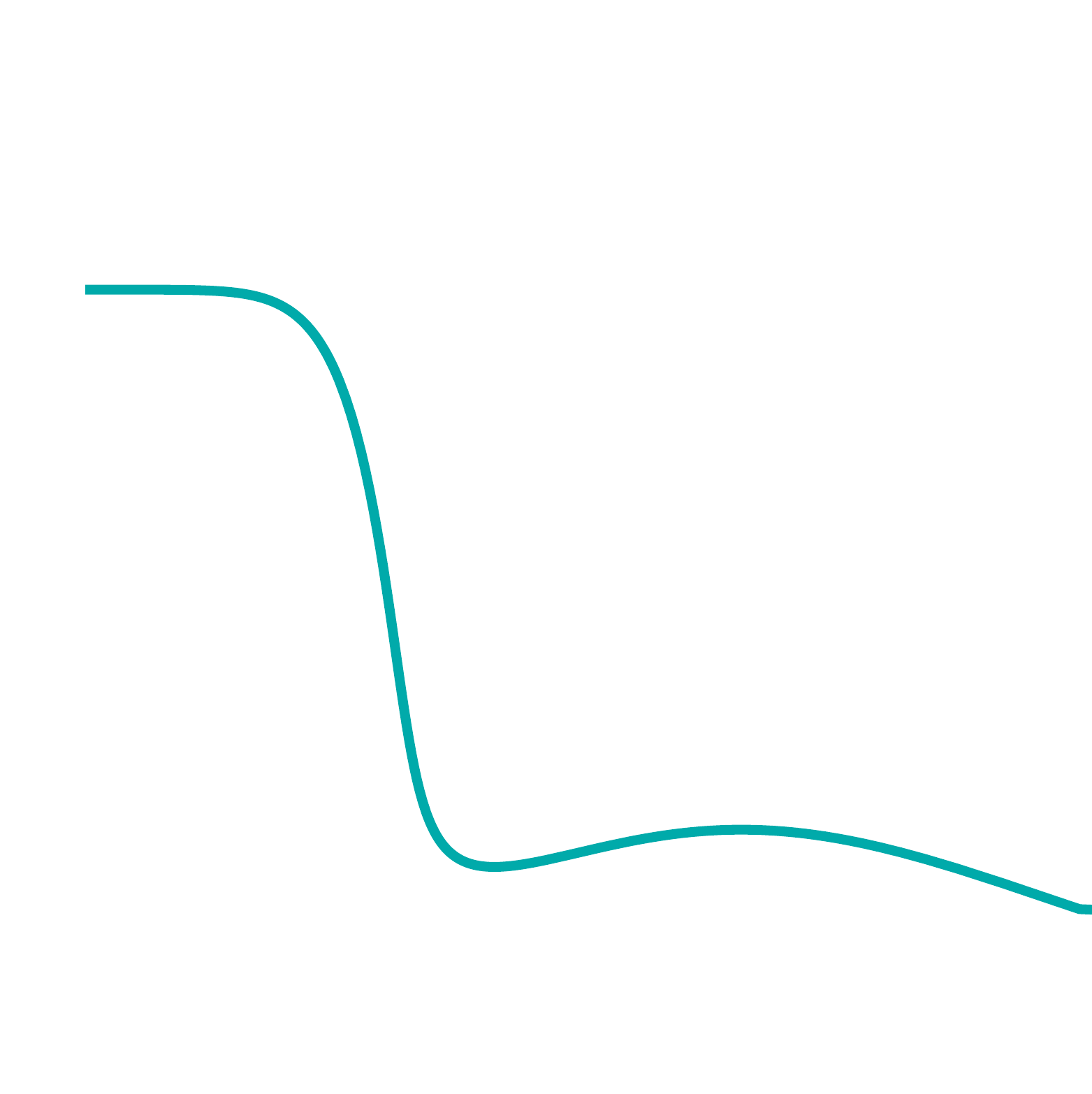}   & \includegraphics[width=0.15\textwidth, height=20mm]{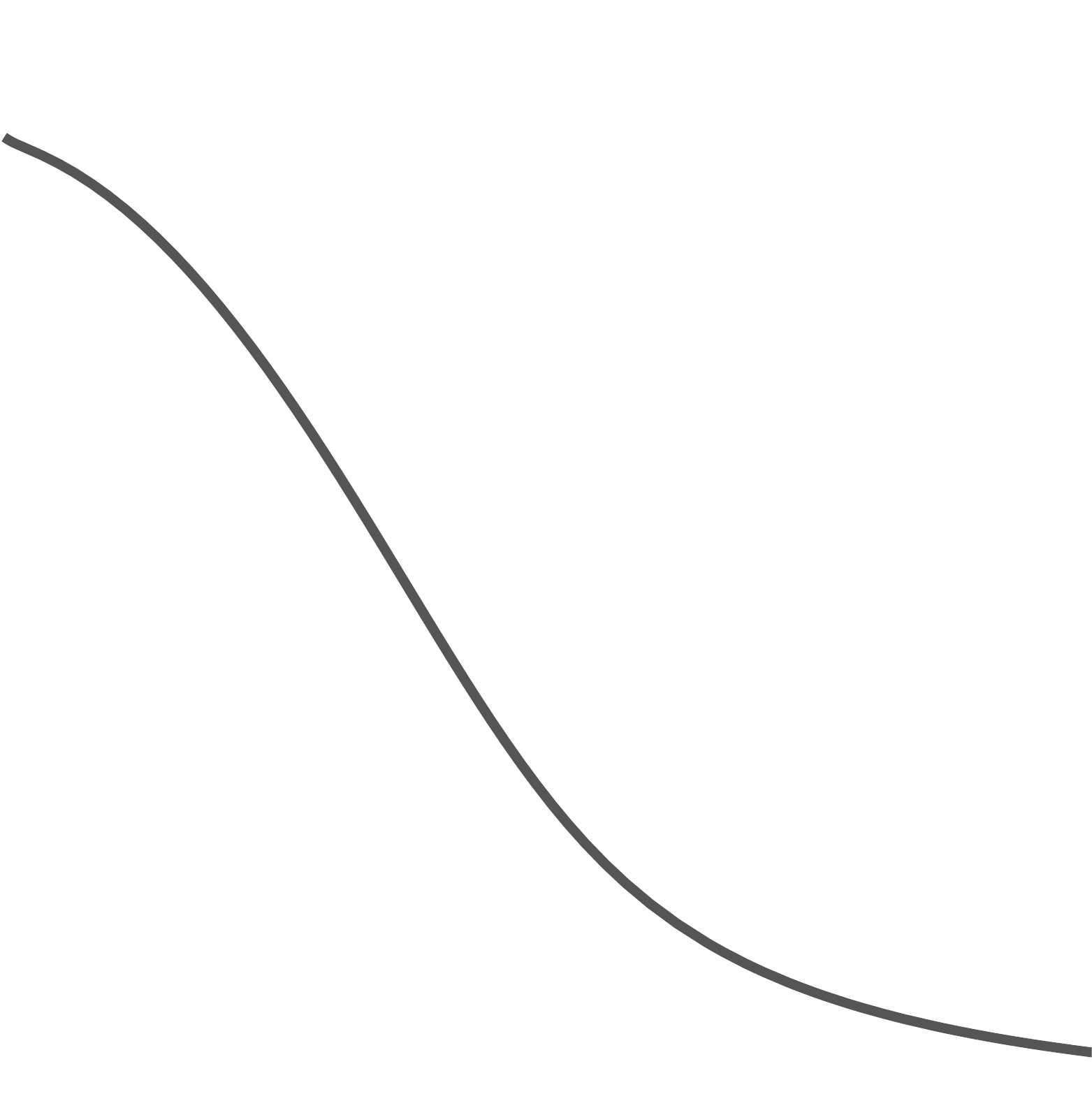}   &  \includegraphics[width=0.15\textwidth, height=20mm]{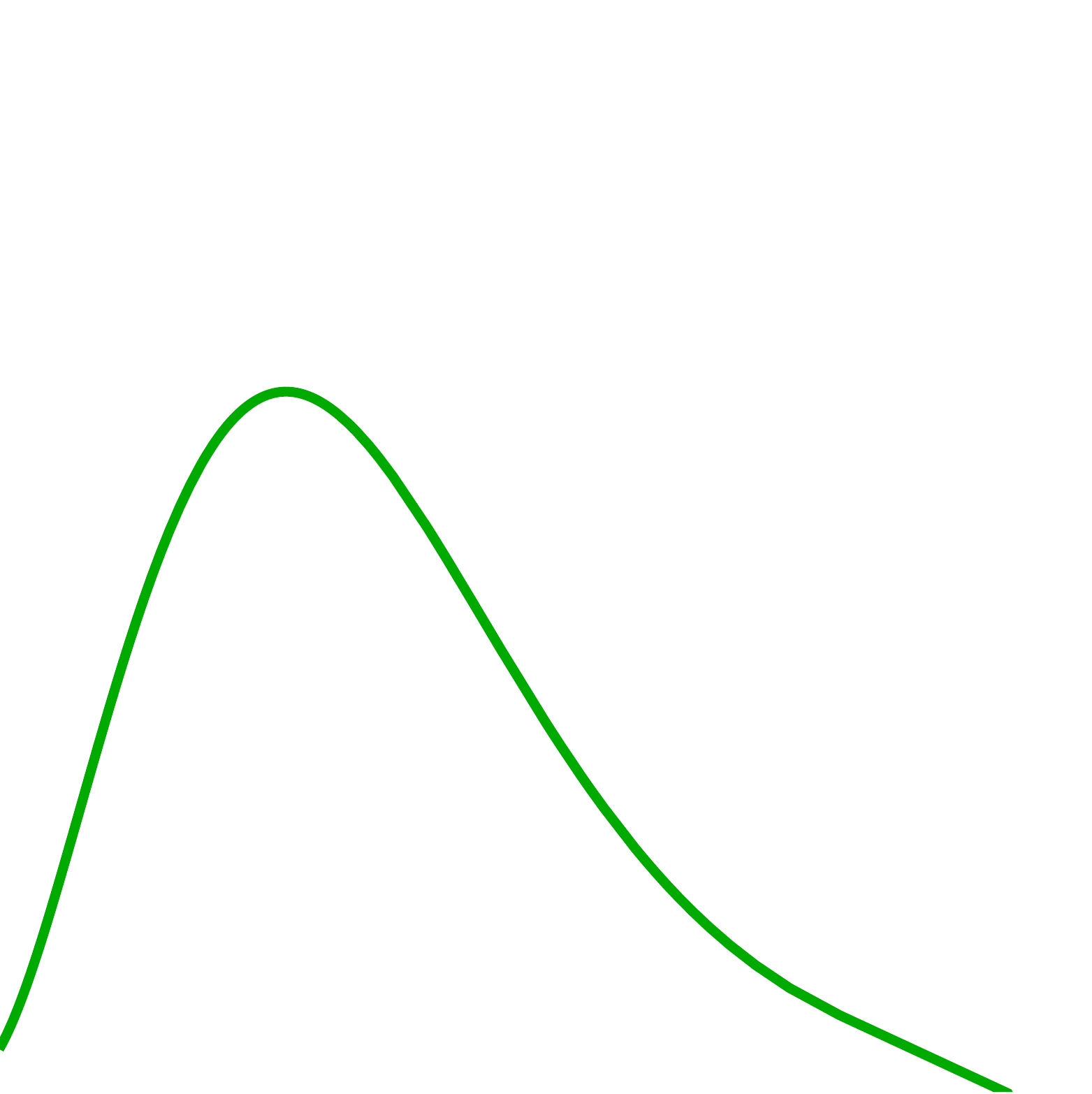}     & \includegraphics[width=0.15\textwidth, height=20mm]{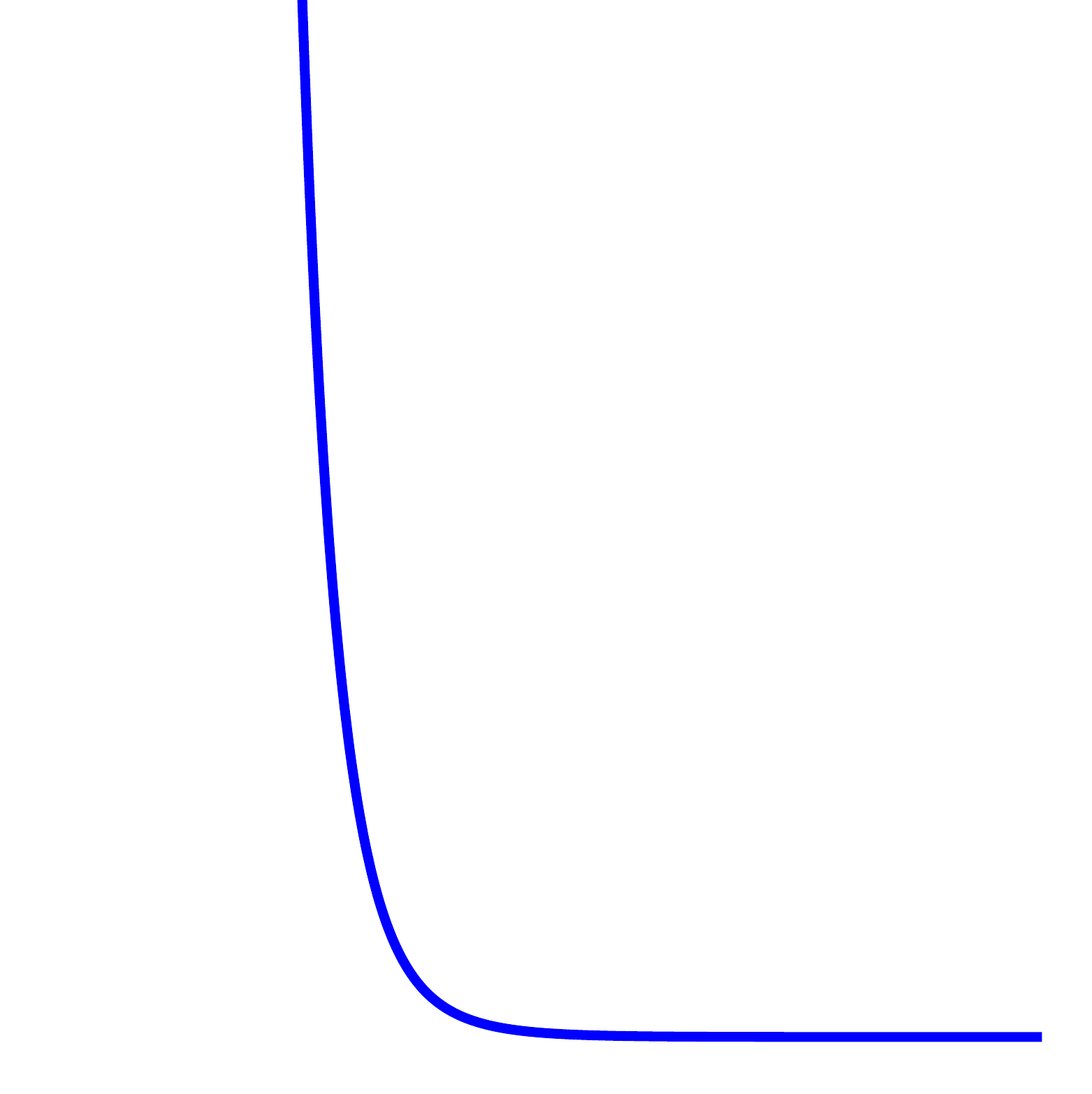}     & \includegraphics[width=0.15\textwidth, height=20mm]{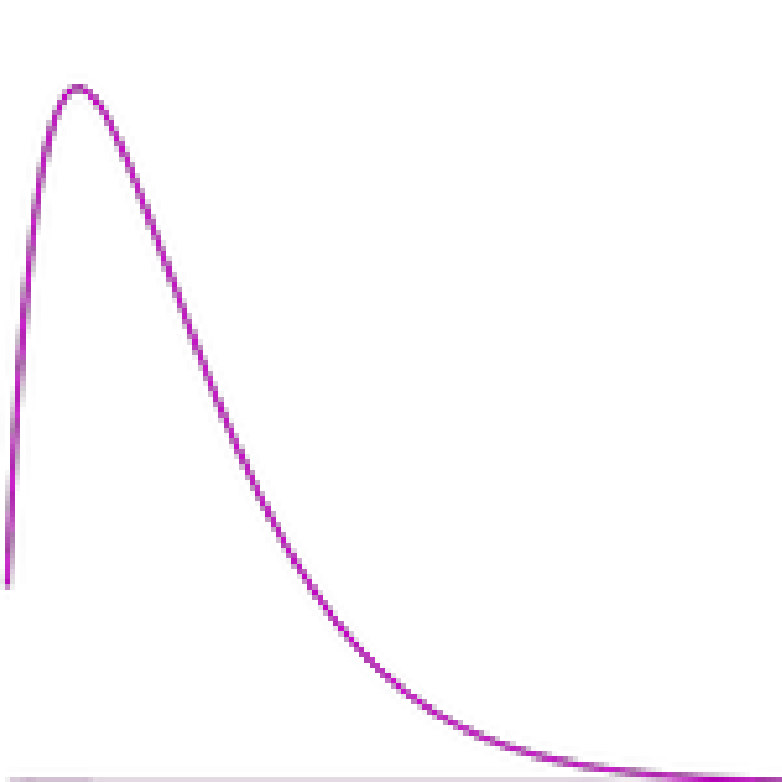}    \\ \cline{2-7}
                           & Fig.\ref{fig:c-b-ISO}    & Fig.\ref{fig:c-b-LT-EF}.A    & Fig.\ref{fig:c-b-LT-EF}.C    & Fig.\ref{fig:c-b-ISO}    & Fig.\ref{fig:c-b-LT-SF}.A     &  Fig.\ref{fig:c-b-LT-SF}.C   \\ \hline
$0.5$
 & \includegraphics[width=0.15\textwidth, height=20mm]{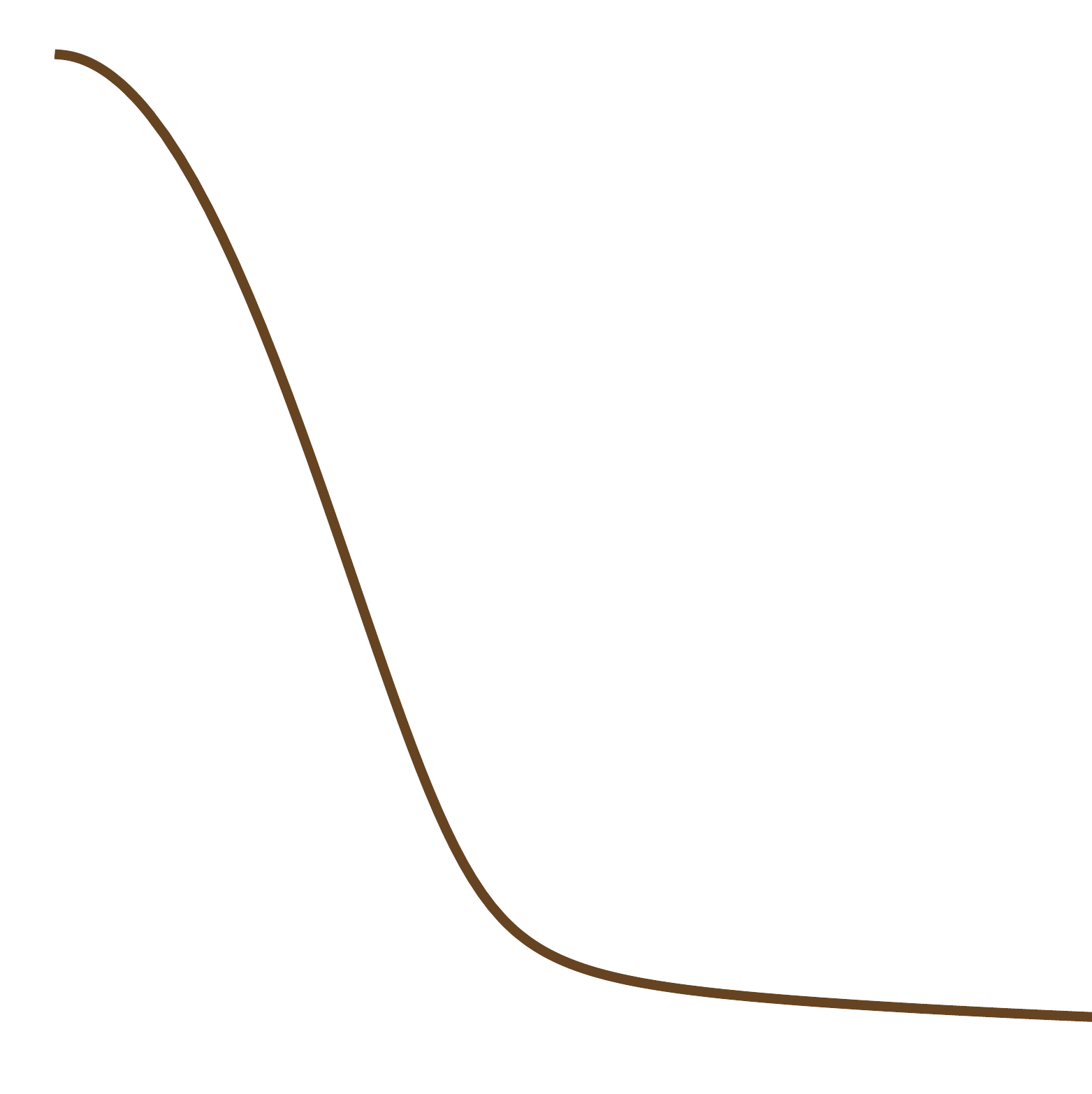}      &  \includegraphics[width=0.15\textwidth, height=20mm]{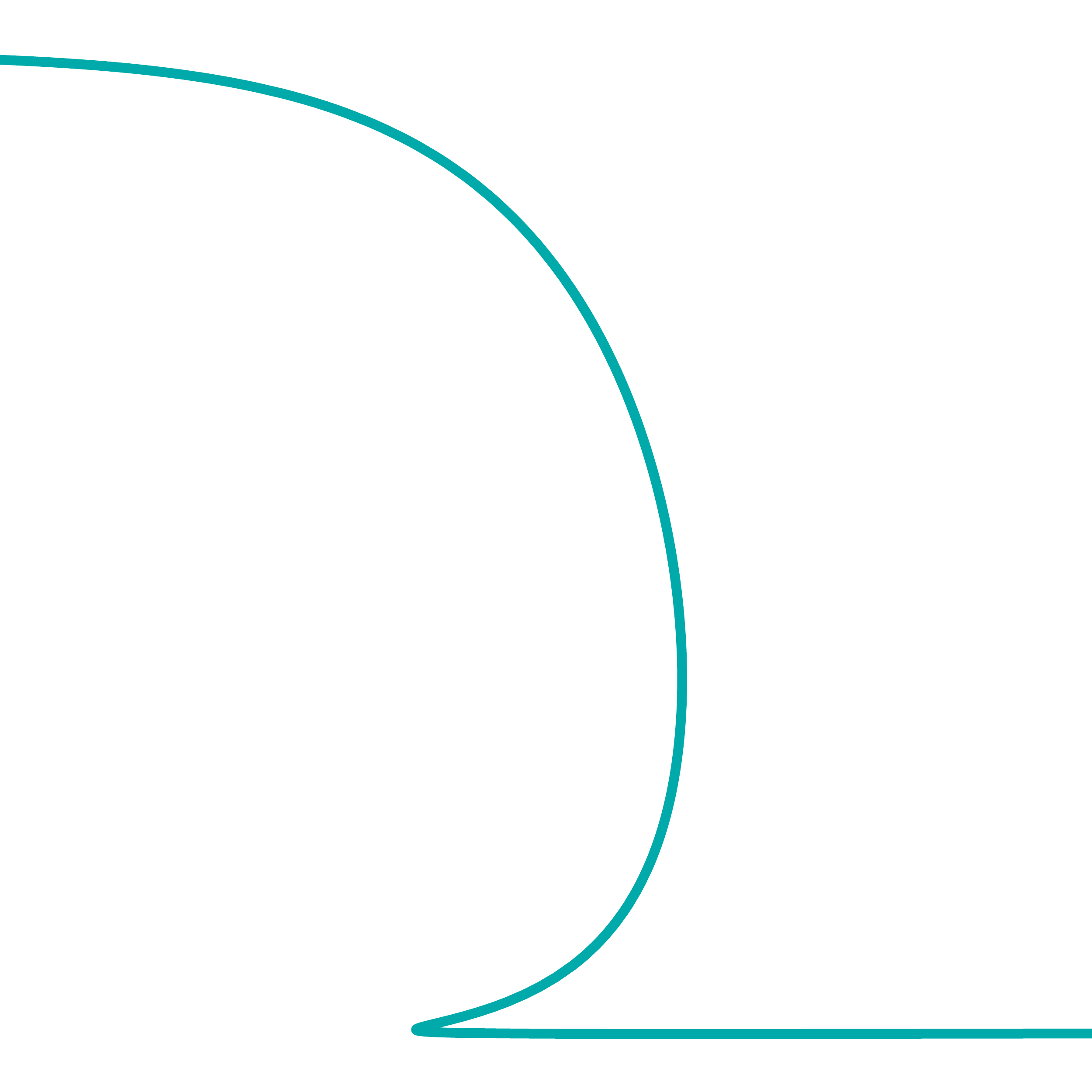}     & \includegraphics[width=0.15\textwidth, height=20mm]{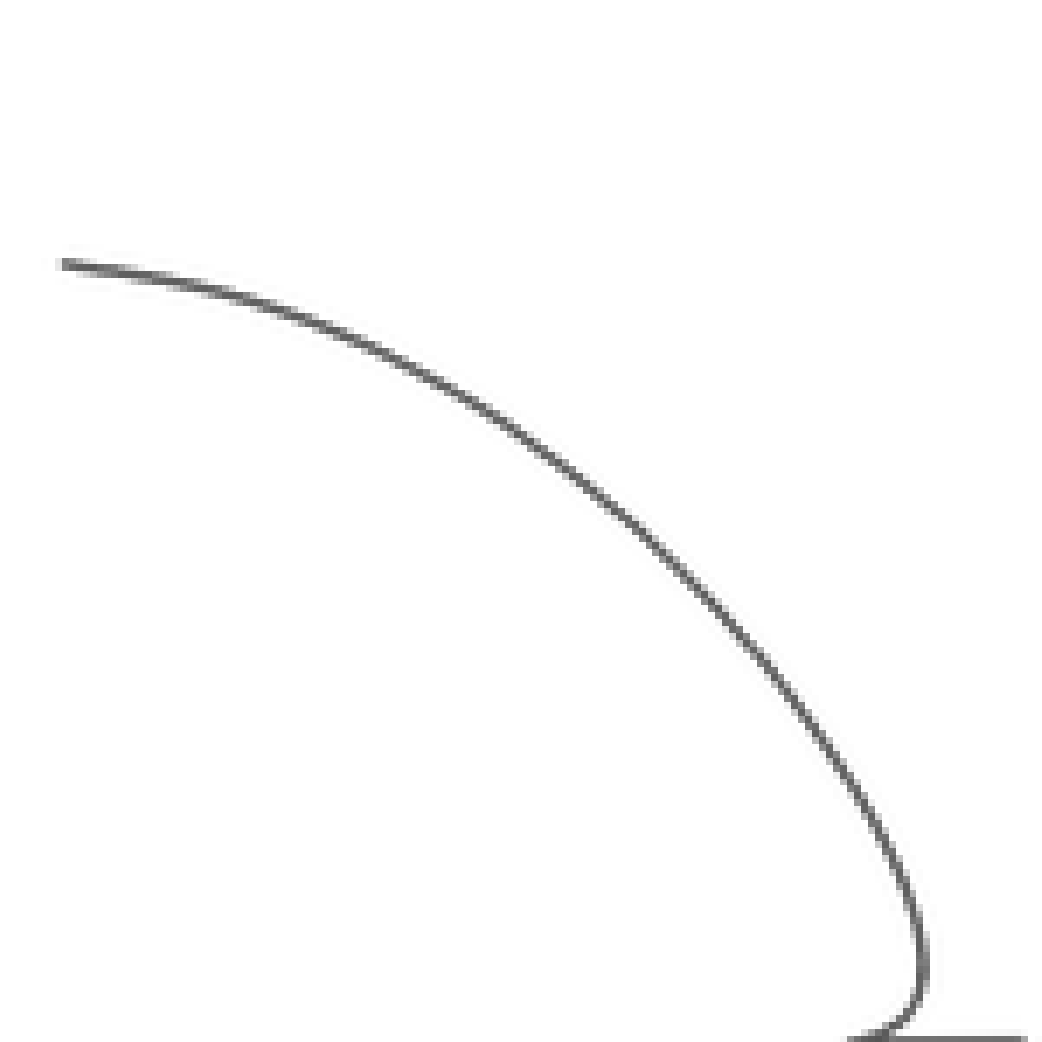}   &
 \includegraphics[width=0.15\textwidth, height=20mm]{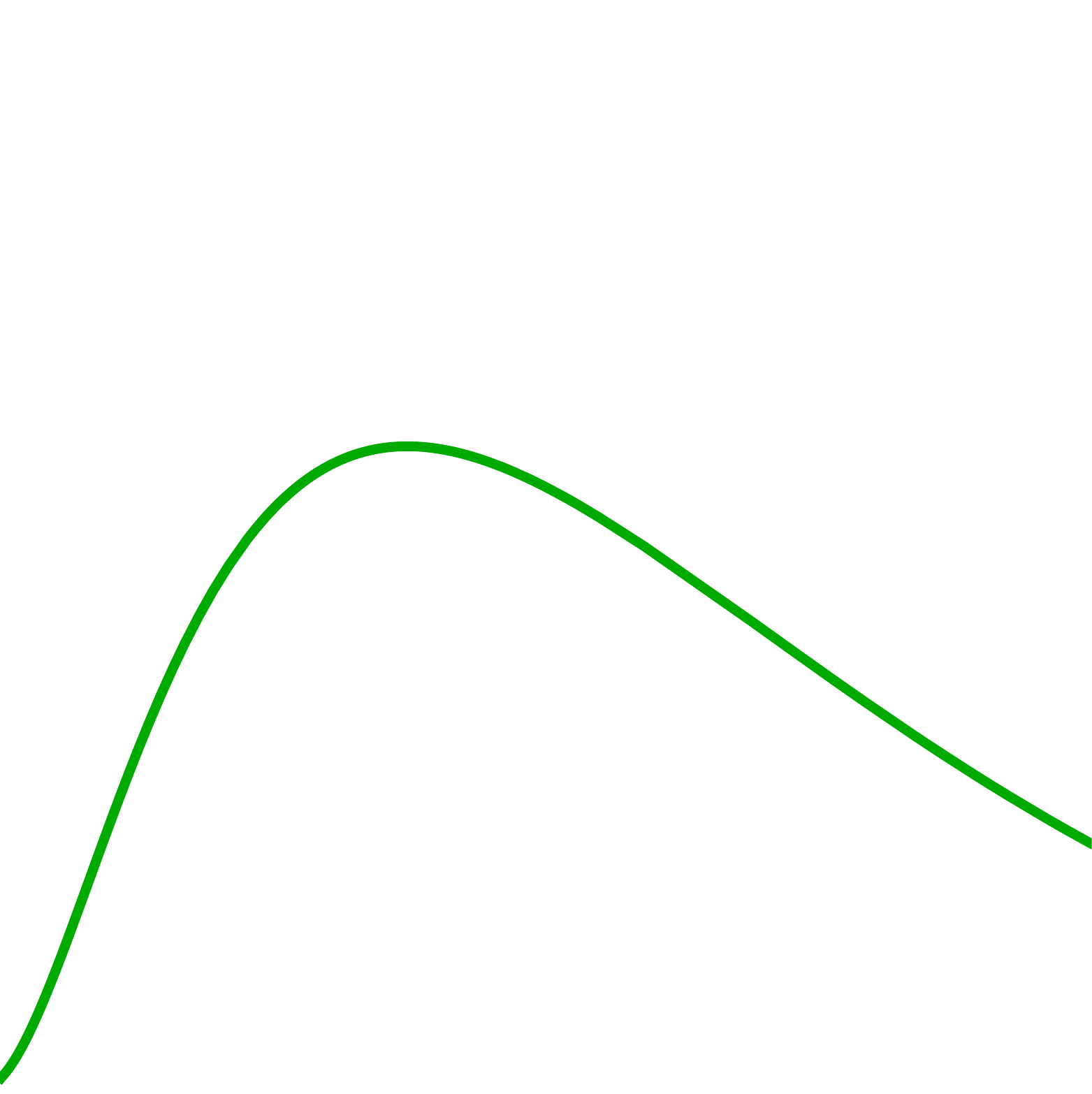}     &  \includegraphics[width=0.15\textwidth, height=20mm]{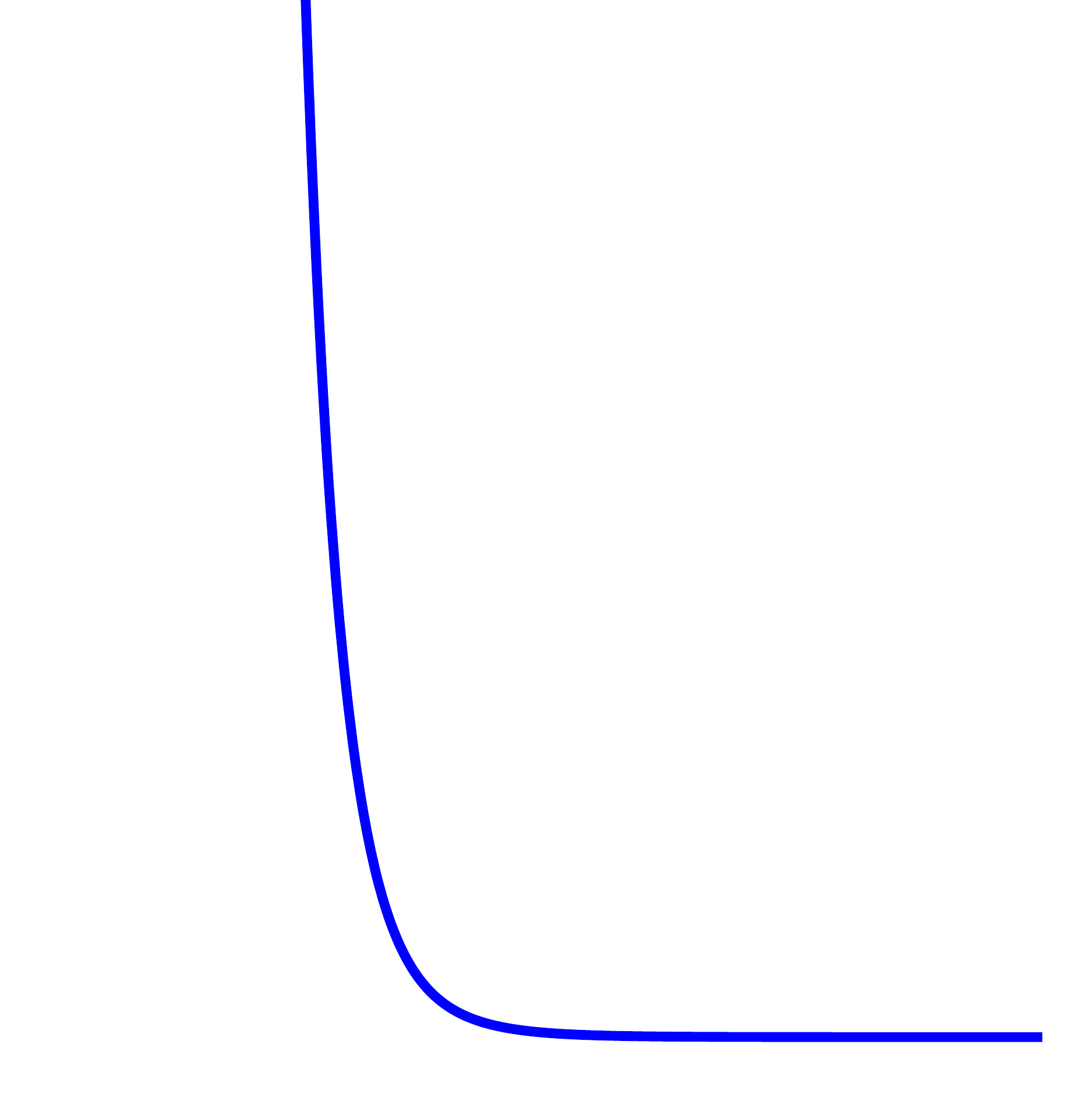}     & \includegraphics[width=0.15\textwidth, height=20mm]{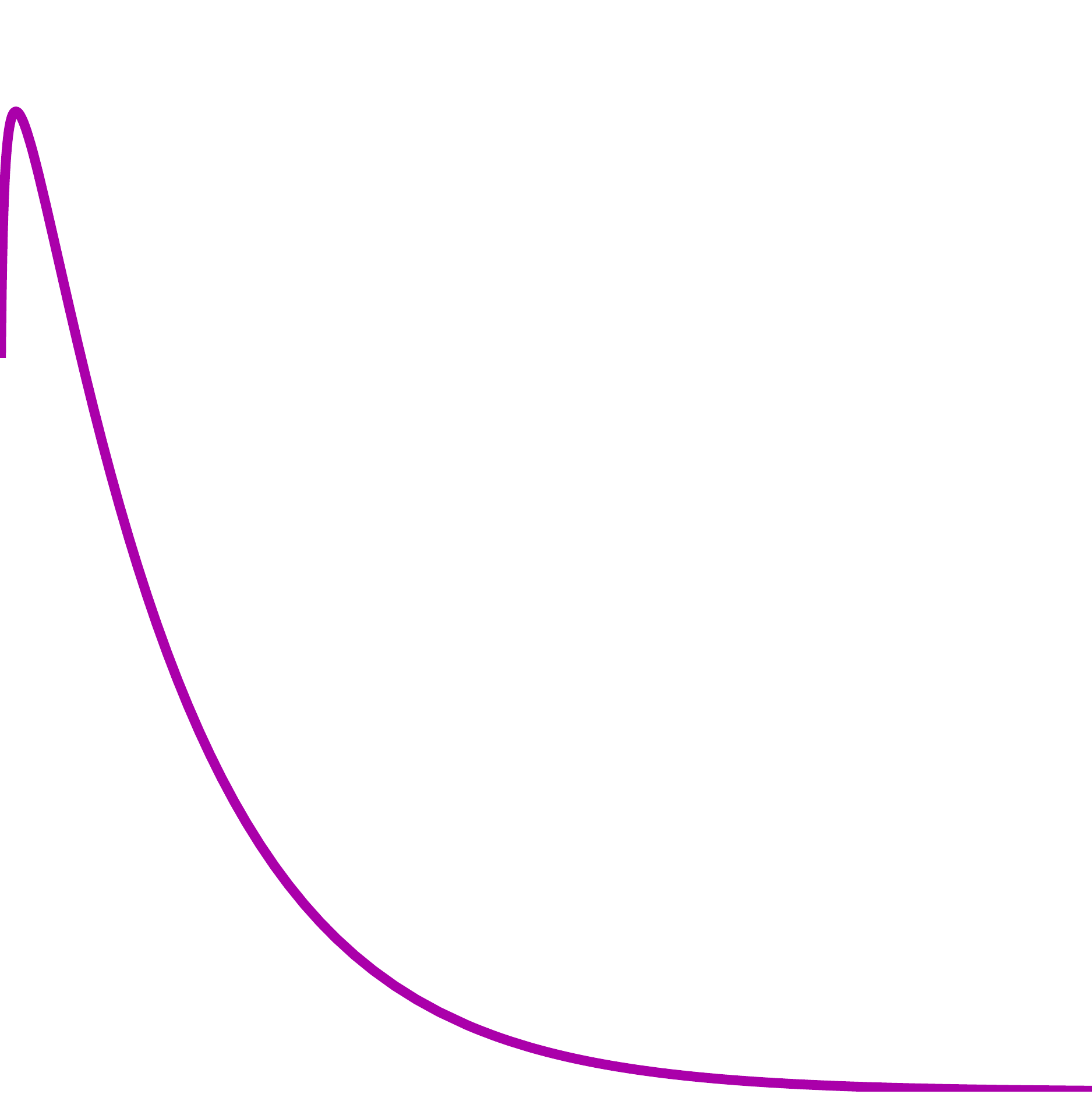}    \\ \cline{2-7}
                        & Fig.\ref{fig:c-b-ISO}    & Fig.\ref{fig:c-b-LT-EF}.B    & Fig.\ref{fig:c-b-LT-EF}.D     & Fig.\ref{fig:c-b-ISO}     &     Fig.\ref{fig:c-b-LT-SF}.B    & Fig.\ref{fig:c-b-LT-SF}.D   \\ \hline
 &  \includegraphics[width=0.15\textwidth, height=20mm]{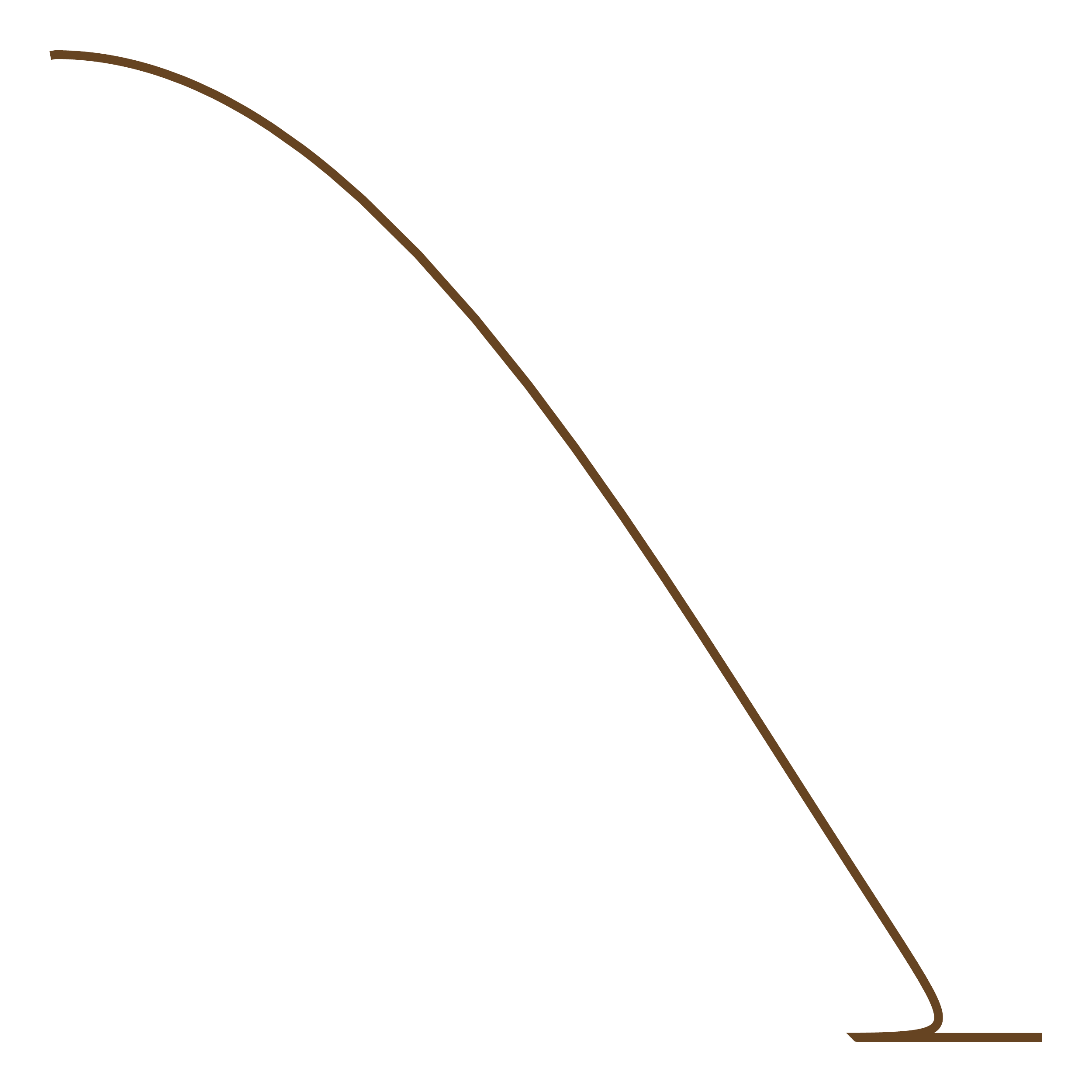}  & \includegraphics[width=0.15\textwidth, height=20mm]{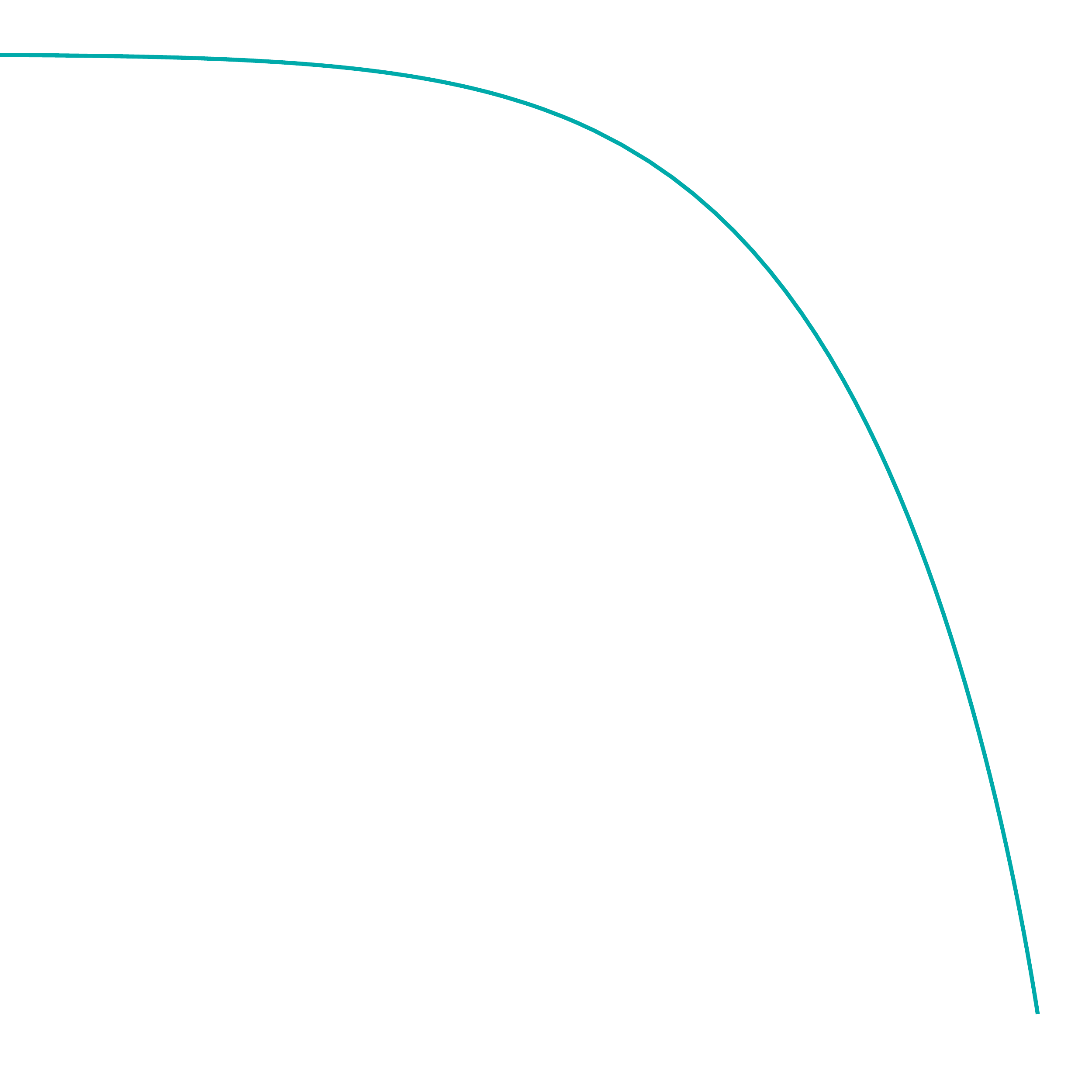}     &  \includegraphics[width=0.15\textwidth, height=20mm]{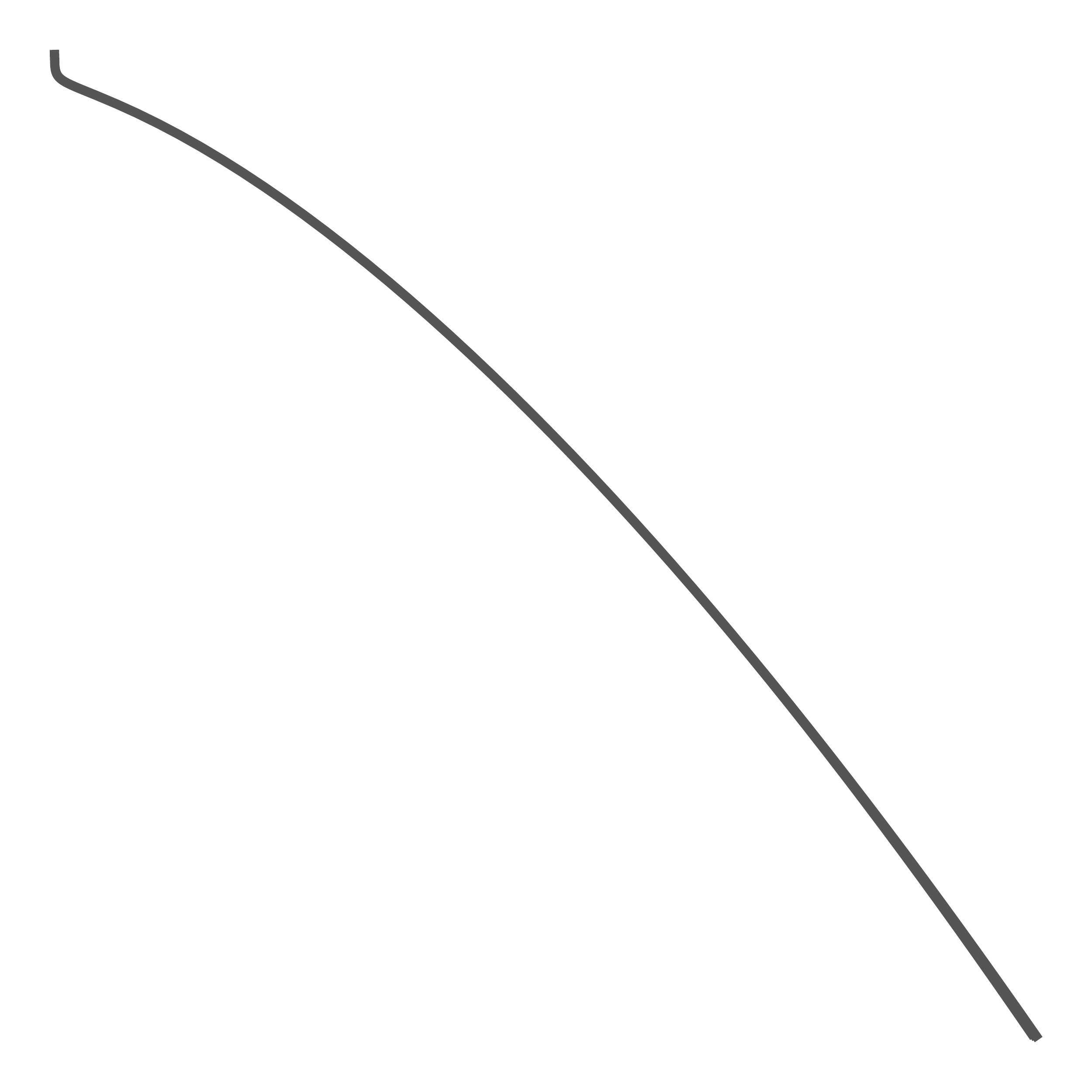}    &  \includegraphics[width=0.15\textwidth, height=20mm]{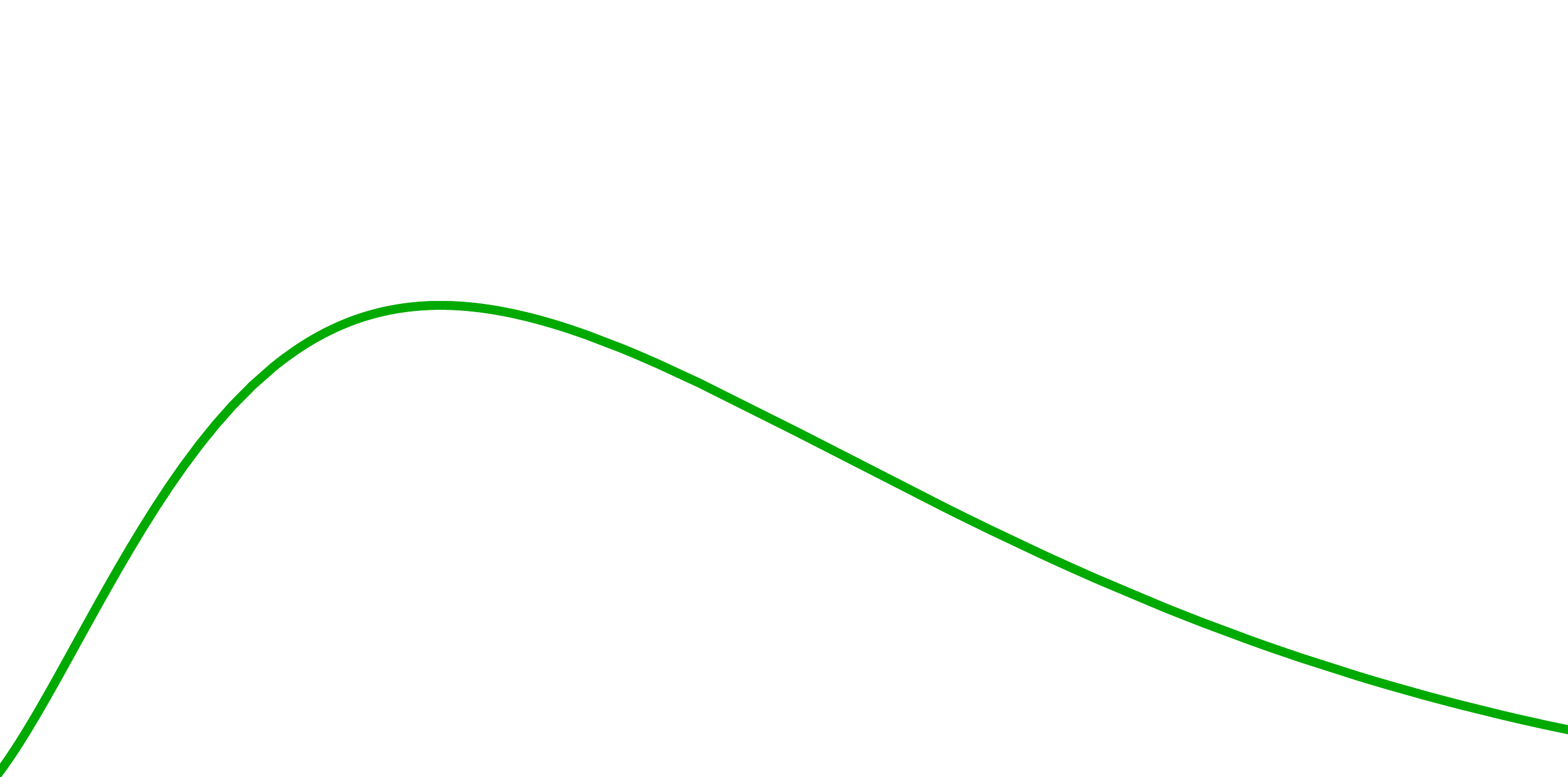}     & \includegraphics[width=0.15\textwidth, height=20mm]{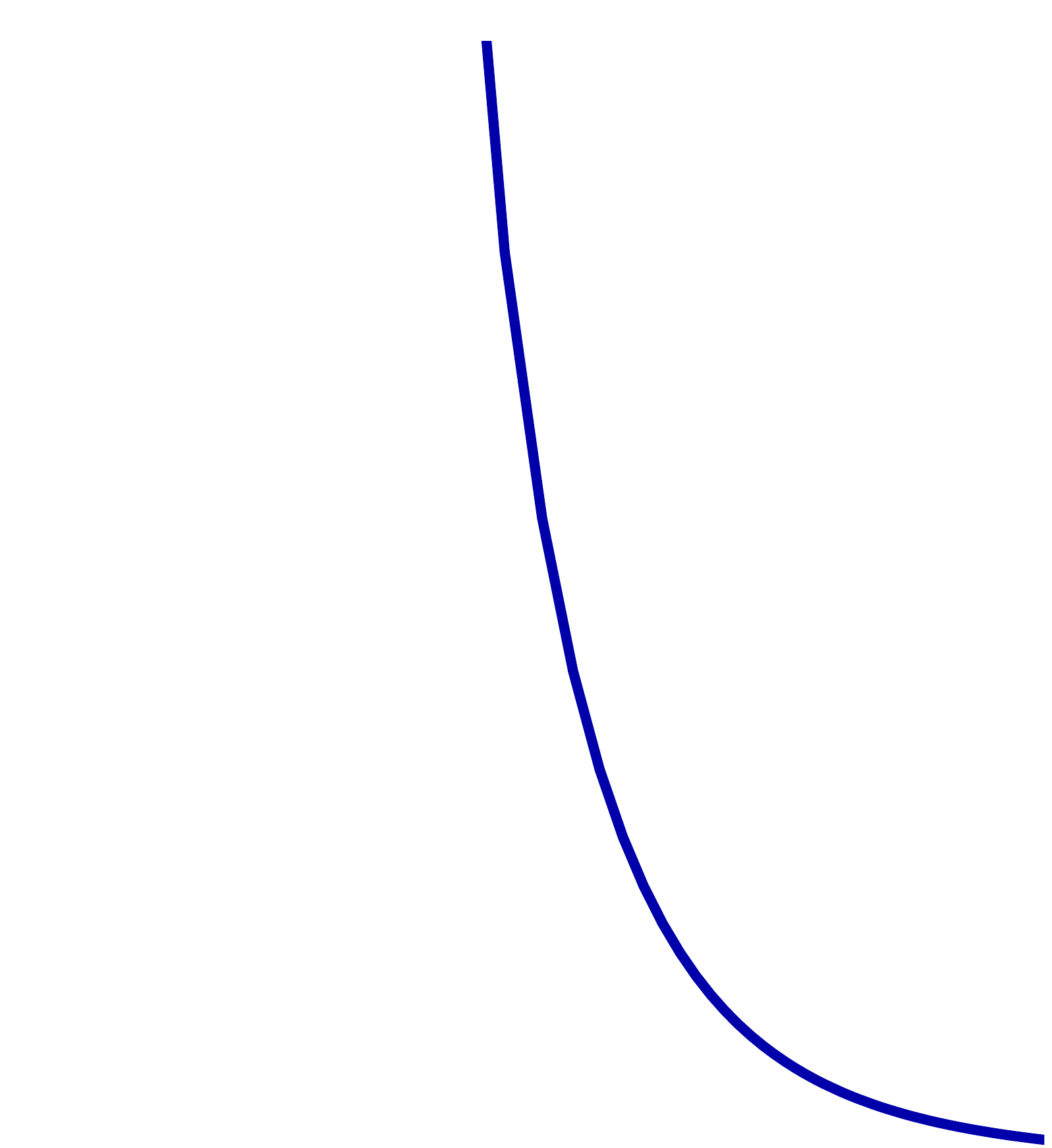}    &  \includegraphics[width=0.15\textwidth, height=20mm]{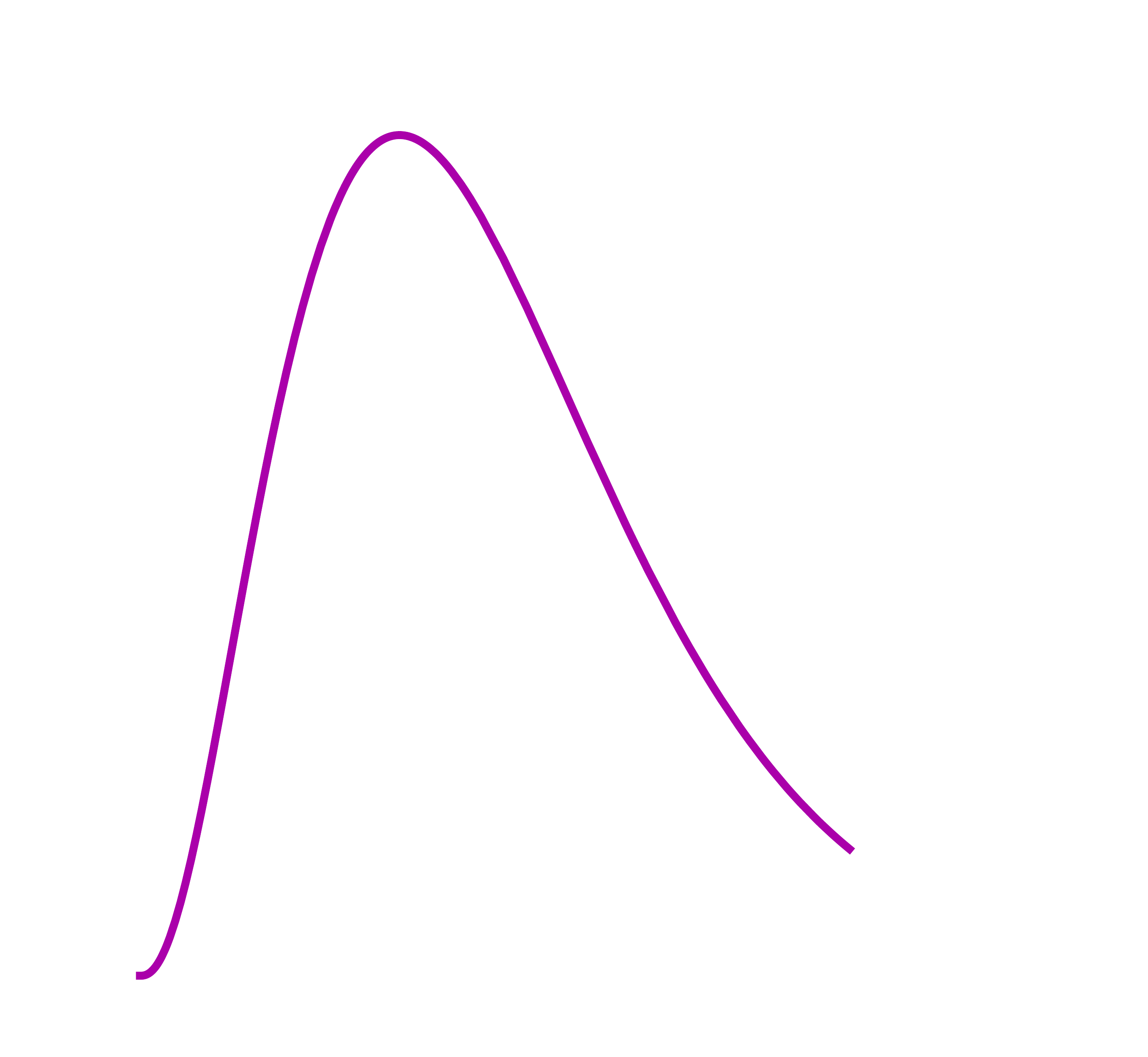}    \\ \cline{2-7}
     PT                      & Fig.\ref{fig:c-b-SLT-PT}.A    & Fig.\ref{fig:c-b-SLT-PT}.B     & Fig.\ref{fig:c-b-SLT-PT}.C    & Fig.\ref{fig:c-b-LT-PT}.A    & Fig.\ref{fig:c-b-LT-PT}.B   & Fig.\ref{fig:c-b-LT-PT}.C  \\ \hline
\end{tabular}
\caption{A summary of the c-function behavior considered in the present work in the EF and the SF based on definition \eqref{GAg}. "ISO" denotes isotropic solution. "L" and "T" denote the longitudinal and transversal orientations of the slab in the anisotropic metric with $\nu=4.5$. "PT" corresponds to  configurations near the background phase transition.}
\end{table}
\label{table:1}

 \begin{itemize}
 \item We have found, that c-functions  in the EF  decrease with increasing $\ell$ for not overly large $\ell$ (see the left part of the Table 2).
\begin{itemize}
 \item 
 For the isotropic case  in regions of $(\mu,T)$-plane far away from the line of the phase transition the c-function decreases with increasing $\ell$ for all  $\ell$.

 \item 
  For the anisotropic case the c-function decreases  while increasing $\ell$ for all transversal and longitudinal orientations for small $\ell$ in all regions of $(\mu,T)$-plane.
  \begin{itemize}
     \item For the transversal orientation different behaviors exhibits in different regions. 
 In the region of small $\mu$ in the transversal case the c-function has a local minimum at $\ell _{min}$ after which it increases up to a local maximum  at $\ell _{max}$ and only then decreases. We get even more interesting behavior with increasing $\mu$. The c-function as a function of $\ell$ becomes multivalued in some interval of $\ell$, $\ell_{left,T}<\ell<\ell_{right,T}$.
 For $\ell>\ell_{right,T}$ crossing the line of the phase transition, the c-function has a jump.

   \item  For longitudinal orientation the c-function for small $\mu$ only decreases with increasing $\ell$. But if we increase $\mu $ we see that the c-function also exhibits multivalued behavior for $\ell_{left,L}<\ell<\ell_{right,L}$ and for $\ell>\ell_{right,T}$ it has jumps as it crosses the phase transition line. 
   \end{itemize}
     \end{itemize}

\item The c-functions calculated in the SF  also  exhibit  non-monotonic behavior in some cases, but there is no multi-valued behavior here (see the right part of  Table 2).
 \begin{itemize}
     \item
In isotropic case in the SF   c-function increases while increasing $\ell$ in UV up to $\ell=\ell_{max}(T,\mu)$. This non-monotony behavior  is related to dilaton behavior near $z=0$.
 
\item  There is different behavior  in anisotropic case in the SF. 
 \begin{itemize}
 \item The c-function exhibits monotonic behavior for the transversal orientation. It decreases with increasing $\ell$ for all $\ell$.

 \item  The c-function increases with increasing $\ell$ in UV up to $\ell=\ell_{max,SF}(T,\mu)$ and when decreases for the longitudinal orientation.
     \end{itemize}
     \end{itemize}   
      \end{itemize}

There are two reasons why we do not need to worry about all these.
 \begin{itemize}
 \item First of all, as has been mentioned in the text,  a non-monotonicity in the anisotropic case is not in contradiction with any of the existing c-theorems as all of them are based on Lorentz invariance.
 \item The saddle points $\ell_{max,SF}(T,\mu)$ as well as the regions of multi-validity of the c-functions,
 $\ell_{left,L}<\ell<\ell_{right,L}$ or $\ell_{left,T}<\ell<\ell_{right,T}$ are located in the regions with large enough values of $\ell$, where the definition  of the c-function using the UV  asymptotics of the solutions can be violated.   \end{itemize}

  \subsection{Entanglement Entropy Phase Transition}\label{Sect:PT}
Let us remind that the criterion for the confinement/deconfinement phase transition in QCD is the behavior of the potential between quarks or the behavior of the temporal Wilson loops. The Wilson loops
can also be computed in HQCD. It turns out that location of  the confinement/deconfinement
line in the $(\mu,T)$ plane  can be close to the background phase transition \cite{AR,
1805.02938,1808.05596}, but not necessary coincide with
 it. For special models the phase transition of the HEE can be used as an indication of the HQCD phase transition \cite{IK,Knaute:2017lll}.

The position of the background phase transition depends on the particular holographic model,  see  \cite{IAGeneral,ARSCornell} and refs therein. It can be also located in the right-bottom part of the $(\mu,T)$ plane starting from a point $(\mu_{cr},T_{cr})$ and going down with increasing $\mu$ till the zero temperature. It also can be located  in the left part of the plane, starting from $T_0$ at $\mu=0$ and going down with increasing $\mu$ till the point $ (\mu_{cr},T_{cr})$.

To find the location of the HEE phase transition on   the $(\mu,T)$ plane one has to find the location of points where the free energy 
$F_A$ corresponding to the reduced $\rho_A$-matrix has a multi-valued behavior.
It turns out that the effective free energy corresponding to the entangled region $A$
 \be dF_A=-S_{A}dT,\ee 
 has a behavior similar to the behavior of the free thermal energy. 
 We can  define the density of the  entanglement effective  free energy as 
  \be df_A=-\eta_A \, dT.\ee 
 The density of the effective free energy as a function of  temperature also has the 
 swallow-tail behavior.

   \begin{figure}[h!]\centering
\includegraphics[width=7cm]{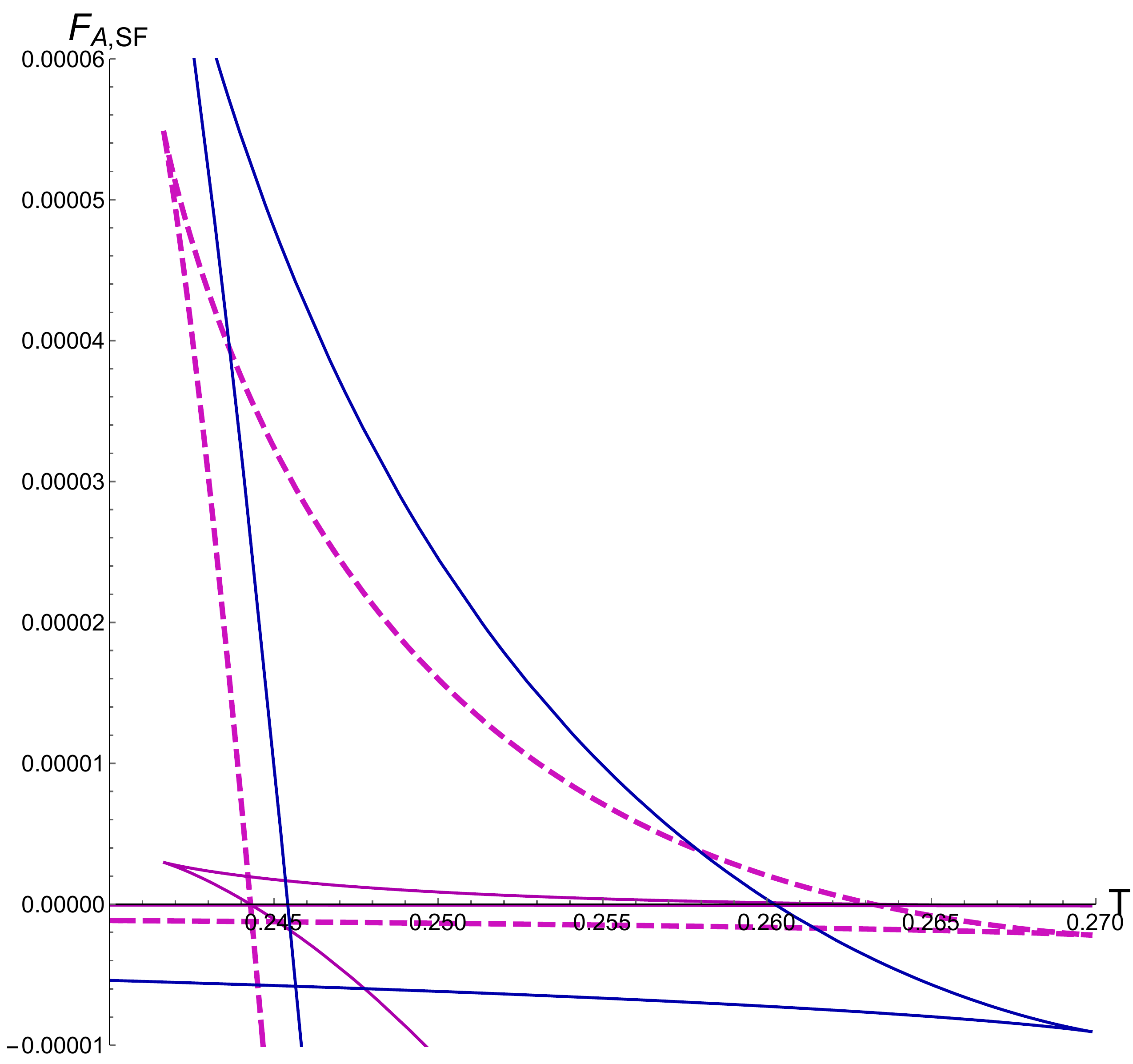}  
 \includegraphics[width=7cm]{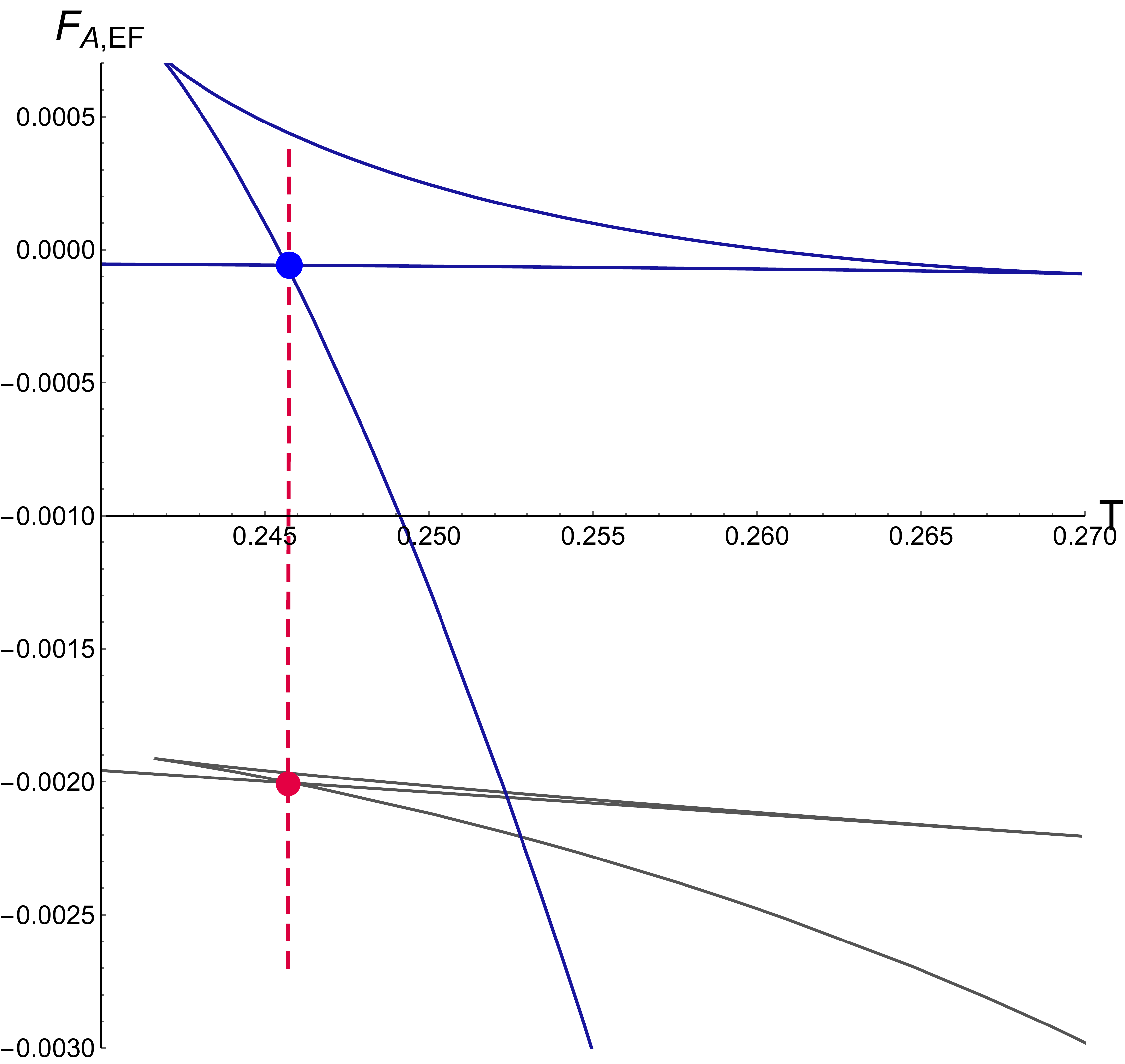}\\
   {\bf A)}\qquad\qquad\qquad\qquad\qquad\qquad\qquad\qquad{\bf B)}\\
   $\,$\\
   \includegraphics[width=7cm]{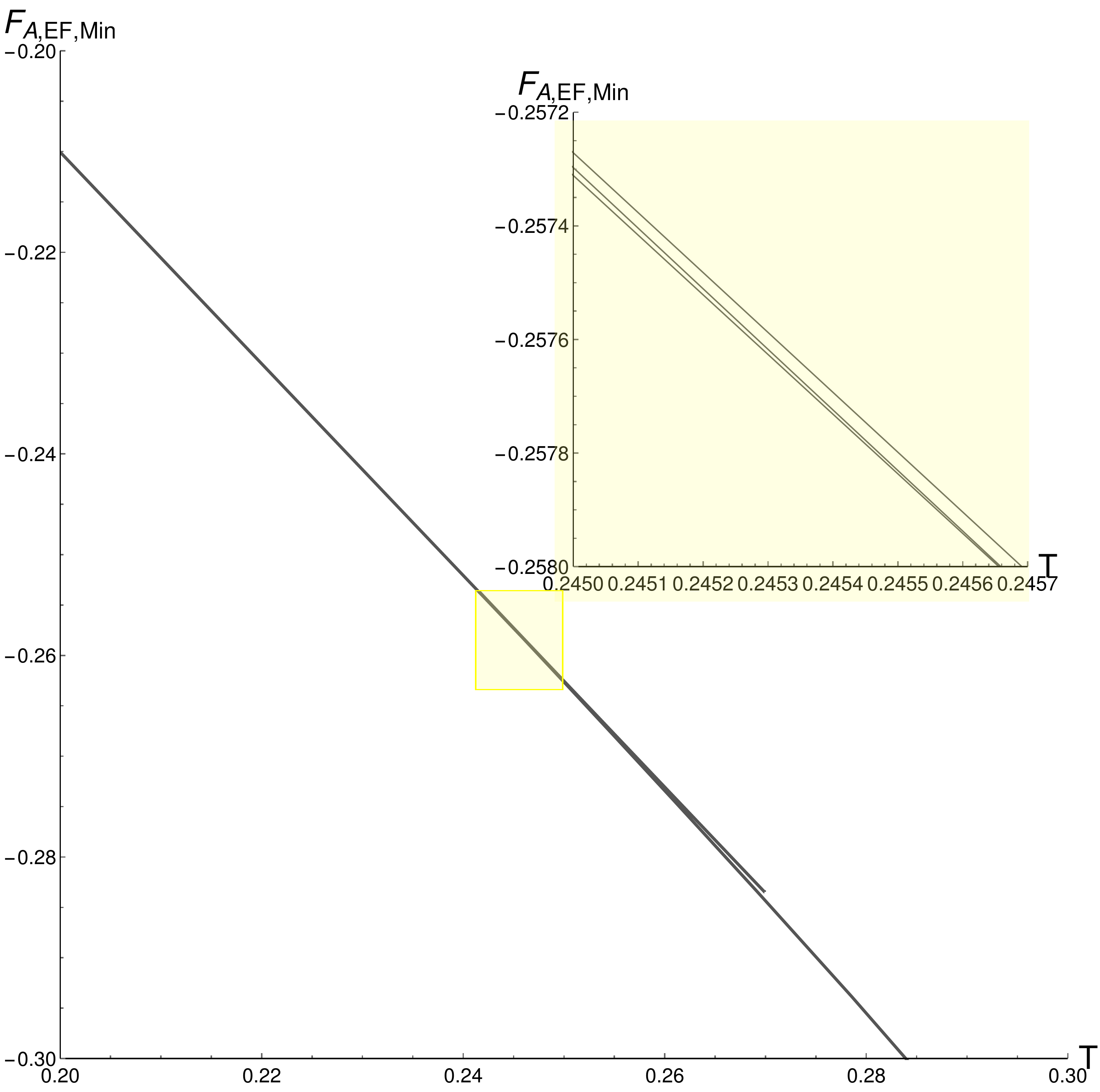}
\\
{\bf C)}
 \caption{{\bf A)} Swallow-tailed temperature dependences of  the thermal free energy density divided by factor 10 (solid blue line) and the entanglement effective free energy density in the SF  (magenta dashed line  based on $\eta$ and  
 magenta solid line  based on $\eta_{CD}$). {\bf B}) The same in the EF (the entanglement free energy density with 
 $\eta_{CD}$ in the EF is shown by gray solid line).  {\bf C}) The entanglement free energy  density based on 
 $\eta$ in the EF. The inset in  {\bf C}) shows that the form of temperature dependence of the 
 entanglement free energy density presented in the main panel is in fact  swallow-tailed. All these plots are done
  for the anisotropic case with $\nu=4.5$ and $\mu =0.2$. 
  }
    \label{fig:phaseHEE-EF-SF}
  \end{figure}
  
This is due to the fact that, on the one hand, slabs that extend infinitely in two directions and being thick enough can  accumulate  enough degrees of freedom to repeat the characteristic form of thermal entropy, and on the other hand,
the three-valued behavior of the function $ z_h = z_h (T) $ is an intrinsic  feature of the background and inevitably leads to
swallow-tail behavior for both ordinary and effective free energy even for thin slabs. However the slab does not include all degrees of freedom, so the  thermal entropy and  the effective entanglement entropy do not coincide exactly, as well as the lines of the thermal and entanglement  phase transitions.

Let us summarize what we found studying behavior of the effective entanglement free energy in different schemes of regularizations and frames.

\begin{itemize}
\item 

We checked the behavior of the effective entanglement free energy density as a function of  temperature in the SF  using different renormalization schemes, see Fig.~\ref{fig:phaseHEE-EF-SF}.{\bf A)}. 
We see that there is no essential dependence on used regularization schemes in the SF. The transition points
for the entanglement free energy in both regularization schemes  almost coincide, also these points are very close  to the transition point obtained from the thermal  free energy.

\item 
We also compared behavior of the effective entanglement free energy and  thermal free energy densities as the functions of  temperature in the EF,   see Fig.~\ref{fig:phaseHEE-EF-SF}.{\bf B)}.  Here we see that the temperatures of  the phase transition points  almost coincide, while the values of the free energy densities at these points do not coincide. 

\item 
Moreover, we analysed  behavior of the effective entanglement free energy defined with the minimal renormalization scheme in EF and found the swallow-tail looking  more flattened, see  Fig.\ref{fig:phaseHEE-EF-SF}.{\bf C)}.

 \item We checked that 
 the locations of the critical points extracted from the HEE density and the HEE itself   almost coincide 
 (numerical calculations were done for different $\ell$ and orientations).  
 
 \end{itemize}

\begin{figure}[h!]\centering
 \includegraphics[width=7cm]{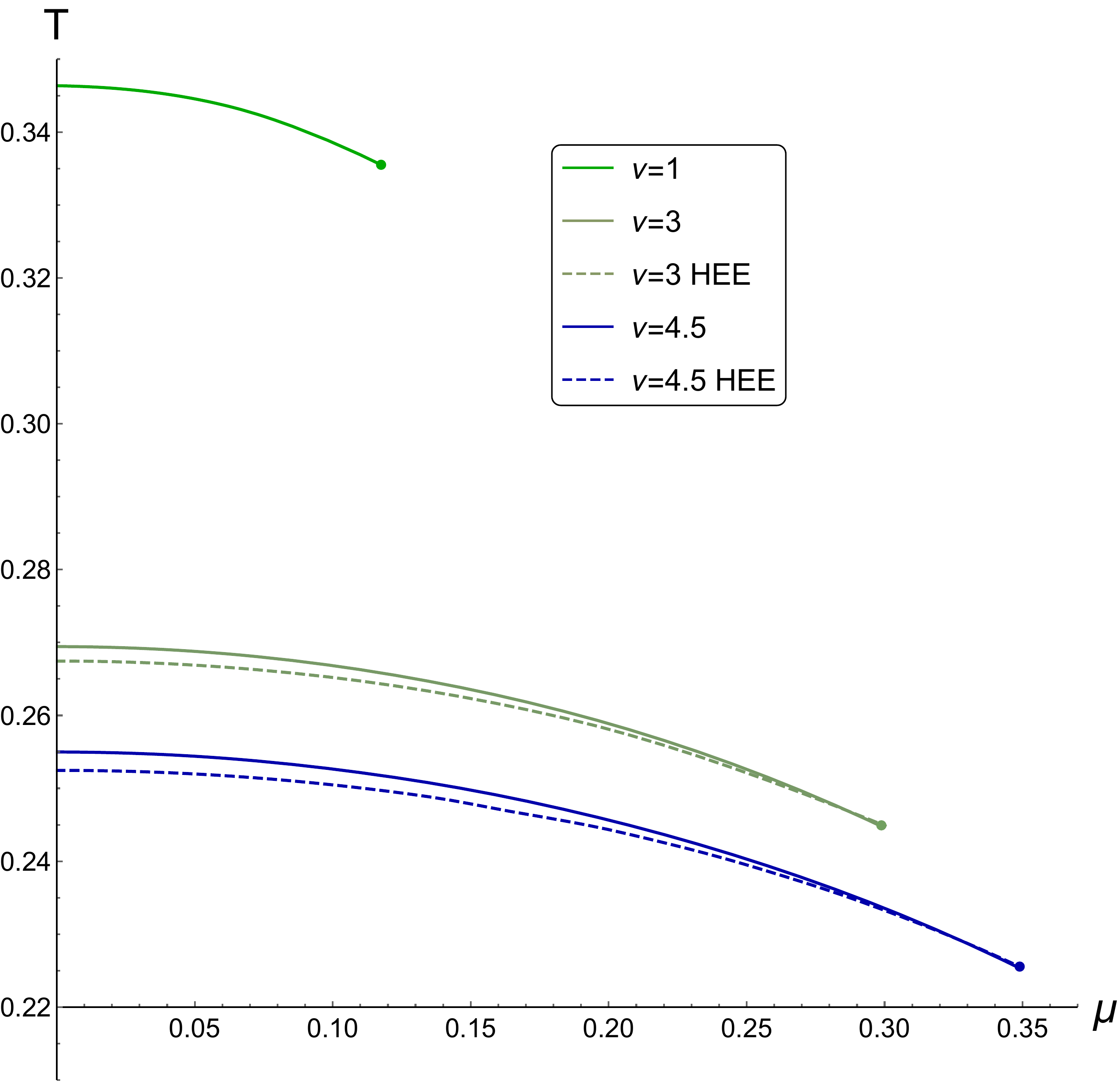}
  \caption{ Phase diagrams for the thermal entropy (solid lines) and   entanglement entropy densities (dashed lines)  for various anisotropy parameters $\nu$, $\nu=1$ (green line), $\nu=3$ (khaki lines) and $\nu=4.5$ (blue lines).}
    \label{fig:phaseHEE}
  \end{figure}

In Fig.\ref{fig:phaseHEE} phase diagrams for the thermal entropy (solid lines) and   entanglement entropy densities (dashed lines)  for various anisotropy parameters $\nu$, $\nu=1$ (green lines), $\nu=3$ (khaki lines) and $\nu=4.5$ (blue lines)
 obtained by numerical calculations based on \eqref{elltheta}, \eqref{Stheta} and \eqref{eta} are presented.

Generally speaking, the location of the phase transition for the HEE is rather close to the background phase transition, see Fig.\ref{fig:phaseHEE}.  Note that in contrast to Wilson loops  behavior in anisotropic background
 \cite{AR}, there is no visible orientation dependence of  HEE phase transition line  in anisotropic cases.

\section{Conclusion}
We have considered  the most general anisotropic holographic model and found the expression for the HEE in terms of the Euler angles defined by the orientation of the slab-shape  area in respect to HIC axes. In a particular case, we have considered the HEE for the model invariant in the transversal directions with the unique anisotropy scaling factor supported by the Einstein-Dilaton-two-Maxwell action \cite{AR}.
The choice of the model \cite{AG} is motivated by agreement of the energy dependence of the produced entropy  with the experimental data for the energy dependence of the total multiplicity of particles produced
in HIC \cite{Alice}  in the anisotropic metric.
This model describes multiplicity and  quark confinement (for heavy quarks),  predicts crossover transition line between confinement/deconfinement phases,
anisotropy in hadron spectrum (for a short time after collisions) \cite{1808.05596,ARSCornell} and phase transition for the spatial Wilson loops \cite{IAQ18}. \\

We have calculated the HEE and its density in the holographic anisotropic model \cite{AR}. We have shown that the  HEE and its density have significant fluctuations near the BB phase transition line in  $(\mu,T)$-plane  for all values of the anisotropy parameter. The lines of thermal and entanglement entropies phase transition in $(\mu,T)$-plane are different in the anisotropic cases  but do not depend on the orientation of the entangling area.
Note that for isotropic case $\nu=1$ the differences between these  lines in the phase diagram plane  are not visible.
We have discussed an application of enormous increasing of the HEE as  an indicator of the background  phase transition.

We have studied the dependence of the c-function on
the thickness $\ell$ of the entanglement  slab. We have  found saddle points of c-function as a function of  $\ell$ as well as its multivalued behaviour. The obtained results are schematically presented in Table 2.
 The c-functions in the EF  decrease while increasing $\ell$ for not too large $\ell$. Moreover,
in the isotropic case  in regions of $(\mu,T)$-plane remote from  the line of the phase transition, the c-function decreases while increasing $\ell$ for all  $\ell$. 
In the anisotropic case, the c-function decreases  with increasing $\ell$ for all transversal and longitudinal orientations for small $\ell$ in all regions of $(\mu,T)$-plane. 
The c-functions calculated in the SF in some cases exhibit  non-monotonic behavior, but there is no multivalued behavior here.
 In isotropic case in the SF   c-function increases with increasing $\ell$ in UV up to $\ell=\ell_{max}(T,\mu)$. This non-monotonic behavior is related with dilaton behavior near $z=0$.
 There is different behavior  in anisotropic case in the SF. 
 It has been mentioned in the text,
 a non-monotonicity in an anisotropic case is not in contradiction with any of the existing c-theorems because  all 
 of them are based on Lorentz invariance.
The saddle points  as well as the regions of multi-validity of the c-functions are located in the regions  where the definition  of the c-function using the UV  asymptotics of the solutions can be violated.

As to  further development, we suppose to study modifications of the model \cite{AR} to include the light quarks following \cite{1703.09184}, incorporate  the chiral phase transition \cite{2002.00075}, and also  perform the numerical calculations in full anisotropic case to incorporate the magnetic field, as has been done in \cite{Gursoy:2018ydr}. 

 We hope that the results presented in this paper, their  interpretations and their further possible adjustment to the phenomenology  data can be of interest for experiments at the future facilities of FAIR \cite{Friman:2011zz}, NICA \cite{nica:whitepaper}, for RHIC's BES II program and CERN, III run.

\section*{Acknowledgments}
 This work is supported by RFBR Grant 18-02-40069 and partially (I.A. and P.S.) by the “BASIS” Science Foundation (grant No. 18-1-1-80-4). We would like to thank  D. Ageev, A. Golubtsova,  
 M. Khramtsov, K. Rannu and 
 I. Volovich for useful discussions.

\end{document}